\documentclass[11pt,a4paper]{article}

\usepackage{amsmath}
\usepackage{amssymb}
\usepackage{graphicx}
\usepackage{color}
\usepackage{rotating}
\usepackage{makeidx}
\usepackage{alltt}
\usepackage{upquote}
\usepackage{longtable}

\usepackage{url}

\usepackage[pdfpagelabels,colorlinks,urlcolor=black,linkcolor=black,citecolor=blue]{hyperref}
\usepackage{amsmath}
\usepackage{amssymb}
\usepackage{dsfont}
\usepackage{amsthm}

\usepackage{bookmark}

\theoremstyle{remark}
\newtheorem{remark}{Remark}
\newtheorem{example}{Example}
\theoremstyle{definition}

\newtheorem{theorem}{Theorem}
\newtheorem{lemma}{Lemma}

\newcommand{\rank}{\mathop{\mathrm{rank}}}
\newcommand{\diag}{\mathop{\mathrm{diag}}}
\newcommand{\spann}{\mathop{\mathrm{span}}}

\renewcommand{\ker}{\mathcal{N}}
\newcommand{\wno}{\mathop{\mathrm{wno}}}

\newcommand{\be}{\begin{equation}}
\newcommand{\ee}{\end{equation}}
\newcommand{\ba}{\left [ \begin{array}}
\newcommand{\ea}{\end{array} \right ]}
\newcommand{\bea}{\begin{eqnarray}}
\newcommand{\eea}{\end{eqnarray}}
\newcommand{\C}{{\mbox{\rm $\scriptscriptstyle ^\mid$\hspace{-0.40em}C}}}

\newcommand{\FF}{{{\rm I \kern -0.2em F}}}
\newcommand{\RR}{{{\rm I \kern -0.2em R}}}
\newcommand{\DD}{{{\rm I \kern -0.2em D}}}
\newcommand{\CC}{{{\mbox{\rm \hspace*{0.05ex}
\rule[.18ex]{.18ex}{1.24ex} \kern -.65em C}}}}

\newcommand{\finr}{{\ \hfill $\Box$}}
\newcommand{\fine}{{\ \hfill $\lozenge$}}
\newcommand{\ii}[1]{{\it #1}}

\makeindex             

\begin{document}
%
\title{\vspace*{-0cm}\LARGE{{\sc Descriptor System Tools (\textbf{DSTOOLS})} } \newline\newline {\LARGE \textsc{User's Guide} } \vspace*{4cm}
}
\author{\Large \textbf{Andreas~Varga}\footnote{Andreas Varga lives in Gilching, Germany. \emph{E-mail address}: \url{varga.andreas@gmail.com}, \emph{URL}: \url{https://sites.google.com/site/andreasvargacontact/}}
}
\date{September 30, 2018}

\maketitle
\vspace*{2cm}
\begin{abstract}
The Descriptor System Tools (\textbf{DSTOOLS}) is a collection of MATLAB functions  for the operation on and manipulation of rational transfer function matrices via their descriptor system realizations. The \textbf{DSTOOLS} collection relies on the Control System Toolbox and several mex-functions based on the Systems and Control Library SLICOT.
Many of the implemented functions are based on the computational procedures described in Chapter 10 of the book: "A. Varga, Solving Fault Diagnosis Problems -- Linear Synthesis Techniques, Springer, 2017". This document is the User's Guide for the version V0.71 of \textbf{DSTOOLS}. First, we present the mathematical background on rational matrices and descriptor systems. Then, we give in-depth information on the
command syntax of the main computational functions. Several examples illustrate the use of the main functions of \textbf{DSTOOLS}.
\end{abstract}
\hypersetup{pageanchor=true}

\vspace*{1.5cm}

\newpage
\pdfbookmark[1]{Contents}{Cont}

\tableofcontents
\hypersetup{linkcolor=blue}

\newpage
\section*{Notations and Symbols}
\addcontentsline{toc}{section}{Notations and Symbols}

%
\begin{longtable}{lp{13cm}}
$\emptyset$ & {empty set} \\
$\mathds{C}$ & {field of complex numbers} \\
$\mathds{R}$ & {field of real numbers} \\
$\mathds{C}_s$ & {stability domain (i.e., open left complex half-plane in continuous-time or open unit disk centered in the origin in discrete-time)}\\
$\partial\mathds{C}_s$ & {boundary of stability domain (i.e., extended imaginary axis with infinity included in continuous-time, or unit circle centered in the origin in discrete-time)} \\
$\overline{\mathds{C}}_s$ & {closure of $\mathds{C}_s$: $\overline{\mathds{C}}_s = \mathds{C}_s \cup \partial\mathds{C}_s$} \\
${\mathds{C}}_u$ & {open instability domain: $\mathds{C}_u := \mathds{C} \setminus \overline{\mathds{C}}_s$}
\\ $\overline{\mathds{C}}_u$ & {closure of ${\mathds{C}}_u$: $\overline{\mathds{C}}_u := \mathds{C}_u \cup \partial\mathds{C}_s$}
\\ $\mathds{C}_g$ & {``good'' domain of $\mathds{C}$}
\\ $\mathds{C}_b$ & {``bad'' domain of $\mathds{C}$: $\mathds{C}_b = \mathds{C} \setminus \mathds{C}_g$}
\\ $s$ & {complex frequency variable in the Laplace transform: $s = \sigma+\mathrm{i}\omega$}
\\ $z$ & {complex frequency variable in the Z-transform: $z = \mathrm{e}^{\,sT}$, $T$ -- sampling time}
\\ $\lambda$ & {complex  frequency variable: $\lambda = s$ in continuous-time or $\lambda = z$ in discrete-time}
\\ $\bar\lambda$ & {complex conjugate of the complex number $\lambda$ }
\\ $\mathds{R}(\lambda)$ & {field of real rational functions  in  indeterminate $\lambda$}
\\ $\mathds{R}(\lambda)$ & {set of rational matrices  in  indeterminate $\lambda$ with real coefficients  and unspecified dimensions}
\\ $\mathds{R}(\lambda)^{p\times m}$ & {set of $p\times m$ rational matrices  in  indeterminate $\lambda$ with real coefficients}
\\ $\mathds{R}{[\lambda]}$ & {ring of real rational polynomials  in  indeterminate $\lambda$}
\\ $\mathds{R}{[\lambda]}$ & {set of polynomial matrices  in  indeterminate $\lambda$ with real coefficients  and unspecified dimensions}
\\ $\mathds{R}{[\lambda]}^{p\times m}$ & {set of $p\times m$ polynomial matrices  in  indeterminate $\lambda$ with real coefficients}
\\ $\delta(G(\lambda))$ & {McMillan degree of the rational matrix $G(\lambda)$}
\\ $G^\sim(\lambda)$ & {Conjugate of $G(\lambda) \in \mathds{R}(\lambda)$: $G^\sim(s) = G^T(-s)$ in continuous-time and $G^\sim(z) = G^T(1/z)$ in discrete-time}
\\ $\ell_2$ & {Banach-space of square-summable sequences}
\\ $\mathcal{L}_2$ & {Lebesgue-space of square-integrable functions}
\\ $\mathcal{H}_2$ & {Hardy-space of square-integrable complex-valued functions analytic in ${\mathds{C}}_u$}
\\  $\mathcal{L}_\infty$ & {Space of complex-valued functions bounded and analytic in ${\partial\mathds{C}}_s$}
\\ $\mathcal{H}_\infty$ & {Hardy-space of complex-valued functions bounded and analytic in ${\mathds{C}}_u$}
\\ $\|G\|_{2}$ & {$\mathcal{H}_2$- or $\mathcal{L}_2$-norm of the transfer function matrix $G(\lambda)$}
\\ $\|G\|_{\infty}$ & {$\mathcal{H}_\infty$- or $\mathcal{L}_\infty$-norm of the transfer function matrix $G(\lambda)$}
\\ $\|G\|_{\infty/2}$ & {either the $\mathcal{H}_\infty$- or $\mathcal{H}_2$-norm of the transfer function matrix $G(\lambda)$}
\\ $\|G\|_H$ & {Hankel norm of the transfer function matrix $G(\lambda)$}
\\ $\delta_\nu(G_1,G_2)$ & $\nu$-gap distance between the transfer function matrices  $G_1(\lambda)$ and $G_2(\lambda)$
\\ col$_i(M)$ & {the $i$-th column of the matrix $M$}
\\ row$_i(M)$ & {the $i$-th row of the matrix $M$}
\\ $M^T$ & {transpose of the matrix $M$}
\\ $M^P$ & {pertranspose of the matrix $M$}
\\ $M^{-1}$ & {inverse of the matrix $M$}
\\ $M^{-T}$ & {transpose of the inverse matrix $M^{-1}$}
\\ $M^{-L}$ & {left inverse of the matrix $M$ (i.e., $M^{-L}M = I$)}
\\ $M^{-R}$ & {right inverse of the matrix $M$ (i.e., $MM^{-R} = I$)}
\\ $M^{\dag}$ & {Moore-Penrose pseudo-inverse of the matrix $M$}
\\ $\overline\sigma(M)$ & {largest singular value of the matrix $M$}
\\ $\underline\sigma(M)$ & {least singular value of the  matrix $M$}
\\ $\ker(M)$ & {kernel (or right nullspace) of the matrix $M$}
\\ $\ker_L(G(\lambda))$ & {left kernel (or left nullspace) of $G(\lambda) \in \mathds{R}(\lambda)$}
\\ $\ker_R(G(\lambda))$ & {right kernel (or right nullspace) of $G(\lambda) \in \mathds{R}(\lambda)$}
\\ $\mathcal{R}(M)$ & {range (or image space) of the matrix $M$}
\\ $\Lambda(A)$ & {set of eigenvalues of the matrix $A$}
\\ $\Lambda(A,E)$ & {set of generalized eigenvalues of the pair $(A,E)$ }
\\ $\Lambda(A-\lambda E)$ & {set of eigenvalues of the pencil $A-\lambda E$}
\\ $\mathbf{u}$ & {unit roundoff of the floating-point representation}
\\ $\mathcal{O}(\epsilon)$ & {quantity of order of $\epsilon$}
\\ $I_n$ or $I$ & {identity matrix of order $n$ or of an order resulting from context}
\\ $e_i$ & {the $i$-th column of the (known size) identity matrix }
\\ $0_{m\times n}$ or $0$ & {zero matrix of size ${m\times n}$ or of a size resulting from context}
\\ $\spann M$ & {span (or linear hull) of the columns of the matrix $M$}
 \end{longtable}

\newpage
\section*{Acronyms}
\addcontentsline{toc}{section}{Acronyms}

%

\begin{longtable}{lp{13cm}}
GCARE&{Generalized continuous-time algebraic Riccati equation}
\\GDARE&{Generalized discrete-time algebraic Riccati equation}
\\GRSD&{Generalized real Schur decomposition}
\\GRSF&{Generalized real Schur form}
\\LCF&{Left coprime factorization}
\\LDP&{Least distance problem}
\\LTI&{Linear time-invariant}
\\MIMO&{Multiple-input multiple-output}
\\RCF&{Right coprime factorization}
\\RSF&{Real Schur form}
\\SISO&{Single-input single-output}
\\SVD&{Singular value decomposition}
\\TFM&{Transfer function matrix}
\end{longtable}

\newpage

\section{Introduction}\label{sec:intro}

The {\sc Descriptor System Tools} (\textbf{DSTOOLS}) is a collection of {MATLAB} functions for the operation on and manipulation of rational transfer function matrices via their descriptor system realizations. The initial version V0.5 of \textbf{DSTOOLS} covers the main computations encountered in the synthesis approaches of linear residual generation filters for continuous- or discrete-time linear systems, described in the Chapter 10 of the author's book \cite{Varg17}: \\

\begin{tabular}{l}Andreas Varga, \emph{Solving Fault Diagnosis Problems - Linear Synthesis Techniques}, \\
vol. 84 of Studies in Systems, Decision and Control, Springer International Publishing, \\
xxviii+394, 2017.
\end{tabular}\\

The functions of the \textbf{DSTOOLS} collection rely on the \emph{Control System Toolbox} \cite{MLCO15} and several mex-functions based on the Systems and Control Library SLICOT \cite{Benn99}.
The current release of \textbf{DSTOOLS} is version V0.71, dated September 30, 2018. \textbf{DSTOOLS} is distributed as a free software via the Bitbucket repository.\footnote{\url{https://bitbucket.org/DSVarga/dstools}} The codes have been developed under MATLAB 2015b and have been also tested with MATLAB 2016a through 2018b. To use the functions of \textbf{DSTOOLS}, the \emph{Control System Toolbox} must be installed in MATLAB running under 64-bit Windows 7, 8, 8.1 or 10.

This document describes version V0.71 of the \textbf{DSTOOLS} collection. It will be continuously extended in parallel with the implementation of new functions. The book \cite{Varg17} represents an important complementary documentation for the \textbf{DSTOOLS} collection: it describes the mathematical background on rational matrices and descriptor systems, and gives detailed descriptions of many of the underlying procedures. Additionally, the M-files of the functions are self-documenting and a detailed documentation can be obtained online by typing help with the M-file name.
Please cite the current version of \textbf{DSTOOLS} as follows: \\

\begin{tabular}{l} A. Varga. DSTOOLS -- The Descriptor System Tools for MATLAB, 2018.\\ \url{https://sites.google.com/site/andreasvargacontact/home/software/dstools}.
\end{tabular}\\

\newpage
\section{Background Material on Generalized System Representations}\label{sec:Basics}

In this section we give background information on two system representations of {linear time-invariant}  systems, namely the input-output representation via rational {transfer function matrices}   and the generalized state-space representation, also  known as descriptor system representation. Since each rational matrix can be interpreted as the transfer function matrix of a descriptor system, the manipulation of rational matrices can be alternatively performed via their descriptor representations, using numerically reliable computational algorithms.

The treatment in depth of most of concepts was not possible in the restricted size of this guide. The equivalence theory of linear matrix pencils is covered in \cite{Gant59}.
The material on rational matrices is covered in several textbooks, of which we mention only the two widely cited books of Kailath \cite{Kail80} and Vidyasagar \cite{Vidy85}. Linear descriptor systems (also known in the literature as linear differential-algebraic-equations-based systems or generalized state-space systems or singular systems), are discussed, to  different depths  and with different focus, in several books \cite{Camp80,Dai89,Kunk06,Duan10}.

\subsection{Rational Transfer Function Matrices}\label{app:TFM}

Transfer functions are used to describe the input-output behaviour of \emph{single-input single-output} (SISO) \emph{linear time-invariant} (LTI) systems by relating the input and output variables via a gain depending on a frequency variable. For a SISO system with input $u(t)$ and output $y(t)$ depending on the continuous time variable $t$, let ${\mathbf{u}}(s) := \mathcal{L}(u(t))$ and $\mathbf{y}(s) := \mathcal{L}(y(t))$ denote the Laplace transformed input and output, respectively. Then, the transfer function of the continuous-time LTI system is defined as
\[ g(s) := \frac{\mathbf{y}(s)}{\mathbf{u}(s)} \]
and relates the input and output in the form
\[ \mathbf{y}(s) = g(s)\mathbf{u}(s) \, .\]
The complex variable  $s = \sigma + j\omega$, has for $\sigma =0$ the interpretation of a complex frequency. If the time variable has a discrete variation with equally spaced values with increments given by a sampling-period $T$, then the transfer function of the discrete-time system is defined using the $\mathcal{Z}$-transforms of the input and output variables $\mathbf{u}(z) := \mathcal{Z}(u(t))$ and $\mathbf{y}(z) := \mathcal{Z}(y(t))$, respectively, as
\[ g(z) := \frac{\mathbf{y}(z)}{\mathbf{u}(z)} \]
and relates the input and output in the form
\[ \mathbf{y}(z) = g(z)\mathbf{u}(z) \, .\]
The complex variable $z$ is related to the complex variable $s$ as $z = e^{sT}$.
We will use the variable $\lambda$ to denote either the $s$ or $z$ complex variables, depending on the context, continuous- or discrete-time, respectively. Throughout this guide, bolded variables as ${\mathbf{u}(\lambda)}$ and ${\mathbf{y}(\lambda)}$ will be used to denote either the Laplace- or $\mathcal{Z}$-transformed quantities of the corresponding time-variables $u(t)$ and $y(t)$.
Furthermore, we will restrict our discussion to rational transfer functions $g(\lambda)$ which can be expressed as a ratio of two polynomials with real coefficients
\be\label{glambda} g(\lambda) = \frac{\alpha(\lambda)}{\beta(\lambda)} = \frac{a_m\lambda^m+a_{m-1}\lambda^{m-1}+\cdots + a_1\lambda+a_0}{b_n\lambda^n+b_{n-1}\lambda^{n-1}+\cdots + b_1\lambda+b_0} \, ,\ee
with $a_m\not = 0$ and $b_n\not = 0$. Thus, $g(\lambda) \in \mathds{R}(\lambda)$, where $\mathds{R}(\lambda)$ is the field of real rational functions.

Transfer function matrices are used to describe the input-output behaviour of \emph{multi-input multi-output} (MIMO) LTI systems by relating the input and output variables via a matrix of gains depending on a frequency variable. Consider a MIMO system with $m$ inputs $u_1(t)$, $\ldots$, $u_m(t)$, which form the $m$-dimensional input vector $u(t) = [\,u_1(t), \ldots, u_m(t)\,]^T$, and $p$ outputs $y_1(t)$, $\ldots$, $y_p(t)$, which form the $p$-dimensional output vector $y(t) = [\,y_1(t), \ldots, y_p(t)\,]^T$. For a continuous  dependence of $u(t)$ and $y(t)$ on the time variable $t$, let $\mathbf{u}(s)$ and $\mathbf{y}(s)$ be the Laplace-transformed input and output vectors, respectively, while in the case of a discrete dependence on $t$, we denote $\mathbf{u}(z)$ and $\mathbf{y}(z)$ the $\mathcal{Z}$-transformed input and output vectors, respectively. We denote with $\lambda$ the frequency variable, which is either $s$ or $z$, depending on the nature of the time variation, continuous or discrete, respectively. Let $G(\lambda)$ be the $p\times m$ \emph{transfer function matrix} (TFM)  defined as
\[ G(\lambda) = \ba{ccc} g_{11}(\lambda) & \cdots & g_{1m}(\lambda) \\
\vdots & \ddots & \vdots \\
g_{p1}(\lambda) & \cdots & g_{pm}(\lambda) \ea \, , \]
which relates relates the $m$-dimensional input vector $u$ to the $p$-dimensional output vector $y$ in the form
\[ \mathbf{y}(\lambda) = G(\lambda)\mathbf{u}(\lambda) \, . \]
The element $g_{ij}(\lambda)$ describes the contribution of the $j$-th input $u_j(t)$ to the $i$-th output $y_i(t)$. We assume that each matrix entry $g_{ij}(\lambda) \in \mathds{R}(\lambda)$ and thus it can be expressed as ratio of two polynomials $\alpha_{ij}(\lambda)$ and $\beta_{ij}(\lambda)$ with real coefficients as $g_{ij}(\lambda) = \alpha_{ij}(\lambda)/\beta_{ij}(\lambda)$ of the form (\ref{glambda}).

Each TFM $G(\lambda)$ belongs to the set of rational matrices with real coefficients, thus having elements in the field of real rational functions $\mathds{R}(\lambda)$. Polynomial matrices, having elements in the ring of polynomials with real coefficients  $\mathds{R}[\lambda]$, can be assimilated in a natural way with special rational matrices with all elements having 1 as denominators. Let $\mathds{R}(\lambda)^{p\times m}$ and $\mathds{R}[\lambda]^{p\times m}$ denote the sets of $p\times m$ rational and polynomial matrices with real coefficients, respectively. To simplify the notation, we will also use $G(\lambda)\in \mathds{R}(\lambda)$ or $G(\lambda)\in \mathds{R}[\lambda]$
if the dimensions of $G(\lambda)$ are not relevant or are clear from the context.

A rational matrix $G(\lambda)\in \mathds{R}(\lambda)$ is called \emph{proper} if $\lim_{\lambda \rightarrow \infty} G(\lambda) = D$, with $D$ having a finite norm. Otherwise, $G(\lambda)$ is called \emph{improper}.
\index{transfer function matrix (TFM)!proper}%
\index{transfer function matrix (TFM)!improper}%
If $D = 0$, then $G(\lambda)$ is \emph{strictly proper}.
\index{transfer function matrix (TFM)!strictly proper}
An invertible $G(\lambda)$ is \emph{biproper} if both $G(\lambda)$ and $G^{-1}(\lambda)$ are proper.
\index{transfer function matrix (TFM)!biproper}
A polynomial matrix $U(\lambda) \in \mathds{R}[\lambda]$ is called \emph{unimodular}  if is invertible and its inverse $U^{-1}(\lambda) \in \mathds{R}[\lambda]$ (i.e., is a polynomial matrix).
\index{polynomial matrix!unimodular}
 The determinant of a unimodular matrix is therefore a constant.

The degree of a rational matrix $G(\lambda)$, also known as the McMillan degree, is defined in Section~\ref{app:tfm_polzer}. We only give here the definition of the degree of a  rational vector $v(\lambda)$. For this, we express first $v(\lambda)$ in the form $v(\lambda) = \widetilde v(\lambda)/d(\lambda)$, where $d(\lambda)$ is the monic least common multiple of all denominator polynomials of the elements of $v(\lambda)$ and $\widetilde v(\lambda)$ is the corresponding polynomial vector $\widetilde v(\lambda) := d(\lambda)v(\lambda)$. Then, $\text{deg } v(\lambda) = \max(\text{deg } \widetilde v(\lambda), \text{deg } v(\lambda))$.

\subsection{Descriptor Systems}\label{app:desc}

A descriptor system is a generalized state-space representation of the form
\be\label{app:dss}
\begin{array}{rcl} E \lambda x(t) &=& Ax(t) + Bu(t) , \\
y(t) &=& Cx(t) + Du(t) ,
\end{array} \ee
where $x(t) \in \mathds{R}^n$ is the state vector, $u(t) \in \mathds{R}^m$ is the input vector, and $y(t) \in \mathds{R}^p$ is the output vector, and where $\lambda$ is either the differential operator $\lambda x(t) = \frac{\text{d}}{\text{d}t}x(t)$ for a continuous-time system or the advance operator
$\lambda x(t) = x(t+1)$ for a discrete-time system. In all what follows, we assume
$E$ is square and possibly singular, and the pencil $A-\lambda E$ is regular (i.e., $\det (A-\lambda E) \not \equiv 0$). If $E = I_n$, we call the representation (\ref{app:dss}) a \emph{standard state-space system}.  The corresponding input-output representation of the descriptor system (\ref{app:dss}) is
\be\label{app:tfm} \mathbf{y}(\lambda) = G(\lambda)\mathbf{u}(\lambda) \, , \ee
where, depending on the system type, $\lambda = s$, the complex variable in the Laplace transform for a continuous-time system, or $\lambda = z$, the complex variable in the $\mathcal{Z}$-transform for a discrete-time system, $\mathbf{y}(\lambda)$ and $\mathbf{u}(\lambda)$ are the Laplace- or $\mathcal{Z}$-transformed output and input vectors,  respectively, and $G(\lambda)$ is the rational TFM of the system, defined as
\be\label{app:dss-tfm} G(\lambda) = C(\lambda E-A)^{-1}B+D \, .\ee
We alternatively denote descriptor systems of the form (\ref{app:dss}) with the quadruple $(A-\lambda E,B,C,D)$ or a standard state-space system with $(A,B,C,D)$ (if $E=I_n$), and use the  notation
\be\label{app:tfm-dss} G(\lambda) := \ba{c|c} A-\lambda E & B \\ \hline C & D \ea \, ,\ee
to relate the  TFM $G(\lambda)$ to a particular  descriptor system realization as in (\ref{app:dss}).

It is well known that a descriptor system representation of the form (\ref{app:dss})
is the most general description for a linear time-invariant system.
Continuous-time descriptor systems arise frequently from modelling interconnected systems containing algebraic loops or constrained mechanical systems which describe contact phenomena. Discrete-time descriptor representations are frequently used to model economic processes.

The manipulation of rational matrices can be easily performed via their descriptor representations.
The main result which allows this is the following \cite{Verg79}:
\begin{theorem}\label{T-dssminreal} For any rational matrix $G(\lambda) \in \mathds{R}(\lambda)^{p\times m}$, there exist $n \geq 0$ and the  real matrices $E, A \in \mathds{R}^{n\times n}$, $B\in \mathds{R}^{n\times m}$, $C\in \mathds{R}^{p\times n}$ and $D\in \mathds{R}^{p\times m}$, with $A-\lambda E$ regular, such that
(\ref{app:dss-tfm}) holds.
\end{theorem}

The descriptor realization $(A-\lambda E,B,C,D)$ of a given rational matrix $G(\lambda)$ is not unique. For example, if $U$ and $V$ are invertible matrices of the size $n$ of the square matrix $E$, then two descriptor realizations $(A-\lambda E,B,C,D)$ and $(\widetilde A-\lambda \widetilde E,\widetilde B,\widetilde C,\widetilde D)$ related by a \emph{system similarity transformation} of the form
\be\label{app:dss-sim} (\widetilde A-\lambda \widetilde E,\widetilde B,\widetilde C,\widetilde D) = (UAV-\lambda UEV,UB,CV,D) \,,\ee
have the same TFM $G(\lambda)$. \index{descriptor system!similarity transformation}
Moreover, among all possible realizations of a given $G(\lambda)$, with different sizes $n$, there exist realizations which have the least dimension.
A descriptor realization $(A-\lambda E,B,C,D)$ of the rational matrix $G(\lambda)$ is called \emph{minimal} if the dimension $n$ of the square matrices $E$ and $A$ is the least possible one. The minimal realization of a given $G(\lambda)$ is also not unique, since two minimal realizations related by a system similarity transformation as in (\ref{app:dss-sim}) correspond to the same $G(\lambda)$.

A minimal descriptor system realization  $(A-\lambda E,B,C,D)$  is characterized by the following five conditions \cite{Verg81}.
\begin{theorem}\label{T-desc-minreal}
A descriptor system realization  $(A-\lambda E,B,C,D)$ of order $n$ is minimal if the following conditions are fulfilled:
\[ \begin{array}{rl} (i) & \rank\ba{cc} A-\lambda E & B \ea = n, \quad \forall \lambda \in \mathds{C} , \\ \\[-3mm]
(ii) & \rank\ba{cc}  E & B \ea = n, \\ \\[-3mm]
(iii) & \rank\ba{c} A-\lambda E \\ C \ea = n, \quad \forall \lambda \in \mathds{C} , \\ \\[-3mm]
(iv) & \rank\ba{c}  E \\ C \ea = n, \\ \\[-3mm]
(v) & A\ker(E) \subseteq \mathcal{R}(E) .
\end{array} \]
\end{theorem}
\index{descriptor system!finite controllability}
\index{descriptor system!infinite controllability}
\index{descriptor system!finite observability}
\index{descriptor system!infinite observability}
\index{descriptor system!irreducible realization}
\index{descriptor system!controllability}
\index{descriptor system!observability}
\index{descriptor system!minimal realization}
The conditions $(i)$ and $(ii)$ are known as \emph{finite} and \emph{infinite controllability}, respectively. A system or, equivalently, the pair $(A-\lambda E,B)$, is called \emph{finite controllable} if it fulfills $(i)$,  \emph{infinite controllable} if it fulfills $(ii)$, and \emph{controllable} if it fulfills   both $(i)$ and $(ii)$. Similarly, the conditions $(iii)$ and $(iv)$ are known as \emph{finite} and \emph{infinite observability}, respectively. A system or, equivalently, the pair $(A-\lambda E,C)$, is called \emph{finite observable} if it fulfills $(iii)$,  \emph{infinite observable} if it fulfills $(iv)$, and \emph{observable} if it fulfills   both $(iii)$ and $(iv)$.
The condition $(v)$ expresses the absence of non-dynamic modes. A descriptor realization which satisfies only $(i)-(iv)$ is called \emph{irreducible} (also weakly minimal). The numerical computation of irreducible realizations is addressed in \cite{Varg90} (see also \cite{Door81} for alternative approaches).

\subsection{Linear Matrix Pencils}\label{sec:matrix-pencils}
Linear matrix pencils of the form $M-\lambda N$, where $M$ and $N$ are $m\times n$ matrices with elements in $\mathds{C}$, play an important role in the theory of generalized LTI systems. In what follows we shortly review the equivalence theory of linear matrix pencils. For more details on this topic see \cite{Gant59}.

The pencil $M-\lambda N$ is called \emph{regular} if $m = n$ and $\det(M-\lambda N) \not\equiv 0$. Otherwise, the pencil is called \emph{singular}.
Two pencils $M-\lambda N$ and $\widetilde M-\lambda \widetilde N$ with $M, N, \widetilde M, \widetilde N \in \mathds{C}^{m\times n}$ are \emph{strictly equivalent} if there exist two invertible matrices $U \in \mathds{C}^{m\times m}$ and $V \in \mathds{C}^{n\times n}$  such that
\index{linear matrix pencil!strict equivalence}
\be\label{pencil-equiv}
U(M-\lambda N)V = \widetilde M -\lambda \widetilde N .\ee

\index{linear matrix pencil!Weierstrass canonical form}
\index{linear matrix pencil!eigenvalues}
\index{linear matrix pencil!finite eigenvalues}
\index{linear matrix pencil!infinite eigenvalues}
\index{canonical form!Weierstrass}
\index{canonical form!Jordan}
For a regular pencil, the strict equivalence leads to the (complex) \emph{Weierstrass canonical form}, which is instrumental to characterize the dynamics of generalized systems.
\begin{lemma}\label{L-WCF}
Let $M-\lambda N$ be an arbitrary regular pencil with $M, N \in \mathds{C}^{n\times n}$. Then, there exist invertible matrices $U \in \mathds{C}^{n\times n}$ and $V \in \mathds{C}^{n\times n}$ such that
\be\label{Weierstrass} U(M-\lambda N)V = \ba{cc} J_f-\lambda I & 0 \\ 0 & I-\lambda J_\infty \ea ,\ee
where $J_f$ is in a (complex) Jordan canonical form
\be\label{Jordan} J_f = \diag \left(J_{s_1}(\lambda_1), J_{s_2}(\lambda_2), \ldots, J_{s_k}(\lambda_k) \right) \, ,\ee
with  $J_{s_i}(\lambda_i)$ an elementary $s_i\times s_i$ Jordan block of the form
\[ J_{s_i}(\lambda_i) = \ba{cccc} \lambda_i & 1  \\ & \lambda_i & \ddots \\ & & \ddots & 1 \\ &&&\lambda_i \ea \]
and $J_\infty$ is nilpotent and has the (nilpotent) Jordan form
\be\label{Jordan-null} J_\infty = \diag \big(J_{s_1^\infty}(0), J_{s_2^\infty}(0), \ldots, J_{s_h^\infty}(0) \big) \, .\ee
\end{lemma}
The Weierstrass canonical form (\ref{Weierstrass}) exhibits the finite and infinite eigenvalues of the pencil $M-\lambda N$. The finite eigenvalues are $\lambda_i$, for $i = 1, \ldots, k$. Overall, by including all multiplicities, there are $n_f = \sum_{i=1}^{k}s_i$ \emph{finite eigenvalues} and $n_\infty = \sum_{i=1}^{h}s_i^\infty$ \emph{infinite eigenvalues}. Infinite eigenvalues with $s_i^\infty =1$ are called \emph{simple infinite eigenvalues}. We can also express the rank of $N$ as
\[ \rank N = n_f+\rank J_\infty = n_f + \sum_{i=1}^{h}(s_i^\infty-1) =  n_f+n_\infty-h = n-h. \]

If $M$ and $N$ are real matrices, then there exist real matrices $U$ and $V$ such that the pencil $U(M-\lambda N)V$ is in a \emph{real} \emph{Weierstrass canonical form}, where  the only difference is that $J_f$ is in a \emph{real} Jordan form \cite[Section 3.4]{Horn13}. In this form, the elementary real Jordan blocks correspond to pairs of complex conjugate eigenvalues.

If $M-\lambda N = A-\lambda I$ (e.g., the pole pencil for a standard state-space system), then all eigenvalues are finite and $J_f$ in the Weierstrass form is simply the (real) Jordan form of $A$. The transformation matrices can be chosen such that $U = V^{-1}$.

The eigenvalue structure of a regular pencil $A-\lambda E$ is completely described by the Weierstrass canonical form (see Lemma~\ref{L-WCF}). However, the computation of this canonical form involves the use of (potentially ill-conditioned) general invertible transformations, and therefore numerical reliability cannot be guaranteed. Fortunately, the computation of Weierstrass canonical form can be avoided in almost all computations, and alternative ``less'' condensed forms can be employed instead, which can be computed by exclusively employing orthogonal similarity transformations. The \emph{generalized real Schur decomposition} (GRSD) of a matrix pair $(A,E)$ reveals the eigenvalues of the regular pencil $A-\lambda E$, by determining the \emph{generalized real Schur form} (GRSF) of the pair $(A,E)$ (a quasi-triangular--triangular form)  using orthogonal similarity transformations on the pencil $A-\lambda E$. The main theoretical result regarding the GRSD is the following theorem.

\index{condensed form!generalized real Schur (GRSF)}
\begin{theorem}\label{L:GRSF} Let $A-\lambda E$ be an $n\times n$ regular pencil, with $A$ and $E$ real matrices. Then, there exist orthogonal transformation matrices $Q$ and $Z$ such that
\be\label{GRSF-gen} S -\lambda T := Q^T(A-\lambda E)Z = \ba{ccc} S_{11} & \cdots & S_{1k} \\
& \ddots & \vdots \\ 0 && S_{kk} \ea - \lambda \ba{ccc} T_{11} & \cdots & T_{1k} \\
& \ddots & \vdots \\ 0 && T_{kk} \ea ,\ee
where each diagonal subpencil $S_{ii}-\lambda T_{ii}$, for $i = 1, \ldots, k$, is either of dimension $1\times 1$ in the case of a finite real or infinite eigenvalue of the pencil $A-\lambda E$ or of dimension $2\times 2$, with $T_{ii}$ upper triangular, in the case of a pair of finite complex conjugate eigenvalues of $A-\lambda E$. 
\end{theorem}
The pair $(S,T)$ in (\ref{GRSF-gen}) is in a GRSF and the eigenvalues of $A-\lambda E$ (or the generalized eigenvalues of the pair $(A,E)$) are given by
\[ \Lambda(A-\lambda E) = \bigcup_{i=1}^k \Lambda(S_{ii}-\lambda T_{ii}) \, .\]
If $E = I$, then we can always choose $Q = Z$, $T = I$ and $S$ is the  \emph{real Schur form} (RSF) of $A$. \index{condensed form!real Schur (RSF)}

The order of eigenvalues (and thus of the associated pairs of diagonal blocks) of the reduced pencil $S-\lambda T$  is arbitrary. The reordering of the pairs of diagonal blocks (thus also of corresponding eigenvalues) can be done by interchanging two adjacent pairs of diagonal blocks of the GRSF. For the swapping of such two pairs of blocks   orthogonal similarity transformations can be used. Thus, any arbitrary reordering of pairs of blocks (and thus of the corresponding eigenvalues) can be achieved in this way. An important application of this fact is the computation of orthogonal bases for the deflating subspaces of the pencil $A-\lambda E$ corresponding to a particular eigenvalue or a particular set of eigenvalues.

\index{linear matrix pencil!Kronecker canonical form}
\index{canonical form!Kronecker}
For a general (singular) pencil, the strict equivalence leads to the (complex) \emph{Kronecker canonical form}, which is instrumental to characterize the zeros and singularities of a descriptor system.
\begin{lemma}\label{L-KCF}
Let $M-\lambda N$ be an arbitrary pencil with $M, N \in \mathds{C}^{m\times n}$. Then, there exist invertible matrices $U \in \mathds{C}^{m\times m}$ and $V \in \mathds{C}^{n\times n}$ such that
\be\label{Kronecker} U(M-\lambda N)V = \ba{ccc} K_r(\lambda) \\ & K_{reg}(\lambda) \\&&K_l(\lambda)\ea ,\ee
where:
\begin{enumerate}
\item[1)] The full row rank pencil $K_r(\lambda)$ has the form
\be\label{kcf-kreps} K_r(\lambda) = \diag \big(L_{\epsilon_1}(\lambda), L_{\epsilon_2}(\lambda), \cdots, L_{\epsilon_{\nu_r}}(\lambda) \big) \, , \ee
with  $L_{i}(\lambda)$  ($i \geq 0$) an $i\times (i+1)$  bidiagonal pencil of form
\be\label{Liblocks} L_i(\lambda) = \ba{cccc} -\lambda & 1 \\ & \ddots & \ddots \\ && -\lambda & 1 \ea  \, ; \ee
\item[2)] The regular pencil $K_{reg}(\lambda)$ is in a Weierstrass canonical form
\be\label{regblocks}  K_{reg}(\lambda) = \ba{cc} \widetilde J_f-\lambda I  \\ & I-\lambda \widetilde J_\infty \ea \, ,\ee
with $\widetilde J_f$ in a (complex) Jordan canonical form as in (\ref{Jordan}) and
with $\widetilde J_\infty$ in a nilpotent Jordan form as in (\ref{Jordan-null});
\item[3)] The full column rank $K_l(\lambda)$ has the form
\be\label{kcf-keta}  K_l(\lambda) = \diag \big(L^T_{\eta_1}(\lambda), L^T_{\eta_2}(\lambda), \cdots, L^T_{\eta_{\nu_l}}(\lambda) \big) \, .\ee
\end{enumerate}
\end{lemma}

As it is apparent from (\ref{Kronecker}), the Kronecker canonical form  exhibits the right and left singular structures of the pencil $M-\lambda N$ via the full row rank block $K_r(\lambda)$ and full column rank block $K_l(\lambda)$, respectively, and the eigenvalue structure via the regular pencil $K_{reg}(\lambda)$. The full row rank pencil $K_r(\lambda)$ is $n_r\times (n_r+\nu_r)$, where $n_r = \sum_{i=1}^{\nu_r} \epsilon_i$, the full column rank pencil $K_l(\lambda)$ is $(n_l+\nu_l)\times n_l$, where $n_l = \sum_{j=1}^{\nu_l} \eta_j$, while the regular pencil $K_{reg}(\lambda)$ is $n_{reg}\times n_{reg}$, with $n_{reg} = \tilde n_f+\tilde n_\infty$, where $\tilde n_f$ is the number of finite eigenvalues in $\Lambda(\widetilde J_f)$ and $\tilde n_\infty$ is the number of infinite eigenvalues in $\Lambda(I-\lambda \widetilde J_\infty)$ (or equivalently the number of null eigenvalues in $\Lambda(\widetilde J_\infty)$). The $\epsilon_i\times (\epsilon_i+1)$ blocks $L_{\epsilon_i}(\lambda)$ with $\epsilon_i \geq 0$
are the right elementary Kronecker blocks, and $\epsilon_i$, for $i = 1, \ldots, \nu_r$, are called the \emph{right Kronecker indices}. The $(\eta_i+1)\times \eta_i$ blocks $L^T_{\eta_i}(\lambda)$ with $\eta_i \geq 0$
are the left elementary Kronecker blocks, and $\eta_i$, for $i = 1, \ldots, \nu_l$,  are called the \emph{left Kronecker indices}. The normal rank $r$ of the pencil $M-\lambda N$ results as
\[  r := \rank (M-\lambda N) = n_r+\tilde n_f+\tilde n_\infty + n_l .\]
If $M-\lambda N$ is regular, then there are no left- and right-Kronecker structures and the Kronecker canonical form is simply the Weierstrass canonical form.
\index{linear matrix pencil!Kronecker indices}

\begin{remark}\label{altKCF}
By additional column permutations of the block $K_r(\lambda)$ and row permutations of the block $K_l(\lambda)$ (which can be included in the left and right transformations matrices $U$ and $V$) we can bring these blocks to the alternative forms
\be\label{altKrKl} K_r(\lambda) = \ba{cc} B_r & A_r-\lambda I_{n_r} \ea, \qquad
K_l(\lambda) = \ba{cc} A_l-\lambda I_{n_l} \\ C_l \ea , \ee
where the pair $(A_r,B_r)$ is in a Brunovsky controllable form
\[ A_r = \ba{cccc} A_{r,1} \\ &A_{r,2}\\ &&\ddots \\&&&A_{r,\nu_r} \ea, \qquad B_r = \ba{cccc} b_{r,1} \\ &b_{r,2}\\ &&\ddots \\&&&b_{r,\nu_r} \ea \, ,\]
with $A_{r,i}$ an $\varepsilon_i\times \varepsilon_i$ matrix and $b_{r,i}$ an $\varepsilon_i \times 1$ column vector  of the forms
\[ A_{r,i} = \ba{ccccc} 0 & I_{\varepsilon_i-1} \\ 0 & 0 \ea = J_{\varepsilon_i}(0), \qquad b_{r,i} = \ba{c} 0 \\ \vdots \\ 0 \\ 1 \ea \, ,\]
and the pair $(A_l,C_l)$ is in a Brunovsky observable form
\be\label{AlCl-bdiag} A_l = \ba{cccc} A_{l,1} \\ &A_{l,2}\\ &&\ddots \\&&&A_{l,\nu_l} \ea, \qquad C_l = \ba{cccc} c_{l,1} \\ &c_{l,2}\\ &&\ddots \\&&&c_{l,\nu_l} \ea \, ,\ee
with $A_{l,i}$ an $\eta_i\times \eta_i$ matrix and $c_{l,i}$ a $1\times \eta_i$ row vector  of the forms
\[ A_{l,i} = \ba{ccccc} 0 & 0 \\ I_{\eta_i-1} & 0 \ea = J^T_{\eta_i}(0), \qquad c_{l,i} = \ba{cccc} 0 & \cdots & 0 & 1 \ea \, .\]

\finr\end{remark}

The computation of the Kronecker-canonical form may involve the use of ill-conditioned transformations and, therefore, is potentially numerically unstable. Fortunately,  alternative staircase forms, called \emph{Kronecker-like forms}, allow to obtain basically the same (or only a part of) structural information on the pencil $M-\lambda N$ by employing exclusively orthogonal transformations (i.e., $U^TU = I$ and $V^TV = I$).

The following result concerns with one of the main Kronecker-like forms.

\index{condensed form!Kronecker-like}
\begin{theorem}\label{T:KLF} Let $M \in \mathds{R}^{m\times n}$ and $N \in \mathds{R}^{m\times n}$ be arbitrary real matrices. Then, there exist orthogonal $U \in \mathds{R}^{m\times m}$ and $V \in \mathds{R}^{n\times n}$, such that
\be\label{Kronecker-like} U(M-\lambda N)V = \ba{ccc} M_r-\lambda N_r & \ast & \ast \\ 0 & M_{reg}-\lambda N_{reg} & \ast\\0&0&M_l-\lambda N_l\ea ,\ee
where
\begin{enumerate}
\item[1)] The $n_r\times (m_r+n_r)$ pencil $M_r-\lambda N_r$ has full row rank, $n_r$, for all $\lambda \in \mathds{C}$ and is in a \emph{controllability staircase form} \index{condensed form!controllability staircase form}
\be\label{pencilright} M_r-\lambda N_r = \ba{cc} B_r & A_r-\lambda E_r \ea \, , \ee
with $B_r \in \mathds{R}^{n_r\times m_r}$, $A_r, E_r \in \mathds{R}^{n_r\times n_r}$, and $E_r$ invertible.
\item[2)] The $n_{reg}\times n_{reg}$ pencil $M_{reg}-\lambda N_{reg}$ is regular and its eigenvalues are the eigenvalues of pencil $M-\lambda N$. The pencil $M_{reg}-\lambda N_{reg}$ may be chosen in a GRSF, with arbitrary ordered diagonal blocks.
\item[3)] The $(p_l+n_l)\times n_l$ pencil $M_l-\lambda N_l$ has full column rank, $n_l$, for all $\lambda \in \mathds{C}$ and is in an \emph{observability staircase form}     \index{condensed form!observability staircase form}
\be\label{pencilleft} M_l-\lambda N_l = \ba{c} A_l-\lambda E_l \\ C_l \ea \, ,\ee
with $C_l \in \mathds{R}^{p_l\times n_l}$, $A_l, E_l \in \mathds{R}^{n_l\times n_l}$, and $E_l$ invertible.
\end{enumerate}

\end{theorem}

The generalized controllability staircase form (\ref{pencilright}) is defined with
\be\label{cscf-defab}
\hspace{-4mm}\ba{c|c}   B_r & A_r \ea =
{\arraycolsep=1mm \ba{c|ccccc}
A_{1,0} & A_{1,1} &  A_{12}  & \cdots &  A_{1,k-1}  & A_{1,k}   \\
0 &  A_{2,1} & A_{22} & \cdots &  A_{2,k-1} & A_{2,k} \\
0 & 0 & A_{32} & \cdots &  A_{3,k-1} &  A_{3,k} \\
\vdots & \vdots & \vdots & \ddots &  \vdots &\vdots \\
0 & 0 & 0 & \cdots &   A_{k,k-1} & A_{k,k}
\ea} \, ,
\ee
where $A_{j,j-1} \in \mathds{R}^{\nu_j\times \nu_{j-1}}$, with $\nu_0 = m_r$,   are full row rank matrices for $j = 1, \ldots, k$, and the resulting upper triangular matrix $E_r$ has a similar block partitioned form
\be\label{cscf-defe} E_r = \ba{ccccc} E_{1,1} & E_{1,2} & \cdots & E_{1,k-1} & E_{1,k} \\
0 & E_{2,2} & \cdots & E_{2,k-1} & E_{2,k} \\
\vdots & \vdots & \ddots & \vdots & \vdots \\
0 & 0 & \cdots & E_{k-1,k-1} & E_{k-1,k} \\
0 & 0 & \cdots & 0 & E_{k,k} \ea \, ,\ee
where $E_{j,j} \in \mathds{R}^{\nu_j\times \nu_j}$.
The resulting block dimensions $\nu_j, \,j = 0, 1, \ldots, k$, satisfy
\[ m_r = \nu_0 \geq \nu_1 \geq \cdots \geq \nu_k > 0 \, .\]
The dimensions $\nu_i$, $i = 1, \ldots, k$, of the diagonal blocks of $A_r-\lambda E_r$ completely determine the right Kronecker structure of $M-\lambda N$ as follows: there are $\nu_{i-1}-\nu_i$ blocks $L_{i-1}(\lambda)$ of size $(i-1)\times i$, for $i = 1, \ldots, k$.

The generalized observability staircase form (\ref{pencilleft}) is
\be\label{oscf-defac}
\ba{c}   A_l \\ \hline C_l \ea =
{\arraycolsep=1mm \ba{ccccc}
 A_{\ell,\ell} &  A_{\ell,\ell-1}  & \cdots &  A_{\ell,2}  & A_{\ell,1}   \\
 A_{\ell-1,\ell} & A_{\ell-1,\ell-1} & \cdots &  A_{\ell-1,2} & A_{\ell-1,1} \\
0 & A_{\ell-2,\ell-1} & \cdots &  A_{\ell-2,2} &  A_{\ell-2,1} \\
\vdots & \vdots & \ddots &  \vdots &\vdots \\
 0 & 0 & \cdots &   A_{1,2} & A_{1,1} \\ \hline
 0 & 0 & \cdots &   0 & A_{0,1 }
\ea} \, ,
\ee
where $A_{j-1,j} \in \mathds{R}^{\mu_{j-1} \times \mu_j}$, with $\mu_0 = p_l$,   are full column rank matrices for $j = 1, \ldots, \ell$, and the resulting upper triangular matrix $E_o$ has a similar block partitioned form
\be\label{oscf-defe} E_l = \ba{ccccc} E_{\ell,\ell} &  E_{\ell,\ell-1}  & \cdots &  E_{\ell,2}  & E_{\ell,1} \\
0 & E_{\ell-1,\ell-1} & \cdots &  E_{\ell-1,2} & E_{\ell-1,1}\\
\vdots & \vdots & \ddots & \vdots & \vdots \\
0 & 0 & \cdots & E_{2,2} & E_{2,1} \\
0 & 0 & \cdots & 0 & E_{1,1} \ea \, ,\ee
with $E_{j,j} \in \mathds{R}^{\mu_j\times \mu_j}$.
The resulting block dimensions $\mu_j, \,j = 0, 1, \ldots, \ell$, satisfy
\[ p_l = \mu_0 \geq \mu_1 \cdots \geq \mu_\ell > 0 \, .\]
The dimensions $\mu_i$, $i = 1, \ldots, \ell$  of the diagonal blocks of $A_l-\lambda E_l$ in the observability staircase form completely determine the left Kronecker structure of $M-\lambda N$ as follows: there are $\mu_{i-1}-\mu_i$ blocks $L^T_{i-1}(\lambda)$ of size $i\times (i-1)$, $i = 1, \ldots, \ell$. We have $n_r = \sum_{i=1}^k\nu_i$ and $n_l = \sum_{i=1}^\ell\mu_i$, and                                                                                 the normal rank of $M-\lambda N$ is $n_r+n_{reg}+n_l$.

Algorithms for the computation of Kronecker-like forms of linear pencils, using SVD-based rank determinations, have been proposed in \cite{Door79a,Demm93}. Albeit numerically reliable, these algorithms have a computational complexity $\mathcal{O}(n^4)$, where $n$ is the minimum of row or column dimensions of the pencil.  More efficient algorithms of complexity $\mathcal{O}(n^3)$ have been proposed in \cite{Beel88,Varg96d,Oara97}, which rely on using QR decompositions with column pivoting for rank determinations.

\subsection{Minimal Nullspace Bases} \label{sec:minbasis}

A $p$-dimensional rational vector $v(\lambda) \in \mathds{R}(\lambda)^p$ can be seen as either a $1\times p$ or a $p\times 1$ rational matrix.
A set of rational vectors $V(\lambda) :=\{v_1(\lambda),\ldots, v_k(\lambda) \}$ is said to be \emph{linearly dependent} over the field  $\mathds{R}(\lambda)$ if there exists $k$ rational functions $\gamma_i(\lambda) \in \mathds{R}(\lambda)$, $i = 1, \ldots, k$, with $\gamma_i(\lambda) \not = 0$ for at least one $i$, such that, the linear combination
\be\label{lindep} \sum_{i=1}^k \gamma_i(\lambda)v_i(\lambda) = 0 \, .\ee
The set of vectors $V(\lambda)$ is \emph{linearly independent} over $\mathds{R}(\lambda)$ if (\ref{lindep}) implies that $\gamma_i(\lambda) = 0$ for each $i = 1, \ldots, k$.
It is important to note, that a linearly dependent set $V(\lambda)$ over  $\mathds{R}(\lambda)$, can be still linearly independent over another field (e.g., the field of reals if $\gamma_i \in \mathds{R}$).

The \emph{normal rank} of a rational matrix $G(\lambda) \in \mathds{R}(\lambda)^{p\times m}$, which we also denote by $\rank G(\lambda)$, is the maximal number of linearly independent rows (or columns) over the field of rational functions $\mathds{R}(\lambda)$.
\index{transfer function matrix (TFM)!normal rank}%
It can be shown that the normal rank of $G(\lambda)$ is the maximally possible rank of the complex matrix $G(\lambda)$ for all values of $\lambda \in \mathds{C}$ such that $G(\lambda)$ has finite norm. This interpretation provides a simple way to determine the normal rank as the maximum of the rank of $G(\lambda)$ for a few random values of the frequency variable $\lambda$.

It is easy to check that the set of $p$-dimensional rational vectors $\mathds{R}(\lambda)^p$ forms a vector space with scalars defined over $\mathds{R}(\lambda)$. If $\mathcal{V}(\lambda) \subset \mathds{R}(\lambda)^p$ is a vector space, then there exists a set of linearly independent rational vectors $V(\lambda) :=\{v_1(\lambda),\ldots, v_{n_b}(\lambda) \} \subset \mathcal{V}(\lambda)$ such that any  vector in $\mathcal{V}(\lambda)$ is a linear combination of the vectors in $V(\lambda)$ (equivalently, any set of $n_b+1$ vectors,  including an arbitrary vector from $\mathcal{V}(\lambda)$ and the $n_b$ vectors in $V(\lambda)$, is linearly dependent). The set $V(\lambda)$ is called a \emph{basis} of the vector space $\mathcal{V}(\lambda)$ and $n_b$ is the dimension of $\mathcal{V}(\lambda)$. With a slight abuse of notation, we denote $V(\lambda)$ the matrix formed of the $n_b$ stacked row vectors
\[ V(\lambda) = \ba{ccc} v_1(\lambda)\\ \vdots \\ v_{n_b}(\lambda) \ea \]
or the $n_b$ concatenated column vectors
\[ V(\lambda) = \ba{ccc} v_1(\lambda)& \cdots & v_{n_b}(\lambda) \ea \, . \]
Interestingly, as basis vectors we can always use polynomial vectors since we can replace each vector $v_i(\lambda)$ of a rational basis, by $v_i(\lambda)$  multiplied with the least common multiple of the denominators of the components of $v_i(\lambda)$.

\index{polynomial basis!minimal}
The use of polynomial bases allows to define the main concepts related to so-called minimal bases.
Let $n_i$ be the degree of the $i$-th polynomial vector $v_i(\lambda)$ of a polynomial basis  $V(\lambda)$ (i.e., $n_i$ is the largest degree of the components of $v_i(\lambda)$). Then, $n := \sum_{i=1}^{n_b}n_{i}$ is, by definition, the \emph{degree} of the polynomial basis $V(\lambda)$. A \emph{minimal polynomial basis} of $\mathcal{V}(\lambda)$ is one for which $n$ has the least achievable value. For a minimal polynomial basis, $n_{i}$, for $i = 1, \ldots, n_b$, are called the \emph{row} or \emph{column} \emph{minimal indices} (also known as \emph{left} or \emph{right} \emph{Kronecker indices}, respectively). Two important examples are the left and right nullspace bases of a rational matrix, which are shortly discussed below.

\index{nullspace!basis}
Let $G(\lambda)$ be a $p\times m$ rational matrix $G(\lambda)$ whose normal rank is $r < \min(p,m)$. It is easy to show that the set
\[ \ker_L(G(\lambda)) := \{ v(\lambda) \in \mathds{R}(\lambda)^{1\times p}\; |\; v(\lambda)G(\lambda) = 0 \} \]
is a linear space called the \index{nullspace!left}%
\emph{left nullspace} of $G(\lambda)$. Analogously,
\[ \ker_R(G(\lambda)) := \{ v(\lambda) \in \mathds{R}(\lambda)^{m\times 1}\; |\; G(\lambda)v(\lambda) = 0 \} \]
is a linear space called the \index{nullspace!right}%
\emph{right nullspace} of $G(\lambda)$.
The dimension of $\ker_L(G(\lambda))$ $\left[\,\ker_R(G(\lambda))\,\right]$ is $p-r$ $[m-r]$, and therefore, there exist  $p-r$ $[m-r]$ linearly independent polynomial vectors which form a minimal polynomial basis for  $\ker_L(G(\lambda))$ $\left[\,\ker_R(G(\lambda))\,\right]$.  Let $n_{l,i}$ $[\,n_{r,i}\,]$ be the \emph{left} [\emph{right}] \emph{minimal indices} and let $n_l := \sum_{i=1}^{p-r}n_{l,i}$ $\left[\,n_r := \sum_{i=1}^{m-r}n_{r,i}\,\right]$ be the least possible degree of the left [right] nullspace basis.  The least left and right degrees $n_l$ and $n_r$, respectively, play an important role in relating the main structural elements of rational matrices (see the discussion of poles and zeros in Section~\ref{app:tfm_polzer}).

Some properties of minimal polynomial bases are summarized below for the left nullspace bases. Similar results can be given for the right nullspace bases.
\index{nullspace!basis!minimal polynomial, left}
\index{minimal basis!polynomial}
\begin{lemma}\label{L-polbasis} Let $G(\lambda)$ be a $p\times m$ rational matrix of normal rank $r$ and let $N_l(\lambda)$ be a $(p-r)\times p$ minimal polynomial basis of the left nullspace $\ker_L(G(\lambda))$ with left minimal (or left Kronecker) indices $n_{l,i}$, $i = 1, \ldots, p-r$. Then the following holds:
\begin{enumerate}
\item The left minimal indices are {unique} up to permutations (i.e.,
if $\widetilde N_l(\lambda)$ is another minimal polynomial basis, then
$N_l(\lambda)$ and $\widetilde N_l(\lambda)$ have the same left minimal
indices).
\item $N_l(\lambda)$ is {irreducible}, having
 full row rank for all $\lambda \in \mathds{C}$ (i.e., $N_l(\lambda)$ has no
finite  zeros, see Section~\ref{app:tfm_polzer}).\index{polynomial basis!irreducible}
\item $N_l(\lambda)$ is {row
reduced} (i.e., the {leading row coefficient matrix} formed
from the coefficients of the highest row degrees has full row
rank.) \index{polynomial basis!row reduced}
\end{enumerate}
\end{lemma}

An irreducible and row-reduced polynomial basis is actually a minimal polynomial basis. Irreducibility implies that any polynomial vector $v(\lambda)$ in the space spanned by the rows of $N_l(\lambda)$, can be expressed as a linear combination of basis vectors $v(\lambda) = \phi(\lambda)N_l(\lambda)$, with $\phi(\lambda)$ being a polynomial vector. In particular, assuming the  rows of $N_l(\lambda)$ are ordered such that $n_{l,i} \leq n_{l,i+1}$ for $i = 1, \ldots, p-r-1$, then for any $v(\lambda)$ of degree $n_{l,i}$, the corresponding $\phi(\lambda)$ has as its $j$-th element a polynomial of degree at most $n_{l,i}-n_{l,j}$ for $j = 1, \ldots, i$, and the rest of components are zero. This property allows to easily generate left polynomial annihilators of given degrees.

\index{rational basis!minimal proper}
\index{rational basis!simple minimal proper}
\index{minimal basis!proper rational}
\index{minimal basis!simple, proper rational}
Minimal polynomial bases allow
to easily build \emph{simple minimal proper rational bases}. These are
proper rational bases having the property that the sum of degrees of the rows [columns] is equal to the least left [right] degree of a minimal polynomial basis $n_l$ [$n_r$]. A
simple minimal proper rational left nullspace basis with arbitrary poles can be
constructed by forming $\widetilde N_l(\lambda) := M(\lambda)N_l(\lambda)$ with
\be\label{M} M(\lambda) = \diag
\big(1/d_1(\lambda), \ldots, 1/d_{p-r}(\lambda)
\big), \ee
where $d_i(\lambda)$ is a polynomial of degree $n_{l,i}$ with arbitrary roots.
Since $N_l(\lambda)$ is row reduced, it follows that $D_l :=
\lim_{\lambda\rightarrow\infty} \widetilde N_l(\lambda)$ has full
row rank (i.e., $\widetilde N_l(\lambda)$ has no infinite zeros, see Section~\ref{app:tfm_polzer}).
A simple minimal proper rational left nullspace basis allows  to generate, in a straightforward manner, left rational annihilators of given McMillan degrees by forming linear combinations of the basis vectors in $\widetilde N_l(\lambda)$ using specially chosen  rational vectors $\phi(\lambda)$ (see Section~\ref{app:tfm_polzer} for the definition of the McMillan degree of a rational matrix). The concept of {simple minimal proper rational basis} has been introduced in
\cite{Vard84} as the natural counterpart of a minimal polynomial basis.

\index{descriptor system!minimal nullspace basis}
\index{nullspace!basis!minimal rational, left}
\index{nullspace!basis!minimal rational, right}

Let $G(\lambda)$ be a $p\times m$ rational matrix of normal rank $r$ and let $(A-\lambda E,B,C,D)$ be an irreducible descriptor realization of $G(\lambda)$.
To determine a basis $N_l(\lambda)$ of the left nullspace of $G(\lambda)$, we can exploit the
simple fact (see \cite[Theorem 2]{Verg79}) that $N_l(\lambda)$ is a left minimal nullspace basis of
$G(\lambda)$ if and only if, for a suitable $M_l(\lambda)$,
\be\label{yl1} Y_l(\lambda) := [\, M_l(\lambda)\; N_l(\lambda)\,] \ee
 is a left minimal
nullspace basis of the associated system matrix pencil
\be\label{systempencil1} S(\lambda) := \ba{cc} A-\lambda E & B\\ C & D \ea .\ee
Thus, to compute $N_l(\lambda)$  we can determine first a
left minimal nullspace basis $Y_l(\lambda)$ for $S(\lambda)$ and then
$N_l(\lambda)$ results as
\[ N_l(\lambda) = Y_l(\lambda)\ba{c} 0 \\ I_{p} \ea . \]
By duality, if $Y_r(\lambda)$ is a right minimal nullspace basis for $S(\lambda)$, then a right minimal nullspace basis of $G(\lambda)$ is given by
\[ N_r(\lambda) = \ba{cc} 0 & I_m \ea Y_r(\lambda). \]

The Kronecker canonical form (\ref{Kronecker}) of the system matrix pencil $S(\lambda)$ in (\ref{systempencil1}) allows to easily determine left and right nullspace bases of $G(\lambda)$.
A numerically reliable computational approach to compute proper minimal null\-space bases of rational matrices is described in \cite{Varg03b} and relies on using the Kronecker-like form (\ref{Kronecker-like}) of the system matrix pencil, which can be determined by using exclusively orthogonal similarity transformations. The computation of simple minimal proper nullspace bases is described in \cite{Varg11e}.
\index{nullspace!basis!minimal proper, right}%
\index{nullspace!basis!simple minimal proper, right}%
\index{nullspace!basis!minimal proper, left}%
\index{nullspace!basis!simple minimal proper, left}%

\subsection{Poles and Zeros} \label{app:tfm_polzer}
\index{transfer function!poles}
\index{transfer function!zeros}
\index{transfer function!poles!finite}
\index{transfer function!poles!infinite}
\index{transfer function!zeros!finite}
\index{transfer function!zeros!infinite}
For a scalar rational function $g(\lambda) \in \mathds{R}(\lambda)$, the values of $\lambda$ for which $g(\lambda)$ is infinite are called the \emph{poles} of $g(\lambda)$. If $g(\lambda) = \alpha(\lambda)/\beta(\lambda)$ has the form in (\ref{glambda}), then the $n$ roots $\lambda^p_i$, $i = 1, \ldots, n$, of $\beta(\lambda)$ are the \emph{finite poles} of $g(\lambda)$, while if $m < n$, there are also, by convention,  $n-m$ \emph{infinite poles}.  The values of $\lambda$ for which $g(\lambda) = 0$ are called the \emph{zeros} of $g(\lambda)$. The $m$ roots $\lambda^z_i$, $i = 1, \ldots, m$, of $\alpha(\lambda)$ are the \emph{finite zeros} of $g(\lambda)$, while if $n < m$, there are also, by convention,  $m-n$ \emph{infinite zeros}. It follows that the number of poles is equal to the number of zeros and is equal to $\max(m,n)$, the degree of $g(\lambda)$.
The rational function $g(\lambda)$ in (\ref{glambda}) can be equivalently expressed in terms of its finite poles and zeros in the factorized form
\be\label{gkpz} g(\lambda) = k_g\frac{\prod_{i=1}^m(\lambda-\lambda^z_i)}{\prod_{i=1}^n(\lambda-\lambda^p_i)} \, ,\ee
where $k_g = a_m/b_n$. If $g(\lambda)$ is the transfer function of a SISO LTI system, then we will always assume that $g(\lambda)$ is in a minimal cancelled (irreducible) form, that is, the polynomials $\alpha(\lambda)$ and $\beta(\lambda)$ in (\ref{glambda}) have 1 as greatest common divisor. Equivalently, the two polynomials have no common roots, and therefore no pole-zero cancellation may occur in (\ref{gkpz}). Two such polynomials are called \emph{coprime}.

\index{transfer function!exponential stability}
\index{transfer function!stable}
\index{transfer function!anti-stable}
\index{transfer function!poles!stable}
\index{transfer function!poles!unstable}
\index{transfer function!minimum-phase}
\index{transfer function!zeros!minimum-phase}
\index{transfer function!zeros!non-minimum-phase}
\index{transfer function!poles!stability degree}

In studying the stability of systems, the poles  play a primordial role. Their real parts, in the case of a continuous-time system, or moduli, in the case of a discrete-time system, determine the asymptotic (exponential) decay or divergence speed of the system output.  A SISO LTI system with the transfer function $g(\lambda)$ is \emph{exponentially stable} (or equivalently $g(\lambda)$ is stable) if $g(\lambda)$ is proper and has all poles in the appropriate \emph{stability domain} $\mathds{C}_s$. The system is \emph{unstable} if it has at least one pole outside of the stability domain and \emph{anti-stable} if all poles lie outside of the stability domain. Poles inside the stability domain are called \emph{stable poles}, while those outside of the stability domain are called \emph{unstable poles}.
For continuous-time systems the stability domain is the open left half complex plane $\mathds{C}_s = \{ s \in \mathds{C} : \Re(s) < 0 \}$, while for discrete-time systems the stability domain is the open unit disk $\mathds{C}_s = \{ z \in \mathds{C} : |z| < 1 \}$. We denote by $\partial \mathds{C}_s$ the boundary of the stability domain. For continuous-time systems, the boundary of the stability domain is the extended imaginary axis (i.e., including the point at infinity)  $\partial \mathds{C}_s = \{\infty \} \cup \{ s \in \mathds{C} : \Re(s) = 0 \}$, while for discrete-time systems the boundary of the stability domain is the unit circle  $\partial \mathds{C}_s = \{ z \in \mathds{C} : |z| = 1 \}$. We denote
$\overline{\mathds{C}}_s = \mathds{C}_s \cup \partial \mathds{C}_s$ the closure of the stability domain.  The \emph{instability domain} of poles we denote by $\overline{\mathds{C}}_u$ and is the complement of $\mathds{C}_s$ in $\mathds{C}$, $\overline{\mathds{C}}_u = \mathds{C} \setminus \mathds{C}_s$. It is also the closure of the set denoted by ${\mathds{C}}_u$, which for a continuous-time system is the open right-half plane ${\mathds{C}}_u = \{ s \in \mathds{C} : \Re(s) > 0 \}$, while for a discrete-time systems is the exterior of the unit circle ${\mathds{C}}_u = \{ z \in \mathds{C} : |z| > 1 \}$. The \emph{stability degree} of poles is defined as the largest real part of the poles in the continuous-time case, or the largest absolute value of the poles in the discrete-time case.

Let  $\mathds{R}_s(\lambda)$ be the set of proper stable transfer functions having poles only in $\mathds{C}_s$.
A transfer function $g(\lambda) \in \mathds{R}_s(\lambda)$ having only zeros in $\mathds{C}_s$ is called \emph{minimum-phase}. Otherwise it is called \emph{non-minimum-phase}. The zeros of $g(\lambda)$ in $\mathds{C}_s$ are called \emph{minimum-phase zeros}, while those outside $\mathds{C}_s$ are called \emph{non-minimum-phase zeros}.

There are no straightforward generalizations of poles and zeros of scalar rational functions to the rational matrix case. Instrumental for a rigorous definition are two canonical forms: the \emph{Smith form} for polynomial matrices and the \emph{Smith-McMillan form} for rational matrices. For polynomial matrices we have the following important result.

\index{polynomial matrix!Smith form}
\index{polynomial matrix!normal rank}
\index{polynomial matrix!invariant polynomials}
\begin{lemma}\label{L-SF} Let $P(\lambda) \in \mathds{R}[\lambda]^{p\times m}$ be any polynomial matrix. Then, there exist unimodular matrices $U(\lambda) \in \mathds{R}[\lambda]^{p\times p}$ and $V(\lambda) \in \mathds{R}[\lambda]^{m\times m}$ such that
\be\label{SF} U(\lambda)P(\lambda)V(\lambda) = S(\lambda) :=
\ba{cccccc} \alpha_1(\lambda) & 0 & \cdots & 0 & \cdots & 0 \\
0 & \alpha_2(\lambda) & \cdots & 0 & \cdots & 0 \\
\vdots & \vdots & \ddots & \vdots & & \vdots \\
0 & 0 & \cdots & \alpha_r(\lambda) & \cdots & 0 \\
\vdots & \vdots &  & \vdots & & \vdots \\
0 & 0 & \cdots & 0 & \cdots & 0\ea \ee
and $\alpha_i(\lambda)$ divides $\alpha_{i+1}(\lambda)$ for $i = 1, \ldots, r-1$.
\end{lemma}
The polynomial matrix $S(\lambda)$ is called the \emph{Smith form} of $P(\lambda)$ and $r$ is the normal rank of $P(\lambda)$. The diagonal elements $\alpha_1(\lambda), \ldots , \alpha_r(\lambda)$ are called the \emph{invariant polynomials} of $P(\lambda)$. The roots of the polynomials $\alpha_i(\lambda)$, for $i = 1, \ldots, r$,
are called the \emph{finite zeros} of the polynomial matrix $P(\lambda)$.
\index{polynomial matrix!zeros}%
\index{polynomial matrix!zeros!finite}%
To each distinct finite zero $\lambda_z$ of $P(\lambda)$, we can associate the  multiplicities $\sigma_i(\lambda_z) \ge 0$ of root $\lambda_z$ in each of the polynomials $\alpha_i(\lambda)$, for $i = 1, \ldots, r$. By convention, $\sigma_i(\lambda_z) = 0$ if $\lambda_z$ is not a root of $\alpha_i(\lambda)$. The divisibility properties of $\alpha_i(\lambda)$ imply that
\[ 0 \leq \sigma_1(\lambda_z) \leq \sigma_2(\lambda_z) \leq \cdots \leq \sigma_r(\lambda_z) \, .\]

Any rational matrix $G(\lambda)$ can be expressed as
\[ G(\lambda) = \frac{P(\lambda)}{d(\lambda)} ,\]
where $d(\lambda)$ is the monic least common multiple of the denominator polynomials of the entries of $G(\lambda)$, and $P(\lambda) := d(\lambda)G(\lambda)$ is a polynomial matrix. Then, we have the following straightforward  extension of Lemma~\ref{L-SF} to rational matrices.
\index{transfer function matrix (TFM)!Smith-McMillan form}
\begin{lemma}\label{L_SMF}
Let $G(\lambda) \in \mathds{R}(\lambda)^{p\times m}$ be any rational  matrix. Then, there exist unimodular matrices $U(\lambda) \in \mathds{R}[\lambda]^{p\times p}$ and $V(\lambda) \in \mathds{R}[\lambda]^{m\times m}$ such that
\be\label{SMF} U(\lambda)G(\lambda)V(\lambda) = H(\lambda) :=
\ba{cccccc} \frac{\alpha_1(\lambda)}{\beta_1(\lambda)} & 0 & \cdots & 0 & \cdots & 0 \\
0 & \frac{\alpha_2(\lambda)}{\beta_2(\lambda)} & \cdots & 0 & \cdots & 0 \\
\vdots & \vdots & \ddots & \vdots & & \vdots \\
0 & 0 & \cdots & \frac{\alpha_r(\lambda)}{\beta_r(\lambda)} & \cdots & 0 \\
\vdots & \vdots &  & \vdots & & \vdots \\
0 & 0 & \cdots & 0 & \cdots & 0\ea \, , \ee
with $\alpha_i(\lambda)$ and $\beta_i(\lambda)$ coprime for $i = 1, \ldots, r$ and $\alpha_{i}(\lambda)$ divides $\alpha_{i+1}(\lambda)$ and $\beta_{i+1}(\lambda)$ divides $\beta_{i}(\lambda)$ for $i = 1, \ldots, r-1$.
\end{lemma}

The rational matrix $H(\lambda)$ is called the \emph{Smith--McMillan form} of $G(\lambda)$ and $r$ is the normal rank of $G(\lambda)$. The Smith-McMillan form is a powerful conceptual tool which allows to define rigorously the notions of poles and zeros of MIMO LTI systems and to establish several basic factorization results of rational matrices.

The roots of the numerator polynomials $\alpha_i(\lambda)$, for $i = 1, \ldots, r$,
are called the \emph{finite zeros} of the rational  matrix $G(\lambda)$ and the roots of the denominator polynomials $\beta_i(\lambda)$, for $i = 1, \ldots, r$,
are called the \emph{finite poles} of the rational  matrix $G(\lambda)$.
To each finite $\lambda_z$, which is a zero or a pole of $G(\lambda)$ (or both), we can associate its  multiplicities $\{\sigma_1(\lambda_z), \ldots, \sigma_r(\lambda_z)\}$, where $\sigma_i(\lambda_z)$ is the multiplicity of $\lambda_z$ either as a pole or a zero of the ratio $\alpha_i(\lambda)/\beta_i(\lambda)$, for $i = 1, \ldots, r$. By convention, we use negative values for poles and positive values for zeros. The divisibility properties of $\alpha_i(\lambda)$  and $\beta_i(\lambda)$ imply that
\[ \sigma_1(\lambda_z) \leq \sigma_2(\lambda_z) \leq \cdots \leq \sigma_r(\lambda_z) \, .\]
The $r$-tuple of multiplicities $\{\sigma_1(\lambda_z), \ldots, \sigma_r(\lambda_z)\}$ completely characterizes the \emph{pole-zero structure} of $G(\lambda)$ in $\lambda_z$.

The relative degrees of  $\alpha_i(\lambda)/\beta_i(\lambda)$ do not provide the correct information on the multiplicity of infinite zeros and poles. This is because the used unimodular transformations may have poles and zeros at infinity. To overcome this, the multiplicity of zeros and poles at infinity are defined in terms of multiplicities of poles and zeros of $G(1/\lambda)$ in $\lambda_z = 0$. This allows to define the multiplicities of \emph{infinite zeros} of a polynomial matrix $P(\lambda)$  as the multiplicities of the null poles in the Smith-McMillan for of $P(1/\lambda)$. \index{polynomial matrix!zeros!infinite}%

The \emph{McMillan degree} of a rational matrix $G(\lambda)$, usually denoted by $\delta(G(\lambda))$, is the number $n_p$ of its poles, both finite and infinite, counting all multiplicities.
\index{transfer function matrix (TFM)!McMillan degree}%
\index{transfer function matrix (TFM)!poles}%
\index{transfer function matrix (TFM)!zeros}%
\index{transfer function matrix (TFM)!poles!infinite}%
\index{transfer function matrix (TFM)!zeros!finite}%
\index{transfer function matrix (TFM)!zeros!infinite}%
If $n_z$ is the number of zeros (finite and infinite, counting all multiplicities), then we have the following important structural relation for any rational matrix \cite{Verg79}
\be\label{pol-zer} n_p = n_z + n_l + n_r ,\ee
where $n_l$ and $n_r$ are the least degrees of the minimal polynomial bases for the left and right nullspaces of $G(\lambda)$, respectively.

For a given rational matrix $G(\lambda)$, any (finite or infinite) pole of its elements $g_{ij}(\lambda)$, is also a pole of $G(\lambda)$. Therefore, many notions related to poles of SISO LTI systems introduced previously can be extended in a straightforward manner to MIMO LTI systems. For example, the notion of properness of $G(\lambda)$ can be equivalently defined as the nonexistence of infinite poles in the elements of $G(\lambda)$ (i.e., $G(\lambda)$ has only finite poles).  The notion of \emph{exponential stability} of a LTI system with a proper TFM $G(\lambda)$ can be defined as the lack of unstable poles in all elements of $G(\lambda)$ (i.e., all poles of $G(\lambda)$ are in the stable domain $\mathds{C}_s$). A TFM $G(\lambda)$ with only stable poles  is called \emph{stable}. Otherwise, $G(\lambda)$ is called \emph{unstable}. A similar definition applies for the \emph{stability degree} of poles.
\index{transfer function matrix (TFM)!stable}
\index{transfer function matrix (TFM)!unstable}

The zeros of $G(z)$ are also called the \emph{transmission zeros} of the corresponding LTI system. A proper and stable $G(z)$ is \emph{minimum-phase} if all its finite zeros are stable. Otherwise, it is called \emph{non-minimum-phase}.
\index{transfer function matrix (TFM)!minimum-phase}
\index{transfer function matrix (TFM)!non-minimum-phase}

\index{linear matrix pencil!strict equivalence}
Consider the irreducible descriptor system $(A-\lambda E,B,C,D)$ with the corresponding TFM $G(\lambda) \in \mathds{R}(\lambda)^{p\times m}$. Two pencils play a fundamental role in defining the main structural elements of the rational matrix $G(\lambda)$. The regular \emph{pole pencil}
\be\label{polepencil} P(\lambda) := A-\lambda E \ee
characterizes the pole structure of $G(\lambda)$, exhibited by the Weierstrass canonical form of the pole pencil $P(\lambda)$.  The (singular)  system matrix pencil
\be\label{systempencil} S(\lambda) := \ba{cc} A-\lambda E & B\\ C & D \ea \ee
characterizes the  zero structure of $G(\lambda)$, as well as the right- and left-singular structures of $G(\lambda)$, which are exhibited by the Kronecker canonical form of the system matrix pencil $S(\lambda)$.

The main structural results for the rational TFM $G(\lambda)$ can be stated in terms of its  irreducible descriptor system realization $(A-\lambda E,B,C,D)$ of order $n$. The following facts rely on the Weierstrass canonical form (\ref{Weierstrass}) of the pole pencil $P(\lambda)$ in (\ref{polepencil})  (see Lemma \ref{L-WCF}) and the Kronecker canonical form (\ref{Kronecker}) of the system matrix pencil $S(\lambda)$ in (\ref{systempencil}) (see Lemma \ref{L-KCF}):
\begin{enumerate}
\item[1)] The \emph{finite poles} of $G(\lambda)$ are the finite eigenvalues of the pole pencil $P(\lambda)$ and are the eigenvalues (counting multiplicities) of the matrix  $J_f$ in the Weierstrass canonical form (\ref{Weierstrass}) of the pencil $P(\lambda)$.
\item[2)] The \emph{infinite poles} of $G(\lambda)$ have multiplicities defined by the multiplicities of the infinite eigenvalues of the pole pencil $P(\lambda)$ minus 1 and are the dimensions minus 1 of the nilpotent Jordan blocks in the matrix $J_\infty$ in the Weierstrass canonical form (\ref{Weierstrass}) of the pencil $P(\lambda)$.
\item[3)] The \emph{finite zeros} of $G(\lambda)$ are the $n_f$ finite eigenvalues (counting multiplicities) of the system matrix pencil $S(\lambda)$ and are the eigenvalues (counting multiplicities) of the matrix  $\widetilde J_f$ in the Kronecker canonical form (\ref{Kronecker}) of the pencil $S(\lambda)$.
\item[4)] The \emph{infinite zeros} of $G(\lambda)$ have multiplicities defined by the multiplicities of the infinite eigenvalues of the system matrix pencil $S(\lambda)$ minus 1 and are the dimensions minus 1 of the nilpotent Jordan blocks in the matrix $\widetilde J_\infty$ in the Kronecker canonical form (\ref{Kronecker}) of the pencil $S(\lambda)$.
\item[5)] The \emph{left minimal indices} of $G(\lambda)$  are pairwise equal to the left Kronecker indices of $S(\lambda)$ and are the row dimensions $\epsilon_i$ of the  blocks $L_{\epsilon_i}(\lambda)$ for $i = 1,\ldots, \nu_r$ in the Kronecker canonical form (\ref{Kronecker}) of the pencil $S(\lambda)$.
\item[6)] The \emph{right minimal indices} of $G(\lambda)$    are pairwise equal to the right Kronecker indices of $S(\lambda)$  and are the column dimensions $\eta_i$ of the  blocks $L^T_{\eta_i}(\lambda)$ for $i = 1,\ldots, \nu_l$ in the Kronecker canonical form (\ref{Kronecker}) of the pencil $S(\lambda)$.
\item[7)] The \emph{normal rank} of $G(\lambda)$ is   $r = \rank S(\lambda)-n = n_r+\tilde n_f+\tilde n_\infty + n_l -n$.
\end{enumerate}
\index{descriptor system!poles}
\index{descriptor system!zeros}
\index{descriptor system!normal rank}

\index{descriptor system!proper}
\index{descriptor system!improper}
\index{descriptor system!polynomial}
These facts allow to formulate simple conditions to characterize some pole-zero related properties, such as, properness, stability or minimum-phase of an irreducible descriptor system $(A-\lambda E,B,C,D)$ in terms of the eigenvalues of the pole and system matrix pencils. The descriptor system $(A-\lambda E,B,C,D)$ is \emph{proper} if all infinite eigenvalues of the regular pencil $A-\lambda E$ are simple (i.e., the system has no infinite poles). It is straightforward to show using the Weierstrass canonical form of the pencil $A-\lambda E$, that any irreducible proper descriptor system can be always reduced to a minimal order descriptor system, with the descriptor matrix $E$ invertible, or even to a standard state-space representation with $E = I$. The irreducible descriptor system  $(A-\lambda E,B,C,D)$ is \emph{improper} if the regular pencil $A-\lambda E$ has at least one infinite eigenvalue which is not simple (i.e, has at least one infinite pole). A \emph{polynomial} descriptor system is one for which $A-\lambda E$ has only infinite eigenvalues of which at least one is not simple (i.e, has only infinite poles). The concept of stability involves naturally the properness of the system. The  irreducible descriptor system $(A-\lambda E,B,C,D)$ is \emph{exponentially stable} if it has only  finite poles and all poles belong to the stable region $\mathds{C}_s$ (the pencil $A-\lambda E$ still can have simple infinite eigenvalues).
\index{descriptor system!exponential stability}
The irreducible descriptor system $(A-\lambda E,B,C,D)$  is \emph{unstable} if it has at least one finite pole outside of the stability domain or at least one infinite pole. The finite poles (or finite eigenvalues) inside the stability domain are called \emph{stable poles} (\emph{stable eigenvalues}), while the poles lying  outside of the stability domain are called \emph{unstable poles}. The irreducible descriptor system $(A-\lambda E,B,C,D)$  is \emph{minimum-phase} if  it has only finite zeros and all finite zeros belong to the stable region $\mathds{C}_s$.

\index{descriptor system!controllable eigenvalue}
\index{descriptor system!uncontrollable eigenvalue}
\index{descriptor system!observable eigenvalue}
\index{descriptor system!unobservable eigenvalue}
\index{descriptor system!finite stabilizable}
\index{descriptor system!finite detectable}
To check the finite controllability condition $(i)$ and finite observability condition $(iii)$ of Theorem~\ref{T-desc-minreal}, it is sufficient to check that
\be\label{eig-controllable} \rank\ba{cc} A-\lambda_i E & B \ea = n \ee
and, respectively,
\be\label{eig-observable} \rank\ba{c} A-\lambda_i E \\ C \ea = n \ee
for all distinct finite eigenvalues $\lambda_i$  of  the regular pencil $A-\lambda E$. A finite eigenvalue $\lambda_i$ is \emph{controllable} if (\ref{eig-controllable}) is fulfilled, and \emph{uncontrollable} otherwise.  Similarly, a finite eigenvalue $\lambda_i$ is \emph{observable} if (\ref{eig-observable}) is fulfilled, and \emph{unobservable} otherwise. If the rank conditions (\ref{eig-controllable}) are fulfilled for all $\lambda_i \in \overline{\mathds{C}}_u$ we call the descriptor system $(A-\lambda E,B,C,D)$ (or equivalently the pair $(A-\lambda E,B)$)  \emph{finite stabilizable}. Finite stabilizability guarantees the existence of a state-feedback matrix $F \in \mathds{R}^{m\times n}$ such that all finite eigenvalues of $A+BF-\lambda E$ lie in $\mathds{C}_s$. If the rank conditions (\ref{eig-observable}) are fulfilled for all $\lambda_i \in \overline{\mathds{C}}_u$ we call the descriptor system $(A-\lambda E,B,C,D)$ (or equivalently the pair $(A-\lambda E,C)$) \emph{finite detectable}. Finite detectability guarantees the existence of an output-injection matrix $K \in \mathds{R}^{n\times p}$ such that all finite eigenvalues of $A+KC-\lambda E$ lie in $\mathds{C}_s$.

\index{descriptor system!strongly stabilizable}
\index{descriptor system!strongly detectable}
The notion of {strong stabilizability} is related to the existence of a state-feedback matrix $F$ such that all finite eigenvalues of $A+BF-\lambda E$ lie in $\mathds{C}_s$ and all infinite eigenvalues of $A+BF-\lambda E$ are simple. The necessary and sufficient conditions for the existence of such an $F$  is the \emph{strong stabilizability} of the pair $(A-\lambda E,B)$, that is: (1)  the finite stabilizability  of the pair $(A-\lambda E,B)$; and (2) $\rank [\, E \; AN_\infty \; B\,] = n$, where the columns of $N_\infty$ form a basis of $\ker(E)$. Similarly, {strong detectability} is related to the existence of an output-injection matrix $K$ such that all finite eigenvalues of $A+KC-\lambda E$ lie in $\mathds{C}_s$ and all infinite eigenvalues of $A+KC-\lambda E$ are simple. The necessary and sufficient conditions for the existence of such a $K$ is the \emph{strong detectability} of the pair $(A-\lambda E,C)$, that is: (1) the finite detectability of the pair $(A-\lambda E,C)$; and (2) $\rank [\, E^T \; A^T\!L_\infty  \;\, C^T\,] = n$, where the columns of $L_\infty$ for a basis of $\ker(E^T)$.

\subsection{Range and Coimage Space Bases} \label{app:tfm_range}
\index{range space!basis!minimal proper}%
\index{range space!basis!inner}%
\index{factorization!full rank}%
For any $p\times m$ real rational matrix $G(\lambda)$ of normal rank $r$, there exists a \emph{full rank factorization} of $G(\lambda)$ of the form
\be\label{full-rank-fac} G(\lambda) = R(\lambda)X(\lambda) , \ee
where $R(\lambda)$ is a $p\times r$ full column rank rational matrix and $X(\lambda)$ is a $r\times m$ full row rank rational matrix. This factorization generalizes the full-rank factorization of constant matrices, and, similarly to the constant case, it is not unique. Indeed, for any $r\times r$ invertible rational matrix $M(\lambda)$, $G(\lambda) = \widetilde R(\lambda)\widetilde X(\lambda)$, with $\widetilde R(\lambda) = R(\lambda)M(\lambda)$ and $\widetilde X = M^{-1}(\lambda)X(\lambda)$, is also a full rank factorization of $G(\lambda)$.

Using (\ref{full-rank-fac}), it is straightforward to show  that $G(\lambda)$ and $R(\lambda)$ have the same range space over the rational functions
\[ \mathcal{R}(G(\lambda)) = \mathcal{R}(R(\lambda)).  \]
For this reason, with a little abuse of language, we will call $R(\lambda)$ the range (or image) matrix of $G(\lambda)$ (or simply the range of $G(\lambda)$). Since $R(\lambda)$ has normal rank $r$, its columns form a set of $r$ basis vectors of $\mathcal{R}(G(\lambda))$.

The poles of $R(\lambda)$ can be chosen (almost) arbitrary, by replacing $R(\lambda)$ with $\widetilde R(\lambda)$, the numerator factor of right coprime factorization $R(\lambda) = \widetilde R(\lambda)M^{-1}(\lambda)$, where $\widetilde R(\lambda)$ and $M(\lambda)$ have only poles in a given complex domain $\mathds{C}_g$.
From the full-rank factorization  (\ref{full-rank-fac}) results that each zero of  $G(\lambda)$ is either a zero of $R(\lambda)$ or of $X(\lambda)$. Therefore, it is possible to construct full-rank factorizations with $R(\lambda)$ having as zeros a symmetric subset of zeros of $G(\lambda)$. A minimal McMillan degree basis results if $R(\lambda)$ has no zeros. Of special interest in some applications are bases which are inner, such that $R(\lambda)$ is stable and satisfies
$R^\sim(\lambda)R(\lambda) = I_r$. Applications of such bases are the computation of inner-outer factorizations (see Section \ref{app_IOF}) and of the normalized coprime factorizations (see Section \ref{appsec:fact}). Methods to compute various full rank factorizations are discussed in \cite{Varg17f}, based on techniques developed in \cite{Oara00} and \cite{Oara05}.

\index{coimage space!basis!minimal proper}%
Similarly, for any $p\times m$ real rational matrix $G(\lambda)$ of normal rank $r$, there exists a \emph{full rank factorization} of $G(\lambda)$ of the form
\be\label{full-rank-fac-alt} G(\lambda) = X(\lambda)R(\lambda) , \ee
where $R(\lambda)$ is a $r\times m$ full row rank rational matrix and $X(\lambda)$ is a $p\times r$ full column rank rational matrix. \index{factorization!full rank}%
Using (\ref{full-rank-fac-alt}), it is straightforward to show  that $G(\lambda)$ and $R(\lambda)$ have the same coimage space (or row space) over the rational functions, which can be also expressed as
\[ \mathcal{R}(G^T(\lambda)) = \mathcal{R}(R^T(\lambda)).  \]
Since $R(\lambda)$ has normal rank $r$, its rows form a set of $r$ basis vectors of the row space of  $G(\lambda)$ (i.e., the coimage of $G(\lambda)$). As in the case of the range space bases, the coimage space basis $R(\lambda)$ of $G(\lambda)$ can be determined to have (almost) arbitrarily chosen poles and the zeros of $R(\lambda)$ include an arbitrary symmetric subset of zeros of $G(\lambda)$. Moreover, $R(\lambda)$ can be chosen stable, without zeros and coinner, satisfying $R(\lambda)R^\sim(\lambda) = I_r$.%
\index{coimage space!basis!coinner}%

\subsection{Additive Decompositions} \label{app:tfm-add}
\index{transfer function matrix (TFM)!additive decomposition}

Let $G(\lambda)$ be the transfer function matrix of a LTI system and let $\mathds{C}_g$ be a domain of interest of the complex plane $\mathds{C}$ for the poles of $G(\lambda)$ (e.g., a certain stability domain). Define
$\mathds{C}_b := \mathds{C}\setminus \mathds{C}_g$, the complement of $\mathds{C}_g$ in $\mathds{C}$ .
Since $\mathds{C}_g$ and $\mathds{C}_b$ are disjoint, each pole of any element $g_{ij}(\lambda)$ of $G(\lambda)$ lies either in $\mathds{C}_g$ or in $\mathds{C}_b$. Therefore, using the well-known partial fraction decomposition results of rational functions,
$G(\lambda)$ can be additively decomposed as
\be\label{tfm_add}  G(\lambda) = G_1(\lambda) + G_2(\lambda) ,\ee
where $G_1(\lambda)$ has only poles in $\mathds{C}_g$, while $G_2(\lambda)$ has only poles in $\mathds{C}_b$. For such a decomposition of $G(\lambda)$ we always have that
\[ \delta\left(G(\lambda)\right) = \delta\left(G_1(\lambda)\right)+\delta\left(G_2(\lambda)\right) \, .\]

For example, if $\mathds{C}_g = \mathds{C}\setminus \{\infty\}$ and $\mathds{C}_b = \{\infty\}$, then (\ref{tfm_add}) represents the additive decomposition of a possibly  improper rational matrix  as the sum of its proper and  polynomial parts. This decomposition, in general, is not unique, because an arbitrary constant term can be always added to one term and subtracted from the other one.  Another frequently used decomposition is the stable-unstable decomposition of proper rational matrices, when
$\mathds{C}_g = \mathds{C}_s$ (stability region) and $\mathds{C}_b = \mathds{C}_u$ (instability region).

\index{descriptor system!additive decomposition}
Let $G(\lambda) = (A-\lambda E,B,C,D)$ be  a descriptor system representation of $G(\lambda)$. Using a general similarity transformation using two invertible matrices $Q$ and $Z$, we can determine an equivalent representation of
$G(\lambda)$ with partitioned system matrices of the form
\be\label{desc-specdec} G(\lambda) =  \ba{c|c} QAZ-\lambda QEZ & QB \\ \hline \\[-3mm] CZ & D \ea
= \ba{cc|c}  A_1 -\lambda E_1 & 0 &  B_{1} \\ 0 &  A_2 -\lambda E_2 &  B_{2} \\ \hline \\[-4mm]  C_1 &  C_2 & D \ea \, ,\ee
where $\Lambda(A_1 -\lambda E_1) \subset \mathds{C}_1$ and $\Lambda(A_2 -\lambda E_2) \subset \mathds{C}_2$.
It follows that $G(\lambda)$
can be additively decomposed as
\be\label{dss_add}  G(\lambda) = G_1(\lambda) + G_2(\lambda) ,\ee
where
\be\label{dss_addterms} G_1(\lambda) = \ba{c|c}  A_1 -\lambda E_1 & B_{1} \\ \hline \\[-4mm]  C_1 &  D \ea, \quad G_2(\lambda) = \ba{c|c}  A_2 -\lambda E_2 & B_{2} \\ \hline \\[-4mm]  C_2 &  0 \ea \, ,\ee
and $G_1(\lambda)$ has only poles in $\mathds{C}_g$, while $G_2(\lambda)$ has only poles in $\mathds{C}_b$.
For the computation of additive spectral decompositions, a numerically reliable procedure is described in \cite{Kags89}.

\subsection{Coprime Factorizations} \label{appsec:fact}
\index{factorization!fractional}
Consider a disjunct partition of the complex plane $\mathds{C}$ as
\be\label{Cgoodbad}  \mathds{C} = \mathds{C}_g \cup \mathds{C}_b, \quad \mathds{C}_g \cap \mathds{C}_b = \emptyset \, ,\ee
where both $\mathds{C}_g$ and $\mathds{C}_b$ are symmetrically located with respect to the real axis, and  such that $\mathds{C}_g$ has at least one point on the real axis. Any rational matrix $G(\lambda)$ can be expressed in a left fractional form
\be\label{lf} G(\lambda) = M^{-1}(\lambda)N(\lambda) \, , \ee
or in a right fractional form
\be\label{rf} G(\lambda) = N(\lambda)M^{-1}(\lambda) \, ,\ee
where both the denominator factor $M(\lambda)$ and the numerator factor $N(\lambda)$ have only poles in $\mathds{C}_g$. These fractional factorizations over a ``good'' domain of poles $\mathds{C}_g$ are important in various observer, fault detection filter, or controller synthesis methods, because they allow to achieve the placement of all poles of a TFM $G(\lambda)$ in the domain $\mathds{C}_g$ simply, by a premultiplication or postmultiplication of $G(\lambda)$ with a suitable $M(\lambda)$.

\index{factorization!left coprime (LCF)}%
\index{factorization!right coprime (RCF)}%
Of special interest are the so-called coprime factorizations, where the factors satisfy additional conditions.  A fractional representation of the form (\ref{lf}) is a \emph{left coprime factorization} (LCF) of $G(\lambda)$ with respect to $\mathds{C}_g$, if there exist $U(\lambda)$ and $V(\lambda)$ with poles only in $\mathds{C}_g$ which satisfy the \emph{Bezout identity}
\[ M(\lambda)U(\lambda)+N(\lambda)V(\lambda) = I \, . \]
A fractional representation of the form (\ref{rf}) is a \emph{right coprime factorization} (RCF) of $G(\lambda)$ with respect to $\mathds{C}_g$, if there exist $U(\lambda)$ and $V(\lambda)$ with poles only in $\mathds{C}_g$ which satisfy
\[ U(\lambda)M(\lambda)+V(\lambda)N(\lambda) = I \, .\]

\index{factorization!left coprime (LCF)!minimum-degree denominator}
\index{factorization!right coprime (RCF)!minimum-degree denominator}
An important class of coprime factorizations is the class of coprime factorizations with minimum-degree denominators. Recall that $\delta\left(G(\lambda)\right)$, the McMillan degree of  $G(\lambda)$,  is defined as the number of poles of $G(\lambda)$, both finite and infinite, counting all multiplicities. It follows that for any $G(\lambda)$ we have $\delta\left(G(\lambda)\right) = n_g + n_b$,
where $n_g$ and $n_b$ are the number of poles of $G(\lambda)$ in $\mathds{C}_g$ and $\mathds{C}_b$, respectively.
The denominator factor $M(\lambda)$ has the minimum-degree property if $\delta\left(M(\lambda)\right) =  n_b$.

\index{factorization!left coprime (LCF)!with inner denominator}
\index{factorization!right coprime (RCF)!with inner denominator}
\index{transfer function matrix (TFM)!conjugate}
\index{transfer function matrix (TFM)!inner}
\index{transfer function matrix (TFM)!coinner}
A square TFM $G(\lambda)$ is \emph{all-pass} if $G^\sim (\lambda)G(\lambda) = I$.
If $G(\lambda)$ is a stable TFM and satisfies $G^\sim (\lambda)G(\lambda) = I$ then it is called an \emph{inner} TFM, while if it satisfies $G(\lambda)G^\sim (\lambda) = I$ it is called a \emph{coinner} TFM. Note that an inner or coinner TFM must not be square, but must have full column rank (injective) or full row rank (surjective), respectively.
It is remarkable, that each proper TFM $G(\lambda)$ without poles on the boundary of stability domain $\partial \mathds{C}_s$ has a stable LCF  of the form (\ref{lf}) or a stable RCF of the form (\ref{rf}) with the denominator factor $M(\lambda)$ inner. As before, the minimum McMillan degree of  the inner denominator $M(\lambda)$ is equal to the number of the unstable poles of $G(\lambda)$.

\index{descriptor system!coprime factorization}

A special class of coprime factorizations consists of the so-called normalized coprime factorizations. For an arbitrary $p\times m$ rational matrix $G(\lambda)$, the
 \emph{normalized left coprime factorization} of $G(\lambda)$ is
\be\label{nlcf} G(\lambda) = M^{-1}(\lambda)N(\lambda), \ee
where $N(\lambda)$ and $M(\lambda)$ are stable TFMs and $[\, N(\lambda) \; M(\lambda)\,]$ is \emph{coinner}, that is
\be\label{nlcf1} N(\lambda)N^\sim(\lambda) + M(\lambda)M^\sim(\lambda) = I_p .\ee
\index{factorization!left coprime (LCF)!normalized}%
Similarly,
the \emph{normalized right coprime factorization} of $G(\lambda)$ is
\be\label{nrcf} G(\lambda) = N(\lambda)M^{-1}(\lambda), \ee
where $N(\lambda)$ and $M(\lambda)$ are stable TFMs and  $\left[\begin{smallmatrix} N(\lambda) \\ M(\lambda) \end{smallmatrix}\right]$ is inner, that is
\be\label{nrcf1} N^\sim(\lambda)N(\lambda) + M^\sim(\lambda)M(\lambda) = I_m .\ee
\index{factorization!right coprime (RCF)!normalized}%

For the computation of  coprime factorizations with minimum degree denominators, descriptor system representation based methods have been proposed  in \cite{Varg98a,Varg17d}, by using iterative pole dislocation techniques developed in the spirit of the approach described in \cite{Door90}. Alternative, non-iterative approaches to compute minimum degree coprime factorizations with inner denominators have been proposed in \cite{Oara99b,Oara05}. For the computation of normalized coprime factorizations computational methods have been proposed in \cite{Oara01} (see also \cite{Varg17f}).

\subsection{Inner-Outer and Spectral Factorizations} \label{app_IOF}
\index{factorization!inner--outer}
\index{factorization!inner--quasi-outer}
\index{factorization!co-outer--coinner}
\index{factorization!quasi-co-outer--coinner}
\index{transfer function matrix (TFM)!outer}
\index{transfer function matrix (TFM)!quasi-outer}
\index{transfer function matrix (TFM)!co-outer}
\index{transfer function matrix (TFM)!quasi-co-outer}
A proper and stable  TFM $G(\lambda)$ is \emph{outer} if it is minimum-phase and full row rank (surjective), and is \emph{co-outer} if it is  minimum-phase and full column rank (injective). A full row rank (full column rank) proper and stable  TFM $G(\lambda)$ is \emph{quasi-outer} (\emph{quasi-co-outer}) if it has only zeros in $\overline{\mathds{C}}_s$ (i.e, in the stability domain and its boundary). Any stable TFM $G(\lambda)$ without zeros in $\partial\mathds{C}_s$  has an \emph{inner--outer factorization}
\be\label{inoutMIMO} G(\lambda) = G_i(\lambda)G_o(\lambda) \, ,\ee
with $G_i(\lambda)$  inner  and $G_o(\lambda)$ outer.
Similarly, $G(\lambda)$ has a \emph{co-outer--coinner factorization}
\be\label{coinoutMIMO} G(\lambda) = G_{co}(\lambda)G_{ci}(\lambda) \, , \ee
with $G_{co}(\lambda)$ co-outer and $G_{ci}(\lambda)$ coinner.
Any stable TFM $G(\lambda)$ has an \emph{inner--quasi-outer factorization} of the form (\ref{inoutMIMO}), where $G_i(\lambda)$ is inner and $G_o(\lambda)$ is quasi-outer, and also has a \emph{quasi-co-outer--coinner factorization} of the form (\ref{coinoutMIMO}), where $G_{ci}(\lambda)$ is coinner and $G_{co}(\lambda)$ is quasi-co-outer.

In some applications, instead of the (compact) inner-outer factorization (\ref{inoutMIMO}), an alternative (extended) factorization with square inner factor is desirable. The \emph{extended inner-outer factorization} and \emph{extended inner--quasi-outer factorization} have the form
\index{factorization!inner--outer!extended}
\index{factorization!inner--quasi-outer!extended}
\be\label{einoutMIMO} G(\lambda) = \ba{cc} G_i(\lambda) & G_i^\bot(\lambda) \ea \ba{c} G_o(\lambda) \\ 0 \ea = U(\lambda) \ba{c} G_o(\lambda) \\ 0 \ea \, , \ee
where $G_i^\bot(\lambda)$ is the inner orthogonal complement of $G_i(\lambda)$ such that $U(\lambda) := \left[\,\!G_i(\lambda) \,\, G_i^\bot(\lambda)\,\!\right]$ is square and inner. Similarly, the  \emph{extended co-outer--coinner factorization} and \emph{extended quasi-co-outer--coinner factorization} have the form
\index{factorization!quasi-co-outer--coinner!extended}
\index{factorization!co-outer--coinner!extended}
\be\label{ecoinoutMIMO} G(\lambda) = \ba{cc} G_{co}(\lambda) & 0 \ea \ba{c} G_{ci}(\lambda) \\ G_{ci}^\bot(\lambda) \ea = \ba{cc} G_{co}(\lambda) & 0 \ea V(\lambda) \, ,\ee
where $G_{ci}^\bot(\lambda)$ is the coinner orthogonal complement of $G_i(\lambda)$ such that
 $V(\lambda) := \ba{c} G_{ci}(\lambda) \\ G_{ci}^\bot(\lambda) \ea$ is square and coinner (thus also inner).

The extended inner-outer factorization (\ref{einoutMIMO}) of a TFM $G(\lambda)$ can be interpreted as a generalization of the orthogonal QR-factorization of a real matrix. The inner factor $U(\lambda) = \left[\,\!G_i(\lambda) \,\, G_i^\bot(\lambda)\,\!\right]$ can be seen as the generalization of an orthogonal  matrix. Its role in an inner-outer factorization is twofold: to compress the given $G(\lambda)$ to a full row rank TFM $G_o(\lambda)$ and to dislocate all zeros of $G(\lambda)$ lying in $\mathds{C}_u$ into positions within $\mathds{C}_s$, which are symmetric (in a certain sense) with respect to $\partial\mathds{C}_s$. A factorization of $G(\lambda)$ as in (\ref{einoutMIMO}), where only the first aspect (i.e., the row compression) is addressed is called an
\emph{extended QR-like factorization} of the rational matrix $G(\lambda)$ and produces a compressed full row rank $G_o(\lambda)$, which contains all zeros of $G(\lambda)$. The similar factorization in (\ref{ecoinoutMIMO}) with $V(\lambda)$ inner, which only addresses the column compression aspect for $G(\lambda)$, is called an \emph{extended RQ-like factorization} of $G(\lambda)$. Applications of these factorizations are in computing the pseudo-inverse of a rational matrix \cite{Oara00} (see also Example \ref{ex:Oara-Varga2}).
\index{factorization!QR-like!extended}%
\index{factorization!RQ-like!extended}%

\index{factorization!spectral}
\index{factorization!spectral!stable left}
\index{factorization!spectral!minimum-phase left}
\index{factorization!spectral!stable right}
\index{factorization!spectral!minimum-phase right}
The outer factor $G_o(\lambda)$ of a TFM $G(\lambda)$ without zeros in $\partial\mathds{C}_s$ satisfies \[  G^\sim (\lambda)G(\lambda) = G_o^\sim (\lambda)G_o(\lambda) \]
and therefore, it is a solution of the \emph{minimum-phase right spectral factorization} problem.
Similarly, the co-outer factor $G_{co}(\lambda)$ of a TFM $G(\lambda)$ without zeros in $\partial\mathds{C}_s$ satisfies
\[  G(\lambda)G^\sim (\lambda) = G_{co}(\lambda)G_{co}^\sim (\lambda) \]
and therefore, it is a solution of the \emph{minimum-phase left spectral factorization} problem.

 Combining the LCF with inner denominator and the inner-outer factorization, we have for an arbitrary $G(\lambda)$, without poles and zeros on the boundary of the stability domain $\partial\mathds{C}_s$, that
\[ G(\lambda) = M_i^{-1}(\lambda)N(\lambda) = M_i^{-1}(\lambda)N_i(\lambda)N_o(\lambda) , \]
where $M_i(\lambda)$ and $N_i(\lambda)$ are inner and $N_o(\lambda)$ is outer. It follows that the outer factor $N_o(\lambda)$ is the solution of the \emph{stable minimum-phase right spectral factorization} problem
\[  G^\sim (\lambda)G(\lambda) = N_o^\sim (\lambda)N_o(\lambda) \, . \]
\index{factorization!spectral!stable minimum-phase right}%
Similarly, by combining the RCF with inner denominator and the co-outer--coinner factorization we obtain
\[ G(\lambda) = N(\lambda)M_i^{-1}(\lambda) = N_{co}(\lambda)N_{ci}(\lambda)M_i^{-1}(\lambda) \, ,\]
with $M_i(\lambda)$ inner, $N_{ci}(\lambda)$ coinner and $N_{co}(\lambda)$ co-outer. Then, $N_{co}(\lambda)$ is the solution of the \emph{stable minimum-phase left spectral factorization} problem
\[  G(\lambda)G^\sim (\lambda) = N_{co}(\lambda)N_{co}^\sim (\lambda) \, . \]
\index{factorization!spectral!stable minimum-phase left}

If $G(\lambda)$ has poles or zeros on the boundary of the stability domain $\partial \mathds{C}_s$, then we can still achieve the above factorizations by including all poles and zeros of $G(\lambda)$ in $\partial \mathds{C}_s$ in the resulting spectral factors $N_o(\lambda)$ or $N_{co}(\lambda)$.

\index{factorization!inner--outer}
\index{factorization!co-outer--coinner}
Assume that the stable proper TFM $G(\lambda)$ has an irreducible descriptor realization
\be\label{Glambda-IOF} G(\lambda) = \ba{c|c} A-\lambda E & B \\ \hline C & D \ea .\ee
For illustration of the inner-outer factorization approach, we only consider in detail the \emph{standard problem} with $E$ invertible and when $G(\lambda)$ has full column rank, in which case explicit formulas for both the inner and outer factors can be derived using the results presented in \cite{Zhou96}. Similar dual results can be obtained if $G(\lambda)$ has full row rank. The factorization procedures to compute inner--quasi-outer factorizations in the most general setting (i.e., for $G(\lambda)$ possibly improper with arbitrary normal rank and arbitrary zeros), are described in \cite{Oara00}
for continuous-time systems and in \cite{Oara05} for discrete-time systems. These methods, essentially reduce the factorization problems to the solution of standard problems, for which straightforward extensions of the results for standard systems in \cite{Zhou96} apply.

We have the following result for a continuous-time system:

\begin{theorem} \label{T:IOF}
 Let $G(s)$ be a  proper and  stable,  full column rank TFM without zeros in $\partial\mathds{C}_s$, with the descriptor system realization $G(s) = (A-sE,B,C,D)$ and $E$ invertible. Then, $G(s)$ has an inner--outer factorization $G(s) = G_i(s)G_o(s)$, with the particular realizations of the factors
\[ G_i(s) = \ba{c|c} A+BF-sE & BH^{-1} \\ \hline \\[-4mm] C+DF & DH^{-1} \ea , \qquad
 G_o(s) = \ba{c|c} A-sE & B \\ \hline \\[-4mm] -HF & H \ea ,\]
where $H$ is an invertible matrix satisfying $D^TD = H^TH$, $F$ is given by
\[ F = -(D^TD)^{-1}(B^TX_sE+D^TC) \, , \]
with $X_s \geq 0$ being the stabilizing solution  of the {generalized continuous-time algebraic Riccati equation} \emph{(GCARE)}
\index{matrix equation!generalized algebraic Riccati!continuous-time (GCARE)}
\[ A^TXE+E^TXA -(E^TXB+C^TD)(D^TD)^{-1}(B^TXE+D^TC)+C^TC = 0 .\]
\end{theorem}
The similar result for a discrete-time system is:
\begin{theorem} \label{T:IOFd}
 Let $G(z)$ be a  proper and  stable,  full column rank TFM without zeros in $\partial\mathds{C}_s$, with the descriptor system realization $G(z) = (A-zE,B,C,D)$ and $E$ invertible. Then,  $G(z)$ has an inner--outer factorization $G(z) = G_i(z)G_o(z)$, with the particular realizations of the factors
\[ G_i(z) = \ba{c|c} A+BF-zE & BH^{-1} \\ \hline \\[-4mm] C+DF & DH^{-1} \ea , \qquad
G_o(z) = \ba{c|c} A-zE & B \\ \hline \\[-4mm] -HF & H \ea ,\]
where $H$ is an invertible matrix satisfying $D^TD+B^TX_sB = H^TH$, $F$ is given by
\[ F = -(H^TH)^{-1}(B^TX_sA+D^TC) \, , \]
with $X_s \geq 0$ being the stabilizing solution  of the {generalized discrete-time algebraic Riccati equation} \emph{(GDARE)}
\index{matrix equation!generalized algebraic Riccati!discrete-time (GDARE)}
\[ A^TXA-E^TXE -(A^TXB+C^TD)(D^TD+B^TXB)^{-1}(B^TXA+D^TC)+C^TC = 0 .\]
\end{theorem}

When computing the \emph{extended inner-outer factorization} (\ref{einoutMIMO}),
the inner orthogonal complement $G_i^\bot(\lambda)$ of $G_i(\lambda)$  is also needed. In the continuous-time case, a descriptor realization of $G_i^\bot(\lambda)$ is given by
\[ G_i^\bot(s) = \ba{c|c} A+BF-sE & -X_s^\dag E^{-T}C^TD^\bot \\ \hline \\[-3.5mm]
C+DF & D^\bot \ea ,\]
where $D^\bot$ is an orthogonal complement chosen such that $\ba{cc} DH^{-1} & D^\bot \ea$ is square and orthogonal. In the discrete-time case we have
\[ G_i^\bot(z) = \ba{c|c} A+BF-zE & Y \\ \hline
C+DF & W \ea ,\]
where $Y$ and $W$  satisfy
\[ \begin{array}{lcl} A^TX_sY + C^TW & = & 0 ,\\
B^TX_sY + D^TW & = & 0 ,\\
W^TW + Y^TX_sY & = & I .
\end{array} \]

The similar results for the co-outer--coinner factorization (or the {extended co-outer--coinner factorization})  can be easily obtained by considering the inner--outer factorization (or its extended version) for the dual system with the TFM $G^T(\lambda)$ having the descriptor realization $(A^T-\lambda E^T,C^T,B^T,D^T)$.


\index{transfer function matrix (TFM)!model-matching problem!approximate}
A special factorization problem encountered when solving the approximate model-matching problem by reducing it to a least distance problem (see equation (\ref{VA:spec}) in Section~\ref{appsec:LDP}) is the following \emph{special left spectral factorization problem}: for a given TFM $G(\lambda)$ without poles in $\partial\mathds{C}_s$ and a given bound $\gamma > \|G(\lambda)\|_\infty$, compute a stable and minimum-phase TFM $G_o(\lambda)$ such that
\[ \gamma^2 I-G(\lambda)G^\sim(\lambda) = G_o(\lambda)G_o^\sim(\lambda) \, .\]
This computation can be addressed in two steps. In the first step, we compute a RCF  $G(\lambda) = N(\lambda)M^{-1}(\lambda)$, with the denominator factor $M(\lambda)$ inner. It follows that
\[ \gamma^2 I - G(\lambda) G^{\sim}(\lambda) = \gamma^2 I - N(\lambda) N^{\sim}(\lambda) ,\]
where $N(\lambda)$ is proper and has only poles in $\mathds{C}_s$. In the second step, we determine the stable and minimum-phase $G_o(\lambda)$ which satisfies
\be\label{special_specfacN} \gamma^2 I - N(\lambda) N^{\sim}(\lambda)  = G_{o}(\lambda) G_{o}^{\sim}(\lambda) \, .\ee
The first step has been already discussed in Section~\ref{appsec:fact}, and therefore we assume that for an irreducible descriptor realization $(A-\lambda E,B,C,D)$ of $G(\lambda)$, we determined a stable $N(\lambda)$ with a descriptor realization $(\widetilde A-\lambda \widetilde E,\widetilde B,\widetilde C,\widetilde D)$.
\index{factorization!spectral!special, stable minimum-phase left}

In the continuous-time case, we can compute the spectral factor $G_o(s)$ by using the following result, which extends to proper descriptor systems the formulas developed in  \cite[Corollary~13.22]{Zhou96}.
\begin{lemma}\label{L:special-specfacc}
Let $N(s)$ be a stable TFM and let $(\widetilde A-s \widetilde E,\widetilde B,\widetilde C,\widetilde D)$ be its  descriptor system realization. For $\gamma > \|N(s)\|_\infty$, a realization of a stable and minimum-phase spectral factor $G_o(s)$, satisfying (\ref{special_specfacN}) for $\lambda=s$,  is given by
\[ G_o(s) = \ba{c|c} \widetilde A -s \widetilde E &  -K_sR^{1/2} \\ \hline \\[-3.5mm]\widetilde C & R^{1/2} \ea \, ,\]
where
\[ \begin{array}{lll} R &=& \gamma^2 I - \widetilde D\widetilde D^T ,\\
K_s &=& (\widetilde EY_s\widetilde C^T+\widetilde B\widetilde D^T)R^{-1} , \end{array} \]
and $Y_s$ is the stabilizing solution of the GCARE
\index{matrix equation!generalized algebraic Riccati!continuous-time (GCARE)}
\[ \widetilde AY\widetilde E^T + \widetilde EY\widetilde A^T + (\widetilde EY\widetilde C^T+\widetilde B\widetilde D^T)R^{-1}(\widetilde CY\widetilde E^T+\widetilde D\widetilde B^T) + \widetilde B\widetilde B^T = 0 \, . \]
\end{lemma}

We have the following analogous result in the discrete-time case, which extends to proper descriptor system the formulas developed in  \cite[Theorem~21.26]{Zhou96} for the dual special right spectral factorization problem (see below).
\begin{lemma}\label{L:special-specfacd} Let $N(z)$ be a stable TFM and let $(\widetilde A-z \widetilde E,\widetilde B,\widetilde C,\widetilde D)$ be its descriptor realization. For $\gamma > \|N(z)\|_\infty$,  a realization of a stable and minimum-phase spectral factor  $G_o(z)$, satisfying (\ref{special_specfacN}) for $\lambda=z$, is given by
\[ G_o(z) = \ba{c|c} \widetilde A -\lambda \widetilde E &  -K_sR^{1/2} \\ \hline \\[-3mm] \widetilde C & R^{1/2} \ea \, , \]
where
\[ \begin{array}{lll} R_D &=& \gamma^2 I - \widetilde D\widetilde D^T ,\\
R &=& R_D-\widetilde CY_s\widetilde C^T  ,\\
K_s &=& (\widetilde AY_s\widetilde C^T+\widetilde B\widetilde D^T)R^{-1} ,\end{array} \]
and $Y_s$ is the stabilizing solution of the GDARE
\index{matrix equation!generalized algebraic Riccati!discrete-time (GDARE)}
\[ \widetilde AY\widetilde A^T - \widetilde EY\widetilde E^T - (\widetilde AY\widetilde C^T+\widetilde B\widetilde D^T)(-R_D+\widetilde CY\widetilde C^T)^{-1}(\widetilde CY\widetilde A^T+\widetilde D\widetilde B^T) + \widetilde B\widetilde B^T = 0 \, . \]
\end{lemma}

Similar formulas to those provided by Lemma~\ref{L:special-specfacc} and
Lemma~\ref{L:special-specfacd} can be derived for the solution of the dual \emph{special right spectral factorization problem}
\[ \gamma^2 I-N^\sim(\lambda)N(\lambda)=G_o^\sim(\lambda)G_o(\lambda) . \].
\index{factorization!spectral!special, stable minimum-phase right}

\subsection{Linear Rational Matrix Equations} \label{appsec:linsys}
\index{transfer function matrix (TFM)!linear rational matrix equation}
For $G(\lambda) \in \mathds{R}(\lambda)^{p\times m}$ and $F(\lambda) \in \mathds{R}(\lambda)^{p\times q}$ consider the solution of the linear rational matrix equation
\be\label{matlinrat} G(\lambda)X(\lambda) = F(\lambda)\, ,\ee
where $X(\lambda) \in \mathds{R}(\lambda)^{m\times q}$ is the solution we seek. The existence of a solution is guaranteed if the compatibility condition for the linear system  (\ref{matlinrat}) is fulfilled.
\begin{lemma} \label{L-EMMP}
The rational equation (\ref{matlinrat}) has a solution  if and only if
\be\label{matcompcond} \rank G(\lambda) = \rank \,[\, G(\lambda) \; F(\lambda)\,]\, .\ee
\end{lemma}
Let $r$ be the rank of $G(\lambda)$. In the most general case, the solution of  (\ref{matlinrat}) (if exists) is not unique and can be expressed as
\be\label{ratsolparam} X(\lambda) = X_0(\lambda) + N_r(\lambda)Y(\lambda) , \ee
where $X_0(\lambda)$ is a particular solution of (\ref{matlinrat}), $N_r(\lambda) \in \mathds{R}(\lambda)^{m\times (m-r)}$ is a rational matrix representing a basis of the right nullspace $\ker_R(G(\lambda))$ (is empty if $r=m$), while $Y(\lambda)\in \mathds{R}(\lambda)^{(m-r)\times q}$ is an arbitrary rational matrix.

An important aspect in control related applications is to establish conditions which ensure the existence of a solution $X(\lambda)$ which has only poles in a ``good'' domain $\mathds{C}_g$, or equivalently, $X(\lambda)$ has no poles in the ``bad`` domain $\mathds{C}_b := \mathds{C}\setminus \mathds{C}_g$.
Such a condition can be obtained in terms of the pole-zero structures of the rational matrices $G(\lambda)$ and $[\, G(\lambda) \; F(\lambda)\,]$ at a particular value $\lambda_z$ of the frequency parameter $\lambda$ (see, for example, \cite{Kail80}).
\begin{lemma} \label{L-SEMMP}
The rational equation (\ref{matlinrat}) has a solution without poles in $\mathds{C}_b$ if and only if (\ref{matcompcond}) is fulfilled and the rational matrices $G(\lambda)$ and $[\, G(\lambda) \; F(\lambda)\,]$ have the same pole-zero structure for all $\lambda_z \in \mathds{C}_b$.
\end{lemma}

\index{transfer function matrix (TFM)!model-matching problem!exact}
The characterization provided by Lemma~\ref{L-SEMMP} is relevant when solving synthesis problems of fault detection filters and controllers using an \emph{exact model-matching} approach, where the physical realizability requires the properness and stability of the solutions (i.e., $\mathds{C}_g = \mathds{C}_s$). For example, if $G(\lambda)$ has unstable zeros in $\lambda_z$, then $F(\lambda)$ must be chosen to have the same (or richer) zero structure in $\lambda_z$, in order to ensure the cancellation of these zeros (appearing now as  unstable poles of any particular solution $X_0(\lambda)$). The fixed poles in $\mathds{C}_b$ of any particular solution $X_0(\lambda)$ correspond to those zeros of $G(\lambda)$ for which the above condition is not fulfilled, and thus no complete cancellations take place.

In the case when $\rank G(\lambda) < m$, an important aspect is the exploitation of the non-uniqueness of the solution in (\ref{matlinrat}) by determining a solution with the least possible McMillan degree. This problem is known in the literature as the \emph{minimum  design problem} (MDP) and primarily targets the  reduction of the complexity of real-time burden when implementing filters or controllers. Of particular importance are proper and stable solutions which are suitable for a physical (causal) realization. If the minimal degree solution is not proper and stable, then it is of interest to find a proper and stable solution with the least McMillan degree.  Surprisingly, this problem does not have a satisfactory procedural solution, most of proposed approaches involves parametric searches using suitably parameterized solutions of given candidate degrees.


\index{descriptor system!linear rational matrix equation}
An equivalent solvability condition to that of Lemma~\ref{L-EMMP} can be derived in terms of descriptor system representations of
$G(\lambda)$ and $F(\lambda)$, which we assume to be of the form
\be\label{GF_rat_sys1} G(\lambda) = \ba{c|c}A-\lambda E & B_G \\ \hline C & D_G \ea, \qquad F(\lambda) = \ba{c|c}A-\lambda E & B_F \\ \hline C & D_F \ea \, .\ee
Such representations, which share the pair $(A-\lambda E,C)$, can be easily obtained by determining a descriptor realization of the compound rational matrix $[\, G(\lambda) \; F(\lambda)\,]$.
It is easy to observe that any solution of (\ref{matlinrat}) is also part of the solution of the linear polynomial equation
\be\label{lineq_pencil1} \ba{cc}A-\lambda E & B_G \\ C & D_G \ea Y(\lambda) = \ba{c} B_F \\ D_F \ea \, ,\ee
where $Y(\lambda) = \left[\begin{smallmatrix} W(\lambda) \\ X(\lambda) \end{smallmatrix}\right]$. Therefore, alternatively to solving (\ref{matlinrat}), we can solve instead (\ref{lineq_pencil1}) for $Y(\lambda)$ and compute $X(\lambda)$ as
\be\label{X_extracted1}  X(\lambda) = [\, 0 \;\; I_m\,]Y(\lambda) \, .\ee

Define the system matrix pencils corresponding to $G(\lambda)$ and the compound $[\, G(\lambda) \; F(\lambda)\,]$ as
\be\label{S_GF}
S_G(\lambda) := \ba{cc}A-\lambda E & B_G \\ C & D_G \ea , \qquad
S_{G,F}(\lambda) := \ba{ccc} A-\lambda E & B_G & B_F \\ C & D_G & D_F \ea. \ee
We have the following result similar to Lemma~\ref{L-EMMP}.

\begin{lemma} \label{L-EMMP-dss}
The rational equation (\ref{matlinrat}) with $G(\lambda)$ and $F(\lambda)$ having the descriptor realizations in (\ref{GF_rat_sys1}) has a solution  if and only if
\be\label{comp_cond_ss1} \rank S_G(\lambda) = \rank S_{G,F}(\lambda) \, .\ee
\end{lemma}

Let $\mathds{C}_b$ be the ``bad`` domain of the complex plane, where the solution $X(\lambda)$ must not have poles.
We have the following result similar to Lemma~\ref{L-SEMMP}.
\begin{lemma} \label{L-SEMMP-dss}
The rational equation (\ref{matlinrat}) with $G(\lambda)$ and $F(\lambda)$ having the descriptor realizations in (\ref{GF_rat_sys1}) has a solution without poles in $\mathds{C}_b$ if and only if the matrix pencils $S_G(\lambda)$ and $S_{G,F}(\lambda)$ defined in (\ref{S_GF}) fulfill (\ref{comp_cond_ss1}) and have the same zero structure for all zeros of $G(\lambda)$ in $\mathds{C}_b$.
\end{lemma}

The general solution of (\ref{matlinrat}) can be expressed as in (\ref{ratsolparam})
where $X_0(\lambda)$ is any particular solution of (\ref{matlinrat}), $N_r(\lambda)$ is a rational  matrix whose columns form a basis for the right nullspace
of $G(\lambda)$, and $Y(\lambda)$ is an arbitrary rational matrix with compatible dimensions. A possible approach
to compute a
solution $X(\lambda)$ of least McMillan degree is to determine
a suitable $Y(\lambda)$ to achieve this goal.
 A computational procedure to determine a least McMillan degree solution $X(\lambda)$  is described in \cite[Section 10.3.7]{Varg17} and relies on the Kronecker-like form of the associated system matrix pencil $S_G(\lambda)$ in (\ref{S_GF}), which is used to determine the particular solution $X_0(\lambda)$, the nullspace basis  $N_r(\lambda)$, and,  combined with
 the
generalized minimum cover algorithm of \cite{Varg04a}, to obtain the least McMillan degree solution.

\subsection{Dynamic Cover-Based Order Reduction}\label{sec:dyncov}
Let $X_1(\lambda)$ and $X_2(\lambda)$ be rational transfer function matrices. For $X_1(\lambda)$ and $X_2(\lambda)$ with the same row dimension, the \emph{right minimal cover problem} is to find $X(\lambda)$ and $Y(\lambda)$ such that
\be\label{rmincov} X(\lambda) = X_1(\lambda) + X_2(\lambda)Y(\lambda) , \ee
and the McMillan degree of $\left[\begin{smallmatrix} X(\lambda) \\ Y(\lambda) \end{smallmatrix}\right]$ is minimal \cite{Anto83}. Similarly, for $X_1(\lambda)$ and $X_2(\lambda)$ with the same column dimension,  the \emph{left minimal cover problem} is to find $X(\lambda)$ and $Y(\lambda)$ such that
\be\label{lmincov} X(\lambda) = X_1(\lambda) + Y(\lambda) X_2(\lambda) , \ee
and the McMillan degree of $[\,X(\lambda) \; Y(\lambda) \,]$ is minimal \cite{Anto83}. Any method to solve a right minimal cover problem can be used to solve the left minimal proper problem, by applying it to the transposed matrices  $X_1^T(\lambda)$ and $X_2^T(\lambda)$ to determine the transposed solution $X^T(\lambda)$ and $Y^T(\lambda)$. Therefore, in what follows we will only discuss the solution of right minimal cover problems.
\index{transfer function matrix (TFM)!left minimal cover problem}%
\index{transfer function matrix (TFM)!right minimal cover problem}%

By imposing additional conditions on $X_1(\lambda)$ and $X_2(\lambda)$, as well as on the class of desired solutions $X(\lambda)$ and $Y(\lambda)$, particular cover problems arise, which may require specific methods for their solution. In what follows, we only discuss the proper case, when $X_1(\lambda)$, $X_2(\lambda)$, $X(\lambda)$ and $Y(\lambda)$ are all proper.   Various applications involving the solution of minimal cover problems are discussed in \cite{Anto83}, as for example---observer and controller synthesis using model-matching based approaches. Other problems, as for example, imposing $X(\lambda) = 0$, are equivalent to solve linear rational equations for the least McMillan degree solutions (see Section~\ref{appsec:linsys}).


\index{descriptor system!left minimal cover problem}%
\index{descriptor system!right minimal cover problem}%
Assume $X_1(\lambda)$  and $X_2(\lambda)$ have the descriptor realizations \be\label{genrealQ} \ba{cc} X_1(\lambda) & X_2(\lambda) \ea = \ba{c|cc} A-\lambda E & B_1 & B_2 \\ \hline C & D_1 & D_2 \ea \, ,\ee
with the descriptor pair $(A-\lambda E, [\,B_1 \; B_2\,])$ controllable and $E$ invertible. The methods we describe, use for $Y(\lambda)$ the descriptor realization
\be\label{Ycover}  Y(\lambda) = \ba{c|c} A+B_2F-\lambda E & B_1+B_2G \\ \hline F & G \ea \, ,\ee
where $F$  and $G$ are state-feedback gain and  feedforward gains, respectively, to be determined. It is straightforward to check that $X(\lambda) = X_1(\lambda) + X_2(\lambda)Y(\lambda)$ has the descriptor system realization
\be\label{Xcover} X(\lambda) := \ba{c|c} A+B_2F-\lambda E & B_1+B_2G \\ \hline C+D_2F & D_1+D_2G \ea \, .\ee
%
%
If the gains $F$ and $G$ are determined such that the pair
 $(A+B_2F-\lambda E,B_1+B_2G)$ is {maximally uncontrollable}, then the resulting realizations of $X(\lambda)$ and $Y(\lambda)$ contain a maximum number of uncontrollable eigenvalues which can be eliminated using minimal realization techniques. In some applications, the use of a strictly proper $Y(\lambda)$ (i.e., $G=0$) is sufficient to achieve the maximal order reduction, therefore, this case will be treated explicitly in what follows.

The problem to determine the matrices $F$ and $G$, which make the descriptor system pair $(A+B_2F-\lambda
E,B_1+B_2G)$ {maximally uncontrollable}, is essentially equivalent \cite{Mors76} to compute a
subspace $\mathcal{V}$ having the least possible dimension and satisfying
\be\label{V2} (\overline A+\overline B_2F)\mathcal{V} \subset \mathcal{V}, \quad \spann\,(\overline B_1+\overline B_2G) \subset \mathcal{V} \, ,\ee
where $\overline A := E^{-1}A$, $\overline B_1 := E^{-1}B_1$, and $\overline B_2 := E^{-1}B_2$.
If we denote $\overline{\mathcal{B}}_1 = \spann\,\overline B_1$ and $\overline{\mathcal{B}}_2 = \spann\,\overline B_2$, then the
above condition for $G = 0$ can be equivalently rewritten as the condition defining a \textit{Type 1}
minimum dynamic cover \cite{Emre77,Kimu77}:
\be\label{cover1}  \overline A\mathcal{V}  \subset \mathcal{V} + \overline{\mathcal{B}}_2 ,\qquad
\overline{\mathcal{B}}_1  \subset \mathcal{V} . \ee
\index{descriptor system!right minimal cover problem!Type 1}%
If we allow for $G \not = 0$, then (\ref{V2}) can be equivalently rewritten as the  condition defining a \textit{Type 2}
minimum dynamic cover \cite{Emre77,Kimu77}:
\be\label{cover2}  \overline A\mathcal{V}  \subset \mathcal{V} + \overline{\mathcal{B}}_2 ,\qquad
\overline{\mathcal{B}}_1  \subset \mathcal{V}+\overline{\mathcal{B}}_2 . \ee
\index{descriptor system!right minimal cover problem!Type 2}%

The underlying computational problems to solve these minimum dynamic cover problems is the
following:
given the controllable descriptor system pair $(A-\lambda E,B)$  with
$A,E \in \mathds{R}^{n\times n}$  and $E$ invertible, $B \in \mathds{R}^{n\times m}$,
and $B$ partitioned as
$B = [\,B_1\; B_2\,]$ with $B_1 \in \mathds{R}^{n\times m_1}$, $B_2 \in \mathds{R}^{n\times m_2}$,
determine the matrices $F$ and $G$ (either with $G = 0$ or $G\not = 0$)
such that the descriptor system pair $(A+B_2F-\lambda E,B_1+B_2G)$ has the maximal
number of uncontrollable eigenvalues. Computational procedures to solve these problems have been proposed in \cite{Varg04a} for descriptor systems and in \cite{Varg03e} for standard systems (see also \textbf{Procedures GRMCOVER1} and \textbf{GRMCOVER2}, described in \cite{Varg17}).

\subsection{Hankel Norm }\label{sec:normTFM}

\index{transfer function matrix (TFM)!Hankel norm}
The \emph{Hankel norm} of a stable TFM $G(\lambda)$ is a measure of the influence of past inputs on future outputs and is denoted by $\|G(\lambda)\|_H$. The precise mathematical definition of the Hankel norm involves some advanced concepts from functional analysis (i.e., it is the norm of the Hankel operator $\Gamma_G$ associated to $G(\lambda)$). For further details, see \cite{Zhou96}.

\index{descriptor system!Hankel norm}
For the evaluation of the Hankel norm of a stable TFM $G(\lambda)$, the state-space representation allows to use explicit formulas.
\begin{lemma}\label{L:Hnormc}
Let $G(s)$ be a proper and stable TFM of a continuous-time system and let $(A-sE,B,C,D)$ be an irreducible descriptor realization with $E$ invertible. Then, the {Hankel norm} of $G(s)$ can be evaluated as
\[ \|G(s)\|_H = \overline\sigma(RES) ,\]
where $P = SS^T$ is the positive semi-definite controllability Gramian and $Q = R^TR$ is the positive semi-definite observability Gramian, which satisfy the following generalized Lyapunov equations \index{matrix equation!generalized Lyapunov}%
\be\label{glyap}  \begin{array}{lll} APE^T +EPA^T + BB^T &=& 0 \, ,\\
A^TQE + E^TQA + C^TC &=&0 \, .\end{array} \ee
\end{lemma}

\begin{lemma}\label{L:Hnormd}
Let $G(z)$ be a proper and stable TFM of a continuous-time system and let $(A-zE,B,C,D)$ be an irreducible descriptor realization with $E$ invertible. Then, the {Hankel norm} of $G(z)$ can be evaluated as
\[ \|G(z)\|_H = \overline\sigma(RES) ,\]
where $P = SS^T$ is the positive semi-definite controllability Gramian and $Q = R^TR$ is the positive semi-definite observability Gramian, which satisfy the following generalized Stein equations
\index{matrix equation!generalized Stein}%
\be\label{gstein} \begin{array}{lll} APA^T -EPE^T+ BB^T &=& 0\, ,\\
A^TQA - E^TQE + C^TC &=&0 \, .\end{array} \ee
\end{lemma}

For the solution of the generalized Lyapunov equations (\ref{glyap}) and the generalized Stein equations  (\ref{gstein}), computational procedures have been proposed in \cite{Penz98a}, which allow to directly determine the Cholesky factors $S$ and $R$ of the Gramians. These methods extend the algorithm of \cite{Hamm82} proposed for standard systems with $E = I$.

\subsection{The Gap Metric}\label{sec:nugap}

\index{transfer function matrix (TFM)!$\nu$-gap metric}

The $\nu$-gap metric $\delta_\nu(G_1,G_2)$ has been introduced in \cite{Vinn93} for standard systems, as a natural measure of the ``distance'' between two proper transfer function matrices $G_1(\lambda)$ and $G_2(\lambda)$ of the same dimensions. The definition of $\delta_\nu(G_1,G_2)$ given in \cite{Vinn93} can be extended to arbitrary transfer function matrices, because this definition involves only the normalized left coprime factorization $G_1(\lambda) = \widetilde M_1^{-1}(\lambda)\widetilde N_1(\lambda)$ and the right coprime factorizations $G_1(\lambda) = N_1(\lambda)M_1^{-1}(\lambda)$ and $G_2(\lambda) = N_2(\lambda)M_2^{-1}(\lambda)$. This definition can also serve as basis for a computational procedure. With
$\widetilde L_2(\lambda) := [\,-\widetilde M_2(\lambda) \; \widetilde N_2(\lambda)\,]$, $R_i(\lambda) := \left[ \begin{smallmatrix} N_i(\lambda) \\ M_i(\lambda) \end{smallmatrix}\right]$ for $i = 1, 2$, and $g(\lambda) := \det\big(R_2^\sim(\lambda) R_1(\lambda)\big)$, we have the following definition:
\be\label{nugap} \delta_\nu(G_1,G_2) := \left\{ \begin{array}{cl} \left\| \widetilde L_2(\lambda)R_1(\lambda) \right\|_\infty & \text{if } g(\lambda) \neq 0 \; \forall \lambda \in \partial\mathds{C}_s \text{ and } \wno(g) = 0 ,\\
1 & \text{otherwise} , \end{array}\right. \ee
where $\wno(g)$ denotes the \emph{winding number} of $g(\lambda)$ about the appropriate critical point for $\lambda$ following the corresponding standard Nyquist contour.  \index{transfer function!winding number}%
The winding number of $g(\lambda)$ can be determined as the difference between the number of unstable zeros of $g(\lambda)$ and the number of unstable poles of $g(\lambda)$ \cite{Vidy11}.
Generally, for any $G_1(\lambda)$ and $G_2(\lambda)$, we have  $0 \leq \delta_\nu(G_1,G_2) \leq 1$.
If $\delta_\nu(G_1,G_2)$ is small, then we can say that
$G_1(\lambda)$ and $G_2(\lambda)$  are close and it is likely that a controller or filter suited for $G_1(\lambda)$ will also work with $G_2(\lambda)$. On the other side, if $\delta_\nu(G_1,G_2)$  is nearly equal to 1, then $G_1(\lambda)$ and $G_2(\lambda)$ are sufficiently distinct, such that an easy discrimination between the two models is possible. A common criticism of the $\nu$-gap metric is that there are many transfer function matrices $G_2(\lambda)$ at a distance $\delta_\nu(G_1,G_2) = 1$ to a given $G_1(\lambda)$, but the metric fails to differentiate between them.

In \cite{Vinn01}, the point-wise $\nu$-gap metric is also defined to evaluate the distance between two models in a single frequency point. If $\lambda_s$ is a fixed complex frequency, then the point-wise $\nu$-gap metric between two transfer-function matrices $G_1(\lambda)$ and $G_2(\lambda)$ at $\lambda_s$ is:
\be\label{nugap-pointwise} \delta_\nu(G_1(\lambda_s),G_2(\lambda_s)) := \left\{ \begin{array}{cl} \big\| \widetilde L_2(\lambda_s)R_1(\lambda_s) \big\|_2 & \text{if } g(\lambda) \neq 0 \; \forall \lambda \in \partial\mathds{C}_s \text{ and } \wno(g) = 0 ,\\
1 & \text{otherwise} , \end{array}\right. \ee

%
\subsection{Balancing-Related Order Reduction } \label{app:balance}

Consider a stable TFM $G(\lambda)$ with a state-space descriptor system realization $(A-\lambda E,B,C,D)$ with $E$ invertible and $\Lambda(A,E) \subset \mathds{C}_s$. The controllability and observability
properties of the state-space realization can be characterized by the controllability and observability gramians. The controllability gramian $P$ and observability gramian $Q$ satisfy appropriate generalized Lyapunov and Stein equations. In the continuous-time case, $P$ and $Q$ satisfy the generalized Lyapunov equations (\ref{glyap}),
while in the discrete-time case, $P$ and $Q$ satisfy the generalized Stein equations (\ref{gstein}).
For a stable system, both gramians $P$ and $Q$ are positive semi-definite matrices and fully characterize the controllability and observability properties. The controllability gramian $P > 0$ if and only if the pair $(A-\lambda E,B)$ is controllable, and the observability gramian $Q > 0$ if and only if the pair $(A-\lambda E,C)$ is observable. Besides these qualitative characterizations, the eigenvalues of $P$ and $Q$ provides quantitative measures of these properties. Thus, the degree of controllability can be defined as the least eigenvalue of $P$ and represents a measure of the nearness of the state-space realization to an uncontrollable one. Analogously, the degree of observability, defined as the least eigenvalue of $Q$, represents a measure of the nearness of the state-space realization to an unobservable one. These measures are \emph{not} invariant to coordinate transformations and, therefore, it is of interest to have state-space representations for which the controllability and observability properties are balanced. In fact, for a controllable and observable system, it is possible to find a particular state-space representation, called a \emph{balanced realization}, for which the two gramians are equal and even diagonal.

The eigenvalues of the gramians of a balanced system are called the \emph{Hankel singular values}. The largest singular value represents the \emph{Hankel norm} $\|G(\lambda)\|_H$ of the corresponding TFM $G(\lambda)$,
\index{transfer function matrix (TFM)!Hankel norm} while the smallest one can be interpreted as a measure of the nearness of the system to a non-minimal one.   Important applications of balanced realizations are to ensure minimum sensitivity to roundoff errors of real-time filter models or to perform model order reduction, by reducing large order models to lower order approximations. The order reduction can be performed by simply truncating the system state to a part corresponding to the ``large'' singular values, which significantly exceed the rest of ``small'' singular values.

A procedure to compute minimal balanced realizations of stable descriptor systems is \textbf{Procedure GBALMR} described in \cite[Section 10.4.4]{Varg17}. This procedure is instrumental in solving the Nehari approximation problem for descriptor systems (see Section~\ref{app:nehari}).
The reduction of a linear state-space model to a balanced minimal realization may involve the usage of ill-conditioned coordinate
transformations (or projections) for systems which are
nearly non-minimal or nearly unstable. This is why, for the computation of minimal realizations or of lower order approximations, the so-called \emph{balancing-free} approaches, as proposed in \cite{Varg91a} for standard systems and in \cite{Styk04} for descriptor systems, are generally more accurate.

\subsection{Solution of the Optimal Nehari Problems} \label{app:nehari}
\index{transfer function matrix (TFM)!Nehari approximation!optimal}
In this section we consider the solution of the following optimal Nehari problem: Given an anti-stable $G(\lambda)$ (i.e., such that $G^\sim(\lambda)$ is stable), find a stable $X(\lambda)$ which is the closest to $G(\lambda)$ and satisfies
\be\label{app:nehari1} \| G(\lambda)-X(\lambda)\|_\infty = \|G^\sim(\lambda)\|_H \,.\ee
This computation is encountered in the solution of the least-distance problem formulated in Section~\ref{appsec:LDP}. As shown in \cite{Glov84}, to solve the optimal Nehari approximation problem (\ref{app:nehari1}), we can solve instead for $X^\sim(\lambda)$ the optimal zeroth-order Hankel-norm approximation problem
\be\label{app:nehari2} \| G^\sim(\lambda)-X^\sim(\lambda)\|_\infty = \|G^\sim(\lambda)\|_H \,.\ee
A computational method for continuous-time systems, can be devised using the method proposed in \cite{Glov84}, which relies on computing first a balanced minimal order state-space realization of $G^\sim(\lambda)$.  The corresponding procedure for  discrete-time systems is much more involved (see \cite{Gu89}) and therefore, a preferred alternative, suggested in \cite{Glov84}, is to use the procedure for continuous-time systems in conjunction with bilinear transformation techniques.
This approach underlies \textbf{Procedure GNEHARI} described in \cite[Section 10.4.5]{Varg17}, which can be employed even for improper discrete-time systems.

If $\rho$ is a given value satisfying $\rho > \|G^\sim(\lambda)\|_H$, then the suboptimal Nehari approximation problem is to determine a stable $X(\lambda)$ such that
\be\label{app:nehari-sub} \| G(\lambda)-X(\lambda)\|_\infty < \rho \,.\ee
To solve this problem, a suboptimal Hankel-norm approximation $X^\sim(\lambda)$ of $G^\sim(\lambda)$ can be computed using the method of  \cite{Glov84}.
\index{transfer function matrix (TFM)!Nehari approximation!suboptimal}

\subsection{Solution of Least-Distance Problems}\label{appsec:LDP}
\index{transfer function matrix (TFM)!model-matching problem!approximate}
A possible solution method of the $\mathcal{H}_\infty$-model-matching problem \cite{Fran87} (see also Section \ref{appsec:AMMP}) is to reduce this problem to a \emph{least-distance problem} (LDP), which can be solved using Nehari-approximation techniques. The optimal LDP is the problem of computing a stable solution $X(\lambda)$ such that
\be\label{glinfldp:optdef}  \|[\, G_1(\lambda)-X(\lambda) \;\; G_2(\lambda)\,] \|_\infty = \min \, , \ee
where $G_1(\lambda)$ and $G_2(\lambda)$ are TFMs without poles in $\partial\mathds{C}_s$. Therefore, the use of the $\mathcal{L}_\infty$-norm in (\ref{glinfldp:optdef}) is assumed.
The suboptimal LDP is to find a stable $X(\lambda)$ such that
\be\label{glinfldp:suboptdef}  \|[\, G_1(\lambda)-X(\lambda) \;\; G_2(\lambda)\,] \|_\infty < \gamma \, , \ee
where $\gamma > \|G_2(\lambda)\|_\infty$.
If $G_2(\lambda)$ is present, we have a
\emph{2-block} LDP, while if $G_2(\lambda)$ is not present we have an \emph{1-block} LDP.
In what follows, we discuss shortly  the solution approaches for the optimal 1- and 2-block problems.

\emph{Solution of the 1-block $\mathcal{H}_\infty$-LDP.} In the case of the $\mathcal{L}_\infty$-norm, the stable optimal solution $X(\lambda)$ of the 1-block problem can be computed by solving an optimal Nehari problem. Let $L_s(\lambda)$ be the stable part and let $L_u(\lambda)$ be the
unstable part in the additive decomposition
\be\label{F1stabUnstab}   G_1 (\lambda) = L_s(\lambda) + L_u(\lambda) \, .\ee
Then, for the optimal solution we have successively
\[  \|  G_1(\lambda)-X(\lambda) \|_\infty = \| L_u(\lambda)- X_s(\lambda) \|_\infty  = \| L_u^\sim(\lambda) \|_H \, ,\]
where $X_s(\lambda)$ is the stable optimal Nehari solution and
\[ X(\lambda) = X_s(\lambda) + L_s(\lambda) \, . \]
\index{transfer function matrix (TFM)!Nehari approximation!optimal}%

\emph{Solution of the 2-block $\mathcal{H}_\infty$-LDP.} A stable optimal solution $X(\lambda)$ of the 2-block LDP can be approximately determined as the solution of the suboptimal 2-block LDP
\be\label{VA:two-blocks_app} \| [\,  G_1(\lambda)-X(\lambda)\;\;   G_2(\lambda) \,] \|_\infty < \gamma ,\ee
where $\gamma_{opt} < \gamma \leq \gamma_{opt}+\varepsilon$, with $\varepsilon$ an arbitrary user specified (accuracy) tolerance for the least achievable value $\gamma_{opt}$ of $\gamma$. The standard solution approach is a bisection-based $\gamma$-iteration method, where the solution of the 2-block problem is approximated by successively computed $\gamma$-suboptimal solutions of appropriately defined 1-block problems \cite{Fran87}.

Let $\gamma_l$ and $\gamma_u$ be lower and upper bounds for $\gamma_{opt}$, respectively. Such bounds can be computed, for example, as
\be\label{VA:gammalu_app} \gamma_l = \|  G_2 (\lambda)
\|_\infty, \quad \gamma_u = \| [\,  G_1(\lambda) \;
G_2(\lambda)\,] \|_\infty \, .\ee
For a given $\gamma > \gamma_l$,
we solve first the stable minimum-phase left spectral factorization problem
\be\label{VA:spec} \gamma^2I -   G_2
(\lambda)  G_2^\sim (\lambda) =
V(\lambda)V^{\sim}(\lambda) ,\ee
where the spectral factor
$V(\lambda)$ is biproper, stable and minimum-phase.
\index{factorization!spectral!special, stable minimum-phase left}%
Further, we
compute the additive decomposition
\be\label{VA:dec} V^{-1}(\lambda)   G_1 (\lambda)  = L_s(\lambda) +
L_u(\lambda),\ee
\index{transfer function matrix (TFM)!additive decomposition}%
where $L_s(\lambda)$ is the stable part and $L_u(\lambda)$ is the
unstable part.
If $\gamma > \gamma_{opt}$, the suboptimal 2-block  problem
(\ref{VA:two-blocks_app}) is equivalent to the suboptimal 1-block problem
\be\label{VA:one-block}
\big\| V^{-1}(\lambda)\big(  G_1
(\lambda)-X(\lambda)\big)   \big \|_\infty \leq 1 \ee
and
$\gamma_H := \|L_u^\sim(\lambda)\|_H < 1$. In this case we readjust the upper bound to $\gamma_u =
\gamma$. If $\gamma \leq \gamma_{opt}$, then $\gamma_H \geq 1$ and we readjust the lower bound to $\gamma_l
= \gamma$. For the bisection value  $\gamma = (\gamma_l +\gamma_u)/2$ we redo the
factorization (\ref{VA:spec}) and decomposition (\ref{VA:dec}). This process is repeated until $\gamma_u-\gamma_l \leq \varepsilon$.

At the end of iterations, we have either $\gamma_{opt} < \gamma \leq \gamma_u$ if $\gamma_H < 1$ or $ \gamma_l < \gamma \leq \gamma_{opt} $ if $\gamma_H \geq 1$, in which case we set $\gamma = \gamma_u$. We compute the stable solution of
(\ref{VA:one-block}) as
\be\label{qbar} X(\lambda) = V(\lambda)(L_s(\lambda) + X_s(\lambda)) , \ee
where, for any $\gamma_1$ satisfying $1 \geq \gamma_1 > \gamma_H$,  $X_s(\lambda)$ is the stable
solution of the optimal Nehari problem
\be\label{VA:neh} \left\| L_u(\lambda)-X_s(\lambda) \right \|_\infty
= \| L_u^\sim(\lambda) \|_H  .\ee
\index{transfer function matrix (TFM)!Nehari approximation!optimal}%

\subsection{Approximate Model Matching}\label{appsec:AMMP}
\index{transfer function matrix (TFM)!model-matching problem!approximate}

The formulation of the approximate {model-matching problems}   is frequently done by formulating suitable error minimization problems in terms of $\mathcal{L}_2$- or $\mathcal{L}_\infty$-norms. For example, the  {approximate} solution of the (left) rational equation $X(\lambda)G(\lambda)=F(\lambda)$ involves the minimization of the norm of the error
$\mathcal{E}(\lambda) := F(\lambda)-X(\lambda)G(\lambda)$, while the  {approximate} solution of the (right) rational equation $G(\lambda)X(\lambda)=F(\lambda)$ involves the minimization of the norm of the error
$\mathcal{E}(\lambda) := F(\lambda)-G(\lambda)X(\lambda)$.  In what follows, we consider either the $\mathcal{L}_\infty$- or $\mathcal{L}_2$-norms of the error, or the $\mathcal{H}_\infty$- or $\mathcal{H}_2$-norms of the error if $\mathcal{E}(\lambda)$ is stable. The corresponding problems are called  $\mathcal{L}_\infty$ \emph{model-matching problem} ($\mathcal{L}_\infty$-MMP), $\mathcal{L}_2$-MMP, $\mathcal{H}_\infty$-MMP or $\mathcal{H}_2$-MMP. We will use $\|\cdot\|_{\infty/2}$ to denote either the $\mathcal{L}_\infty$- or $\mathcal{L}_2$-norm, or the $\mathcal{H}_\infty$- or $\mathcal{H}_2$-norm, depending on the context. In practical applications, further conditions may be imposed on the desired solution $X(\lambda)$, as for example stability or properness, in conjunction with a stable $F(\lambda)$.

For the existence of a solution the error norm must be finite.
Surprisingly, necessary and sufficient condition for the existence of an optimal solution of the $\mathcal{L}_\infty$-MMP are unknown (at least to the author), but several practice relevant sufficient conditions can be easily established. For example, an optimal solution of the $\mathcal{L}_\infty$-MMP exists if $F(\lambda)$ has no poles in $\partial\mathds{C}_s$ (e.g., it is stable). Also, an optimal solution of the $\mathcal{L}_2$-MMP exists if $F(\lambda)$ has no poles in $\partial\mathds{C}_s$ (e.g., it is stable) and additionally it is strictly proper.
This result holds for arbitrary $G(\lambda)$ (e.g., even a improper one).

For filter and feedforward controller design applications, a standard formulation of the $\mathcal{H}_\infty$ \emph{model-matching problem} ($\mathcal{H}_\infty$-MMP) is: given $G(\lambda), F(\lambda) \in \mathcal{H}_\infty$,  find $X(\lambda) \in \mathcal{H}_\infty$ which minimizes $\|\mathcal{E}(\lambda)\|_\infty$. The $\mathcal{H}_2$ \emph{model-matching problem} ($\mathcal{H}_2$-MMP) has a similar formulation.
The following results, taken from \cite{Fran87}, provide sufficient conditions for the solvability of the $\mathcal{H}_\infty$-MMP and $\mathcal{H}_2$-MMP in the standard case.
\begin{lemma} \label{L:AMMP-Hinf} An optimal solution $X(\lambda)$ of the $\mathcal{H}_\infty$-MMP exists if $G(\lambda)$  has no zeros in $\partial\mathds{C}_s$.
\index{model-matching problem!approximate (AMMP), $\mathcal{H}_\infty$-norm!solvability|ii}
\end{lemma}
\begin{lemma} \label{L:AMMP-H2} An optimal solution $X(\lambda)$ of the $\mathcal{H}_2$-MMP exists if $G(\lambda)$  has no zeros in $\partial\mathds{C}_s$ and, additionally, in the continuous-time
\be\label{ammp-h2} \begin{array}{ll}
\rank G(\infty) = \rank \ba{c}G(\infty) \\ F(\infty) \ea , & \text{for a left equation} \, , \\[4mm]
\rank G(\infty) = \rank [\,G(\infty) \; F(\infty) \,] , & \text{for a right equation}  \, . \end{array} \ee
\index{model-matching problem!approximate (AMMP), $\mathcal{H}_2$-norm!solvability|ii}
\end{lemma}
\noindent The conditions of Lemma \ref{L:AMMP-Hinf} and Lemma \ref{L:AMMP-H2}  are clearly not necessary.  For example, in the case of solving the equations $X(\lambda)G(\lambda) = G(\lambda)$ or $G(\lambda)X(\lambda) = G(\lambda)$, the trivial optimal (exact) solution $X(\lambda) = I$ exists for any stable $G(\lambda)$.
The rank conditions (\ref{ammp-h2}) merely ensures that the optimal solution $X(\lambda)$ of the $\mathcal{H}_2$-MMP can be chosen such that the resulting $\mathcal{E}(\lambda)$ is strictly proper and therefore $\| \mathcal{E}(\lambda)\|_2$ is finite. For the equations $X(\lambda)G(\lambda) = G(\lambda)$  or $G(\lambda)X(\lambda) = G(\lambda)$, they are fulfilled for arbitrary $G(\lambda)$, and the trivial optimal (exact) solution $X(\lambda) = I$ exists for arbitrary stable $G(\lambda)$.

In what follows, we will sketch a general approach for solving approximate MMPs based on transforming the error minimization problems into appropriate \emph{least distance problems} (LDPs), by using a special factorization of $G(\lambda)$. For convenience, we will further assume $F(\lambda)$ stable, but will relax the condition on $G(\lambda)$, where for the stability of the solution, it is only assumed that $G(\lambda)$ has no zeros in $\partial\mathds{C}_s$ (i.e., it can be unstable and even improper). This condition is sufficient for the existence of a stable optimal solution of the $\mathcal{L}_\infty$-MMP, as it will be apparent from the constructive solution provided below. In what follows, this will be called the \emph{standard case}, while the \emph{non-standard case} is when $G(\lambda)$ has zeros in $\partial\mathds{C}_s$. In this case, the resulting optimal solution $X(\lambda)$ may have poles in $\partial\mathds{C}_s$, which represents a subset of the zeros of $G(\lambda)$ lying in $\partial\mathds{C}_s$.

We only present a (constructive) computational approach to determine the optimal approximate solution of the left rational equation $X(\lambda)G(\lambda) = F(\lambda)$. However, the same procedure is also applicable to compute the approximate solution of the right rational equation $G(\lambda)X(\lambda) = F(\lambda)$, by applying it to  the transposed equation $\widetilde X(\lambda)G^T(\lambda) = F^T(\lambda)$ to compute $\widetilde X(\lambda)$ and obtain $X(\lambda) = \widetilde X^T(\lambda)$.

The first step of the proposed constructive solution of both $\mathcal{L}_\infty$- and $\mathcal{L}_2$-MMPs is the computation of the (extended) quasi-co-outer--coinner factorization of $G(\lambda)$ to reduce these problems to  $\mathcal{L}_\infty$- or $\mathcal{L}_2$-LDPs, respectively. Consider the factorization
\be\label{amm:iofac}
 G(\lambda) = \ba{cc} G_o(\lambda) & 0 \ea G_i(\lambda) = \ba{cc} G_o(\lambda) & 0 \ea \ba{c} G_{i,1}(\lambda)\\ G_{i,2}(\lambda) \ea = G_o(\lambda)G_{i,1}(\lambda), \ee
 where $G_i(\lambda) := \left[\begin{smallmatrix} G_{i,1}(\lambda)\\ G_{i,2}(\lambda) \end{smallmatrix}\right]$ is square and inner and $G_o(\lambda)$ is full column rank (i.e., is left invertible). In the standard case (i.e.,  when $G(\lambda)$ has no zeros in $\partial\mathds{C}_s$),  $G_o(\lambda)$ has only stable zeros and $G_o(\lambda)$ has a stable left inverse $G_o^{-L}(\lambda)$. In the non-standard case (i.e.,  when $G(\lambda)$ has zeros in $\partial\mathds{C}_s$), $G_o(\lambda)$ has unstable zeros (which are the zeros of $G(\lambda)$ in $\partial\mathds{C}_s$) and any left inverse $G_o^{-L}(\lambda)$ has these zeros as poles. In the standard case, if  $G(\lambda)$  is additionally stable, then the resulting $G_o(\lambda)$ is outer. The factorization (\ref{amm:iofac}) can be computed using algorithms proposed in \cite{Oara00} in the continuous-time case and \cite{Oara05} for the discrete-time case.

The factorization (\ref{amm:iofac}) allows to write successively
\[
\begin{array}{ll}
\|\mathcal{E}(\lambda)\|_{\infty/2} &= \|F(\lambda) -X(\lambda)G(\lambda)\|_{\infty/2} \\ & = \left\|\left(F(\lambda) G_i^{\sim}(\lambda)-X(\lambda)\ba{cc} G_o(\lambda) & 0 \ea\right)  G_i(\lambda)\right\|_{\infty/2} \\ & = \big\|\big[\, \widetilde F_1(\lambda)-Y(\lambda) \; \widetilde F_2(\lambda)\, \big]\big\|_{\infty/2} \, ,\end{array}\]
where $Y(\lambda) := X(\lambda)G_o(\lambda) \in \mathcal{H}_\infty$ and
\[ F(\lambda) G_i^{\sim}(\lambda) = {\arraycolsep=1mm\ba{c|c} F(\lambda) G_{i,1}^{\sim}(\lambda) & F(\lambda) G_{i,2}^{\sim}(\lambda) \ea := \ba{c|c} \widetilde F_1(\lambda) & \widetilde F_2(\lambda) \ea \, .} \]
Thus, the problem of computing an approximate solution $X(\lambda)$ which minimizes the error norm $\|\mathcal{E}(\lambda)\|_{\infty/2}$ has been reduced to a LDP to compute the stable solution $Y(\lambda)$ which minimizes the norm $\big\|\big[\, \widetilde F_1(\lambda)-Y(\lambda) \; \widetilde F_2(\lambda)\, \big]\big\|_{\infty/2}$. The solution of the original MMP is given by
\[ X(\lambda) = Y(\lambda)G_o^{-L}(\lambda)\, , \]
where $G_o^{-L}(\lambda)$ is particular left inverse of $G_o(\lambda)$, determined such that its only unstable poles are the unstable zeros of $G_o(\lambda)$ lying in $\partial\mathds{C}_s$. It follows, that in the standard case $X(\lambda)$ results always stable, while in the non-standard case  $X(\lambda)$ may result unstable, having possibly some poles lying in $\partial\mathds{C}_s$. However, $X(\lambda)$ may result stable even in the non-standard case if unstable pole-zero cancellations take place when computing $X(\lambda)$ as a consequence of the presence of such zeros in the transfer function matrix $F(\lambda)$.

In general, we have that $\big[\, \widetilde F_1(\lambda) \; \widetilde F_2(\lambda) \,\big] \not \in \mathcal{H}_\infty$.
If $\widetilde F_2(\lambda)$ is present (i.e., $G(\lambda)$ has no full column rank), we have a
\emph{2-block} LDP, while if $G(\lambda)$ has full column rank, then $\widetilde F_2(\lambda)$ is not present and we have an \emph{1-block} LDP. The solution of the $\mathcal{H}_\infty$-LDP has been discussed in Section \ref{appsec:LDP}. Therefore, in what follows, we only shortly discuss  the solution approach of the $\mathcal{H}_2$-LDP.

Let $L_s(\lambda)$ be the stable part and let $L_u(\lambda)$ be the
unstable part in the additive decomposition
\be\label{F2stabUnstab} \widetilde  F_1 (\lambda) = L_s(\lambda) + L_u(\lambda) \, .\ee
In the case of $\mathcal{H}_2$-norm, the solution of the LDP  is
\[ Y(\lambda) = L_{s}(\lambda) ,\]
where $L_{s}(\lambda)$ is the stable projection in (\ref{F2stabUnstab}). In the continuous-time case, we take the unstable projection $L_u(s)$  strictly proper.
With the above choice, the achieved minimum $\mathcal{H}_2$-norm of ${\mathcal{E}}(\lambda)$ is
\[ \| {\mathcal{E}}(\lambda)\|_2 =  \|[\,L_{u}(\lambda) \;\; \widetilde  F_2(\lambda)\,]\|_2
 \, .
\]
Since the underlying TFMs are unstable, the $\mathcal{L}_2$-norm is used in the last equation. In the continuous-time case, the error norm $\| {\mathcal{E}}(s)\|_2$ is finite only if $\widetilde F_2(s)$ is strictly proper.

In some applications, it is sufficient to determine a so-called $\gamma$-suboptimal solution $X(\lambda)$, which, for a given sub-optimality degree $\gamma$, satisfies
\[ \|  {\mathcal{E}}(\lambda)\|_{\infty} \le \gamma .\]
The choice of $\gamma$ must satisfy $\gamma > \gamma_{opt}$, where $\gamma_{opt}$ is the optimal (minimal) value of $\|{\mathcal{E}}(\lambda)\|_{\infty}$. The solution approach is similar as in the optimal case, but instead of solving an optimal $\mathcal{H}_\infty$-LDP, a $\gamma$-suboptimal $\mathcal{H}_\infty$-LDP is solved, by determining a stable $Y(\lambda)$ which satisfies
\[ \big\|\big[\, \widetilde F_1(\lambda)-Y(\lambda) \; \widetilde F_2(\lambda)\, \big]\big\|_{\infty} < \gamma .\]
Since no $\gamma$-iteration is needed to be performed, the solution of the sub-optimal problem is significantly easier than the solution of the optimal problem.

\newpage
\section{Description of DSTOOLS} \label{sec:userguide}
This user's guide is intended to provide  basic information on the \textbf{DSTOOLS} collection for the operation on and manipulation of rational transfer function matrices via their descriptor system realizations as described in Section~\ref{sec:Basics}.
The notations and terminology used throughout this guide have been introduced and extensively discussed in Chapter 9 of the accompanying book \cite{Varg17}, while Chapter 10 also represents the main reference for the implemented  computational methods in \textbf{DSTOOLS}. Information on the requirements for installing \textbf{DSTOOLS} are given in Appendix \ref{appA}.

In this section, we present first a short overview of the existing functions of \textbf{DSTOOLS} and then, we give in-depth information on the command syntax of the functions of the \textbf{DSTOOLS} collection.
To execute the examples presented in this guide, simply copy and paste the presented code sequences into the MATLAB command window.

\subsection{Quick Reference Tables}
The current release of \textbf{DSTOOLS} is version V0.71, dated September 30, 2018. The corresponding \texttt{Contents.m} file is listed in Appendix \ref{app:contents}.
This section contains quick reference tables for the functions of the \textbf{DSTOOLS} collection.
The main  M- and MEX-files available in the current version of \textbf{DSTOOLS} are listed below by category, with short descriptions.

\begin{center}
\begin{longtable}{|p{2.2cm}|p{12.9cm}|} \hline
\multicolumn{2}{|c|}{\textbf{Demonstration}} \\ \hline
  \texttt{\bfseries DSToolsdemo}& Demonstration of Descriptor System Tools\\
  \hline
\multicolumn{2}{c}{} \\ \hline
\multicolumn{2}{|c|}{\textbf{System analysis}} \\ \hline
\texttt{\bfseries gpole}& Poles of a LTI descriptor system.\\ \hline
\texttt{\bfseries gzero} &Zeros of a LTI descriptor system. \\ \hline
\texttt{\bfseries gnrank}& Normal rank of a transfer function matrix of a LTI system.\\   \hline
\texttt{\bfseries ghanorm}& Hankel norm of a proper and stable LTI descriptor system.\\   \hline
\texttt{\bfseries gnugap}& $\nu$-gap distance between two LTI systems.
\\   \hline
\multicolumn{2}{c}{} \\ \hline
\multicolumn{2}{|c|}{\textbf{Order reduction}} \\ \hline
  \texttt{\bfseries gir}& Reduced order realizations of a LTI descriptor system.\\ \hline
  \texttt{\bfseries gminreal}&  Minimal realization of a LTI descriptor system.\\
  \hline
  \texttt{\bfseries gbalmr}&  Balancing-based model reduction of a stable LTI descriptor system.\\
  \hline
  \texttt{\bfseries gss2ss}&  Conversions to SVD-like coordinate forms without non-dynamic modes.\\
  \hline
\multicolumn{2}{c}{} \\ \hline
\multicolumn{2}{|c|}{\textbf{Operations on transfer function matrices}} \\ \hline
  \texttt{\bfseries grnull}&  Right nullspace basis of a transfer function matrix.\\  \hline
  \texttt{\bfseries glnull}&  Left nullspace basis of a transfer function matrix.\\
  \hline
  \texttt{\bfseries grange}&  Range space basis of a transfer function matrix.\\
  \hline
  \texttt{\bfseries gcrange}&  Coimage space basis of a transfer function matrix.\\
  \hline
  \texttt{\bfseries grsol}&  Solution of the linear rational matrix equation $G(\lambda)X(\lambda)=F(\lambda)$.\\  \hline
  \texttt{\bfseries glsol}&  Solution of the linear rational matrix equation $X(\lambda)G(\lambda)=F(\lambda)$.\\
  \hline
  \texttt{\bfseries gsdec}&  Generalized additive spectral decompositions.\\   \hline
  \texttt{\bfseries grmcover1}&  Right minimum dynamic cover of Type 1 based order reduction.\\  \hline
  \texttt{\bfseries glmcover1}&  Left minimum dynamic cover of Type 1 based order reduction.\\
  \hline
  \texttt{\bfseries grmcover2}&  Right minimum dynamic cover of Type 2 based order reduction.\\
  \hline
  \texttt{\bfseries glmcover2}&  Left minimum dynamic cover of Type 2 based order reduction.\\   \hline
  \texttt{\bfseries gbilin}&  Generalized bilinear transformation.\\
  \hline
  \texttt{\bfseries gbilin1}&  Transfer functions of commonly used bilinear transformations.\\
  \hline
\multicolumn{2}{c}{} \\ \hline
\multicolumn{2}{|c|}{\textbf{Factorizations of transfer function matrices}} \\ \hline
  \texttt{\bfseries grcf}&  Right coprime factorization with proper and stable factors.\\  \hline
  \texttt{\bfseries glcf }&  Left coprime factorization with proper and stable factors.\\
  \hline
  \texttt{\bfseries grcfid}&  Right coprime factorization with inner denominator.\\
  \hline
  \texttt{\bfseries glcfid}&  Left coprime factorization with inner denominator.\\ \hline
  \texttt{\bfseries gnrcf}&  Normalized right coprime factorization.\\  \hline
  \texttt{\bfseries gnlcf }&  Normalized left coprime factorization.\\
  \hline
  \texttt{\bfseries giofac}&  Inner-outer and QR-like factorizations of a transfer function matrix.\\
  \hline
  \texttt{\bfseries goifac}&  Co-outer--coinner and RQ-like factorizations of a transfer function matrix.\\   \hline
  \texttt{\bfseries grsfg}&  Right spectral factorization of $\gamma^2I-G^\sim(\lambda)G(\lambda)$.\\
  \hline
  \texttt{\bfseries glsfg}&  Left spectral factorization of $\gamma^2I-G(\lambda)G^\sim(\lambda)$.\\ \hline
\multicolumn{2}{c}{} \\ \hline
\multicolumn{2}{|c|}{\textbf{Model-matching problems}} \\ \hline
\texttt{\bfseries grasol}&  Approximate solution of the linear rational matrix equation $G(\lambda)X(\lambda)\!=\!F(\lambda)$.\\  \hline
  \texttt{\bfseries glasol}&  Approximate solution of the linear rational matrix equation $X(\lambda)G(\lambda)\!=\!F(\lambda)$.\\
  \hline
  \texttt{\bfseries glinfldp}&  Solution of the least distance problem $\min_{X(\lambda)}\|[\, G_1(\lambda)-X(\lambda)\; G_2(\lambda)\,]\|_\infty$.\\   \hline
  \texttt{\bfseries gnehari}&  Generalized Nehari approximation.\\
  \hline
\multicolumn{2}{c}{} \\ \hline
\multicolumn{2}{|c|}{\textbf{Matrix pencils and stabilization}} \\ \hline
  \texttt{\bfseries gklf}&  Kronecker-like staircase forms of a linear matrix pencil.\\
  \hline
  \texttt{\bfseries gsklf}&  Special Kronecker-like form of a system matrix pencil.\\
  \hline
  \texttt{\bfseries gsorsf}&  Specially ordered generalized real Schur form.\\   \hline
  \texttt{\bfseries gsfstab}&  Generalized state-feedback stabilization.\\
  \hline
\multicolumn{2}{c}{} \\ \hline
\multicolumn{2}{|c|}{\textbf{SLICOT-based MEX-files}} \\ \hline
  \texttt{\bfseries \url{sl_gstra}}&  Descriptor system coordinate transformations.\\
  \hline
\texttt{\bfseries \url{sl_gminr}}&  Minimal realization of descriptor systems.\\  \hline
  \texttt{\bfseries \url{sl_gsep}}&  Descriptor system spectral separations.\\
  \hline
  \texttt{\bfseries \url{sl_gzero}}&  Computation of system zeros and Kronecker structure.\\  \hline
  \texttt{\bfseries \url{sl_klf}}&  Pencil reduction to Kronecker-like forms.\\   \hline
  \texttt{\bfseries \url{sl_glme}}&  Solution of generalized linear matrix equations.\\
  \hline
\end{longtable}
\end{center}

\subsection{Getting Started}

In this section  we shortly recall how to construct generalized LTI system objects using commands of the  Control System Toolbox (CST) of MATLAB \cite{MLCO15} and discuss some basic model conversion techniques and operations with rational matrices via their descriptor system representations. We also illustrate and compare the functionality of some functions available in the CST and \textbf{DSTOOLS} to perform model conversions and manipulations.

\subsubsection{Building Generalized LTI Models}

To describe generalized LTI systems via their TFM representations, the CST supports two model objects (or classes) called \texttt{\bfseries tf} and \texttt{\bfseries zpk}.   The model class \texttt{\bfseries tf} represents the elements of the TFM  as ratios of two polynomials, in the form (\ref{glambda}). The corresponding constructor command \texttt{\bfseries tf} can be used to build transfer functions for SISO or TFMs for MIMO LTI systems. The model class \texttt{\bfseries zpk}  represents the elements of the TFM in a zero-pole-gain (factorized) form as in (\ref{gkpz}) and the corresponding constructor command is \texttt{\bfseries zpk}.

A straightforward method to enter TFM based models relies on a powerful feature of the CST to connect subsystems. For example, to enter the continuous-time improper TFM
\[ G_c(s) = \left[\begin{array}{cc} s^2 & \displaystyle\frac{s}{s +1}\\[2mm]0 & \displaystyle\frac{1}{s} \end{array}\right],
\]
the following commands can be used
\begin{verbatim}
s = tf('s');                  % define the complex variable s
Gc = [s^2 s/(s+1); 0 1/s]     % define the 2-by-2 improper Gc(s)
\end{verbatim}
Similarly, the discrete-time improper TFM
\[ G_d(z) = \left[\begin{array}{cc} z^2 & \displaystyle\frac{z}{z - 2}\\[2mm] 0 & \displaystyle\frac{1}{z} \end{array}\right] , \]
with sampling time equal to 0.5, can be entered using the commands
\begin{verbatim}
z = tf('z',0.5);              % define the complex variable z
Gd = [z^2 z/(z-2); 0 1/z]     % define the 2-by-2 improper Gd(z)
\end{verbatim}

To describe LTI systems in state-space form, the model class \texttt{ss} is provided jointly with the class constructor commands \texttt{\bfseries dss}, for descriptor systems and \texttt{\bfseries ss} for standard state-space systems (with $E = I$).  For example, a descriptor system model of the form (\ref{app:dss}) can be constructed by entering the system matrices $E$, $A$, $B$, $C$ and $D$ and using the command \texttt{\bfseries dss}. The descriptor system with the state-space realization
\[ \ba{c|c}A-s E & B\\ \hline C & D \ea =
\left[\begin{array}{ccccc|cc} 1 & - s & 0 & 0 & 0 & 0 & 0\\ 0 & 1 & - s & 0 & 0 & 0 & 0\\ 0 & 0 & 1 & 0 & 0 & -1 & 0\\ 0 & 0 & 0 &  -1 -s & 0 & 0 & 1\\ 0 & 0 & 0 & 0 & - s & 0 & 1\\ \hline 1 & 0 & 0 & -1 & 0 & 0 & 1\\ 0 & 0 & 0 & 0 & 1 & 0 & 0 \end{array}\right]
\]
can be entered using the following commands:
\begin{verbatim}
A = [ 1 0 0 0 0; 0 1 0 0 0; 0 0 1 0 0; 0 0 0 -1 0; 0 0 0  0 0];
B = [ 0 0 -1 0 0; 0 0 0 1 1]';
C = [ 1 0 0 -1 0; 0 0 0 0 1];
D = [ 0 1; 0 0 ];
E = [ 0 1 0 0 0; 0 0 1 0 0; 0 0 0 0 0; 0 0 0 1 0; 0 0 0 0 1];
sys = dss(A,B,C,D,E);
\end{verbatim}
If $E = I$, the system model is a standard state-space model, which can be constructed using the class constructor command \texttt{\bfseries ss}.

A less known aspect of building descriptor system models is the available ``freedom''  in the CST to allow the construction of descriptor system models with a singular pole pencil $A-\lambda E$. This ``freedom'' is very questionable and may lead to potential conceptual and computational difficulties. This  can be easily illustrated with the following trivial non-regular descriptor system, for which many of the functions of the CST produce misleading warnings or questionable results, as shown by the following sequence of commands:
\begin{verbatim}
A = 0; E = 0; B = 1; C = 1; D = 0;
syst = dss(A,B,C,D,E);
pole(syst)       % the pole must be NaN, similar to eig(A,E), and not empty!
tf(syst)         % the transfer function is infinite and not NaN!
evalfr(syst,1)   % this evaluation of an infinite frequency response is correct
isproper(syst)   % this test is wrong, because the system is not proper
\end{verbatim}
The function \texttt{\bfseries gpole} of \textbf{DSTOOLS} can be used to check the regularity of the pole pencil. Both  commands \texttt{\bfseries gpole} and \texttt{\bfseries eig} below
\begin{verbatim}
gpole(syst)
eig(A,E)
\end{verbatim}
compute the ``correct'' value of the pole, which, in this case, is \texttt{NaN}.

The \textbf{DSTOOLS} collection exclusively deals with regular descriptor models, for which the pole pencil $A-\lambda E$ is regular. All functions of \textbf{DSTOOLS} guarantee that the computed results are regular descriptor systems, provided the input systems are regular.  We have to stress, that for efficiency reasons, in most of functions of \textbf{DSTOOLS}, the regularity condition for the input system data is only \emph{assumed}, but it is not explicitly checked. Therefore, it is likely that some functions, even if they perform without issuing error messages, may still deliver nonsense results if the regularity assumption is not fulfilled by the input system descriptions.

\subsubsection{Conversions between LTI Model Representations}

Most of functions of \textbf{DSTOOLS} accept only state-space system objects as inputs and therefore the input-output models must be converted to a standard or descriptor state-space form. For the above defined TFMs $G_c(s)$ and $G_d(z)$, this conversion can be done simply using
\begin{verbatim}
sysc = ss(Gc)
sysd = ss(Gd)
\end{verbatim}
The resulting state-space realization are usually non-minimal. Incidentally, both of the resulting state-space realizations \texttt{sysc} and \texttt{sysd} have order 5 and are minimal.

For visualization purposes, often the more compact TFM models are better suited than state-space models. Both class constructor commands \texttt{\bfseries tf} and \texttt{\bfseries zpk} can also serve to explicitly convert a LTI model to \texttt{tf} or \texttt{zpk} forms, respectively. The resulting transfer-function models computed from state-space models, usually contains uncancelled common factors in the numerator and denominator polynomials of the rational matrix elements, and, therefore, are non-minimal.

To compute minimal realizations, the function  \texttt{\bfseries minreal} is available in the CST. This function, applied to transfer-function models, enforces the cancellation of common factors in the numerator and denominator polynomials of each element.
However, this function is \emph{only} applicable to proper descriptor systems (also including the case of systems with singular $E$). When applied
to an improper descriptor system, an error message is issued:
\begin{verbatim}
order(minreal(sysc))
Error using DynamicSystem/minreal (line 53)
The "minreal" command cannot be used for models with more zeros than poles.
\end{verbatim}
Unfortunately, the above error message is misleading, since for the system \texttt{sysc} with an invertible TFM $G_c(s)$, the number of poles and zeros (counting also the infinite poles and zeros), must coincide in accordance with (\ref{pol-zer}). This can be checked with the functions \texttt{\bfseries gpole} and \texttt{\bfseries gzero}  of \textbf{DSTOOLS}:
\begin{verbatim}
POLES = gpole(sysc)
ZEROS = gzero(sysc)
\end{verbatim}
which produce the following results:
\begin{verbatim}
POLES =
   0.0000 + 0.0000i
  -1.0000 + 0.0000i
      Inf + 0.0000i
      Inf + 0.0000i

ZEROS =
  -1.0000 + 0.0000i
   0.0000 + 0.0000i
   0.0000 + 0.0000i
      Inf + 0.0000i
\end{verbatim}
These results also indicate that the McMillan degree of the system is 4 (recall that the order of the minimal descriptor state-space realization is 5).
In contrast, the functions \texttt{\bfseries pole} and \texttt{\bfseries tzero} of the CST, compute only the finite poles and finite zeros, respectively, and provide no hint on the actual McMillan degree.

Alternative functions to compute irreducible and minimal realizations, are, respectively, the functions \texttt{\bfseries gir} and \texttt{\bfseries gminreal} available in \textbf{DSTOOLS}. For example, the minimality of the above computed realizations can be checked with the function \texttt{\bfseries gminreal} of \textbf{DSTOOLS}:
\begin{verbatim}
order(gminreal(sysc))  % computes the order of the minimal realization
\end{verbatim}

\subsubsection{Conversion to Standard State-Space Form} \label{gss2ss}

A useful conversion which we discuss separately is the conversion of a descriptor system model of a proper system into a standard state-space model.
Specifically, we discuss  the conversion of a proper descriptor system of the form
\be\label{dssmodel} \begin{array}{rcl} E\lambda x(t) &=& Ax(t)+Bu(t) \, ,\\ y(t) &=& Cx(t)+Du(t) \, ,\end{array} \ee
with $x(t) \in \mathds{R}^n$,  to a standard state-space system of the form
\be\label{ssmodel1} \begin{array}{rcl} \lambda{\widetilde x}(t) &=& \widetilde A\widetilde x(t)+\widetilde Bu(t) \, ,\\ y(t) &=& \widetilde C\widetilde x(t)+\widetilde Du(t) \, ,\end{array} \ee
with $\widetilde x(t) \in \mathds{R}^{\tilde n}$ and $\tilde n \leq n$, and such that
the two systems have the same transfer function matrices, i.e.
\[ C(\lambda E-A)^{-1} B + D = \widetilde C(\lambda I_{\tilde n}-\widetilde A)^{-1} \widetilde B + \widetilde D .\]
This conversion is usually necessary, to obtain for the designed controllers and filters simpler representations, which are better suited for real-time processing. However, we cautiously recommend to avoid such conversions at early steps of the synthesis procedures, unless it is possible to guarantee that no significant loss of accuracy takes place due to ill-conditioned transformations.

For simplicity, we consider only the case when the descriptor realization $(A-\lambda E,B,C,D)$ is already irreducible. Such realizations can be obtained, for example, using the \textbf{DSTOOLS} function \texttt{\bfseries gir}. We further assume that the regular pencil $A-\lambda E$ has $r$ finite eigenvalues and  $n-r$ simple eigenvalues at infinity, where $r$ is the rank of $E$.
When $E$ is nonsingular, we can simply choose $\widetilde x(t) = x(t)$ and
\[ \widetilde A = E^{-1}A, \quad \widetilde B = E^{-1}B, \quad
\widetilde C = C, \quad \widetilde D = D , \]
or alternatively choose $\widetilde x(t) = Ex(t)$ and
\[ \widetilde A = AE^{-1}, \quad \widetilde B = B, \quad
\widetilde C = CE^{-1}, \quad \widetilde D = D \, .\]
In these conversion formulas, the inverse of $E$ is explicitly involved and, therefore, severe loss of accuracy can occur if the condition number $\kappa(E) := \big\|E\big\|_2\big\|E^{-1}\big\|_2$ is large.

A numerically better conversion approach is to use the \emph{singular value decomposition} (SVD)
\( E = U \Sigma V^T \), with $U$ and $V$ orthogonal matrices and $\Sigma$ a diagonal matrix whose diagonal elements are the decreasingly ordered singular values $\sigma_1 \geq \sigma_2 \geq \cdots \geq \sigma_n > 0$. We can choose $\widetilde x(t) = \Sigma^{\frac{1}{2}}V^Tx(t)$ and
\be\label{dss2ss-svd}  \widetilde A =  \Sigma^{-\frac{1}{2}}U^TAV\Sigma^{-\frac{1}{2}}, \quad \widetilde B = \Sigma^{-\frac{1}{2}}U^TB, \quad
\widetilde C = CV\Sigma^{-\frac{1}{2}}, \quad \widetilde D = D \, .\ee
From the SVD of $E$, we can easily compute the condition number $\kappa(E) = \sigma_1/\sigma_n$, and thus have a rough estimation of potential loss of accuracy induced by using the above transformation.

It is also possible to use the QR-decomposition $E = QR$, with $Q$ orthogonal and $R$ upper-triangular and with positive diagonal elements (this can always be arranged using an orthogonal diagonal scaling matrix). Let $R^{\frac{1}{2}}$ be the upper-triangular square root of $R$, which can be computed using the method described in \cite[Algorithm 6.7]{High08}, which is also implemented in the MATLAB function \texttt{\bfseries sqrtm}.
We can choose $\widetilde x(t) = R^{\frac{1}{2}}x(t)$ and
\be\label{dss2ss-qr} \widetilde A =  R^{-\frac{1}{2}}Q^TAR^{-\frac{1}{2}}, \quad \widetilde B = R^{-\frac{1}{2}}Q^TB, \quad
\widetilde C = CR^{-\frac{1}{2}}, \quad \widetilde D = D \, .\ee
From the QR-decomposition of $E$, we can easily compute the condition number $\kappa(E) = \kappa(R)$ (e.g., by using \texttt{rcond(R)} to estimate the reverse condition number $1/\kappa(R)$). In this way, we can have a rough estimation of potential loss of accuracy induced by using the above transformation. Although this method can be generally used, its primary use is when the original pair $(A,E)$ is already in a GRSF, with $A$ upper quasi-triangular and $E$ upper triangular. In this case, the resulting $\widetilde A$ is in a RSF.

More involved transformation is necessary when $E$ is singular, with $\rank\,E = r < n$. In this case, we can employ the singular value decomposition of  $E$ in the form
\be\label{dss2ss-svd1}  E = U \Sigma V^T := \ba{cc} U_1 & U_2 \ea \ba{cc}\widetilde\Sigma & 0 \\ 0 & 0 \ea  \ba{cc} V_1 & V_2 \ea^T  ,\ee
where $\widetilde\Sigma$ is a nonsingular diagonal matrix of order $\tilde n := r$ with the nonzero singular values of $E$ on the diagonal, and $U$ and $V$ are compatibly partitioned orthogonal matrices. If we apply a system similarity transformation with the transformation matrices
\[ \widetilde U = \diag\big(\widetilde \Sigma^{-\frac{1}{2}}, I_{n-r}\big)U^T , \quad \widetilde V= V\diag\big(\widetilde \Sigma^{-\frac{1}{2}}, I_{n-r}\big) \] we obtain
\be\label{dss2ss-svd2} \widetilde U (A-\lambda E) \widetilde V = \ba{cc} A_{11}-\lambda I_{r} & A_{12}\\ A_{21} & A_{22} \ea, \quad \widetilde U B = \ba{c} B_1\\ B_2 \ea, \quad C\widetilde V = \ba{cc} C_1 & C_2 \ea \, , \ee
where $A_{22}$ is nonsingular, due to the assumption of only simple infinite eigenvalues of the regular pencil $A-\lambda E$.
The above transformed matrices correspond to the coordinate transformation $\overline x = \widetilde V^{-1}x(t)$ and lead to the partitioned system representation
\[ \begin{array}{ccl} \lambda{\overline x}_1(t) & = & A_{11}\overline x_1(t) + A_{12}\overline x_2(t) + B_1 u(t) \, ,\\
0 & = & A_{21}\overline x_1(t) + A_{22}\overline x_2(t) + B_2 u(t) \, ,\\
y(t) &=& C_1\overline x_1(t)+ C_2\overline x_2(t) + Du(t)\, , \end{array}
\]
where $\overline x(t) = \ba{c}\overline x_1(t) \\ \overline x_2(t)\ea$ is partitioned such that $\overline x_1(t) \in \mathds{R}^{r}$ and $\overline x_2(t) \in \mathds{R}^{n-r}$. We can solve the second (algebraic) equation for $\overline x_2(t)$ to obtain
\[  \overline x_2(t) = -A^{-1}_{22}A_{21}^{}\overline x_1^{}(t) -A^{-1}_{22}B_2^{}u(t) \]
and arrive to a standard system representation with $\widetilde x(t) = \overline x_1(t)$ and the corresponding matrices
\be\label{dss2ss-svd3} \begin{array}{ll} \widetilde A = A_{11}^{}-A_{12}^{}A^{-1}_{22}A_{21}^{}, & \quad\widetilde B = B_1^{}- A_{12}^{}A^{-1}_{22}B_2^{}, \\ \\[-1mm] \widetilde C = C_1^{}- C_2^{} A^{-1}_{22}A_{21}^{}, & \quad\widetilde D = D -C_2^{}A^{-1}_{22}B_2^{} .\end{array}\ee
In this case, if any of the condition numbers $\kappa(\widetilde \Sigma)$ or $\kappa(A_{22})$ is large, potential accuracy losses can be induced by the conversion to a standard state-space form.

The conversion formulas (\ref{dss2ss-svd}) and (\ref{dss2ss-svd1})-(\ref{dss2ss-svd3}) underly the implementation of the function \texttt{\bfseries gss2ss} of \textbf{DSTOOLS} (used as default options). The formulas
(\ref{dss2ss-qr}) are used, by default, if the input pair $(A,E)$ is in a GRSF, with $E$ invertible. The CST function \texttt{\bfseries isproper} has a hidden input flag (\texttt{'explicit'}), which allows to convert proper descriptor models to a standard state-space form.

\subsubsection{Sensitivity Issues for Polynomial-Based Representations}
The ill-conditioning of polynomial-based representations was a constant discussion subject in the control literature to justify the advantage of using state-space realization-based models for numerical computations. For an authoritative  discussion see \cite{Door81}.  More details on these issues are provided in Chapter 6 of \cite{Door03}.

The extreme sensitivity of roots of polynomials with respect to small variations in the coefficients, illustrated in the following example, is well known in the literature and is inherent for polynomial-based representations above a certain degree (say $n > 10$). Therefore, all algorithms which involve rounding errors are doomed to fail by giving results of extremely poor accuracy when dealing with an ill-conditioned polynomial. This potential loss of accuracy
is one of the main reasons why polynomial-based system representations with rational or polynomial matrices are generally not suited for numerical computations.

It is well known that polynomials with multiple roots are very sensitive to small variations in the coefficients. However, it is less known that this large sensitivity may be present even in the case of polynomials with well separated roots, if the order of the polynomial is sufficiently large.
This will be illustrated by the following example.
\begin{example}\label{ex:wilkinson}
The simple transfer function
\[ g(s) = \frac{1}{(s+1)(s+2)\cdots(s+25)} = \frac{1}{s^{25}+325s^{24} + \cdots + 25! } \]
has the exact poles $\{-1, -2, \ldots, -25\}$. The denominator is a modification coined by Daniel Kressner (private communication) of the famous Wilkinson polynomial analyzed in \cite{Wilk84} (originally of order 20 and with positive roots). This polynomial has been used in many works to illustrate the pitfalls of algorithms for computing eigenvalues of a matrix by computing the roots of its characteristic polynomial.

If we explicitly construct the transfer function $g(s)$ and compute its poles using the MATLAB commands
\begin{verbatim}
g = tf(1,poly(-25:1:-1));
sys = ss(g);
pole(sys)
\end{verbatim}
inaccurate poles with significant imaginary parts result, as can be observed from Fig.~\ref{FigEx_Poles}.
For example, instead the poles at $-19$ and $-20$, two complex conjugate poles at $-19.8511 \pm 3.2657\mathrm{i}$ result.\footnote{Computed with \textsc{Matlab} R2015b Version, running under 64-Bit Microsoft Windows 10}

\begin{figure}[h]
\begin{center}
\includegraphics[width=10cm]{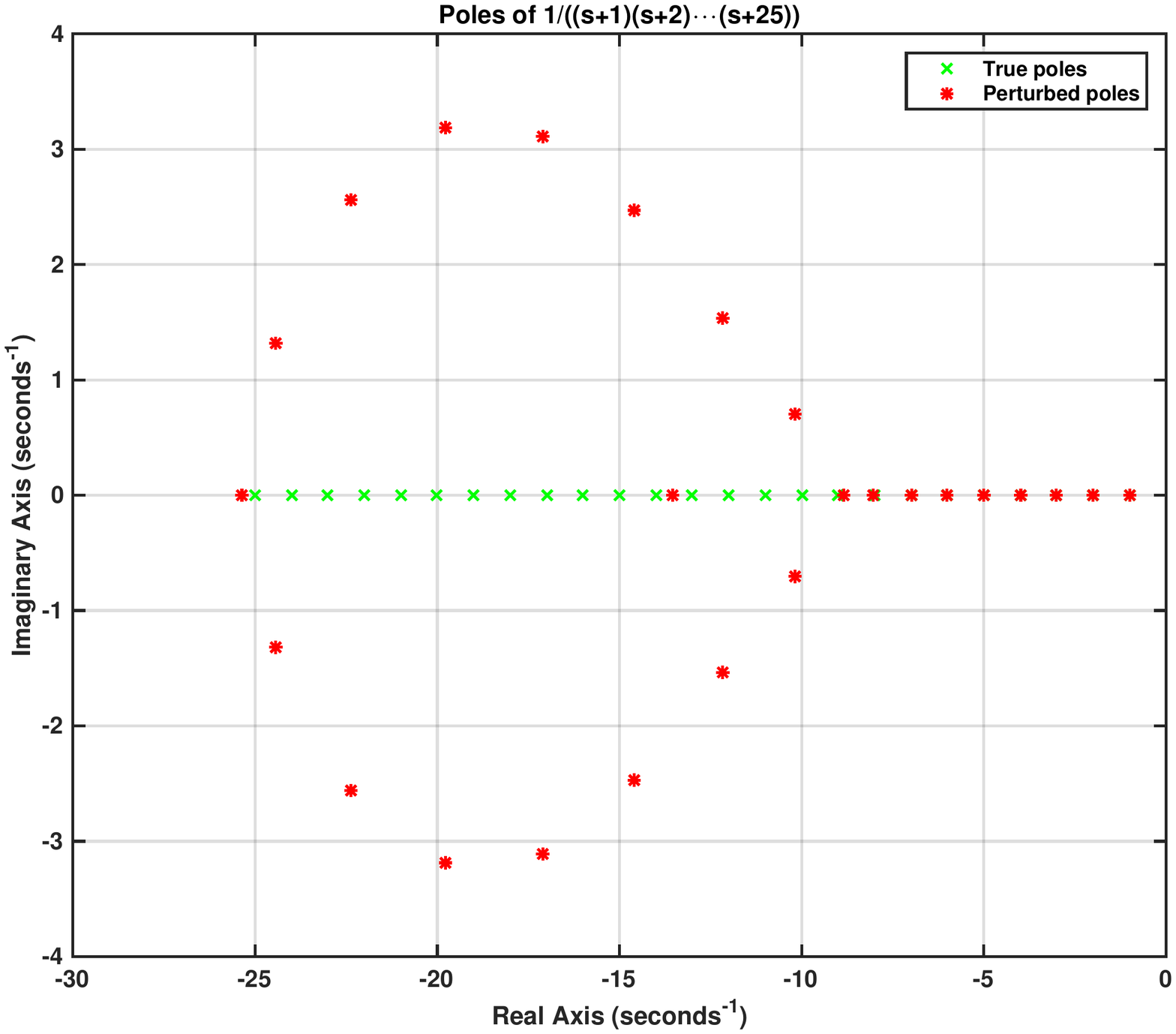} \caption{Example for high sensitivity of polynomial poles } 
 \label{FigEx_Poles}
\end{center}
\end{figure}

The main reason for these inaccurately computed  poles is the high sensitivity of the polynomial roots to small variations in the coefficients. In this case, inherent truncations take place in representing the large integer coefficients due to the finite representation with 16 accurate digits of double-precision floating-point numbers. For example, the constant term in the denominator
$25! \approx 1.55\cdot 10^{25}$ has 25 decimal digits, thus can not be exactly represented with 16 digits precision. While the relative error in representing $25!$ is of the order of the machine-precision $\varepsilon_M \approx 10^{-16}$, the absolute error is of the order of $10^9$!

For this particular example, the zero-pole-gain based representation is possibly better suited as starting point for building a state-space realization. For example, a state-space realization of $g(s)$ can be constructed, which still preserves the full accuracy of poles, as shown in the following example:
\begin{verbatim}
s = zpk('s'); g = 1; for i=1:20, g = g/(s+i); end
sys = ss(g);
pole(sys)
\end{verbatim}

Interestingly, the following command sequence builds a 60-th order descriptor system realization of $g(s)$, which contains 40 non-dynamics modes. After eliminating the non-dynamic modes and converting the realization to a standard state-space model, the full accuracy of poles is still preserved to machine precision.
\begin{verbatim}
s = ss(tf('s')); g = 1; for i=1:20, g = g/(s+i); end
sys = gss2ss(g);
pole(sys)
\end{verbatim}
\fine
\end{example}

This example illustrates that using descriptor system models for model building may occasionally alleviate the numerical difficulties which are inherent when using polynomial based models.

\subsubsection{Operations with Rational Matrices}
In this section we succinctly discuss the operations with rational matrices via their descriptor system realizations. These operations play an important role in implementing many of the functions available in \textbf{DSTOOLS} and therefore, we present some illustrations of the usage of these operations. Let $G(\lambda)$ be a rational TFM having the descriptor system realization $(A-\lambda E,B,C,D)$. We assume that two system objects have been defined to represent $G(\lambda)$: the transfer-function model \texttt{g}  (e.g., defined with either the \texttt{tf} or  \texttt{zpk} constructors)  and the state-space model \texttt{sys}  (e.g., defined either with the \texttt{ss} or the \texttt{dss} constructors).

The \emph{transpose} $G^T(\lambda)$ and the corresponding \emph{dual} descriptor system realization $G^T(\lambda)  = (A^T-\lambda E^T,C^T,B^T,D^T)$ can be simply computed as \texttt{g.'} and \texttt{sys.'}, respectively. This operation can be useful to implement, for example, a left oriented function by using an already available right oriented function. For example, to compute a left coprime factorization of $G(\lambda)$ as $G(\lambda) = M^{-1}(\lambda)N(\lambda)$, the function \texttt{\bfseries glcf} calls the function \texttt{\bfseries grcf} to compute a right coprime factorization of the transpose as $G^T(\lambda) = \widetilde N(\lambda)\widetilde M^{-1}(\lambda)$ and sets $N(\lambda) = \widetilde N^T(\lambda)$ and $M(\lambda) = \widetilde M^T(\lambda)$. An interesting aspect of computing right coprime factorizations with the function \texttt{\bfseries grcf} is that both resulting descriptor realizations of $\widetilde N(\lambda) = (\widetilde A_N-\lambda \widetilde E_N,\widetilde B_N,\widetilde C_N,\widetilde D_N) $ and $\widetilde M(\lambda) = (\widetilde A_M-\lambda \widetilde E_M,\widetilde B_M,\widetilde C_M,\widetilde D_M)$ have the resulting pairs $(\widetilde A_N, \widetilde E_N)$ and $(\widetilde A_M, \widetilde E_M)$ in a GRSF (i.e., with the pole pencils upper quasi-triangular). This condensed form of the pole pencils may be useful for further computations, where this structure can be efficiently exploited (e.g., the eigenvalues can be computed at no cost using the function \texttt{\bfseries ordeig} or the function \texttt{\bfseries grcf} can be applied a second time, with a significantly less computational burden). Therefore, it is highly desirable that the left oriented function \texttt{\bfseries glcf} preserves the upper quasi-triangular shape of the pole pencils of the factors. However,  by simply forming the transposed matrices of the dual descriptor systems, this useful property is lost. To preserve the upper quasi-triangular form of the pole pencil of the dual system, an alternative realization of the dual system can be constructed as $G^T(\lambda) = (PA^TP-\lambda PE^TP,PC^T,B^TP,D^T)$, where $P$ is the (orthogonal) permutation matrix with ones down on the secondary diagonal
\[ P = {\arraycolsep=0.5mm\def\arraystretch{0.1}\ba{rcc} ~0&& 1\\[-1mm] & \reflectbox{$\ddots$} \\ 1 &&0~\ea} \, . \]

The following sequence of commands is used in the function \texttt{\bfseries glcf} to efficiently build permuted dual realizations using the function \texttt{\bfseries xperm} of the CST:
\begin{verbatim}
% apply grcf to the dual system, by trying to preserve upper-triangular shapes
[sysn,sysm] = grcf(xperm(sys,order(sys):-1:1).',options);
% build dual factors, by preserving upper-triangular shapes
sysn = xperm(sysn,order(sysn):-1:1).';
sysm = xperm(sysm,order(sysm):-1:1).';
\end{verbatim}

If $G(\lambda)$ is invertible, then an inversion free realization of the \emph{inverse} TFM $G^{-1}(\lambda)$ is given by
\index{descriptor system!inverse}
\be\label{invsys-dss} G^{-1}(\lambda) = \ba{cc|c} A-\lambda E & B & 0\\C & D & I \\ \hline 0 & -I & 0 \ea \, .\ee
This realization is not minimal, even if the original realization is minimal.
If $D$ is invertible, then an alternative realization of the inverse is
\be\label{invsys-ss} G^{-1}(\lambda) = \ba{c|c} A-BD^{-1}C-\lambda E & -BD^{-1} \\ \hline \\[-3.5mm] D^{-1}C & D^{-1} \ea ,\ee
which is minimal if the original realization is minimal.

To compute inverse systems, the overloaded function \texttt{inv} can be used to compute the inverse systems as \texttt{inv(g)} or \texttt{inv(sys)}. The function \texttt{inv} applied to a descriptor state-space model always builds the realization (\ref{invsys-dss}) of the inverse   (even if $E = I$), while for standard state-space systems  (with $E = I$), the realization (\ref{invsys-ss}) is used, provided $D$ is reasonably well-conditioned with respect to the inversion. For transfer-function models, an automatic conversion to state-space form is performed, and the computed inverse is converted back to the transfer-function form. These conversions often lead to non-minimal representations containing uncancelled factors between the numerator and denominators of the matrix elements.

Operations involving inverses can often be performed using the overloaded matrix operators \url{\ } (left divide) or \texttt{/} (right divide), as for example, to compute \url{sys1\ sys2} or \texttt{sys1/sys2} for two systems \texttt{sys1} and \texttt{sys2}. However, these operation do not avoid the explicit building of the inverses, as can be seen by performing \texttt{sys1/sys1} or \url{sys2\ sys2}, which, instead of producing a non-dynamic systems with  the direct feedthrough gain equal to the identity matrix,  determine realizations of orders (at least) double of the orders of \texttt{sys1} or \texttt{sys2}. Alternative computation of \url{sys1\ sys2} can be done using the function \texttt{\bfseries glsol} of \textbf{DSTOOLS}, while for the evaluation of \texttt{sys1/sys2}, the function \texttt{\bfseries grsol} of \textbf{DSTOOLS} can be used. These functions are generally applicable to solve compatible systems of linear rational matrix equations and always determine minimal order realizations of the solutions. %

\index{descriptor system!conjugate}
The \emph{conjugate} (or \emph{adjoint}) TFM $G^\sim(\lambda)$ is defined in the continuous-time case as $G^\sim(s) = G^T(-s)$ and has the realization
\[ G^\sim(s) = \ba{c|c} -A^T-s E^T & C^T \\ \hline \\[-3mm] -B^T & D^T \ea \, , \]
while in the discrete-time case $G^\sim(z) = G^T(1/z)$ and has the realization
\[ G^\sim(z) = \ba{cc|c} E^T -z A^T & 0 & -C^T \\
z B^T & I & D^T \\ \hline 0 & I & 0 \ea \, .\]
If $G(z)$ has a standard state-space realization $(A,B,C,D)$ with $A$ invertible, then an alternative realization of $G^\sim(z)$ is
\[ G^\sim(z) = \ba{c|c} A^{-T} -z I &  -A^{-T}C^T \\ \hline \\[-3mm]
B^TA^{-T} &  D^T -B^TA^{-T}C^T \ea \, .\]
This realization is only recommended if $A$ is well-conditioned with respect to the inversion.

The {conjugate} of a transfer function model \texttt{g} can be simply computed as \texttt{g'}, while for a state-space model \texttt{sys} with \texttt{sys'}. For a discrete-time state-space model, the conjugate system is always a descriptor system, which is usually non-minimal (frequently contains non-dynamic modes).

\subsection{Functions for System Analysis} \label{dstools:analysis}

The system analysis functions cover the computation of poles and zeros, of normal rank and Hankel norm of the transfer function matrix of a LTI descriptor system.

\subsubsection{\texttt{\bfseries gpole}}

\subsubsection*{Syntax}
\index{M-functions!\texttt{\bfseries gpole}}
\begin{verbatim}
[POLES,INFO] = gpole(SYS)
[POLES,INFO] = gpole(SYS,TOL)
[POLES,INFO] = gpole(SYS,TOL,OFFSET)
\end{verbatim}

\subsubsection*{Description}

\noindent \texttt{gpole} computes for a LTI descriptor system, the finite and infinite zeros of the pole pencil, and provides information related to the pole pencil structure.
\index{transfer function matrix (TFM)!poles}%
\index{transfer function matrix (TFM)!zeros}%
\index{transfer function matrix (TFM)!poles!finite}%
\index{transfer function matrix (TFM)!poles!infinite}%
\index{transfer function matrix (TFM)!zeros!finite}%
\index{transfer function matrix (TFM)!zeros!infinite}%
\index{linear matrix pencil!zeros}%
\index{linear matrix pencil!zeros!finite}%
\index{linear matrix pencil!zeros!infinite}%

\subsubsection*{Input data}

\begin{description}
\item
\texttt{SYS} is a LTI system in a descriptor system state-space form
\be\label{gpole:sysss}
\begin{aligned}
E\lambda x(t)  &=   Ax(t)+ B u(t) ,\\
y(t) &=  C x(t)+ D u(t) .
\end{aligned}
\ee
\item
\texttt{TOL} is a relative tolerance used for rank determinations. If \texttt{TOL} is not specified as input or if \texttt{TOL} = 0, an internally computed default value is used.
\item
 \texttt{OFFSET}  is the stability
 boundary offset $\beta$, to be used  to assess the finite eigenvalues which belong to $\partial\mathds{C}_s$ (the boundary of the stability domain) as follows: in the
 continuous-time case these are the finite
  eigenvalues having real parts in the interval $[-\beta, \beta]$, while in the
 discrete-time case these are the finite eigenvalues having moduli in the
 interval $[1-\beta, 1+\beta]$. (Default: $\beta = 1.4901\cdot 10^{-08}$). \item
\end{description}

\subsubsection*{Output data}

\begin{description}
\item
\texttt{POLES} is a complex column vector which contains the zeros (finite and infinite) of the \emph{pole pencil} $A-\lambda E$. These are the poles of the TFM of \texttt{SYS}
    if the pencil $A-\lambda E$ is regular and the descriptor realization $(A-\lambda E,B,C,D)$ is irreducible. If the pencil $A-\lambda E$ is not regular, a number of components of \texttt{POLES} equal to the rank deficiency of $A-\lambda E$ are set to \texttt{NaN}.
\item
 \texttt{INFO} is a MATLAB structure containing additional structural information related to the pole pencil $A-\lambda E$, as follows:\\
\pagebreak[2]
{\begin{longtable}{|l|p{12cm}|} \hline
\textbf{\texttt{INFO} fields} & \textbf{Description} \\ \hline
 \texttt{nfev}   & number of finite eigenvalues of the pencil $A-\lambda E$,
           and also the number of finite poles of \texttt{SYS} if \texttt{SYS} is irreducible; \\ \hline
 \texttt{niev}   & number of infinite eigenvalues of the pencil $A-\lambda E$;\\ \hline
 \texttt{nisev}   & number of simple infinite eigenvalues of the
          pencil $A-\lambda E$ (also the number of non-dynamic modes); \\ \hline
 \texttt{nip}   & number of infinite zeros of the pencil $A-\lambda E$, and also the number of infinite poles of the system \texttt{SYS}  if \texttt{SYS} is irreducible;  \\ \hline
 \texttt{nfsev}   & number of (stable) finite  eigenvalues in $\mathds{C}_s$ (the stability  domain);  \\ \hline
\texttt{nfsbev}   & number of finite eigenvalues in $\partial\mathds{C}_s$ (the boundary of the stability  domain);  \\ \hline
 \texttt{nfuev}   & number of (unstable) finite  eigenvalues in $\mathds{C}_u$ (the open instability  domain without including infinity);  \\ \hline
 \texttt{nhev}   & number of hiden eigenvalues of the pencil $A-\lambda E$ (can be nonzero only if the pencil $A-\lambda E$  is singular, see \textbf{Method});  \\ \hline
 \texttt{nrank}   & normal rank of the pencil $A-\lambda E$;  \\ \hline
\texttt{miev}   & integer row vector, which contains the multiplicities of the infinite eigenvalues of the pencil $A-\lambda E$  (also the dimensions of the
           elementary infinite blocks in the Kronecker form of $A-\lambda E$);  \\ \hline
 \texttt{mip}   & integer row vector, which contains the information on the
           multiplicities of the infinite zeros of the pencil $A-\lambda E$ as follows:
           $A-\lambda E$ has \texttt{INFO.mip$(i)$} infinite zeros of multiplicity $i$.
           \texttt{INFO.mip} is empty if $A-\lambda E$ has no infinite zeros.  \\ \hline
 \texttt{kr}   & integer row vector, which contains the right Kronecker
           indices of the pencil $A-\lambda E$.  For a regular pencil, this
           vector is empty.   \\ \hline
 \texttt{kl}   & integer row vector, which contains the left Kronecker
           indices of the pencil $A-\lambda E$.  For a regular pencil, this
           vector is empty.   \\ \hline
 \texttt{regular}   & logical value, which is set to \texttt{true} if the pencil $A-\lambda E$ is regular, or to  \texttt{false}, if the pencil $A-\lambda E$ is singular. \\ \hline
 \texttt{proper}   & logical value, which is set to \texttt{true} if the pencil $A-\lambda E$ is regular and all its infinite eigenvalues are simple (has only non-dynamic modes), or to  \texttt{false}, if the pencil $A-\lambda E$ is singular or it is regular,
                 but has non-simple infinite eigenvalues. \\ \hline
\texttt{stable}   & logical value, which is set to \texttt{true} if the pencil $A-\lambda E$ is regular and has only stable finite  eigenvalues and all its infinite eigenvalues are simple (has only non-dynamic modes), or to  \texttt{false}, if the pencil $A-\lambda E$ is singular or it is regular,
                 but has unstable finite eigenvalues or  non-simple infinite eigenvalues. \\ \hline
\end{longtable}}

\end{description}

\subsubsection*{Method}
Let $G(\lambda)$ be the TFM $G(\lambda) = C(\lambda E-A)^{-1}B+D$ of the LTI system \texttt{SYS}.
For the definition of the poles of $G\lambda)$ in terms of the descriptor realization (\ref{gpole:sysss}), see Section \ref{app:tfm_polzer}. If the descriptor system realization
 $(A-\lambda E,B,C,D)$ of \texttt{SYS} has a regular pole pencil $A-\lambda E$ and is irreducible (i.e., controllable and observable), then the computed finite poles in \texttt{POLES} are simply the finite generalized eigenvalues of the pair $(A,E)$, and the multiplicities of the infinite generalized eigenvalues of $(A,E)$ are in excess with one with respect to the multiplicities of the infinite poles. For the computation of the eigenvalues of $A-\lambda E$ and the information related to its Kronecker structure, the zeros computation algorithm of \cite{Misr94} (see also \cite{Svar85}) is applied to the particular system matrix pencil $S(\lambda) := A-\lambda E$ (i.e., of a system without inputs and outputs), by calling the MEX-function \url{sl_gzero}. This algorithm determines the following structural information related to the pencil $A-\lambda E$:
 \begin{itemize}
 \item $n_f$ finite zeros of $A-\lambda E$, $\lambda_i$, $i = 1, \ldots, n_f$ ($n_f$ is provided in \texttt{INFO.nfev});
  \item the dimensions $s_i^\infty$, for $i = 1, \ldots, h$, of the elementary infinite Jordan blocks in (\ref{Jordan-null}) in the Weierstrass canonical form of the regular part (\ref{regblocks}), representing the multiplicities of infinite eigenvalues; the number of infinite eigenvalues is $n_\infty = \sum_{i=1}^{h}s_i^\infty$, of which $n^0_\infty$ are simple infinite eigenvalues  for which $ s_i^\infty = 1$ ($n_\infty$ is provided in \texttt{INFO.niev}, $n^0_\infty$  is provided in \texttt{INFO.nsiev}, and the multiplicities of the infinite eigenvalues are provided in \texttt{INFO.miev});
 \item the numbers $m_i^\infty$, for $i = 1, \ldots, k$, where $m_i^\infty$ is the number of the elementary infinite Jordan blocks of order $i+1$ in (\ref{Jordan-null}) in the Weierstrass canonical form of the regular part (\ref{regblocks}); the number of infinite zeros is  $n_{p,\infty} = \sum_{i=1}^{k}im_i^\infty$ (also equal to $n_\infty-n^0_\infty$)  ($n_{p,\infty} $ is provided in \texttt{INFO.nip} and the numbers $m_i^\infty$, for $i = 1, \ldots, k$,  are provided in \texttt{INFO.mip});
 \item $\nu_r$ right Kronecker indices $\epsilon_i = 0$, for $i = 1, \ldots, \nu_r$ of the pencil $A-\lambda E$, corresponding to the $\nu_r$ Kronecker blocks $L_{\epsilon_i}(\lambda)$  of the form $\epsilon_i\times(\epsilon_i+1)$, which are part of the Kronecker-form of the pencil $A-\lambda E$, as in (\ref{kcf-kreps}) (the right Kronecker indices are provided in \texttt{INFO.kr});
 \item $\nu_l$ left Kronecker indices $\eta_i = 0$, for $i = 1, \ldots, \nu_l$ of the pencil $A-\lambda E$, corresponding to the $\nu_l$ Kronecker blocks $L_{\eta_i}^T(\lambda)$  of the form $(\eta_i+1)\times\eta_i$, which are part of the Kronecker-form of the pencil $A-\lambda E$, as in (\ref{kcf-keta}) (the left Kronecker indices are provided in \texttt{INFO.kl});
 \item $\rho := n_f+n_\infty + \sum_{i=1}^{\nu_r}\epsilon_i + \sum_{i=1}^{\nu_l} \eta_i$, the normal rank of the $n$-th order pencil $A-\lambda E$, where $n-\rho$ is the rank deficiency of $A-\lambda E$ ($\rho$ is provided in \texttt{INFO.nrank}).
\end{itemize}
The multiplicities of the infinite zeros are related to the orders of the elementary infinite blocks, as follows: to each elementary infinite block of order $s_i^\infty > 1$ corresponds an infinite zero of multiplicity $s_i^\infty-1$.

The presence of the right or left Kronecker indices indicates that the pencil $A-\lambda E$ is singular. In this case, a number of $\nu_r+\nu_l$ generalized eigenvalues of the pair $(A,E)$ are \emph{hidden} (their values depend on the employed transformation matrices) and a number of eigenvalues corresponding to the rank deficiency of $A-\lambda E$  are not defined, because would involve forming the fractions $\frac{0}{0}$. Therefore, the components of the vector \texttt{POLES} consist in general of: $n_f$ finite values $\lambda_i$, $i = 1, \ldots, n_f$; $n_\infty$ infinite values; and $n-\rho$ values set to \texttt{NaN}. The hidden eigenvalues are not included in \texttt{POLES}. \emph{Note: The term hidden eigenvalue is not standard in the literature and has been introduced only for convenience. Some hidden eigenvalues may explicitly appear when directly using the function} \texttt{eig} \emph{as} \texttt{eig(SYS.a,SYS.e)}. \emph{Interestingly,} \texttt{eig(SYS.a.',SYS.e.')} \emph{may produce different hidden eigenvalues!}

The number of finite stable eigenvalues of $A-\lambda E$ lying in $\mathds{C}_s$ (the appropriate open stability domain) is provided in \texttt{INFO.nfsev}.
The number of finite unstable eigenvalues of $A-\lambda E$ lying in $\mathds{C}_u$ (the appropriate open instability domain without including infinity) is provided in \texttt{INFO.nfuev}.
The number of finite eigenvalues of $A-\lambda E$ lying in $\partial\mathds{C}_s$ (the boundary of the appropriate stability domain) is provided in \texttt{INFO.nfsbev}. If these eigenvalues have multiplicity one, they are also called the \emph{marginally stable poles} of the system \texttt{SYS}. The stability boundary offset $\beta$ specified in \texttt{OFFSET} is used to numerically assess  if an eigenvalue belongs or not to $\mathds{C}_s$, $\partial\mathds{C}_s$ or $\mathds{C}_u$.

Additional qualitative information is provided in the \texttt{INFO} structure, to characterize some properties related to system poles:
\begin{itemize}
\item
\texttt{INFO.regular} is set to \texttt{true} if $A-\lambda E$ is a regular pencil, and is set to \texttt{false} for a singular $A-\lambda E$ (in which case $\rho < n$ and $\det (A-\lambda E) \equiv 0$).
\item
\texttt{INFO.proper} is set to \texttt{true} if $A-\lambda E$ is a regular pencil and all its infinite eigenvalues are simple. This property characterizes a \emph{proper} system \texttt{SYS}, provided the descriptor realization $(A-\lambda E,B,C,D)$ is irreducible. Otherwise, \texttt{INFO.proper} is set to \texttt{false} (i.e., for a singular $A-\lambda E$ or if non-simple infinite eigenvalues are present).
\item
\texttt{INFO.stable} is set to \texttt{true} for an exponentially stable system, for which $A-\lambda E$ is a regular pencil, with all its finite eigenvalues belonging to the stability domain $\mathds{C}_s$, and with only simple infinite eigenvalues. Otherwise, \texttt{INFO.stable} is set to \texttt{false}.
\end{itemize}
\index{MEX-functions!\url{sl_gzero}}

\subsubsection{\texttt{\bfseries gzero}}

\subsubsection*{Syntax}
\index{M-functions!\texttt{\bfseries gzero}}
\begin{verbatim}
[Z,INFO] = gzero(SYS)
[Z,INFO] = gzero(SYS,TOL)
[Z,INFO] = gzero(SYS,TOL,OFFSET)
\end{verbatim}

\subsubsection*{Description}

\noindent \texttt{gzero} computes for a LTI descriptor system, the finite and infinite zeros of the system matrix pencil, and provides information related to the system matrix pencil structure.

\index{transfer function matrix (TFM)!zeros}%
\index{transfer function matrix (TFM)!zeros!finite}%
\index{transfer function matrix (TFM)!zeros!infinite}%
\index{transfer function matrix (TFM)!normal rank}%

\subsubsection*{Input data}

\begin{description}
\item
\texttt{SYS} is a LTI system in a descriptor system state-space form
\be\label{gzero:sysss}
\begin{aligned}
E\lambda x(t)  &=   Ax(t)+ B u(t) ,\\
y(t) &=  C x(t)+ D u(t) .
\end{aligned}
\ee
\item \texttt{TOL} is a relative tolerance used for rank determinations. If \texttt{TOL} is not specified as input or if \texttt{TOL} = 0, an internally computed default value is used.
\item
 \texttt{OFFSET}  is the stability
 boundary offset $\beta$, to be used  to assess the finite zeros which belong to $\partial\mathds{C}_s$ (the boundary of the stability domain) as follows: in the
 continuous-time case these are the finite
  zeros having real parts in the interval $[-\beta, \beta]$, while in the
 discrete-time case these are the finite zeros having moduli in the
 interval $[1-\beta, 1+\beta]$. \newline (Default: $\beta = 1.4901\cdot 10^{-08}$). \item
\end{description}

\subsubsection*{Output data}

\begin{description}
\item
\texttt{Z} is a complex column vector which contains the invariant zeros (finite and infinite) of the system matrix pencil
\be\label{gzero-syspencil} S(\lambda) = \ba{cc} A-\lambda E & B \\ C & D \ea . \ee
These are also called the transmission zeros of the TFM $G(\lambda)$ of \texttt{SYS}
    if the descriptor realization $(A-\lambda E,B,C,D)$ is irreducible.
\item
 \texttt{INFO} is a MATLAB structure containing additional structural information related to the system matrix pencil $S(\lambda)$, as follows:\\
\pagebreak[2]
{\begin{longtable}{|l|p{12cm}|} \hline
\textbf{\texttt{INFO} fields} & \textbf{Description} \\ \hline
 \texttt{nfz}   & number of finite zeros of the pencil $S(\lambda)$; \\ \hline
 \texttt{niev}   & number of infinite eigenvalues of the pencil $S(\lambda)$;\\ \hline
 \texttt{nisev}   & number of simple infinite eigenvalues of the
          pencil $S(\lambda)$; \\ \hline
 \texttt{niz}   & number of infinite zeros of the pencil $S(\lambda)$;  \\ \hline
 \texttt{nfsz}   & number of finite stable zeros in $\mathds{C}_s$ (the stability  domain);  \\ \hline
\texttt{nfsbz}   & number of finite zeros in $\partial\mathds{C}_s$ (the boundary of the stability  domain);  \\ \hline
 \texttt{nfuz}   & number of finite unstable zeros in $\mathds{C}_u$ (the open instability  domain without including infinity);  \\ \hline
 \texttt{nrank}   & normal rank of the pencil $S(\lambda)$;  \\ \hline
\texttt{miev}   & integer row vector, which contains the multiplicities of the infinite eigenvalues of the pencil $S(\lambda)$  (also the dimensions of the
           elementary infinite blocks in the Kronecker form of $S(\lambda)$);  \\ \hline
 \texttt{miz}   & integer row vector, which contains the information on the
           multiplicities of the infinite zeros of the pencil $S(\lambda)$ as follows:
           $S(\lambda)$ has \texttt{INFO.miz$(i)$} infinite zeros of multiplicity $i$.
           \texttt{INFO.miz} is empty if $S(\lambda)$ has no infinite zeros.  \\ \hline
 \texttt{kr}   & integer row vector, which contains the right Kronecker
           indices of the pencil $S(\lambda)$.     \\ \hline
 \texttt{kl}   & integer row vector, which contains the left Kronecker
           indices of the pencil $S(\lambda)$.     \\ \hline
 \texttt{minphase}   & logical value, which is set to \texttt{true} if the pencil $S(\lambda)$ has only stable finite  eigenvalues and all its infinite eigenvalues are simple, or to  \texttt{false}, if the pencil $S(\lambda)$ has unstable finite eigenvalues or  infinite zeros. \\ \hline
\end{longtable}}

\end{description}

\subsubsection*{Method}
Let $G(\lambda)$ be the TFM $G(\lambda) = C(\lambda E-A)^{-1}B+D$ of the LTI system \texttt{SYS}.
For the definition of the zeros of $G\lambda)$ in terms of the descriptor realization (\ref{gzero:sysss}), see Section \ref{app:tfm_polzer}.
If the descriptor system realization
 $(A-\lambda E,B,C,D)$ of \texttt{SYS} is irreducible (i.e., controllable and observable), then the computed finite zeros in \texttt{Z} are simply the finite generalized eigenvalues of the system matrix pencil $S(\lambda)$, and the multiplicities of the infinite generalized eigenvalues of $S(\lambda)$ are in excess with one with respect to the multiplicities of the infinite zeros.
For the computation of the eigenvalues of $S(\lambda)$ and its Kronecker structure, the zeros computation algorithm of \cite{Misr94} is applied to the system matrix pencil $S(\lambda)$ in (\ref{systempencil}), by calling the MEX-function \url{sl_gzero}. \index{MEX-functions!\url{sl_gzero}}%
In the case of a standard state-space model with $E = I$, \url{sl_gzero} uses the algorithm of \cite{Emam82} in conjunctions with the extension proposed in \cite{Svar85}.
These algorithms determine:
 \begin{itemize}
 \item the $n_f$ finite zeros of $S(\lambda)$, $\lambda_i$, $i = 1, \ldots, n_f$ ($n_f$ is provided in \texttt{INFO.nfz});
  \item the dimensions $s_i^\infty$, for $i = 1, \ldots, h$, of the elementary infinite Jordan blocks in (\ref{Jordan-null}) in the Weierstrass canonical form of the regular part (\ref{regblocks}), representing the multiplicities of infinite eigenvalues; the number of infinite eigenvalues is $n_\infty = \sum_{i=1}^{h}s_i^\infty$, of which $n^0_\infty$ are simple infinite eigenvalues  for which $s_i^\infty = 1$ ($n_\infty$ is provided in \texttt{INFO.niev}, $n^0_\infty$  is provided in \texttt{INFO.nsiev}, and the multiplicities of the infinite eigenvalues are provided in \texttt{INFO.miev});
 \item the numbers $m_i^\infty$, for $i = 1, \ldots, k$, where $m_i^\infty$ is the number of the elementary infinite Jordan blocks of order $i+1$ in (\ref{Jordan-null}) in the Weierstrass canonical form of the regular part (\ref{regblocks}); the number of infinite zeros is  $n_{z,\infty} = \sum_{i=1}^{k}im_i^\infty$ (also equal to $n_\infty-n^0_\infty$) ($n_{z,\infty} $ is provided in \texttt{INFO.niz} and the numbers $m_i^\infty$, for $i = 1, \ldots, k$, are provided in \texttt{INFO.miz});
  \item the $\nu_r$ right Kronecker indices $\epsilon_i$, for $i = 1, \ldots, \nu_r$, of the pencil $S(\lambda)$ (corresponding to the $\nu_r$ Kronecker blocks $L_{\epsilon_i}(\lambda)$  of the form $\epsilon_i\times(\epsilon_i+1)$ which are part of the Kronecker canonical form of the pencil $S(\lambda)$, as in (\ref{kcf-kreps})) (the right Kronecker indices are provided in \texttt{INFO.kr});
 \item the $\nu_l$ left Kronecker indices $\eta_i$, for $i = 1, \ldots, \nu_l$, of the pencil $S(\lambda)$ (corresponding to the $\nu_l$ Kronecker blocks $L_{\eta_i}(\lambda)$  of the form $\eta_i\times(\eta_i+1)$ which are part of the Kronecker canonical form of the pencil $S(\lambda)$, as in (\ref{kcf-keta})) (the left Kronecker indices are provided in \texttt{INFO.kl});
 \item $\rho := n_f+n_\infty + \sum_{i=1}^{\nu_r}\epsilon_i + \sum_{i=1}^{\nu_l} \eta_i$, the normal rank of the pencil $S(\lambda)$ ($\rho$ is provided in \texttt{INFO.nrank}).
 \end{itemize}
The multiplicities of the infinite zeros are related to the orders of the elementary infinite blocks, as follows: to each elementary infinite block of order $s_i^\infty > 1$ corresponds an infinite zero of multiplicity $s_i^\infty-1$.

The number of finite stable zeros of $S(\lambda)$ lying in $\mathds{C}_s$ (the appropriate open stability domain) is provided in \texttt{INFO.nfsz}.
The number of finite unstable zeros of $S(\lambda)$ lying in $\mathds{C}_u$ (the appropriate open instability domain without including infinity) is provided in \texttt{INFO.nfuz}.
The number of finite zeros of $S(\lambda)$ lying in $\partial\mathds{C}_s$ (the boundary of the appropriate stability domain) is provided in \texttt{INFO.nfsbz}.  The stability boundary offset $\beta$ specified in \texttt{OFFSET} is used to numerically assess  if an eigenvalue belongs or not to $\mathds{C}_s$, $\partial\mathds{C}_s$ or $\mathds{C}_u$.

The components of the vector \texttt{Z} are the $n_f$ finite values $\lambda_i$, $i = 1, \ldots, n_f$ and the $n_\infty$ infinite values.
Additional qualitative information is provided in the \texttt{INFO} structure, to characterize the minimum-phase property:
\texttt{INFO.minphase} is set to \texttt{true} if all finite eigenvalues of $S(\lambda)$ are stable and $S(\lambda)$ has only simple infinite eigenvalues. Otherwise, \texttt{INFO.minphase} is set to \texttt{false}.

\subsubsection*{Application examples}

The \emph{input decoupling zeros} are defined as the zeros of the particular system matrix pencil
\[ S(\lambda) := [\, A-\lambda E\; B\,] ,\]
which corresponds to a system with empty outputs.
The finite zeros of $S(\lambda)$ are also known as the uncontrollable finite  eigenvalues of the pencil $A-\lambda E$, while the infinite zeros of $S(\lambda)$ corresponds to uncontrollable infinite eigenvalues of the pencil $A-\lambda E$ (their multiplicities exceeds with 1 the multiplicities of the infinite zeros). The function \texttt{\bfseries gzero} can be used to compute the input decoupling zeros of a LTI state-space system \texttt{SYS} using

\begin{verbatim}
[Z,INFO] = gzero(SYS([],:),TOL)
\end{verbatim}
For a controllable pair $(A-\lambda E,B)$, \texttt{Z} results empty.

Similarly, the \emph{output decoupling zeros} are defined as the invariant zeros of the particular system matrix
\[ S(\lambda) := \ba{c} A-\lambda E\\ C \ea ,\]
which corresponds to a system with empty inputs.
The finite zeros of $S(\lambda)$ are also known as the unobservable finite eigenvalues of the pencil $A-\lambda E$, while the infinite zeros of $S(\lambda)$ corresponds to unobservable infinite eigenvalues of the pencil $A-\lambda E$ (their multiplicities exceeds with 1 the multiplicities of the infinite zeros).  The function \texttt{\bfseries gzero} can be used to compute the output decoupling zeros of a LTI state-space system \texttt{SYS} using

\begin{verbatim}
[Z,INFO] = gzero(SYS(:,[]),TOL)
\end{verbatim}
For an observable pair $(A-\lambda E,C)$, \texttt{Z} results empty.

Finally, the zeros of the pole pencil can be computed as the invariant zeros of the particular system matrix
\[ S(\lambda) := A-\lambda E,\]
which corresponds to a system with empty inputs and empty outputs. The function \texttt{\bfseries gzero} can be used to compute the zeros of the pole pencil of a LTI state-space system \texttt{SYS} using

\begin{verbatim}
[Z,INFO] = gzero(SYS([],[]),TOL)
\end{verbatim}
For an irreducible realization $(A-\lambda E,B,C,D)$, \texttt{Z} contains the poles (finite and infinite) of the system \texttt{SYS}.

\subsubsection{\texttt{\bfseries gnrank}}

\subsubsection*{Syntax}
\index{M-functions!\texttt{\bfseries gnrank}}
\begin{verbatim}
NR = gnrank(SYS)
NR = gnrank(SYS,TOL)
\end{verbatim}

\subsubsection*{Description}

\noindent \texttt{gnrank} computes for the LTI system \texttt{SYS}, the normal rank of its transfer function matrix.
\index{transfer function matrix (TFM)!normal rank}%

\subsubsection*{Input data}

\begin{description}
\item
\texttt{SYS} is a LTI system, which can be specified in a descriptor system state-space form
\be\label{gnrank:sysss}
\begin{aligned}
E\lambda x(t)  &=   Ax(t)+ B u(t) ,\\
y(t) &=  C x(t)+ D u(t)
\end{aligned}
\ee
or in an input-output form
\be\label{gnrank:systf} \mathbf{y}(\lambda) = G(\lambda)\mathbf{u}(\lambda) \, , \ee
where $G(\lambda)$ is the rational transfer function matrix of the system.
\item \texttt{TOL} is a relative tolerance used for rank determinations. If \texttt{TOL} is not specified as input or if \texttt{TOL} = 0, an internally computed default value is used.
\end{description}

\subsubsection*{Output data}

\begin{description}
\item
\texttt{NR} is the normal rank of the transfer function matrix $G(\lambda)$ of the LTI system \texttt{SYS}.
\end{description}

\subsubsection*{Method}
If \texttt{SYS} is specified in the descriptor form (\ref{gnrank:sysss}), then its transfer-function matrix is $G(\lambda) = C(\lambda E-A)^{-1}B+D$. If \texttt{SYS} is specified in the input-output form (\ref{gnrank:systf}), then a (possibly non-minimal) state-space realization of the form (\ref{gnrank:sysss}) is automatically constructed.
For the definition of the normal rank $r$ of a rational TFM $G(\lambda)$, see Section \ref{sec:minbasis}. For the calculation of the normal rank $r$ of $G\lambda)$ in terms of the descriptor representation, we use the relation
\[ r = \rank S(\lambda)-n ,\]
where $\rank S(\lambda)$ is the normal rank of the system matrix pencil $S(\lambda)$ defined as
\be\label{gnrank-syspencil} S(\lambda) = \ba{cc} A-\lambda E & B \\ C & D \ea  \ee
and $n$ is the order of the descriptor state-space realization.
For the computation of the normal rank of $S(\lambda)$, the structural elements of its Kronecker structure are determined using the zeros computation algorithm of \cite{Misr94}, by calling the MEX-function \url{sl_gzero}. \index{MEX-functions!\url{sl_gzero}}%
In the case of a standard state-space model with $E = I$, \url{sl_gzero} uses the algorithm of \cite{Emam82} in conjunctions with the extension proposed in \cite{Svar85}.
These algorithms determine:
 \begin{itemize}
 \item $n_f$, the number of finite eigenvalues of $S(\lambda)$;
 \item the $\nu_r$ right Kronecker indices $\epsilon_i$, for $i = 1, \ldots, \nu_r$ of the pencil $S(\lambda)$ (corresponding to the $\nu_r$ Kronecker blocks $L_{\epsilon_i}(\lambda)$  of the form $\epsilon_i\times(\epsilon_i+1)$ which are part of the Kronecker canonical form of the pencil $S(\lambda)$, as in (\ref{kcf-kreps}));
 \item the $\nu_l$ left Kronecker indices $\eta_i$, for $i = 1, \ldots, \nu_l$ of the pencil $S(\lambda)$ (corresponding to the $\nu_l$ Kronecker blocks $L_{\eta_i}(\lambda)$  of the form $\eta_i\times(\eta_i+1)$ which are part of the Kronecker canonical form of the pencil $S(\lambda)$, as in (\ref{kcf-keta}));
 \item the orders of the elementary infinite blocks $s_i^\infty$ , for $i = 1, \ldots, h$, (i.e., the dimensions of the elementary infinite Jordan blocks in (\ref{Jordan-null}) in the Weierstrass canonical form of the regular part (\ref{regblocks})).
 \end{itemize}
The normal rank of $S(\lambda)$ is determined as \index{transfer function matrix (TFM)!normal rank}%
\[ \rank S(\lambda) = n_f+\sum_{i=1}^{h}s_i^\infty + \sum_{i=1}^{\nu_r}\epsilon_i + \sum_{i=1}^{\nu_l} \eta_i .\]

\subsubsection*{Alternative computation of normal rank}

The normal rank of a LTI system \texttt{SYS} can be alternatively computed by evaluating the rank of the TFM $G(\lambda)$ at a random frequency (or taking the maximum rank for a few random frequencies). The following command can be used for this purpose:
\begin{verbatim}
nr = rank(evalfr(SYS,rand),TOL)
\end{verbatim}

\subsubsection{\texttt{\bfseries ghanorm}}

\subsubsection*{Syntax}
\index{M-functions!\texttt{\bfseries ghanorm}}
\begin{verbatim}
[HANORM,HS] = ghanorm(SYS)
\end{verbatim}

\subsubsection*{Description}

\noindent \texttt{ghanorm} computes for a proper and stable LTI state-space system \texttt{SYS} with the transfer function matrix $G(\lambda)$, the Hankel norm $\|G(\lambda)\|_H$ and the Hankel singular values of the system.
\index{transfer function matrix (TFM)!Hankel norm}
\index{descriptor system!Hankel norm}

\subsubsection*{Input data}

\begin{description}
\item
\texttt{SYS} is a LTI system,  in a descriptor system state-space form
\be\label{ghanorm:sysss}
\begin{aligned}
E\lambda x(t)  &=   Ax(t)+ B u(t) ,\\
y(t) &=  C x(t)+ D u(t) .
\end{aligned}
\ee
\end{description}

\subsubsection*{Output data}

\begin{description}
\item
\texttt{HANORM} is the Hankel norm $\|G(\lambda)\|_H$ of  the transfer function matrix $G(\lambda)$ of the LTI system \texttt{SYS}.
\item
\texttt{HS} is a column vector which contains the decreasingly ordered Hankel singular values of \texttt{SYS}.
\end{description}

\subsubsection*{Method}
Let $G(\lambda)$ be the TFM of the LTI \texttt{SYS}.
For the definition and computation of the Hankel norm of a proper and stable rational TFM $G(\lambda)$ see Section \ref{sec:normTFM} and for the definition and computation of the  Hankel singular values, see Section \ref{app:balance}. If the original descriptor system is proper, but $E$ is singular, then an automatic conversion  is performed using the function \texttt{gss2ss}, to a reduced order descriptor state-space representation with $E$ invertible and upper triangular. For the solution of the intervening generalized Lyapunov equation (\ref{glyap}) or  generalized Stein equation (\ref{gstein}), the MEX-function \url{sl_glme} is called. \index{MEX-functions!\url{sl_glme}}%

\subsubsection{\texttt{\bfseries gnugap}}

\subsubsection*{Syntax}
\index{M-functions!\texttt{\bfseries gnugap}}
\begin{verbatim}
[NUGAP,FPEAK] = gnugap(SYS1,SYS2,TOL)
[NUGAP,FPEAK] = gnugap(SYS1,SYS2,TOL,FREQ)
[NUGAP,FPEAK] = gnugap(SYS1,SYS2,TOL,FREQ,OFFSET)
\end{verbatim}

\subsubsection*{Description}

\noindent \texttt{gnugap} computes for two LTI systems \texttt{SYS1} and \texttt{SYS2}  with the transfer function matrices $G_1(\lambda)$ and $G_2(\lambda)$, respectively, the $\nu$-gap distance $\delta_\nu\big(G_1(\lambda),G_2(\lambda)\big)$  between the two systems.
\index{transfer function matrix (TFM)!$\nu$-gap metric}
\index{descriptor system!$\nu$-gap metric}

\subsubsection*{Input data}

\begin{description}
\item
\texttt{SYS1} is a LTI system, which can be specified in a descriptor system state-space form
\be\label{gnugap:sysss1}
\begin{aligned}
E_1\lambda x_1(t)  &=   A_1x_1(t)+ B_1 u(t) ,\\
y_1(t) &=  C_1 x_1(t)+ D_1 u(t)
\end{aligned}
\ee
or in an input-output form
\be\label{gnugap:systf1} \mathbf{y}_1(\lambda) = G_1(\lambda)\mathbf{u}(\lambda) \, , \ee
where $G_1(\lambda)$ is the rational transfer function matrix of the system \texttt{SYS1}.
\item
\texttt{SYS2} is a LTI system, which can be specified in a descriptor system state-space form
\be\label{gnugap:sysss2}
\begin{aligned}
E_2\lambda x_2(t)  &=   A_2x_2(t)+ B_2 u(t) ,\\
y_2(t) &=  C_2 x_2(t)+ D_2 u(t)
\end{aligned}
\ee
or in an input-output form
\be\label{gnugap:systf2} \mathbf{y}_2(\lambda) = G_2(\lambda)\mathbf{u}(\lambda) \, , \ee
where $G_2(\lambda)$ is the rational transfer function matrix of the system \texttt{SYS2}.
\item \texttt{TOL} is a relative tolerance used for rank determinations. If \texttt{TOL} is not specified as input or if \texttt{TOL} = 0, an internally computed default value is used.
\item \texttt{FREQ} is a real vector with nonnegative elements, which contains a set of frequency values to be used for the point-wise evaluation of the distances. If \texttt{FREQ = []} or not specified, no point-wise evaluation is performed.
\item
 \texttt{OFFSET}  is the stability
 boundary offset $\beta$, to be used  to assess the finite zeros which belong to $\partial\mathds{C}_s$ (the boundary of the stability domain) as follows: in the
 continuous-time case these are the finite
  zeros having real parts in the interval $[-\beta, \beta]$, while in the
 discrete-time case these are the finite zeros having moduli in the
 interval $[1-\beta, 1+\beta]$. \newline (Default: $\beta = 1.4901\cdot 10^{-08}$). \end{description}

\subsubsection*{Output data}

\begin{description}
\item
\texttt{NUGAP}, if \texttt{FREQ = []} or not specified, is the $\nu$-gap distance $\delta_\nu\big(G_1(\lambda),G_2(\lambda)\big)$  between the two systems. If  \texttt{FREQ} is a non-empty vector with $n_f$ elements, then \texttt{NUGAP} is an $n_f$-dimensional vector, where \texttt{NUGAP($i$)} contains the point-wise distance at $i$-th frequency value  \texttt{FREQ($i$)}.
\item
\texttt{FPEAK} is the frequency value where the peak value of $\delta_\nu\big(G_1(\lambda),G_2(\lambda)\big)$  is achieved. \texttt{FPEAK = []} if \texttt{NUGAP} = 1.
\end{description}

\subsubsection*{Method}
The $\nu$-gap metric is defined in Section \ref{sec:nugap} for arbitrary transfer function matrices. The critical aspect of the computation of $\delta_\nu\big(G_1(\lambda),G_2(\lambda)\big)$, using as computational basis the definition (\ref{nugap}), is the reliable determination of the winding number of $g(\lambda) := \det\big(R_2^\sim(\lambda) R_1(\lambda)\big)$, using the computed poles and zeros of $R_2^\sim(\lambda) R_1(\lambda)$, and the checking of the associated conditions on its determinant. For the computation of the poles and zeros, an irreducible realization of   $R_2^\sim(\lambda) R_1(\lambda)$ is first computed, and then the poles are computed using the function \texttt{gpole} and the zeros using the function \texttt{gzero}.

\subsection{Functions for System Order Reduction}

The system order reduction functions cover the computation of irreducible or minimal realizations, the balancing-related model reduction, and the conversion of descriptor systems representation to various SVD-like coordinate forms without non-dynamic modes, including the conversion to standard state-space forms.


\subsubsection{\texttt{\bfseries gir}}
\index{M-functions!\texttt{\bfseries gir}}
\subsubsection*{Syntax}
\begin{verbatim}
SYSR = gir(SYS)
SYSR = gir(SYS,TOL)
SYSR = gir(SYS,TOL,JOBOPT)
\end{verbatim}

\subsubsection*{Description}

\noindent \texttt{gir} computes for a LTI descriptor state-space system $(A-\lambda E,B,C,D)$, a reduced order (e.g., controllable, observable, or irreducible) descriptor realization $(\widetilde A-\lambda \widetilde E,\widetilde B,\widetilde C,D)$, such that the corresponding transfer function matrices are equal.
\index{descriptor system!irreducible realization}

\subsubsection*{Input data}

\begin{description}
\item
\texttt{SYS} is a LTI system,  in a  descriptor system state-space form
\be\label{gir:sysss}
\begin{aligned}
E\lambda x(t)  &=   Ax(t)+ B u(t) ,\\
y(t) &=  C x(t)+ D u(t) .
\end{aligned}
\ee
\item
 \texttt{TOL} is a relative tolerance used for rank determinations. If \texttt{TOL} is not specified as input or if \texttt{TOL} = 0, an internally computed default value is used. \\
\item
 \texttt{JOBOPT}  is a character option variable to specify various order reduction  options, as follows: {\tabcolsep=1mm
\begin{longtable}{lcp{10.4cm}}
   ~~~~~~~~\texttt{'irreducible'} &--& compute an irreducible
                                             descriptor realization (default); \\
   ~~~~~~~~\texttt{'finite'}&--& compute a finite controllable and
                              finite observable realization \\
   ~~~~~~~~\texttt{'infinite'}&--& compute an infinite controllable and
                              infinite observable realization\\
   ~~~~~~~~\texttt{'contr'}&--& compute a controllable realization \\
   ~~~~~~~~\texttt{'obs'}&--& compute an observable realization \\
   ~~~~~~~~\url{'finite_contr'}&--& compute a finite controllable realization \\
   ~~~~~~~~\url{'infinite_contr'}&--& compute an infinite controllable realization \\
   ~~~~~~~~\url{'finite_obs'}&--& compute a finite observable realization \\
   ~~~~~~~~\url{'infinite_obs'}&--& compute an infinite observable realization \\
\end{longtable}
}
\end{description}

\subsubsection*{Output data}

\begin{description}
\item
\texttt{SYSR} contains the resulting reduced order system in a descriptor system state-space form
\be\label{gir:sysssred}
\begin{aligned}
\widetilde E\lambda \widetilde x(t)  &=   \widetilde A\widetilde x(t)+ \widetilde B u(t) ,\\
y(t) &=  \widetilde C x(t)+ D u(t) .
\end{aligned}
\ee
\texttt{SYSR} has the same TFM as \texttt{SYS} and the resulting order of \texttt{SYSR} depends on the order reduction option selected via \texttt{JOBOPT}. If no order reduction takes place, then \texttt{SYSR} has the same realization as \texttt{SYS}.
\end{description}

\subsubsection*{Remark on input and output data} The function \texttt{\bfseries gir} accepts an array of LTI systems \texttt{SYS} as input parameter. In this case, \texttt{SYSR} is also an array of LTI systems of the same size as \texttt{SYS}. To each component system in \texttt{SYS(:,:,i,j)} corresponds a component system \texttt{SYSR(:,:,i,j)} with a reduced order.

\subsubsection*{Method}
Let $G(\lambda)$ be the TFM of the LTI \texttt{SYS} with the descriptor realization (\ref{gir:sysss}). The conditions for finite and infinite controllability and observability are given by Theorem \ref{T-desc-minreal}, as conditions $(i)-(iv)$. The concepts of finite controllable and observable eigenvalues of the pencil $A-\lambda E$ are discussed in Section \ref{app:tfm_polzer}. The elimination of uncontrollable or unobservable eigenvalues is done by determining the so-called Kalman controllability or Kalman observability forms, which exhibit explicitly these eigenvalues.
For the computation of the controllability Kalman form for a descriptor system representation, the \textbf{Procedure GCSF} described in \cite{Varg17} is employed, which employs the orthogonal reduction technique proposed in \cite{Varg90}. This algorithm computes for a pair $(A-\lambda E,B)$, orthogonal transformation matrices $Q$ and $Z$ such that the matrices of the transformed triple $(\widetilde A-\lambda \widetilde E,\widetilde B,\widetilde C) := (Q^T AZ-\lambda Q^T EZ,Q^T B,CZ)$ have the form
\[ \widetilde A -\lambda \widetilde E = \ba{cc}A_c-\lambda E_c & \ast \\ 0 & A_{\bar c}-\lambda E_{\bar c} \ea, \quad \widetilde B = \ba{c} B_c\\ 0 \ea, \quad  \widetilde C = [\, C_c \; \ast\,], \]
where  the pair $(A_c-\lambda E_c,B_c)$ is finite controllable, and $\Lambda(A_{\bar c}-\lambda E_{\bar c})$ contains the finite uncontrollable eigenvalues (as well as possible some infinite uncontrollable eigenvalues too).
The reduced order system $(A_c-\lambda E_c,B_c,C_c,D)$ is finite controllable and has the same TFM $G(\lambda)$ as the original system. In this way, all uncontrollable finite eigenvalues contained in $A_{\bar c}-\lambda E_{\bar c}$ have been removed from the system model. By applying the same algorithm to the dual realization $(A^T-\lambda E^T,C^T,B^T,D^T)$, the finite unobservable eigenvalues can be removed (as the finite uncontrollable eigenvalues of the dual system). If the matrices $A$ and $E$ are interchanged, by forming a model with $(E-\lambda A,B,C,D)$, then the uncontrollable or unobservable null eigenvalues can be removed using the same algorithms. However, by removing the uncontrollable or unobservable null eigenvalues of this model, we remove in fact the infinite uncontrollable or unobservable eigenvalues of the original model. An irreducible (controllable and observable) realization can be thus computed in four steps, which form the \textbf{Procedure GIR} described in \cite{Varg17}.  The computational method to compute an irreducible realization or to perform only a specific step of the \textbf{Procedure GIR} is implemented in
the MEX-function \url{sl_gminr}, which is called, with appropriately set options, by the function \texttt{\bfseries gir}. \index{MEX-functions!\url{sl_gminr}}%

\subsubsection*{Application example}
The function \texttt{\bfseries gir} displays messages indicating the number of uncontrollable and the number of unobservable eigenvalues removed from the model. This verbose output (especially when applied to array of systems)  can be disabled as shown in the example below:
\begin{verbatim}
warning off
sysr = gir(sys);
warning on
\end{verbatim}

\subsubsection{\texttt{\bfseries gminreal}}
\index{M-functions!\texttt{\bfseries gminreal}}
\subsubsection*{Syntax}
\begin{verbatim}
[SYSMIN,INFO] = gminreal(SYS)
[SYSMIN,INFO] = gminreal(SYS,TOL)
[SYSMIN,INFO] = gminreal(SYS,TOL,NDMONLYFLAG)
\end{verbatim}

\subsubsection*{Description}

\noindent \texttt{gminreal} computes for a LTI descriptor state-space system $(A-\lambda E,B,C,D)$, a minimal order  descriptor realization $( A_m - \lambda E_m, B_m, C_m, D_m )$ (i.e., controllable, observable, without non-dynamic modes), such that the corresponding transfer function matrices are equal.
\index{descriptor system!irreducible realization}
\index{descriptor system!minimal realization}

\subsubsection*{Input data}
\begin{description}
\item
\texttt{SYS} is a LTI system,  in a  descriptor system state-space form
\be\label{gminreal:sysss}
\begin{aligned}
E\lambda x(t)  &=   Ax(t)+ B u(t) ,\\
y(t) &=  C x(t)+ D u(t) .
\end{aligned}
\ee
\item
\texttt{TOL} is a relative tolerance used for rank determinations. If \texttt{TOL} is not specified as input or if \texttt{TOL} = 0, an internally computed default value is used.
\item
\texttt{NDMONLYFLAG} is a character option variable to be set to \texttt{'ndmonly'}  to remove only the non-dynamic modes (i.e., simple infinite eigenvalues). By default, the non-dynamic modes are jointly removed  with the uncontrollable and unobservable eigenvalues.
\end{description}

\subsubsection*{Output data}
\begin{description}
\item
\texttt{SYSMIN} contains the resulting minimal order system in a descriptor system state-space form
\be\label{gminreal:sysssred}
\begin{aligned}
E_m\lambda  x_m(t)  &=    A_m x_m(t)+ B_m u(t) ,\\
y_m(t) &=  C_m x_m(t)+ D_m u(t) .
\end{aligned}
\ee
\texttt{SYSMIN} has the same TFM as \texttt{SYS}. If \texttt{NDMONLYFLAG = 'ndmonly'} is specified, then the resulting \texttt{SYSMIN} contains a reduced order system without non-dynamic modes. If no order reduction takes place, then \texttt{SYSMIN} has the same realization as \texttt{SYS}.
\item
\texttt{INFO(1:3)} contains information on the
number of removed eigenvalues, as follows:\\
\hspace*{-2mm}
{\tabcolsep=1mm\begin{tabular}{lll}
    \texttt{INFO(1)} & -- &the number of removed uncontrollable eigenvalues; \\
    \texttt{INFO(2)}& -- &the number of removed unobservable eigenvalues; \\
    \texttt{INFO(3)}& -- &the number of removed non-dynamic (infinite) eigenvalues.
    \end{tabular}}
\end{description}

\subsubsection*{Method}
The conditions for minimality are given by Theorem \ref{T-desc-minreal}, as conditions $(i)-(v)$. The concepts of finite controllable and observable eigenvalues of the pencil $A-\lambda E$ are discussed in Section \ref{app:tfm_polzer}. The elimination of uncontrollable and unobservable eigenvalues is done in several steps, by determining appropriate Kalman controllability and observability forms, which explicitly exhibit these eigenvalues. The basic computation is the reduction of the descriptor system matrices to the controllability Kalman form using the orthogonal transformation based reduction technique proposed in \cite{Varg90} (see also \textbf{Procedure GCSF} described in \cite{Varg17}).
 For more details on the computation of irreducible realizations, see the description of \textbf{Method} for the function  \texttt{\bfseries gir}.

To remove the non-dynamic infinite eigenvalues of a descriptor system $(A-\lambda E,B,C,D)$, the system matrices are reduced to a SVD-like coordinate form
         \[ (\widetilde A-\lambda \widetilde E,\widetilde B,\widetilde C,D) := ( Q^T A Z - \lambda Q^T E Z, Q^T B, C Z, D), \]
where
\[  \begin{array}{rll}
\widetilde A-\lambda \widetilde E & = &    \ba{ccc} A_{11}-\lambda  E_{11} & A_{12} & A_{13} \\
                            A_{21} & A_{22} &     0  \\
                            A_{31} &   0    &     0  \ea \, , \quad
        \widetilde  B = \ba{c} B_1 \\ B_2 \\ B_3 \ea \, , \\ \\[-3mm]
       \widetilde C & = &  \ba{ccc}  C_1 & C_2 & C_3 \ea \, ,
       \end{array}
\]
with $E_{11}$ and $A_{22}$ upper triangular invertible matrices.
Then, the reduced descriptor system without non-dynamic modes $( A_m - \lambda E_m, B_m, C_m, D_m )$ is computed as
\[
 \begin{array}{rll} A_m -\lambda E_m & = &  \ba{cc} A^{}_{11} - A^{}_{12} A_{22}^{-1} A^{}_{21}  -\lambda  E^{}_{11} & A^{}_{13}\\
                                         A_{31}                             &   0    \ea \, , \\ \\[-3mm]
     B_m           & = &            \ba{c}  B^{}_1 - A^{}_{12} A_{22}^{-1} B^{}_2 \\ B_3 \ea \, , \\ \\[-3mm]
     C_m  & = &  \ba{cc} C_1 - C^{}_2 A_{22}^{-1} A^{}_{21}  & C^{}_3 \ea \, , \\ \\[-3mm]
     D_m  & = &  D - C^{}_2 A_{22}^{-1} B^{}_2 \, .
     \end{array} \]

The computational method to determine a minimal order descriptor system realization or a reduced order realization without non-dynamic modes is implemented in the MEX-function \url{sl_gminr}, which is called, with appropriately set options, by the function \texttt{\bfseries gminreal}. \index{MEX-functions!\url{sl_gminr}}%

\subsubsection*{Application example}
The function \texttt{\bfseries gminreal} displays messages indicating the number of eigenvalues (uncontrollable, unobservable, non-dynamic) removed from the model. This verbose output can be disabled as shown in the example below:
\begin{verbatim}
warning off
sysmin = gminreal(sys);
warning on
\end{verbatim}

\subsubsection{\texttt{\bfseries gbalmr}}
\index{M-functions!\texttt{\bfseries gbalmr}}
\subsubsection*{Syntax}
\begin{verbatim}
[SYSR,HS] = gbalmr(SYS)
[SYSR,HS] = gbalmr(SYS,TOL)
[SYSR,HS] = gbalmr(SYS,TOL,BALANCE)
\end{verbatim}

\subsubsection*{Description}
\texttt{\bfseries gbalmr} performs model reduction of a stable LTI state-space system using balancing-related methods.

\subsubsection*{Input data}
\begin{description}
\item
\texttt{SYS} is a stable LTI system,  in a  descriptor system state-space form
\be\label{gbalmr:sysss}
\begin{aligned}
E\lambda x(t)  &=   Ax(t)+ B u(t) ,\\
y(t) &=  C x(t)+ D u(t) .
\end{aligned}
\ee
\item
\texttt{TOL} is a relative tolerance used to determine the order of the reduced model. If \texttt{TOL} is not specified as input, or if \texttt{TOL} is empty, or if \texttt{TOL} = 0, then the value \texttt{TOL = sqrt(eps)} is internally used.
\item
\texttt{BALANCE} is a character option variable to be set to \texttt{'balance'} to compute a balanced realization of the reduced order model. By default, the state-space realization of the computed reduced order model is not balanced.
\end{description}

\subsubsection*{Output data}
\begin{description}
\item
\texttt{SYSR} contains the minimal realization of the resulting reduced order system in a descriptor system state-space form
\be\label{gbalmr:sysssred}
\begin{aligned}
E_r\lambda  x_r(t)  &=    A_r x_r(t)+ B_r u(t) ,\\
y_r(t) &=  C_r x_r(t)+ D_r u(t) .
\end{aligned}
\ee
The realization of \texttt{SYSR} is balanced (i.e., $E_r = I$ and the controllability and observability Gramians are equal and diagonal) if the balancing option \texttt{BALANCE = 'balance'} has been specified. The order of \texttt{SYSR} is the number of Hankel singular values of \texttt{SYS}, which are greater than \texttt{TOL}$\,\|G(\lambda)\|_H$, where $G(\lambda)$ is the TFM of the system (\ref{gbalmr:sysss}).
\item
\texttt{HS} is a column vector which contains the decreasingly ordered Hankel singular values of \texttt{SYS}.
\end{description}

\subsubsection*{Method}

For the order reduction of a standard system (i.e., with $E = I$),  the
balancing-free method of \cite{Varg91} or the balancing-based method of \cite{Tomb87} are used. For a descriptor system the balancing related order reduction methods of \cite{Styk04} are used. For the solution of the intervening Lyapunov and Stein equation for the Cholesky factors of the solution, the mex-function \url{sl_glme} is used.
\index{MEX-functions!\url{sl_glme}}%

\subsubsection{\texttt{\bfseries gss2ss}}
\index{M-functions!\texttt{\bfseries gss2ss}}
\subsubsection*{Syntax}
\begin{verbatim}
[SYSR,RANKE] = gss2ss(SYS)
[SYSR,RANKE] = gss2ss(SYS,TOL)
[SYSR,RANKE] = gss2ss(SYS,TOL,ESHAPE)
\end{verbatim}

\subsubsection*{Description}
\texttt{\bfseries gss2ss} performs the conversion of a LTI descriptor state-space systems $(A-\lambda E,B,C,D)$ to a SVD-like coordinate form $(A_r - \lambda E_r, B_r, C_r, D_r)$  without non-dynamic modes, such that the corresponding transfer function matrices are equal.

\subsubsection*{Input data}
\begin{description}
\item
\texttt{SYS} is a LTI system,  in a  descriptor system state-space form
\be\label{gss2ss:sysss}
\begin{aligned}
E\lambda x(t)  &=   Ax(t)+ B u(t) ,\\
y(t) &=  C x(t)+ D u(t) .
\end{aligned}
\ee
\item
\texttt{TOL} is a relative tolerance used for rank determinations. If \texttt{TOL} is not specified as input or if \texttt{TOL} = 0, an internally computed default value is used.
\item
\texttt{ESHAPE} is a character option variable to specify the shape of the leading invertible diagonal block $E_{11}$ of the resulting descriptor matrix $E_r = \diag (E_{11},0)$ (see \textbf{Method}). The following options can be used for \texttt{ESHAPE}: \\[1mm]
{\tabcolsep=1mm\begin{tabular}{lcp{12.5cm}}
\texttt{'diag'} &--& diagonal (the nonzero diagonal elements are the
                            decreasingly ordered nonzero singular values
                             of $E$); \\
\texttt{'triu'} &--& upper triangular; \\
\texttt{'ident'} &--& identity (default).
\end{tabular}}
\end{description}

\subsubsection*{Output data}
\begin{description}
\item
\texttt{SYSR} contains the resulting reduced order system without non-dynamic modes, in a descriptor system state-space form
\be\label{gss2ss:sysssred}
\begin{aligned}
E_r\lambda  x_r(t)  &=    A_r x_r(t)+ B_r u(t) ,\\
y_r(t) &=  C_r x_r(t)+ D_r u(t) ,
\end{aligned}
\ee
where $E_r$ has a block-diagonal form $E_r = \diag (E_{11},0)$, with $E_{11}$ invertible. The resulting shape of $E_{11}$ is in accordance with the specified option by \texttt{ESHAPE}.
\texttt{SYSR} has the same TFM as \texttt{SYS} and the resulting order of
 \texttt{SYSR} is the order of \texttt{SYS} minus the number of simple infinite (non-dynamic) eigenvalues of the pole pencil $A-\lambda E$.
\item
\texttt{RANKE} is the rank of $E$ (and also the order of $E_{11}$).
\end{description}

\subsubsection*{Method}
To remove the non-dynamic eigenvalues of the descriptor system $(A-\lambda E,B,C,D)$, the system matrices are first reduced using non-orthogonal transformation matrices $Q$ and $Z$ to a SVD-like coordinate form
         \[ (\widetilde A-\lambda \widetilde E,\widetilde B,\widetilde C,D) := ( QA Z - \lambda QE Z, QB, C Z, D), \]
where
\[  \begin{array}{rll}
\widetilde A-\lambda \widetilde E & = &    \ba{ccc} A_{11}-\lambda  E_{11} & A_{12} & A_{13} \\
                            A_{21} & I &     0  \\
                            A_{31} &   0    &     0  \ea \, , \quad
        \widetilde  B = \ba{c} B_1 \\ B_2 \\ B_3 \ea \, , \\ \\[-3mm]
       \widetilde C & = &  \ba{ccc}  C_1 & C_2 & C_3 \ea \, ,
       \end{array}
\]
with $E_{11}$ invertible and either upper triangular (if \texttt{ESHAPE = 'triu'}) or diagonal (if \texttt{ESHAPE = 'diag'} or \texttt{ESHAPE = 'identity'}). To compute the above  SVD-like form, the MEX-function \url{sl_gstra} is called by the function \texttt{\bfseries gss2ss}, with appropriately set options. \index{MEX-functions!\url{sl_gstra}}%

If \texttt{ESHAPE = 'triu'} or \texttt{ESHAPE = 'diag'}, the reduced descriptor system without non-dynamic modes $(A_r - \lambda E_r, B_r, C_r, D_r )$ is computed as
\be\label{gss2ss:triu}
 \begin{array}{rll} A_r -\lambda E_r & = &  \ba{cc} A^{}_{11} - A^{}_{12} A^{}_{21}  -\lambda  E^{}_{11} & A^{}_{13}\\
                                         A_{31}                             &   0    \ea \, , \\ \\[-3mm]
     B_r           & = &            \ba{c}  B^{}_1 - A^{}_{12} B^{}_2 \\ B_3 \ea \, , \\ \\[-3mm]
     C_r  & = &  \ba{cc} C_1 - C^{}_2 A_{21} & C^{}_3 \ea \, , \\ \\[-3mm]
     D_r  & = &  D - C^{}_2 B^{}_2 \, .
     \end{array} \ee
If \texttt{ESHAPE = 'identity'}, then the above matrices are computed as
\[
 \begin{array}{rll} A_r -\lambda E_r & = &  \ba{cc} E_{11}^{-1/2}(A^{}_{11} - A^{}_{12} A^{}_{21})E_{11}^{-1/2}  -\lambda  I & E_{11}^{-1/2}A^{}_{13}\\
                                         A_{31}E_{11}^{-1/2}                             &   0    \ea \, , \\ \\[-3mm]
     B_r           & = &            \ba{c}  E_{11}^{-1/2}(B^{}_1 - A^{}_{12} B^{}_2) \\ B_3 \ea \, , \\ \\[-3mm]
     C_r  & = &  \ba{cc} (C_1 - C^{}_2 A_{21})E_{11}^{-1/2} & C^{}_3 \ea \, , \\ \\[-3mm]
     D_r  & = &  D - C^{}_2 B^{}_2 \, .
     \end{array} \]
The particular case of an invertible $E$ and with the pair $(A,E)$ in a generalized Hessenberg form (i.e., with $A$ in upper Hessenberg form and $E$ upper triangular) is handled separately, in order to preserve the Hessenberg form of $A$. In this case, with the additional assumption that all diagonal elements of $E$ are positive (this can be easily arranged by changing the signs of the corresponding rows of $E$, $A$ and $B$),  the matrices of the standard state-space realization $(A_r-\lambda I,B_r,C_r,D)$ are computed with
\be\label{gss2ss-qr} A_r =  E^{-1/2}AE^{-1/2}, \quad B_r = E^{-1/2}B, \quad
C_r = CE^{-1/2} \, ,\ee
where $E^{-1/2}$ is the square root of $E$ computed using the method described in \cite[Algorithm 6.7]{High08} (implemented in the MATLAB function \texttt{\bfseries sqrtm}).

\subsection{Functions for Operations on Generalized LTI Systems}\label{dstools:operations}
These functions cover the computation of rational nullspace and range space bases, the solution of linear rational equations, the computation of additive spectral decompositions and order reductions using minimal dynamic cover based techniques.

\subsubsection{\texttt{\bfseries grnull}}
\index{M-functions!\texttt{\bfseries grnull}}
\subsubsection*{Syntax}
\begin{verbatim}
[SYSRNULL,INFO] = grnull(SYS,OPTIONS)
\end{verbatim}

\subsubsection*{Description}
\texttt{\bfseries grnull} computes a proper rational  basis $N_r(\lambda)$ of the right nullspace of the transfer function matrix $G_1(\lambda)$ of a LTI descriptor system,
such that
\[ G_1(\lambda)N_r(\lambda) = 0 \]
and determines $G_2(\lambda)N_r(\lambda)$, where $G_2(\lambda)$ is a TFM having the same number of columns as $G_1(\lambda)$.
\index{descriptor system!minimal nullspace basis}
\index{nullspace!basis!minimal proper, right}%
\index{nullspace!basis!simple minimal proper, right}

\subsubsection*{Input data}
\begin{description}
\item
\texttt{SYS} is an output concatenated compound LTI system, \texttt{SYS = [ SYS1; SYS2 ]},  in a  descriptor system state-space form
\be\label{grnull:sysss}
\begin{aligned}
E\lambda x(t)  &=   Ax(t)+ B u(t)  ,\\
y_1(t) &=  C_1 x(t)+ D_1 u(t)  , \\
y_2(t) &=  C_2 x(t)+ D_2 u(t)  ,
\end{aligned}
\ee
where \texttt{SYS1} has the transfer function matrix  $G_1(\lambda)$ with the descriptor system realization $(A-\lambda E,B,C_1,D_1)$,
\texttt{SYS2} has the transfer function matrix  $G_2(\lambda)$ with the descriptor system  realization $(A-\lambda E,B,C_2,D_2)$, and $y_1(t) \in \mathds{R}^{p_1}$ and $y_2(t) \in \mathds{R}^{p_2}$ are the outputs of \texttt{SYS1} and \texttt{SYS2}, respectively.
\item
 \texttt{OPTIONS} is a MATLAB structure to specify user options and has the following fields:\\
{\tabcolsep=1mm\setlength\LTleft{30pt}
\begin{longtable}{|l|lcp{10cm}|} \hline
\textbf{\texttt{OPTIONS} fields} & \multicolumn{3}{l|}{\textbf{Description}} \\ \hline
 \texttt{tol}   & \multicolumn{3}{p{10cm}|}{relative tolerance for rank computations \newline (Default: internally computed)} \\ \hline
 \texttt{p2}   & \multicolumn{3}{l|}{$p_2$, the number of outputs of \texttt{SYS2} (Default: $p_2 = 0$) } \\ \hline
 \texttt{simple}   & \multicolumn{3}{l|}{option to compute a simple proper basis:}\\
                 &  \texttt{true} &--& compute a simple basis; the orders of the
                            basis vectors are provided in \texttt{INFO.degs}; \\
                 &  \texttt{false}&--& no simple basis computed (default)  \\
                                        \hline
 \texttt{inner}   & \multicolumn{3}{p{10cm}|}{option to compute an inner basis:}\\
                  &  \texttt{true} &--& compute an inner basis; if the option for
                             simple basis has been selected then each
                             basis vector results inner and the orders of
                             the basis vectors are provided in \texttt{INFO.deg}
                             (the resulting basis may not be inner) \\
                  &  \texttt{false}&--& no inner basis is computed (default)  \\
                                        \hline
\texttt{offset}   & \multicolumn{3}{p{11.5cm}|}{stability
 boundary offset $\beta$, to be used  to assess the finite zeros which belong to $\partial\mathds{C}_s$ (the boundary of the stability domain) as follows: in the
 continuous-time case these are the finite
  zeros having real parts in the interval $[-\beta, \beta]$, while in the
 discrete-time case these are the finite zeros having moduli in the
 interval $[1-\beta, 1+\beta]$ \newline (Default: $\beta = 1.4901\cdot 10^{-08}$). } \\ \hline
 \texttt{tcond}   & \multicolumn{3}{l|}{maximum allowed value for the condition numbers of the  employed}\\
     & \multicolumn{3}{l|}{non-orthogonal transformation matrices (Default: $10^4$)}\\
    & \multicolumn{3}{l|}{(only used if \texttt{OPTIONS.simple = true}) } \\
                    \hline
\texttt{sdeg}   & \multicolumn{3}{p{11cm}|}{prescribed stability degree for the resulting right nullspace basis
                    (Default: \texttt{[ ]}) } \\ \hline
 \texttt{poles}   & \multicolumn{3}{p{10cm}|}{a complex conjugated set of desired poles  to be assigned for the resulting right nullspace basis
                     (Default: \texttt{[ ]})}\\
                                        \hline
\end{longtable}}
\end{description}

\subsubsection*{Output data}
\begin{description}
\item
\texttt{SYSRNULL} contains  the output concatenated compound system \texttt{[ NR;  SYS2*NR ]}, in a descriptor system state-space form
\be\label{grnull:rnullsysss}
\begin{aligned}
\widetilde E_r\lambda  x_r(t)  &=    \widetilde A_r x_r(t)+ \widetilde B_{r} v(t) ,\\
y_{r,1}(t) &=  \widetilde C_{r,1} x_r(t)+ \widetilde D_{r,1} v(t) , \\
y_{r,2}(t) &=  \widetilde C_{r,2} x_r(t)+ \widetilde D_{r,2} v(t)  ,
\end{aligned}
\ee
where \texttt{NR} is the descriptor realization $(\widetilde A_r-\lambda \widetilde E_r,\widetilde B_{r},\widetilde C_{r,1},\widetilde D_{r,1})$ of the right nullspace basis $N_r(\lambda)$ of the transfer function matrix $G_1(\lambda)$, and \texttt{SYS2*NR} (the series coupling of \texttt{SYS2} and \texttt{NR}) is the descriptor system realization $(\widetilde A_r-\lambda \widetilde E_r,\widetilde B_{r},\widetilde C_{r,2},\widetilde D_{r,2})$ of the transfer function matrix $G_2(\lambda)N_r(\lambda)$.
\item
 \texttt{INFO} is a MATLAB structure containing additional information, as follows:\\
 \pagebreak[4]
{\setlength\LTleft{30pt}\begin{longtable}{|l|p{12cm}|} \hline
\textbf{\texttt{INFO} fields} & \textbf{Description} \\ \hline
 \texttt{nrank}   & normal rank of the transfer function matrix of \texttt{SYS1}; \\ \hline \\[-4.5mm]
 \texttt{stdim}   & dimensions of the diagonal blocks of $\widetilde A_r-\lambda \widetilde E_r$: \\
                &if \texttt{OPTIONS.simple = false}, these are the row dimensions
                of the full row rank subdiagonal blocks of the pencil
                $[\, \widetilde B_r\; \widetilde A_r-\lambda \widetilde E_r\,]$ in controllability staircase form; \\
                &if \texttt{OPTIONS.simple = true}, these are the orders of the
                state-space realizations of the proper rational vectors of
                the computed simple proper rational right nullspace
                basis of \texttt{SYS1}; \\ \hline
 \texttt{degs}   & increasingly ordered degrees of the vectors of a
                polynomial right nullspace basis of the transfer function matrix $G_1(\lambda)$ of \texttt{SYS1}, representing
                the right Kronecker indices of $G_1(\lambda)$;
                also the orders of the realizations of the proper rational
                vectors of a simple proper rational right nullspace
                basis. If \texttt{OPTIONS.simple = true}, \texttt{INFO.deg}$(i)$ is
                the dimension of the $i$-th diagonal blocks of $ \widetilde A_r$ and $ \widetilde E_r$. \\ \hline
 \texttt{tcond}   & maximum of the condition numbers of the employed
                non-orthogonal transformation matrices; a warning is
                issued if \texttt{INFO.tcond} $\geq$ \texttt{OPTIONS.tcond}.\\ \hline
\texttt{fnorm}   & the norm of the employed state-feedback used for
                stabilization; is zero if both \texttt{OPTION.sdeg} and \texttt{OPTIONS.pole} are empty.  \\ \hline
\end{longtable}}
\end{description}

\subsubsection*{Method}
For the definitions related to minimal nullspace bases see Section \ref{sec:minbasis}. In what follows, we sketch the approach to compute minimal proper rational right nullspace bases proposed in \cite{Varg03b} (see also \cite[Section 10.3.2]{Varg17} for more details). This approach is employed if \texttt{OPTIONS.simple = false}.

Let $G_1(\lambda)$ be the $p_1\times m$ TFM of \texttt{SYS1} and let $G_2(\lambda)$ be the $p_2\times m$ TFM of  \texttt{SYS2}. Assume $r = \rank G_1(\lambda)$ is the normal rank of $G_1(\lambda)$.   Let $N_{r}(\lambda)$ be a $m\times (m-r)$ rational left nullspace basis of $G_1(\lambda)$ satisfying
\[ G_1(\lambda)N_{r}(\lambda) = 0 .\]
Such a basis can be computed as
\be\label{grnull:rns}                N_{2}(\lambda) = [\, 0 \;\; I_m\,]Y_r(\lambda), \ee
  where $Y_r(\lambda)$ is a rational basis of the right nullspace of the
  system matrix pencil
\be\label{grnull:syspen}            S(\lambda) = \ba{cc} A-\lambda E &  B \\
                          C_1 &       D_1 \ea .\ee
The right nullspace $Y_r(\lambda)$ is determined using the Kronecker-like
  staircase form of $S(\lambda)$ computed as
 \be\label{grnull:klf}  Q^TS(\lambda)Z =  \ba{ccc} B_r & A_r-\lambda E_r & \ast \\
                               0 &  0 &A_l-\lambda E_l \ea \, ,\ee
  where $Q$ and $Z$ are orthogonal transformation matrices,
  the subpencil $A_l-\lambda E_l$ contains the right Kronecker structure
  and the regular part of $S(\lambda)$, and the subpencil
  $[\, B_r\; A_r-\lambda E_r \,]$
  contains the right Kronecker structure of $S(\lambda)$.
  $[\, B_r\; A_r-\lambda E_r \,]$ is obtained in an controllability staircase form with $[\, B_r\; A_r\,]$ as in (\ref{cscf-defab}) and $E_r$ upper triangular and nonsingular, as in (\ref{cscf-defe}). $Y_r(\lambda)$ results as
  \[ Y_r(\lambda) = Z \ba{c} I \\   (\lambda E_r-A_r-B_rF)^{-1}B_r \\  0
\ea , \]
where $F$ is a stabilizing state feedback ($F = 0$ if both \texttt{OPTION.sdeg} and \texttt{OPTIONS.pole} are empty).
The rational basis $N_{r}(\lambda)$ results using (\ref{grnull:rns}) with the controllable
  state-space realization
  \[ (\widetilde A_r-\lambda \widetilde E_r,\widetilde B_{r},\widetilde C_{r,1},\widetilde D_{r,1}) := (A_r+B_rF-\lambda E_r,B_{r},C_{r,1}+D_{r,1}F,D_{r,1}), \] where
  \[ [\; \ast \;\; C_{r,1} \;\; D_{r,1}\,] := [\, 0\;\; I_m\,]Z.  \]
The resulting basis is column proper, that is,
 $D_{r,1}$ has full column rank. The descriptor realization of $G_2(\lambda)N_{r}(\lambda)$ is obtained as
 \[ (\widetilde A_r-\lambda \widetilde E_r,\widetilde B_{r},\widetilde C_{r,2},\widetilde D_{r,2}) := (A_r+B_rF-\lambda E_r,B_{r},C_{r,2}+D_{r,2}F,D_{r,2}), \] where
  \[ [\; \ast \;\; C_{r,2} \;\; D_{r,2}\,] := [\,C_2\;\; D_2\,]Z  .  \]
If \texttt{OPTIONS.inner = true}, an inner basis is determined as the inner factor $N_{ri}(\lambda)$ of the QR-like factorization $N_r(\lambda) = N_{ri}(\lambda)N_{ro}(\lambda)$, with $N_{ro}(\lambda)$ invertible and with all its zeros stable.  $N_{ro}(\lambda)$ has the realization $(A_r+B_rF-\lambda E_r,B_r,-HF,H)$,
 where $F$ is  the stabilizing state feedback computed by solving an appropriate
      control Riccati equation and $H$ is a suitable invertible feedthrough matrix
      (see Theorem \ref{T:IOF} and Theorem \ref{T:IOFd} for details). Accordingly, the realizations for the inner basis $N_{ri}(\lambda)$ and $G_2(\lambda)N_{ri}(\lambda)$ can be explicitly computed as
 \[ N_{ri}(\lambda) = \ba{c|c} A_r+B_rF-\lambda E_r & B_rH^{-1} \\ \hline C_{r,1}+D_{r,1}F & D_{r,1}H^{-1}\ea ,\quad  G_2(\lambda)N_{ri}(\lambda) = \ba{c|c} A_r+B_rF-\lambda E_r & B_rH^{-1} \\ \hline C_{r,2}+D_{r,2}F & D_{r,2}H^{-1} \ea . \]

 \begin{remark} The existence of an inner basis $N_{ri}(\lambda)$ requires that $N_r(\lambda)$ has no zeros on the boundary of the stability region. This condition is ensured, for example, if $N_r(\lambda)$ is minimal, which is guaranteed if the realization $(A-\lambda E,B,C_1,D_1)$ of \texttt{SYS1} is minimal. More generally, \texttt{SYS2} must not have poles on the boundary of the stability domain, which are uncontrollable eigenvalues of the realization of \texttt{SYS1}. In this case, there is no inner basis $N_{ri}(\lambda)$ such that $G_2(\lambda)N_{ri}(\lambda)$ is stable too. Therefore, the condition on zeros is only checked if \texttt{SYS2} is provided, and, if not fulfilled, the resulting $N_r(\lambda)$  and $G_2(\lambda)N_{r}(\lambda)$ are returned. \finr
 \end{remark}

For the computation of the Kronecker-like form (\ref{grnull:klf}), the mex-function
\url{sl_klf}, based on the algorithm proposed in \cite{Beel87}, is called  by the function \texttt{\bfseries grnull}. \index{MEX-functions!\url{sl_klf}}%
\texttt{INFO.stdim} contains the dimensions $\nu_j, \,j = 1, \ldots, k$ of the diagonal blocks in the staircase form (\ref{cscf-defab}). \texttt{INFO.degs} contains the degrees of a minimal polynomial left nullspace basis. These are the right Kronecker indices of the system matrix pencil $S(\lambda)$ in (\ref{grnull:syspen}) and are determined as follows:  there are $\nu_{i-1}-\nu_i$ vectors of degree $i-1$, for $i = 1, \ldots, k$, where $\nu_0 := m-r$.

If \texttt{OPTIONS.simple = true}, a simple proper left nullspace basis is computed, using the method of \cite{Varg08a} to determine a simple basis from a
      proper basis as computed above. The employed dynamic covers based algorithm relies on performing non-orthogonal similarity transformations. The estimated maximum condition number used in these computations is provided in \texttt{INFO.tcond}.
For a simple proper basis, $\widetilde A_r$, $\widetilde E_r$ and $\widetilde B_r$ are block
   diagonal
\[ \widetilde A_r-\lambda \widetilde E_r = \diag (A_r^{1}-\lambda E_r^{1}, \cdots, A_r^{k}-\lambda E_r^{k}), \quad  \widetilde B_r = \diag (B_r^{1}, \cdots, B_r^{k}), \] with $\widetilde E_r$ upper triangular. The state-space realization of the $i$-th basis (column) vector $v_i(\lambda)$ can
   be explicitly constructed as $(A_r^{i}-\lambda E_r^{i},B_{r}^{i},\widetilde C_{r,1},D_{r,1}^{i})$, where $D_{r,1}^{i}$ is the $i$-th column of $D_{r,1}$. \texttt{INFO.stdim} contains the dimensions of the diagonal blocks $A_r^{i}-\lambda E_r^{i}$, $i = 1, \ldots, k$ and are equal to \texttt{INFO.degs}.
   The corresponding realization for $G_2(\lambda)v_i(\lambda)$ is constructed as $(A_r^{i}-\lambda E_r^{i},B_{r}^{i},\widetilde C_{r,2},D_{r,2}^{i})$, where $D_{r,2}^{i}$ is the $i$-th column of $D_{r,2}$. If \texttt{OPTIONS.inner = true}, then each basis vector is determined inner, by applying  the above approach  separately to each   basis vector.

The resulting realization of \texttt{SYSRNULL} is minimal provided the
   realization of \texttt{SYS} is minimal. However, \texttt{NR} is a minimal proper basis
   only if the realization $(A-\lambda E,B,C_1,D_1)$ of \texttt{SYS1} is minimal. In
   this case, \texttt{INFO.degs} are the degrees of the vectors of a minimal
   polynomial basis or, if \texttt{OPTIONS.simple = true}, of the resulting
   minimal simple proper basis.

\subsubsection*{Example}

\begin{example}\label{ex:grnull}
Consider the transfer function matrix used in \cite[p.~459]{Kail80}
\be\label{ex:grnull_TFM} G(s) = {\def\arraystretch{1.4}
\left[\begin{array}{cccc} \displaystyle\frac{1}{s} & 0 &  \displaystyle\frac{1}{s} & s\\ 0 & {\left(s + 1\right)}^2 & {\left(s + 1\right)}^2 & 0\\ -1 & {\left(s + 1\right)}^2 & s^2 + 2\, s & - s^2 \end{array}\right]} ,
\ee
which has normal rank $r = 2$ and the right Kronecker indices $\nu_1 = 0$ and $\nu_2 = 2$. A simple proper minimal right nullspace basis $N_r(s)$ with the poles assigned in $-1$ has been computed with the function \texttt{grnull} as
\[ N_r(s) = {\def\arraystretch{2}\ba{cc}
-0.5547 & \displaystyle\frac{0.70165s^2}{(s+1)^2} \\
-0.5547 & \displaystyle\frac{-0.43853 s^2}{(s+1)^2}\\
\phantom{-}0.5547 & \displaystyle\frac{0.43853 s^2}{(s+1)^2} \\
0 & \displaystyle\frac{-1.14}{(s+1)^2} \ea} \]
and has two basis vectors of McMillan degrees 0 and 2. A polynomial basis results simply by taking the numerator polynomial vectors
\[ \widetilde N_r(s) = {\def\arraystretch{2}\ba{cc}
-0.5547 & \phantom{-}0.70165s^2 \\
-0.5547 & -0.43853 s^2\\
\phantom{-}0.5547 & \phantom{-}0.43853 s^2 \\
0 & -1.14 \ea} \]

To compute $N_r(s)$ and $\widetilde N_r(s)$, the following sequence of commands can be used:
\begin{verbatim}


% Kailath (1980), page 459: rank 2 matrix
s = tf('s');
G = [1/s 0 1/s s;
    0 (s+1)^2 (s+1)^2 0;
    -1 (s+1)^2 s^2+2*s -s^2];
sys = gir(ss(G),1.e-7);

% set options for simple basis and pole assignment
options = struct('tol',1.e-7,'simple',true,'poles',[-1,-1]);

% compute a simple right nullspace basis Nr(s)
[Nr,info] = grnull(sys,options);
tf(Nr), rki = info.degs  % right Kronecker indices

% check nullspace condition G(s)*Nr(s) = 0
gminreal(sys*Nr,1.e-7)

% minimal polynomial basis computation
Nrtf = tf(Nr);
Nrp = tf(Nrtf.num,1)

% check nullspace condition G(s)*Nrp(s) = 0
gminreal(sys*Nrp,1.e-7)
\end{verbatim}
\fine\end{example}

For examples illustrating the computational details of determining simple and polynomial bases, see Example \ref{ex:simple-basis} and  Example \ref{ex:polynomial-basis}, respectively.

\subsubsection{\texttt{\bfseries glnull}}
\index{M-functions!\texttt{\bfseries glnull}}
\subsubsection*{Syntax}
\begin{verbatim}
[SYSLNULL,INFO] = glnull(SYS,OPTIONS)
\end{verbatim}

\subsubsection*{Description}
\texttt{\bfseries glnull} computes a proper rational  basis $N_l(\lambda)$ of the left nullspace of the transfer function matrix $G_1(\lambda)$ of a LTI descriptor system,
such that
\[ N_l(\lambda)G_1(\lambda) = 0 \]
and determines $N_l(\lambda)G_2(\lambda)$, where $G_2(\lambda)$ is a TFM having the same number of rows as $G_1(\lambda)$.
\index{descriptor system!minimal nullspace basis}
\index{nullspace!basis!minimal proper, left}
\index{nullspace!basis!simple minimal proper, left}

\subsubsection*{Input data}
\begin{description}
\item
\texttt{SYS} is an input concatenated compound LTI system, \texttt{SYS = [ SYS1 SYS2 ]},  in a  descriptor system state-space form
\be\label{glnull:sysss}
\begin{aligned}
E\lambda x(t)  &=   Ax(t)+ B_1 u_1(t) + B_2 u_2(t) ,\\
y(t) &=  C x(t)+ D_1 u_1(t) + D_2 u_2(t) ,
\end{aligned}
\ee
where \texttt{SYS1} has the transfer function matrix  $G_1(\lambda)$ with the descriptor system realization $(A-\lambda E,B_1,C,D_1)$,
\texttt{SYS2} has the transfer function matrix  $G_2(\lambda)$ with the descriptor system  realization $(A-\lambda E,B_2,C,D_2)$, and $u_1(t) \in \mathds{R}^{m_1}$ and $u_2(t) \in \mathds{R}^{m_2}$ are the inputs of \texttt{SYS1} and \texttt{SYS2}, respectively.
\item
 \texttt{OPTIONS} is a MATLAB structure to specify user options and has the following fields:\\
{\tabcolsep=1mm
\begin{longtable}{|l|lcp{10cm}|} \hline
\textbf{\texttt{OPTIONS} fields} & \multicolumn{3}{l|}{\textbf{Description}} \\ \hline
 \texttt{tol}   & \multicolumn{3}{p{10cm}|}{relative tolerance for rank computations \newline (Default: internally computed);} \\ \hline
 \texttt{m2}   & \multicolumn{3}{l|}{$m_2$, the number of inputs of \texttt{SYS2} (Default: $m_2 = 0$); } \\ \hline
 \texttt{simple}   & \multicolumn{3}{l|}{option to compute a simple proper basis:}\\
                 &  \texttt{true} &--& compute a simple basis; the orders of the
                            basis vectors are provided in \texttt{INFO.degs}; \\
                 &  \texttt{false}&--& no simple basis computed (default);  \\
                                        \hline
 \texttt{coinner}   & \multicolumn{3}{p{10cm}|}{option to compute a coinner basis:}\\
                  &  \texttt{true} &--& compute a coinner basis; if the option for
                             simple basis has been selected then each
                             basis vector results coinner and the orders of
                             the basis vectors are provided in \texttt{INFO.deg}
                             (the resulting basis may not be coinner) \\
                  &  \texttt{false}&--& no coinner basis is computed (default)  \\
                                        \hline
\texttt{offset}   & \multicolumn{3}{p{11.5cm}|}{stability
 boundary offset $\beta$, to be used  to assess the finite zeros which belong to $\partial\mathds{C}_s$ (the boundary of the stability domain) as follows: in the
 continuous-time case these are the finite
  zeros having real parts in the interval $[-\beta, \beta]$, while in the
 discrete-time case these are the finite zeros having moduli in the
 interval $[1-\beta, 1+\beta]$ \newline (Default: $\beta = 1.4901\cdot 10^{-08}$). } \\ \hline
 \texttt{tcond}   & \multicolumn{3}{l|}{maximum allowed value for the condition numbers of the  employed}\\
     & \multicolumn{3}{l|}{non-orthogonal transformation matrices (Default: $10^4$)}\\
    & \multicolumn{3}{l|}{(only used if \texttt{OPTIONS.simple = true}); } \\
                    \hline
\texttt{sdeg}   & \multicolumn{3}{p{11cm}|}{prescribed stability degree for the resulting left nullspace basis
                    (Default: \texttt{[ ]}) } \\ \hline
 \texttt{poles}   & \multicolumn{3}{p{11cm}|}{a complex conjugated set of desired poles  to be assigned for the resulting left nullspace basis
                     (Default: \texttt{[ ]}).}\\
                                        \hline
\end{longtable}}
\end{description}

\subsubsection*{Output data}
\begin{description}
\item
\texttt{SYSLNULL} contains  the input concatenated compound system \texttt{[ NL  NL*SYS2 ]}, in a descriptor system state-space form
\be\label{glnull:lnullsysss}
\begin{aligned}
\widetilde E_l\lambda  x_l(t)  &=    \widetilde A_l x_l(t)+  \widetilde B_{l,1} v_1(t)+  \widetilde B_{l,2} v_2(t) ,\\
y_l(t) &=   \widetilde C_l x_l(t)+  \widetilde D_{l,1} v_1(t)+  \widetilde D_{l,2} v_2(t)  ,
\end{aligned}
\ee
where \texttt{NL} is the descriptor system realization $( \widetilde A_l-\lambda \widetilde E_l, \widetilde B_{l,1}, \widetilde C_l, \widetilde D_{l,1})$ of the left nullspace basis $N_l(\lambda)$ of the transfer function matrix $G_1(\lambda)$, and \texttt{NL*SYS2} (the series coupling of \texttt{NL} and \texttt{SYS2}) is  the descriptor system realization $( \widetilde A_l-\lambda  \widetilde E_l, \widetilde B_{l,2}, \widetilde C_l, \widetilde D_{l,2})$ of the transfer function matrix $N_l(\lambda)G_2(\lambda)$.
\item
 \texttt{INFO} is a MATLAB structure containing additional information, as follows:\\
{\begin{longtable}{|l|p{12cm}|} \hline
\textbf{\texttt{INFO} fields} & \textbf{Description} \\ \hline
 \texttt{nrank}   & normal rank of the transfer function matrix of \texttt{SYS1}; \\ \hline \\[-4mm]
 \texttt{stdim}   & dimensions of the diagonal blocks of $ \widetilde A_l-\lambda \widetilde E_l$: \\
                &if \texttt{OPTIONS.simple = false}, these are the column dimensions
                of the full column rank subdiagonal blocks of the pencil
                $\left[\begin{smallmatrix}  \widetilde A_l-\lambda \widetilde E_l\\ \widetilde C_l\end{smallmatrix}\right]$ in observability staircase form; \\
                &if \texttt{OPTIONS.simple = true}, these are the orders of the
                state-space realizations of the proper rational vectors of
                the computed simple proper rational left nullspace
                basis of \texttt{SYS1}; \\ \hline
 \texttt{degs}   & increasingly ordered degrees of the vectors of a
                polynomial left nullspace basis of the transfer function matrix $G_1(\lambda)$ of \texttt{SYS1}, representing
                the left Kronecker indices of $G_1(\lambda)$;
                also the orders of the realizations of the proper rational
                vectors of a simple proper rational left nullspace
                basis. If \texttt{OPTIONS.simple = true}, \texttt{INFO.deg}$(i)$ is
                the dimension of the $i$-th diagonal blocks of $ \widetilde A_l$ and $ \widetilde E_l$. \\ \hline
 \texttt{tcond}   & maximum of the condition numbers of the employed
                non-orthogonal transformation matrices; a warning is
                issued if \texttt{INFO.tcond} $\geq$ \texttt{OPTIONS.tcond}.\\ \hline
\end{longtable}}
\end{description}

\subsubsection*{Method}
For the definitions related to minimal nullspace bases see Section \ref{sec:minbasis}. In what follows, we sketch the approach to compute minimal proper rational left nullspace bases, which is the dual version of the method proposed in \cite{Varg03b} (see also \cite[Section 10.3.2]{Varg17} for more details). This approach is employed if \texttt{OPTIONS.simple = false}.

Let $G_1(\lambda)$ be the $p\times m_1$ TFM of \texttt{SYS1} and let $G_2(\lambda)$ be the $p\times m_2$ TFM of  \texttt{SYS2}. Assume $r = \rank G_1(\lambda)$ is the normal rank of $G_1(\lambda)$.   Let $N_{l}(\lambda)$ be a $(p-r)\times p$ rational left nullspace basis of $G_1(\lambda)$ satisfying
\[ N_{l}(\lambda)G_1(\lambda) = 0 .\]
Such a basis can be computed as
\be\label{glnull:lns}                N_{l}(\lambda) = Y_l(\lambda)\ba{c}0\\ I_p\ea, \ee
  where $Y_l(\lambda)$ is a rational basis of the left nullspace of the
  system matrix pencil
\be\label{glnull:syspen}            S(\lambda) = \ba{cc} A-\lambda E &  B_1 \\
                          C &       D_1 \ea .\ee
The left nullspace $Y_l(\lambda)$ is determined using the Kronecker-like
  staircase form of $S(\lambda)$ computed as
 \be\label{glnull:klf}  Q^TS(\lambda)Z =  \ba{cc} A_r-\lambda E_r & \ast \\
                               0 &  A_l-\lambda E_l \\
                               0 &      C_l \ea \, ,\ee
  where $Q$ and $Z$ are orthogonal transformation matrices,
  the subpencil $A_r-\lambda E_r$ contains the right Kronecker structure
  and the regular part of $S(\lambda)$, and the subpencil $\left[\begin{smallmatrix} A_l-\lambda E_l\\C_l\end{smallmatrix}\right]$
  contains the left Kronecker structure of $S(\lambda)$.
  $\left[\begin{smallmatrix} A_l-\lambda E_l\\C_l\end{smallmatrix}\right]$ is obtained in an observability staircase form with $\left[\begin{smallmatrix} A_l\\C_l\end{smallmatrix}\right]$ as in (\ref{oscf-defac}) and $E_l$ upper triangular and nonsingular, as in (\ref{oscf-defe}). $Y_l(\lambda)$ results as
  \[ Y_l(\lambda) = {\arraycolsep=2mm\ba{c|c|c} 0 &  C_l(\lambda E_l-A_l-FC_l)^{-1} & I
\ea} Q^T , \]
where $F$ is a stabilizing output injection matrix. The rational basis $N_{l}(\lambda)$ results using (\ref{glnull:lns}) with the observable
  state-space realization
  \[ ( \widetilde A_l-\lambda  \widetilde E_l, \widetilde B_{l,1}, \widetilde C_l, \widetilde D_{l,1}) := (A_l+FC_l-\lambda E_l,B_{l,1}+FD_{l,1},C_l,D_{l,1}), \] where
  \[ \ba{c} \ast \\ B_{l,1} \\ D_{l,1} \ea := Q^T \ba{c}0\\ I_p\ea.  \]
The resulting basis is row proper, that is,
 $D_{l,1}$ has full row rank. The descriptor realization of $N_{l}(\lambda)G_2(\lambda)$ is obtained as
 \[ ( \widetilde A_l-\lambda  \widetilde E_l, \widetilde B_{l,2}, \widetilde C_l, \widetilde D_{l,2}) := (A_l+FC_l-\lambda E_l,B_{l,2}+FD_{l,2},C_l,D_{l,2}), \] where
  \[ \ba{c} \ast \\ B_{l,2} \\ D_{l,2} \ea := Q^T \ba{c}B_2\\ D_2\ea.  \]

If \texttt{OPTIONS.coinner = true}, a coinner basis is determined as the coinner factor $N_{li}(\lambda)$ of the RQ-like factorization $N_l(\lambda) = N_{li}(\lambda)N_{lo}(\lambda)$, with $N_{lo}(\lambda)$ invertible and with all its zeros stable.  $N_{lo}(\lambda)$ has the realization $(A_l+FC_l-\lambda E_l,-FH,C_l,H)$,
 where $F$ is  the stabilizing output injection computed by solving an appropriate
      filter Riccati equation and $H$ is a suitable invertible feedthrough matrix
      (for details, see Theorem \ref{T:IOF} and Theorem \ref{T:IOFd} applied to the dual system). Accordingly, the realizations for the coinner basis $N_{li}(\lambda)$ and $N_{li}(\lambda)G_2(\lambda)$ can be explicitly computed as
 \[ N_{li}(\lambda) = {\arraycolsep=1mm\ba{c|c} A_l+FC_l-\lambda E_l & B_{l,1}+FD_{l,1}  \\ \hline H^{-1}C_l & H^{-1}D_{l,1}\ea ,\quad  N_{li}(\lambda)G_2(\lambda) = \ba{c|c} A_l+FC_l-\lambda E_l & B_{l,2}+FD_{l,2} \\ \hline H^{-1}C_l & H^{-1}D_{l,2} \ea} . \]

 \begin{remark} The existence of a coinner basis $N_{li}(\lambda)$ requires that $N_l(\lambda)$ has no zeros on the boundary of the stability region. This condition is ensured, for example, if $N_l(\lambda)$ is minimal, which is guaranteed if the realization $(A-\lambda E,B_1,C,D_1)$ of \texttt{SYS1} is minimal. More generally, \texttt{SYS2} must not have poles on the boundary of the stability domain, which are unobservable eigenvalues of the realization of \texttt{SYS1}. In this case, there is no coinner basis $N_{li}(\lambda)$ such that $N_{li}(\lambda)G_2(\lambda)$ is stable too. Therefore, the condition on zeros is only checked if \texttt{SYS2} is provided, and, if not fulfilled, the resulting $N_l(\lambda)$  and $N_{l}(\lambda)G_2(\lambda)$ are returned. \finr
 \end{remark}

For the computation of the Kronecker-like form (\ref{glnull:klf}), the mex-function
\url{sl_klf}, based on the algorithm proposed in \cite{Beel87}, is called  by the function \texttt{\bfseries glnull}. \index{MEX-functions!\url{sl_klf}}%
\texttt{INFO.stdim} contains the dimensions $\mu_j, \,j = 1, \ldots, \ell$ of the diagonal blocks in the staircase form (\ref{oscf-defac}). \texttt{INFO.degs} contains the degrees of a minimal polynomial left nullspace basis. These are the left Kronecker indices of the system matrix pencil $S(\lambda)$ in (\ref{glnull:syspen}) and are determined as follows:  there are $\mu_{i-1}-\mu_i$ vectors of degree $i-1$, for $i = 1, \ldots, \ell$, where $\mu_0 := p-r$.

If \texttt{OPTIONS.simple = true}, a simple proper left nullspace basis is computed, using the method of \cite{Varg08a} to determine a simple basis from a
       proper basis as computed above. The employed dynamic covers based algorithm relies on performing non-orthogonal similarity transformations. The estimated maximum condition number used in these computations is provided in \texttt{INFO.tcond}.
For a simple proper basis, $A_l$, $E_l$ and $C_l$ are block
   diagonal
\[ \widetilde A_l-\lambda \widetilde E_l = \diag (A_l^{1}-\lambda E_l^{1}, \cdots, A_l^{\ell}-\lambda E_l^{\ell}), \quad  \widetilde C_l = \diag (C_l^{1}, \cdots, C_l^{\ell}), \] with $E_l$ upper triangular. The state-space realization of the $i$-th basis (row) vector $v_i(\lambda)$ can
   be explicitly constructed as $(A_l^{i}-\lambda E_l^{i},\widetilde B_{l,1},C_l^{i},D_{l,1}^{i})$, where $D_{l,1}^{i}$ is the $i$-th row of $D_{l,1}$. \texttt{INFO.stdim} contains the dimensions of the diagonal blocks $A_l^{i}-\lambda E_l^{i}$, $i = 1, \ldots, \ell$ and are equal to \texttt{INFO.degs}.
   The corresponding descriptor system realization for $v_i(\lambda)G_2(\lambda)$ is constructed as $(A_l^{i}-\lambda E_l^{i},\widetilde B_{l,2},C_l^{i},D_{l,2}^{i})$, where $D_{l,2}^{i}$ is the $i$-th row of $D_{l,2}$. If \texttt{OPTIONS.coinner = true}, then each basis vector is determined coinner, by applying  the above approach  separately to each   basis vector.

The resulting realization of \texttt{SYSLNULL} is minimal provided the
   realization of \texttt{SYS} is minimal. However, \texttt{NL} is a minimal proper basis
   only if the realization $(A-\lambda E,B_1,C,D_1)$ of \texttt{SYS1} is minimal. In
   this case, \texttt{INFO.degs} are the degrees of the vectors of a minimal
   polynomial basis or, if \texttt{OPTIONS.simple = true}, of the resulting
   minimal simple proper basis.

\subsubsection*{Example}
\begin{example}
Consider the $3\times 4$ transfer function matrix (\ref{ex:grnull_TFM}) from \cite[p.~459]{Kail80} used in Example \ref{ex:grnull}.
$G(s)$ has normal rank $r = 2$ and a left Kronecker index $\mu_1 = 1$. A proper minimal left nullspace basis $N_l(s)$ with the poles assigned in $-1$ has been computed with the function \texttt{glnull} as
\[ N_l(s) = \ba{ccc}
-\displaystyle\frac{s}{s+1} &  \displaystyle\frac{1}{s+1} & -\displaystyle\frac{1}{s+1}  \ea \]
and consists of a single basis vector of McMillan degree 1. To compute $N_l(s)$, the following sequence of commands can be used:
\newpage
\begin{verbatim}
% Kailath (1980), page 459: rank 2 matrix
s = tf('s');
G = [1/s 0 1/s s;
    0 (s+1)^2 (s+1)^2 0;
    -1 (s+1)^2 s^2+2*s -s^2];
sys = gir(ss(G),1.e-7);

% set options for simple basis and pole assignment
options = struct('tol',1.e-7,'simple',true,'poles',[-1]);

% compute a simple left nullspace basis Nl(s)
[Nl,info] = glnull(sys,options);
minreal(tf(Nl))

% check the nullspace condition Nl(s)*G(s) = 0
gminreal(Nl*sys,1.e-7)
\end{verbatim}
\fine\end{example}

\subsubsection{\texttt{\bfseries grange}}
\index{M-functions!\texttt{\bfseries grange}}

\subsubsection*{Syntax}
\begin{verbatim}
[SYSR,SYSX,INFO] = grange(SYS,OPTIONS)
\end{verbatim}

\subsubsection*{Description}
\texttt{\bfseries grange} computes a proper rational  basis $R(\lambda)$ of the range space of the transfer function matrix $G(\lambda)$ of a LTI descriptor system, and a full row rank $X(\lambda)$ such that
\be\label{grange:def} G(\lambda) = R(\lambda) X(\lambda)  \ee
is a full-rank factorization of $G(\lambda)$.
\index{factorization!full rank}%

\subsubsection*{Input data}
\begin{description}
\item
\texttt{SYS} is a LTI system in a  descriptor system state-space form
\be\label{grange:sysss}
\begin{aligned}
E\lambda x(t)  &=   Ax(t)+ Bu(t) ,\\
y(t) &=  C x(t)+ D u(t) ,
\end{aligned}
\ee
whose transfer function matrix is $G(\lambda)$.
\item
\pagebreak[4]
 \texttt{OPTIONS} is a MATLAB structure to specify user options and has the following fields:\\
{\tabcolsep=1mm
\begin{longtable}{|l|lcp{10cm}|} \hline
\textbf{\texttt{OPTIONS} fields} & \multicolumn{3}{l|}{\textbf{Description}} \\ \hline
 \texttt{tol}   & \multicolumn{3}{l|}{relative tolerance for rank computations (Default: internally computed)} \\ \hline
\texttt{offset}   & \multicolumn{3}{p{12.5cm}|}{stability
 boundary offset $\beta$, to be used  to assess the finite zeros which belong to $\partial\mathds{C}_s$ (the boundary of the stability domain) as follows: in the
 continuous-time case these are the finite
  zeros having real parts in the interval $[-\beta, \beta]$, while in the
 discrete-time case these are the finite zeros having moduli in the
 interval $[1-\beta, 1+\beta]$ (Default: $\beta = 1.4901\cdot 10^{-08}$). } \\ \hline
 \texttt{zeros}   & \multicolumn{3}{p{12.5cm}|}{option for the selection of zeros to be included in  the computed range space basis:}\\
                 &  \texttt{'none'}&--& include no zeros (default)   \\
                 &  \texttt{'all'}&--& include all zeros of \texttt{SYS} \\
                 &  \texttt{'unstable'}&--& include all unstable zeros of \texttt{SYS} \\
                 &  \texttt{'s-unstable'}&--& include all strictly unstable zeros of \texttt{SYS}, both finite and infinite \\
                 &  \texttt{'stable'} &--& include all stable zeros of \texttt{SYS} \\
                 &  \texttt{'finite'} &--& include all finite zeros of \texttt{SYS} \\
                 &  \texttt{'infinite'}&--& include all infinite zeros of \texttt{SYS}  \\
                                        \hline
 \texttt{inner}   & \multicolumn{3}{l|}{option to compute an inner basis:}\\
                  &  \texttt{true} &--& compute an inner basis (only if  \texttt{OPTIONS.zeros = 'none'} or \texttt{OPTIONS.zeros = 'unstable'}); \\
                  &  \texttt{false}&--& no inner basis is computed (default)  \\
                                        \hline
 \texttt{balance}   & \multicolumn{3}{p{12.5cm}|}{balancing option for the Riccati
                    equation solvers (see functions \texttt{care} and \texttt{dare} of the Control System Toolbox):}\\
                  &  \texttt{true} &--& perform balancing (default); \\
                  &  \texttt{false}&--& disable balancing.   \\
                                        \hline
\end{longtable}}
\end{description}

\subsubsection*{Output data}
\begin{description}
\item
\texttt{SYSR} contains  the descriptor system state-space realization of the full column rank proper range basis matrix  $R(\lambda)$ in the form
\be\label{grange:sysi}
\begin{aligned}
 E_R\lambda x_R(t)  &=   A_R x_R(t)+ B_R v(t)  ,\\
y_R(t) &=  C_R x_R(t)+ D_R v(t)  ,
\end{aligned}
\ee
where $E_R$ is invertible. The resulting $R(\lambda)$ contains the selected zeros of $G(\lambda)$ via the option parameter \texttt{OPTIONS.zeros}.  $R(\lambda)$ is inner if \texttt{OPTIONS.inner = true} was selected. The dimension $r$ of the input vector $v(t)$ is the normal rank of $G(\lambda)$.
\item
\texttt{SYSX} contains  the descriptor system state-space realization of the full row rank transfer function matrix $X(\lambda)$ in the form
\be\label{grange:syso}
\begin{aligned}
 E\lambda \widetilde x(t)  &=   A \widetilde x(t)+ B u(t)  ,\\
\widetilde y(t) &=  \widetilde C \widetilde x(t)+ \widetilde D u(t)  ,
\end{aligned}
\ee
where the dimension $r$ of the output vector $\widetilde y(t)$ is the normal rank of $G(\lambda)$.
\item
 \texttt{INFO} is a MATLAB structure containing additional information, as follows:\\
{\begin{longtable}{|l|p{12cm}|} \hline
\textbf{\texttt{INFO} fields} & \textbf{Description} \\ \hline
 \texttt{nrank}   & normal rank of the transfer function matrix $G(\lambda)$; \\ \hline
 \texttt{nfuz}   & number of finite unstable zeros of
              \texttt{SYS} lying on the boundary of the stability region $\partial\mathds{C}_s$ within the offset specified
by \texttt{OPTION.offset};\\ \hline
 \texttt{niuz}   & number of infinite zeros of
              \texttt{SYS} in the continuous-time case
              and 0 in the discrete-time case\\ \hline
 \texttt{ricrez}   & diagnosis flag, as provided provided by the generalized
               Riccati equation solvers \texttt{care} and \texttt{dare}; if non-negative,
               this value represents the Frobenius norm of relative
               residual of the Riccati equation, while a negative value
               indicates failure of solving the Riccati equation.  \\ \hline
\end{longtable}}
\end{description}

\subsubsection*{Method}
Consider a disjunct partition of the complex plane $\mathds{C}$ as
\be\label{grange:Cgoodbad}  \mathds{C} = \mathds{C}_g \cup \mathds{C}_b, \quad \mathds{C}_g \cap \mathds{C}_b = \emptyset \, ,\ee
where $\mathds{C}_g$ and $\mathds{C}_b$ are symmetric with respect to the real axis.
$\mathds{C}_g$ and $\mathds{C}_b$ are associated with the ``good'' and ``bad'' domains of the complex plane $\mathds{C}$ for the poles and zeros of $G(\lambda)$.
Assume $G(\lambda)$ is a $p\times m$ real rational matrix of normal rank $r$, with a $\mathds{C}_b$-{stabilizable} descriptor system realization (\ref{grange:sysss}). Then, there exist two orthogonal matrices $U$ and $Z$ such that
\be\label{spec-klf}
\ba{cc} U & 0 \\ 0 & I \ea \ba{cc} A - \lambda E & B \\ \hline C & D \ea Z =
\ba{cccc} A_{rg}-\lambda E_{rg} & \ast & \ast & \ast \\
0 & A_{b\ell}-\lambda E_{b\ell} & B_{b\ell} & \ast \\
0 & 0 & 0 & B_n \\ \hline
0 & C_{b\ell} & D_{b\ell} & \ast \ea , \ee
where
\begin{itemize}
\item[(a)] The pencil $A_{rg}-\lambda E_{rg}$ has full row rank for $\lambda \in \mathds{C}_g$ and $E_{rg}$ has full row rank.
\item[(b)]  $E_{b\ell}$ and $B_n$ are invertible, the pencil
\be\label{syspencil} \ba{cc}   A_{b\ell}-\lambda E_{b\ell} & B_{b\ell} \\ C_{b\ell} & D_{b\ell} \ea \ee
has full column rank $n_{b\ell}+r$ for $\lambda \in\mathds{C}_g$ and the pair $(A_{b\ell}-\lambda E_{b\ell}, B_{b\ell})$ is $\mathds{C}_b$-stabilizable.
\end{itemize}
The range matrix of $G(\lambda)$, which includes the zeros of $G(\lambda)$ in $\mathds{C}_b$, has the proper descriptor system realization
\be\label{range} R(\lambda) = \ba{c|c} A_{b\ell}-\lambda E_{b\ell} & B_{b\ell} \\ \hline C_{b\ell} & D_{b\ell} \ea . \ee
If  \texttt{OPTIONS.inner = true} was selected, the inner range matrix is determined in the form
\be\label{range_stab} R(\lambda) = \ba{c|c} A_{b\ell}+B_{b\ell}F-\lambda E_{b\ell} & B_{b\ell}W \\ \hline C_{b\ell}+D_{b\ell}F & D_{b\ell}W \ea , \ee
where $W$ is a suitable invertible matrix and $F$ is a stabilizing state-feedback matrix.
$X(\lambda)$ is determined with a descriptor realization of the form
\be\label{full-row-rank} X(\lambda) = \ba{c|c} A-\lambda E & B \\ \hline \\[-3.5mm] \widetilde C & \widetilde D \ea , \ee
where $[\,\widetilde C \; \widetilde D\, ] = W^{-1}[\, 0 \;\; -F \;\; I_r \;\; 0 \,]Z^T$.

The overall factorization approach is described in \cite{Varg17f}.
The reduction of the system matrix pencil to the special Kronecker-like form (\ref{spec-klf}) is described in \cite{Oara05} and involves the use of the mex-function \url{sl_klf} to compute the appropriate Kronecker-like form.
\index{MEX-functions!\url{sl_klf}}%
For the computation of an inner basis, extensions of the standard inner-outer factorization methods of \cite{Zhou96} are used. These methods involve the solution of appropriate (continuous- or discrete-time) generalized algebraic Riccati equations. For additional details, see \cite{Oara00} for continuous-time systems and \cite{Oara05} for discrete-time systems.

\subsubsection*{Example}
\begin{example}\label{ex:Oara-Varga}
This is \textbf{Example 1} from \cite{Oara00} of the transfer function matrix of a continuous-time proper system:
\be\label{grange:ex1} G(s) = {\def\arraystretch{2}
\left[\begin{array}{ccc} \displaystyle\frac{s - 1}{s + 2} & \displaystyle\frac{s}{s + 2} & \displaystyle\frac{1}{s + 2}\\ 0 & \displaystyle\frac{s - 2}{{\left(s + 1\right)}^2} & \displaystyle\frac{s - 2}{{\left(s + 1\right)}^2}\\ \displaystyle\frac{s - 1}{s + 2} & \displaystyle\frac{s^2 + 2\, s - 2}{\left(s + 1\right)\, \left(s + 2\right)} & \displaystyle\frac{2\, s - 1}{\left(s + 1\right)\, \left(s + 2\right)} \end{array}\right]} \, .
\ee
$G(s)$ has zeros at $\{1, 2, \infty\}$, poles at $\{-1, -1, -2, -2 \}$, and normal rank $r = 2$.

A minimum proper basis of $\mathcal{R}(G(s))$, computed with \texttt{\bfseries grange}, is
\[ R(s) = \frac{1}{s + 1.374}
\ba{rr}
1.552 s + 2.124 & 1.314 s + 1.817\\
0.593 s + 1.186 & -0.758 s - 1.516 \\
2.145 s + 2.717 & 0.5558 s + 1.059
\ea , \]
has McMillan-degree 1 and no zeros. The full row rank factor $X(s)$, satisfying $G(s) = R(s)X(s)$, has McMillan degree 4, and zeros at $\{1, 2, -1.374, \infty\}$. The zero at  $-1.374$ is equal to the pole of $R(s)$.

The full-rank factorization of $G(s)$ has been computed using the following sequence of commands:\\
\begin{verbatim}
% Oara and Varga (2000), Example 1
s = tf('s');  % define the complex variable s
% enter G(s) and determine a minimal state-space realization
G = [(s-1)/(s+2) s/(s+2) 1/(s+2);
    0 (s-2)/(s+1)^2 (s-2)/(s+1)^2;
    (s-1)/(s+2) (s^2+2*s-2)/(s+1)/(s+2) (2*s-1)/(s+1)/(s+2)];
sys = minreal(ss(G));
gpole(sys)    % the system is stable
gzero(sys)    % the system has 2 unstable zeros and an infinite zero
nrank(sys)    % the normal rank of G(s) is 2

% compute the full-rank factorization G(s) = R(s)*X(s)
[sysr,sysx] = grange(sys,struct('tol',1.e-7));

% check the factorization
norm(sysr*sysx-sys,inf)   %  ||R(s)*X(s)-G(s)||_inf = 0

gzero(sysr)               % R(s) has no zeros
gzero(sysx)               % X(s) has all zeros of G(s)
\end{verbatim}

\fine\end{example}

\subsubsection{\texttt{\bfseries gcrange}}
\index{M-functions!\texttt{\bfseries gcrange}}

\subsubsection*{Syntax}
\begin{verbatim}
[SYSR,SYSX,INFO] = gcrange(SYS,OPTIONS)
\end{verbatim}

\subsubsection*{Description}
\texttt{\bfseries grange} computes a proper rational  basis $R(\lambda)$ of the coimage space of the transfer function matrix $G(\lambda)$ of a LTI descriptor system, and a full column rank $X(\lambda)$ such that
\be\label{gcrange:def} G(\lambda) = X(\lambda)R(\lambda)   \ee
is a full-rank factorization of $G(\lambda)$.
\index{factorization!full rank}%

\subsubsection*{Input data}
\begin{description}
\item
\texttt{SYS} is a LTI system in a  descriptor system state-space form
\be\label{gcrange:sysss}
\begin{aligned}
E\lambda x(t)  &=   Ax(t)+ Bu(t) ,\\
y(t) &=  C x(t)+ D u(t) ,
\end{aligned}
\ee
whose transfer function matrix is $G(\lambda)$.
\item
 \texttt{OPTIONS} is a MATLAB structure to specify user options and has the following fields:\\
{\tabcolsep=1mm
\begin{longtable}{|l|lcp{10cm}|} \hline
\textbf{\texttt{OPTIONS} fields} & \multicolumn{3}{l|}{\textbf{Description}} \\ \hline
 \texttt{tol}   & \multicolumn{3}{l|}{relative tolerance for rank computations (Default: internally computed)} \\ \hline
\texttt{offset}   & \multicolumn{3}{p{12.5cm}|}{stability
 boundary offset $\beta$, to be used  to assess the finite zeros which belong to $\partial\mathds{C}_s$ (the boundary of the stability domain) as follows: in the
 continuous-time case these are the finite
  zeros having real parts in the interval $[-\beta, \beta]$, while in the
 discrete-time case these are the finite zeros having moduli in the
 interval $[1-\beta, 1+\beta]$ (Default: $\beta = 1.4901\cdot 10^{-08}$). } \\ \hline
 \texttt{zeros}   & \multicolumn{3}{p{12cm}|}{option for the selection of zeros to be included in  the computed coimage space basis:}\\
                 &  \texttt{'none'}&--& include no zeros (default)   \\
                 &  \texttt{'all'}&--& include all zeros of \texttt{SYS} \\
                 &  \texttt{'unstable'}&--& include all unstable zeros of \texttt{SYS} \\
                 &  \texttt{'s-unstable'}&--& include all strictly unstable zeros of \texttt{SYS}, both finite and infinite \\
                 &  \texttt{'stable'} &--& include all stable zeros of \texttt{SYS} \\
                 &  \texttt{'finite'} &--& include all finite zeros of \texttt{SYS} \\
                 &  \texttt{'infinite'}&--& include all infinite zeros of \texttt{SYS}  \\
                                        \hline
 \texttt{coinner}   & \multicolumn{3}{l|}{option to compute a coinner basis:}\\
                  &  \texttt{true} &--& compute a coinner basis (only if  \texttt{OPTIONS.zeros = 'none'} or \texttt{OPTIONS.zeros = 'unstable'}); \\
                  &  \texttt{false}&--& no coinner basis is computed (default)  \\
                                        \hline
 \texttt{balance}   & \multicolumn{3}{p{12cm}|}{balancing option for the Riccati
                    equation solvers (see functions \texttt{care} and \texttt{dare} of the Control System Toolbox):}\\
                  &  \texttt{true} &--& perform balancing (default); \\
                  &  \texttt{false}&--& disable balancing.   \\
                                        \hline
\end{longtable}}
\end{description}

\subsubsection*{Output data}
\begin{description}
\item
\texttt{SYSR} contains  the descriptor system state-space realization of the full row rank proper coimage basis matrix  $R(\lambda)$ in the form
\be\label{gcrange:sysi}
\begin{aligned}
 E_R\lambda x_R(t)  &=   A_R x_R(t)+ B_R u(t)  ,\\
w(t) &=  C_R x_R(t)+ D_R u(t)  ,
\end{aligned}
\ee
where $E_R$ is invertible. The resulting $R(\lambda)$ contains the selected zeros of $G(\lambda)$ via the option parameter \texttt{OPTIONS.zeros}.  $R(\lambda)$ is coinner if \texttt{OPTIONS.inner = true} was selected. The dimension $r$ of the output vector $w(t)$ is the normal rank of $G(\lambda)$.
\item
\texttt{SYSX} contains  the descriptor system state-space realization of the full column rank transfer function matrix $X(\lambda)$ in the form
\be\label{gcrange:syso}
\begin{aligned}
 E\lambda \widetilde x(t)  &=   A \widetilde x(t)+ \widetilde B w(t)  ,\\
\widetilde y(t) &=   C \widetilde x(t)+ \widetilde D w(t)  ,
\end{aligned}
\ee
where the dimension $r$ of the input vector $w(t)$ is the normal rank of $G(\lambda)$.
\item
 \texttt{INFO} is a MATLAB structure containing additional information, as follows:\\
{\begin{longtable}{|l|p{12cm}|} \hline
\textbf{\texttt{INFO} fields} & \textbf{Description} \\ \hline
 \texttt{nrank}   & normal rank of the transfer function matrix $G(\lambda)$; \\ \hline
 \texttt{nfuz}   & number of finite unstable zeros of
              \texttt{SYS} lying on the boundary of the stability region $\partial\mathds{C}_s$ within the offset specified
by \texttt{OPTION.offset};\\ \hline
 \texttt{niuz}   & number of infinite zeros of
              \texttt{SYS} in the continuous-time case
              and 0 in the discrete-time case\\ \hline
 \texttt{ricrez}   & diagnosis flag, as provided provided by the generalized
               Riccati equation solvers \texttt{care} and \texttt{dare}; if non-negative,
               this value represents the Frobenius norm of relative
               residual of the Riccati equation, while a negative value
               indicates failure of solving the Riccati equation.  \\ \hline
\end{longtable}}
\end{description}

\subsubsection*{Method}
The coimage computation is performed by applying the range computation method described in \cite{Varg17f}, to the
 dual descriptor system realization corresponding to the transposed
 rational matrix $G^T(\lambda)$ to obtain the full rank factorization
 \[ G^T(\lambda) = \widetilde R(\lambda)\widetilde X(\lambda) , \]
 where $\widetilde R(\lambda)$ is full column rank and contains those zeros of $G(\lambda)$ which have been selected via the \texttt{OPTIONS.zeros} field, while $\widetilde X(\lambda)$ has full row rank. The coimage basis of $G(\lambda)$ is given by $R(\lambda) = \widetilde R^T(\lambda)$  and the corresponding $X(\lambda)$ in (\ref{gcrange:def}) is given by $X(\lambda) = \widetilde X^T(\lambda)$.

The factorization approach of  \cite{Varg17f} is based  on the reduction of the system matrix pencil to a special Kronecker-like form (see (\ref{spec-klf})) and  is described in \cite{Oara05}. This reduction involves the use of the mex-function \url{sl_klf} to compute the appropriate Kronecker-like form.
\index{MEX-functions!\url{sl_klf}}%
For the computation of an coinner basis, extensions of the standard inner-outer factorization methods of \cite{Zhou96} are used. These methods involve the solution of appropriate (continuous- or discrete-time) generalized algebraic Riccati equations. For additional details, see \cite{Oara00} for continuous-time systems and \cite{Oara05} for discrete-time systems.

\subsubsection*{Example}
\begin{example}\label{ex:Oara-Varga2}
This is \textbf{Example 2} from \cite{Oara00}, and concerns with the transfer function matrix (\ref{grange:ex1}) of a continuous-time proper system, already considered in Example \ref{ex:Oara-Varga}. The Moore-Penrose pseudo-inverse $G^\dag(s)$ of the rational matrix $G(s)$ can be computed in three steps, using a simplified version of the approach described in \cite{Oara00}:
\begin{enumerate}
\item Compute a  full-rank factorization $G(s) =  U(s)G_1(s)$,
with $U(s)$, a minimal inner range matrix, and $G_1(s)$ full row rank.
\item Compute the dual full-rank factorization $G_1(s) = G_2(s)V(s)$,
with $V(s)$, a minimal coinner coimage (i.e., $V(s)V^\sim(s) = I$), and $G_2(s)$ invertible.
\item Compute
\[ G^\dag(s) = V^\sim(s) G_2^{-1}(s) U^\sim(s). \]
\end{enumerate}

These computational steps are implemented in the following MATLAB code:
\pagebreak[4]
\begin{verbatim}

%  Oara and Varga (2000), Example 2
s = tf('s');  % define the complex variable s
% enter G(s) and determine a minimal state-space realization
Gs = [(s-1)/(s+2) s/(s+2) 1/(s+2);
    0 (s-2)/(s+1)^2 (s-2)/(s+1)^2;
    (s-1)/(s+2) (s^2+2*s-2)/(s+1)/(s+2) (2*s-1)/(s+1)/(s+2)];
G = minreal(ss(Gs));

% use tolerance 1.e-7 for rank determinations
opt = struct('tol',1.e-7,'inner',true,'coinner',true);

% compute the full-rank factorization G(s) = U(s)*G1(s) with U(s) inner
[U,G1] = grange(G,opt);

% compute the full-rank factorization G1(s) = G2(s)*V(s) with V(s) coinner
[V,G2] = gcrange(G1,opt); 

% compute the pseudo-inverse Gpinv
Gpinv = gir(V'*(G2\(U')));

% check the four axioms defining the Moore-Penrose pseudo-inverse
norm(G*Gpinv*G-G,inf)               % (i)   ||G*Gpinv*G-G||_inf = 0
norm(Gpinv*G*Gpinv-Gpinv,inf)       % (ii)  ||Gpinv*G*Gpinv-Gpinv||_inf = 0
norm(G*Gpinv-(G*Gpinv)',inf)        % (iii) ||G*Gpinv-(G*Gpinv)'||_inf = 0
norm(Gpinv*G-(Gpinv*G)',inf)        % (iv)  ||Gpinv*G-(Gpinv*G)'||_inf = 0
\end{verbatim}

\fine\end{example}

\subsubsection{\texttt{\bfseries grsol}}
\index{M-functions!\texttt{\bfseries grsol}}

\subsubsection*{Syntax}
\begin{verbatim}
[SYSX,INFO,SYSGEN] = grsol(SYSG,SYSF,OPTIONS)
[SYSX,INFO,SYSGEN] = grsol(SYSGF,MF,OPTIONS)
\end{verbatim}

\subsubsection*{Description}
\texttt{\bfseries grsol} computes the solution $X(\lambda)$ of the linear rational matrix equation
\be\label{grsol:system} G(\lambda)X(\lambda) = F(\lambda), \ee
where $G(\lambda)$ and $F(\lambda)$ are the (rational) transfer function matrices of  LTI descriptor systems.
\index{transfer function matrix (TFM)!linear rational matrix equation}%
\index{descriptor system!linear rational matrix equation}%

\subsubsection*{Input data}
For the usage with
\begin{verbatim}
[SYSX,INFO,SYSGEN] = grsol(SYSG,SYSF,OPTIONS)
\end{verbatim}
the input parameters are as follows:
\begin{description}
\item
\texttt{SYSG} is a LTI system, whose transfer function matrix is $G(\lambda)$, and is in   a descriptor system state-space form
\be\label{grsol:sysssG}
\begin{aligned}
E_G\lambda x_G(t)  &=   A_Gx_G(t)+ B_G u(t)  ,\\
y_G(t) &=  C_G x_G(t)+ D_G u(t)  ,
\end{aligned}
\ee
where $y_G(t) \in \mathds{R}^{p}$.
\item
\texttt{SYSF} is a LTI system, whose transfer function matrix is $F(\lambda)$, and is in a  descriptor system state-space form
\be\label{grsol:sysssF}
\begin{aligned}
E_F\lambda x_F(t)  &=   A_Fx_F(t)+ B_F v(t)  ,\\
y_F(t) &=  C_F x_F(t)+ D_F v(t)  ,
\end{aligned}
\ee
where $y_F(t) \in \mathds{R}^{p}$.
\item
 \texttt{OPTIONS} is a MATLAB structure to specify user options and has the following fields:\\
{\tabcolsep=1mm
\begin{longtable}{|l|lcp{10cm}|} \hline
\textbf{\texttt{OPTIONS} fields} & \multicolumn{3}{l|}{\textbf{Description}} \\ \hline
 \texttt{tol}   & \multicolumn{3}{l|}{relative tolerance for rank computations (Default: internally computed);} \\ \hline
 \texttt{sdeg}   & \multicolumn{3}{p{11cm}|}{prescribed stability degree for the free
                    poles of the solution $X(\lambda)$
                    (Default: \texttt{[ ]}, i.e., no stabilization performed); } \\ \hline
 \texttt{poles}   & \multicolumn{3}{p{11cm}|}{a complex conjugated set of desired poles  to be assigned for the free poles of the solution $X(\lambda)$
                     (Default: \texttt{[ ]});}\\
                                        \hline
 \texttt{mindeg}   & \multicolumn{3}{p{11cm}|}{option to compute a minimum degree solution:}\\
                 &  \texttt{true} &--& determine a minimum order solution; \\
                 &  \texttt{false}&--& determine a particular solution which has
                             possibly non-minimal order (default).  \\
                                        \hline
\end{longtable}}
\end{description}
For the usage with
\begin{verbatim}
[SYSX,INFO,SYSGEN] = grsol(SYSGF,MF,OPTIONS)
\end{verbatim}
the input parameters are as follows:
\begin{description}
\item
\texttt{SYSGF} is an input concatenated compound LTI system, \texttt{SYSGF = [ SYSG SYSF ]},  in a  descriptor system state-space form
\be\label{grsol:sysssGF}
\begin{aligned}
E\lambda x(t)  &=   Ax(t)+ B_G u(t)+ B_F v(t)  ,\\
y(t) &=  Cx(t)+ D_G u(t)+ D_F v(t)  ,
\end{aligned}
\ee
where \texttt{SYSG} has the transfer function matrix $G(\lambda)$, with the descriptor system realization $(A-\lambda E,B_G,C,D_G)$, and
\texttt{SYSF} has the transfer function matrix $F(\lambda)$, with the descriptor system realization $(A-\lambda E,B_F,C,D_F)$.
\item
\texttt{MF} is the dimension of the input vector $v(t)$ of the system \texttt{SYSF}. \item
\texttt{OPTIONS} is a MATLAB structure to specify user options and has the same fields as described previously.
\end{description}

\subsubsection*{Output data}
\begin{description}
\item
\texttt{SYSX} contains  the descriptor system state-space realization of the solution $X(\lambda)$ in the form
\be\label{grsol:sysssX}
\begin{aligned}
\widetilde E\lambda \widetilde x(t)  &=   \widetilde A \widetilde x(t)+ \widetilde B v(t)  ,\\
u(t) &=  \widetilde C \widetilde x(t)+ \widetilde D v(t)  .
\end{aligned}
\ee
\item
 \texttt{INFO} is a MATLAB structure containing additional information, as follows:\\
{\begin{longtable}{|l|p{12cm}|} \hline
\textbf{\texttt{INFO} fields} & \textbf{Description} \\ \hline
 \texttt{nrank}   & normal rank of the transfer function matrix $G(\lambda)$; \\ \hline
 \texttt{rdeg}   & vector which contains the relative column degrees of
       $X(\lambda)$ (i.e., the numbers of integrators/delays needed to make
       each column of $X(\lambda)$ proper);\\ \hline
 \texttt{tcond}   & maximum of the condition numbers of the employed
                non-orthogonal transformation matrices (large values indicate possible  loss of numerical stability); \\ \hline
 \texttt{fnorm}   & the norm of the employed state-feedback/feedforward used
       for dynamic cover computation if \texttt{OPTIONS.mindeg = true}, or for
       stabilization of free poles if \texttt{OPTION.sdeg} is not empty (large values indicate possible  loss of numerical stability);  \\ \hline
 \texttt{nr}   & the order of $A_r-\lambda E_r$, also the row dimension of $B_r$ and also the number of freely assignable poles of the solution (see \textbf{Method});   \\ \hline
 \texttt{nf}   & the order of $A_f-\lambda E_f$ (see \textbf{Method});   \\ \hline
 \texttt{ninf}   & the order of $A_\infty-\lambda E_\infty$ (see \textbf{Method}).   \\ \hline
\end{longtable}}
\item
\texttt{SYSGEN} contains  the input concatenated compound system
\texttt{[ SYSX0  SYSNR ]}, in a descriptor system state-space form
\be\label{grsol:sysssgen}
\begin{aligned}
E_g\lambda  x_g(t)  &=    A_g x_g(t)+ B_{0} v_1(t)+ B_{N} v_2(t) ,\\
y_g(t) &=  C_g x_g(t)+ D_{0} v_1(t)+ D_{N} v_2(t)  ,
\end{aligned}
\ee
where the transfer function matrix $X_0(\lambda)$ of \texttt{SYSX0}, with the descriptor system realization $(A_g-\lambda E_g,B_{0},C_g,D_{0})$, is a particular solution satisfying $G(\lambda)X_0(\lambda) = F(\lambda)$, and the transfer function matrix $N_r(\lambda)$ of \texttt{SYSNR}, with the descriptor system realization $(A_g-\lambda E_g,B_{N},C_g,D_{N})$, is a proper right nullspace basis of $G(\lambda)$, satisfying $G(\lambda)N_r(\lambda) = 0$. The transfer function matrices $X_0(\lambda)$ and $N_r(\lambda)$ can be used to generate all solutions of the system (\ref{grsol:system}) as $X(\lambda) = X_0(\lambda)+N_r(\lambda)Y(\lambda)$, where
$Y(\lambda)$ is an arbitrary rational matrix with suitable dimensions.

\end{description}

\subsubsection*{Method}
The employed solution approach of the linear rational matrix equation (\ref{grsol:system}) is sketched in Section \ref{appsec:linsys}. This approach is based on a realization of the form (\ref{grsol:sysssGF}) of the input concatenated compound system \texttt{SYSGF = [ SYSG SYSF ]}. A detailed computational procedure to solve (\ref{grsol:system}) is described in  \cite{Varg04b} (see also \cite[Section 10.3.7]{Varg17} for more details). For the intervening computation of the Kronecker-like form of the system matrix pencil of the descriptor system \texttt{SYSG}, the mex-function \url{sl_klf}, based on the algorithm proposed in \cite{Beel87}, is employed. \index{MEX-functions!\url{sl_klf}}%
If the descriptor realizations of \texttt{SYSG} and \texttt{SYSF} are separately provided, then an irreducible realization of the  input concatenated compound system \texttt{[ SYSG SYSF ]} is internally computed. For orthogonal transformation based order reduction purposes, the mex-function \url{sl_gminr} is employed. \index{MEX-functions!\url{sl_gminr}}%
Furthermore, the mex-function \url{sl_gstra} is employed to count controllable infinite poles (needed to determine the relative column degrees) and for reduction to SVD-like form (needed for the elimination of  non-dynamic modes) \index{MEX-functions!\url{sl_gstra}}%

For the computation of the solution, a so-called \emph{generator} of all solutions is determined in the form
(\ref{grsol:sysssgen}), where the resulting matrices have the forms
\[   {\arraycolsep=1mm \begin{array}{rclrcl}   A_g-\lambda E_g &=& \ba{ccc} A_r+B_rF-\lambda E_r & \ast & \ast \\ 0 & A_f-\lambda E_f & \ast \\ 0 & 0 & A_\infty-\lambda E_\infty\ea , \quad
[\;B_0 \mid  B_N\, ] &=&  \ba{c|c} B_1 & B_r \\ B_2 & 0 \\ B_3 & 0 \ea , \\ \\[-2mm]
          C_g  &=& \;[\quad     C_r+D_NF  \quad \ast \quad \ast \quad ] \, ,
          \end{array}}
\]
with the descriptor pair $(A_r-\lambda E_r,B_r)$ controllable and $E_r$ invertible,  $A_f-\lambda E_f$ regular and $E_f$ invertible (i.e., $A_f-\lambda E_f$ has only finite eigenvalues), and $A_\infty-\lambda E_\infty$ regular and upper triangular with $A_\infty$ invertible and $E_\infty$ nillpotent (i.e. $A_\infty-\lambda E_\infty$ has only infinite eigenvalues), and $D_N$ full row rank. If $G(\lambda)$ is a $p\times m$ TFM of normal rank $r$, then the matrices
$B_r$ and $D_N$ have $m-r$ columns. The pair $(A_r-\lambda E_r,B_r)$ being controllable, the eigenvalues of $A_r+B_rF-\lambda E_r$ can be freely assigned by using a suitably chosen state-feedback matrix $F$.    The descriptor system realization (\ref{grsol:sysssgen}) is usually not minimal, being uncontrollable, or having non-dynamic modes,  or both. The generator contains the particular solution $X_0(\lambda)$ with the descriptor realization $(A_g-\lambda E_g,B_{0},C_g,D_{0})$ and
a right nullspace basis $N_r(\lambda)$ of $G(\lambda)$ with the (non-minimal) descriptor system realization $(A_g-\lambda E_g,B_{N},C_g,D_{N})$.  A minimal order descriptor
system realization of $N_r(\lambda)$ is $(A_r+B_rF-\lambda E_r,B_r,C_r+D_NF,D_N)$.
All solutions of the system (\ref{grsol:system}) can be expressed as $X(\lambda) = X_0(\lambda)+N_r(\lambda)Y(\lambda)$, where
$Y(\lambda)$ is an arbitrary rational matrix with suitable dimensions.

The resulting generator \texttt{SYSGEN} contains the descriptor system realization
\[ [\, X_0(\lambda) \; N_r(\lambda)\,] =
(A_g-\lambda E_g,[\,B_{0}\;B_N\,],C_g,[\,D_{0}\;D_N\,]) . \]
The orders of the diagonal blocks of $A_g-\lambda E_g$ are provided in the \texttt{INFO} structure as follows:
\texttt{INFO.nr} contains the order of $A_r-\lambda E_r$, \texttt{INFO.nf} contains the order of $A_f-\lambda E_f$, and \texttt{INFO.ninf} contains the order of $A_\infty-\lambda E_\infty$. \texttt{INFO.nrank} contains the normal rank $r$ of $G(\lambda)$. The relative column degrees, provided in \texttt{INFO.rdeg}, are the numbers of infinite poles of the successive columns of $X_0(\lambda)$.

If \texttt{OPTIONS.mindeg = false}, the computed solution \texttt{SYSX} represents a minimal realization of the particular solution $X_0(\lambda)$, computed by eliminating the uncontrollable eigenvalues and non-dynamic modes of the realization $(A_g-\lambda E_g,B_{0},C_g,D_{0})$, where $F$ is determined such that the (free) eigenvalues of $A_r+B_rF-\lambda E_r$ are moved to the stability domain specified via \texttt{OPTIONS.sdeg} or to locations specified in \texttt{OPTIONS.poles}. If both \texttt{OPTIONS.sdeg} and \texttt{OPTIONS.poles} are empty, then $F = 0$ is used.

If \texttt{OPTIONS.mindeg = true}, the computed solution \texttt{SYSX}  represents a minimal realization of $X(\lambda) = X_0(\lambda)+N_r(\lambda)Y(\lambda)$, where $Y(\lambda)$ is determined such that $X(\lambda)$ has the least achievable McMillan degree. For this computation, order reduction based on computing minimum dynamic covers is employed (see \textbf{Procedure GRMCOVER2} in \cite[Section 10.4.3]{Varg17}).

\subsubsection*{Example}

\begin{example}\label{ex:Gao-Antsaklis1989}
This example is taken from \cite{Gao89}, where an one-sided model matching problem is solved, which involves  the computation of a stable and proper solution of  $G(s)X(s) = F(s)$ with
\[ G(s) =
\left[\begin{array}{cc} \displaystyle\frac{s - 1}{s\,  \left(s + 1\right)} &  \displaystyle\frac{s - 1}{s\, \left(s + 2\right)} \end{array}\right], \quad
F(s) =
\left[\begin{array}{cc}  \displaystyle\frac{s - 1}{\left(s + 1\right)\, \left(s + 3\right)} & \displaystyle\frac{s - 1}{\left(s + 1\right)\, \left(s + 4\right)} \end{array}\right] .
\]
Both $G(s)$ and $[\, G(s)\; F(s)\,]$ have rank equal to 1 and zeros $\{1,\infty\}$. It follows, according to Lemma \ref{L-SEMMP}, that the linear rational matrix equation $G(s)X(s) = F(s)$ has a stable and proper solution. A fourth order stable and proper solution has been computed in \cite{Gao89}. A least order solution, with McMillan degree equal to 2, has been computed with \texttt{grsol} as
\[ X(s) = {\def\arraystretch{2}\ba{cc} \displaystyle\frac{0.48889 (s-1.045)}{s+3} & \displaystyle\frac{0.41333 (s-1.419)}{s+4} \\
\displaystyle\frac{0.51111 (s+2)}{s+3} & \displaystyle\frac{0.58667 (s+2)}{s+4} \ea } .\]

To compute a least order solution $X(s)$, the following sequence of commands can be used:
\begin{verbatim}
% Gao & Antsaklis (1989)
s = tf('s');
G = [(s-1)/(s*(s+1))  (s-1)/(s*(s+2))];
F = [(s-1)/((s+1)*(s+3))  (s-1)/((s+1)*(s+4))];

% build a minimal realization of [G(s) F(s)]
sysgf = minreal(ss([G F]));

% solve G(s)*X(s) = F(s) for the least order solution
[X,info] = grsol(sysgf,2,struct('mindeg',true)); info
minreal(zpk(X))

% check solution
minreal(G*X-F)
\end{verbatim}
\fine\end{example}

\subsubsection{\texttt{\bfseries glsol}}
\index{M-functions!\texttt{\bfseries glsol}}

\subsubsection*{Syntax}
\begin{verbatim}
[SYSX,INFO,SYSGEN] = glsol(SYSG,SYSF,OPTIONS)
[SYSX,INFO,SYSGEN] = glsol(SYSGF,MF,OPTIONS)
\end{verbatim}

\subsubsection*{Description}
\texttt{\bfseries glsol} computes the solution $X(\lambda)$ of the linear rational matrix equation
\be\label{glsol:system} X(\lambda)G(\lambda) = F(\lambda), \ee
where $G(\lambda)$ and $F(\lambda)$ are the (rational) transfer function matrices of  LTI descriptor systems.
\index{transfer function matrix (TFM)!linear rational matrix equation}%
\index{descriptor system!linear rational matrix equation}%

\subsubsection*{Input data}
For the usage with
\begin{verbatim}
[SYSX,INFO,SYSGEN] = glsol(SYSG,SYSF,OPTIONS)
\end{verbatim}
the input parameters are as follows:
\begin{description}
\item
\texttt{SYSG} is a LTI system, whose transfer function matrix is $G(\lambda)$, and is in  a  descriptor system state-space form
\be\label{glsol:sysssG}
\begin{aligned}
E_G\lambda x_G(t)  &=   A_Gx_G(t)+ B_G u(t)  ,\\
y_G(t) &=  C_G x_G(t)+ D_G u(t)  ,
\end{aligned}
\ee
where $u(t) \in \mathds{R}^{m}$.
\item
\texttt{SYSF} is a LTI system, whose transfer function matrix is $F(\lambda)$, and is in a descriptor system state-space form
\be\label{glsol:sysssF}
\begin{aligned}
E_F\lambda x_F(t)  &=   A_Fx_F(t)+ B_F v(t)  ,\\
y_F(t) &=  C_F x_F(t)+ D_F v(t)  ,
\end{aligned}
\ee
where $v(t) \in \mathds{R}^{m}$.
\item
 \texttt{OPTIONS} is a MATLAB structure to specify user options and has the following fields:\\
{\tabcolsep=1mm
\begin{longtable}{|l|lcp{10cm}|} \hline
\textbf{\texttt{OPTIONS} fields} & \multicolumn{3}{l|}{\textbf{Description}} \\ \hline
 \texttt{tol}   & \multicolumn{3}{l|}{relative tolerance for rank computations (Default: internally computed);} \\ \hline
 \texttt{sdeg}   & \multicolumn{3}{p{11cm}|}{prescribed stability degree for the free
                    poles of the solution $X(\lambda)$
                    (Default: \texttt{[ ]}, i.e., no stabilization performed); } \\ \hline
 \texttt{poles}   & \multicolumn{3}{p{11cm}|}{a complex conjugated set of desired poles  to be assigned for the free poles of the solution $X(\lambda)$
                     (Default: \texttt{[ ]});}\\
                                        \hline
 \texttt{mindeg}   & \multicolumn{3}{p{11cm}|}{option to compute a minimum degree solution:}\\
                 &  \texttt{true} &--& determine a minimum order solution; \\
                 &  \texttt{false}&--& determine a particular solution which has
                             possibly non-minimal order (default).  \\
                                        \hline
\end{longtable}}
\end{description}
For the usage with
\begin{verbatim}
[SYSX,INFO,SYSGEN] = glsol(SYSGF,MF,OPTIONS)
\end{verbatim}
the input parameters are as follows:
\begin{description}
\item
\texttt{SYSGF} is an output concatenated compound LTI system, \texttt{SYSGF = [ SYSG; SYSF ]},  in a  descriptor system state-space form
\be\label{glsol:sysssGF}
\begin{aligned}
E\lambda x(t)  &=   Ax(t)+ B u(t)  ,\\
y_G(t) &=  C_Gx(t)+ D_G u(t)  , \\
y_F(t) &=  C_Fx(t)+ D_F u(t)  ,
\end{aligned}
\ee
where \texttt{SYSG} has the transfer function matrix $G(\lambda)$, with the descriptor system realization $(A-\lambda E,B,C_G,D_G)$, and
\texttt{SYSF} has the transfer function matrix $F(\lambda)$, with the descriptor system realization $(A-\lambda E,B,CF,D_F)$.
\item
\texttt{MF} is the dimension of the output vector $y_F(t)$ of the system \texttt{SYSF}.
\item
\texttt{OPTIONS} is a MATLAB structure to specify user options and has the same fields as described previously.
\end{description}

\subsubsection*{Output data}
\begin{description}
\item
\texttt{SYSX} contains  the descriptor system state-space realization of the solution $X(\lambda)$ in the form
\be\label{glsol:sysssX}
\begin{aligned}
\widetilde E\lambda \widetilde x(t)  &=   \widetilde A \widetilde x(t)+ \widetilde B y_G(t)  ,\\
y(t) &=  \widetilde C \widetilde x(t)+ \widetilde D y_G(t)  .
\end{aligned}
\ee
\item
 \texttt{INFO} is a MATLAB structure containing additional information, as follows:\\
{\begin{longtable}{|l|p{12cm}|} \hline
\textbf{\texttt{INFO} fields} & \textbf{Description} \\ \hline
 \texttt{nrank}   & normal rank of the transfer function matrix $G(\lambda)$; \\ \hline
 \texttt{rdeg}   & vector which contains the relative row degrees of
       $X(\lambda)$ (i.e., the numbers of integrators/delays needed to make
       each row of $X(\lambda)$ proper);\\ \hline
 \texttt{tcond}   & maximum of the condition numbers of the employed
                non-orthogonal transformation matrices (large values indicate possible  loss of numerical stability); \\ \hline
 \texttt{fnorm}   & the norm of the employed state-feedback/feedforward used
       for dynamic cover computation if \texttt{OPTIONS.mindeg = true}, or for
       stabilization of free poles if \texttt{OPTION.sdeg} is not empty (large values indicate possible  loss of numerical stability);  \\ \hline
 \texttt{ninf}   & the order of $A_\infty-\lambda E_\infty$ (see \textbf{Method});   \\ \hline
  \texttt{nf}   & the order of $A_f-\lambda E_f$ (see \textbf{Method});   \\ \hline
\texttt{nl}   & the order of $A_l-\lambda E_l$, also the column dimension of $C_l$, and also the number of freely assignable poles of the solution (see \textbf{Method}), if \texttt{SYSGEN} is computed; otherwise, empty.   \\ \hline
\end{longtable}}
\item
\texttt{SYSGEN} contains  the output concatenated compound system
\texttt{[ SYSX0;  SYSNL ]}, in a descriptor system state-space form
\be\label{glsol:sysssgen}
\begin{aligned}
E_g\lambda  x_g(t)  &=    A_g x_g(t)+ B_{g} v(t) ,\\
y_0(t) &=  C_0 x_g(t)+ D_{0} v(t)  , \\
y_N(t) &=  C_N x_g(t)+ D_{N} v(t)  ,
\end{aligned}
\ee
where the transfer function matrix $X_0(\lambda)$ of \texttt{SYSX0}, with the descriptor system realization $(A_g-\lambda E_g,B_{g},C_0,D_{0})$, is a particular solution satisfying $X_0(\lambda)G(\lambda) = F(\lambda)$, and the transfer function matrix $N_l(\lambda)$ of \texttt{SYSNL}, with the descriptor system realization $(A_g-\lambda E_g,B_{g},C_N,D_{N})$, is a proper left nullspace basis of $G(\lambda)$, satisfying $N_l(\lambda)G(\lambda) = 0$. The transfer function matrices $X_0(\lambda)$ and $N_l(\lambda)$ can be used to generate all solutions of the system (\ref{glsol:system}) as $X(\lambda) = X_0(\lambda)+Y(\lambda)N_l(\lambda)$, where
$Y(\lambda)$ is an arbitrary rational matrix with suitable dimensions.

\end{description}

\subsubsection*{Method}
To solve the linear rational equation (\ref{glsol:system}), the function \texttt{\bfseries glsol} calls \texttt{\bfseries grsol} to solve the dual system
$G^T(\lambda)\widetilde X(\lambda) = F^T(\lambda)$
and obtain the solution as $X(\lambda) = \widetilde X^T(\lambda)$.
\index{M-functions!\texttt{\bfseries grsol}}%
Thus, the employed solution method corresponds to the dual approach sketched in Section \ref{sec:minbasis} (see also \cite{Varg04b} and \cite[Section 10.3.7]{Varg17} for more details). The function \texttt{\bfseries grsol} relies on the mex-functions
\url{sl_klf}, \url{sl_gminr}, and \url{sl_gstra}.
 \index{MEX-functions!\url{sl_klf}}%
\index{MEX-functions!\url{sl_gminr}}%
\index{MEX-functions!\url{sl_gstra}}%

For the computation of the solution, a so-called \emph{generator} of all solutions is determined in the form (\ref{glsol:sysssgen}), where the resulting matrices have the forms
\[   {\arraycolsep=1mm \begin{array}{rcl}   A_g-\lambda E_g &=& \ba{ccc} A_\infty-\lambda E_\infty & \ast & \ast \\ 0 & A_f-\lambda E_f & \ast \\ 0 & 0 & A_l+KC_l-\lambda E_l\ea , \quad
B_g =  \ba{c} \ast \\ \ast \\ B_l+KD_N \ea , \\ \\[-2mm]
          \ba{c} C_0 \\  C_N\ea   &=& \ba{ccc}C_1 & C_2 & C_3\\ 0 & 0 & C_l \ea \, ,
          \end{array} }
\]
with $A_\infty-\lambda E_\infty$ regular and upper triangular with $A_\infty$ invertible and $E_\infty$ nillpotent (i.e. $A_\infty-\lambda E_\infty$ has only infinite eigenvalues), $A_f-\lambda E_f$ regular and $E_f$ invertible (i.e., $A_f-\lambda E_f$ has only finite eigenvalues), the descriptor pair $(A_l-\lambda E_l,C_l)$ observable and $E_l$ invertible,   and $D_N$ full column rank. If $G(\lambda)$ is a $p\times m$ TFM of normal rank $r$, then the matrices
$C_l$ and $D_N$ have $p-r$ rows. The pair $(A_l-\lambda E_l,C_l)$ being observable, the eigenvalues of $A_l+KC_l-\lambda E_l$ can be freely assigned by using a suitably chosen output-injection matrix $K$.

The descriptor system realization (\ref{glsol:sysssgen}) is usually not minimal, being unobservable, or having non-dynamic modes,  or both. The generator contains the particular solution $X_0(\lambda)$ with the descriptor realization $(A_g-\lambda E_g,B_{g},C_0,D_{0})$ and
a left nullspace basis $N_l(\lambda)$ of $G(\lambda)$ with the (non-minimal) descriptor system realization $(A_g-\lambda E_g,B_{g},C_N,D_{N})$.  A minimal order descriptor
system realization of $N_r(\lambda)$ is $(A_l+KC_l-\lambda E_l,B_l+KD_N,C_l,D_N)$.
All solutions of the system (\ref{grsol:system}) can be expressed as $X(\lambda) = X_0(\lambda)+Y(\lambda)N_l(\lambda)$, where
$Y(\lambda)$ is an arbitrary rational matrix with suitable dimensions.

The resulting generator \texttt{SYSGEN} contains the descriptor system realization
\[ \ba{c} X_0(\lambda) \\ N_l(\lambda)\ea =
\left(A_g-\lambda E_g,B_g,\left[\begin{matrix}C_{0}\\C_N\end{matrix}\right],\left[\begin{matrix}D_{0}\\D_N\end{matrix}\right]\right) . \]
The orders of the diagonal blocks of $A_g-\lambda E_g$ are provided in the \texttt{INFO} structure as follows: \texttt{INFO.ninf} contains the order of $A_\infty-\lambda E_\infty$, \texttt{INFO.nf} contains the order of $A_f-\lambda E_f$, and
\texttt{INFO.nl} contains the order of $A_l-\lambda E_l$. \texttt{INFO.nrank} contains the normal rank $r$ of $G(\lambda)$. The relative row degrees, provided in \texttt{INFO.rdeg}, are the numbers of infinite poles of the successive rows of $X_0(\lambda)$.

If \texttt{OPTIONS.mindeg = false}, the computed solution \texttt{SYSX} represents a minimal realization of the particular solution $X_0(\lambda)$, computed by eliminating the unobservable eigenvalues and non-dynamic modes of the realization $(A_g-\lambda E_g,B_g,C_{0},D_{0})$, where $K$ is determined such that the (free) eigenvalues of $A_l+KC_l-\lambda E_l$ are moved to the stability domain specified via \texttt{OPTIONS.sdeg} or to locations specified in \texttt{OPTIONS.poles}. If both \texttt{OPTIONS.sdeg} and \texttt{OPTIONS.poles} are empty, then $K = 0$ is used.

If \texttt{OPTIONS.mindeg = true}, the computed solution \texttt{SYSX}  represents a minimal realization of $X(\lambda) = X_0(\lambda)+Y(\lambda)N_l(\lambda)$, where $Y(\lambda)$ is determined such that $X(\lambda)$ has the least achievable McMillan degree. For this computation, order reduction based on computing minimum dynamic covers is employed (see \textbf{Procedure GRMCOVER2} in \cite[Section 10.4.3]{Varg17}).

\subsubsection*{Examples}

\begin{example}\label{ex:Wang-Davison}
This example illustrates the computation of a left inverse of least McMillan degree used in \cite{Wang73}.
The transfer function matrices $G(s)$ and $F(s)$ are
\[ G(s) = \frac{1}{s^2+3s+2}\ba{cc} s+1 & s+2\\ s+3 & s^2+2s\\ s^2+3s & 0 \ea, \quad F(s) = \ba{cc} 1 & 0 \\ 0 & 1 \ea. \]
The solution $X(s)$ of the rational equation $X(s)G(s) = I$ is a left inverse of $G(s)$.
A third order minimal state-space realization of $G(s)$ can be computed using the function \texttt{\bfseries gir} (or \texttt{\bfseries minreal}). To compute a least McMillan order proper left inverse, the following MATLAB commands can be used:
\begin{verbatim}
% Wang and Davison Example (1973)
s = tf('s');
g = [ s+1 s+2; s+3 s^2+2*s; s^2+3*s 0 ]/(s^2+3*s+2); f = eye(2);
sysg = gir(ss(g)); sysf = ss(f);

% compute a least order solution of X(s)G(s) = I
[sysx,info] = glsol(sysg,sysf,struct('mindeg',true)); info
gpole(sysx)   % the left inverse is unstable

% check solution
gir(sysx*sysg-sysf)
\end{verbatim}
The computed solution is unstable. In fact, it turns out that all least order left inverses are unstable.  \fine

\end{example}

\begin{example}\label{ex:Wang-Davison-stable}
This example illustrates the computation of a stable left inverse for the example of \cite{Wang73}, used previously. Since both $G(s)$ and $\left[\begin{smallmatrix} G(s)\\I_2\end{smallmatrix}\right]$ have no zeros (in particular no unstable zeros), a stable solution of equation $X(s)G(s) = I$ exists, in accordance with Lemma~\ref{L-SEMMP}. In Example \ref{ex:Wang-Davison}, we computed \texttt{INFO.nl} = 3, and, therefore, we can assign three free poles of the inverse to $\{-1,-2,-3\}$ to obtain a stable left inverse.
This can be done by calling \texttt{\bfseries glsol} as follows:
\begin{verbatim}
sysx = glsol(sysg,sysf,struct('poles',[-1 -2 -3]));
gpole(sysx)
\end{verbatim}
The resulting minimal realization of left inverse \texttt{sysx} has the poles equal to $\{-1,-2,-3\}$. \fine
\end{example}

\subsubsection{\texttt{\bfseries gsdec}}
\index{M-functions!\texttt{\bfseries gsdec}}

\subsubsection*{Syntax}
\begin{verbatim}
[SYS1,SYS2,Q,Z] = gsdec(SYS,OPTIONS)
\end{verbatim}

\subsubsection*{Description}
\texttt{\bfseries gsdec} computes, for the transfer function matrix $G(\lambda)$ of a LTI descriptor system, additive spectral decompositions in the form
\be\label{gsdec:tfm_add}  G(\lambda) = G_1(\lambda) + G_2(\lambda) ,\ee
\index{transfer function matrix (TFM)!additive decomposition}%
\index{descriptor system!additive decomposition}%
where $G_1(\lambda)$ has only poles in a certain domain of interest $\mathds{C}_g \subset \mathds{C}$ and $G_2(\lambda)$ has only poles in the complementary domain $\mathds{C}\setminus\mathds{C}_g$.

\subsubsection*{Input data}
\begin{description}
\item
\texttt{SYS} is a LTI system, whose transfer function matrix is $G(\lambda)$, and is in a  descriptor system state-space form
\be\label{gsdec:sysssG}
\begin{aligned}
E\lambda x(t)  &=   Ax(t)+ Bu(t)  ,\\
y(t) &=  Cx(t)+ Du(t)  .
\end{aligned}
\ee
\item
 \texttt{OPTIONS} is a MATLAB structure to specify user options and has the following fields:
{\tabcolsep=1mm
\begin{longtable}{|l|lcp{10cm}|} \hline
\textbf{\texttt{OPTIONS} fields} & \multicolumn{3}{l|}{\textbf{Description}} \\ \hline
 \texttt{tol}   & \multicolumn{3}{l|}{relative tolerance for rank computations (Default: internally computed)} \\ \hline
 \texttt{smarg}   & \multicolumn{3}{p{12cm}|}{stability margin for the
                  stable poles of the transfer function matrix $G(\lambda)$ of \texttt{SYS}, such that,
                  in the continuous-time case, the stable eigenvalues
                  have real parts less than or equal to \texttt{OPTIONS.smarg}, and
                  in the discrete-time case, the stable eigenvalues
                  have moduli less than or equal to \texttt{OPTIONS.smarg}. \newline
                  (Default: \texttt{-sqrt(eps)} for a continuous-time system \texttt{SYS}; \newline
                  \phantom{(Default:} \texttt{1-sqrt(eps}) for a discrete-time system \texttt{SYS}.)} \\ \hline
 \texttt{job}   & \multicolumn{3}{p{11cm}|}{option for specific spectral separation tasks:}\\
                 &  \texttt{'finite'} &--& $G_1(\lambda)$ has only finite poles and
                              $G_2(\lambda)$ has only infinite poles (default); \\
                 &  \texttt{'infinite'}&--& $G_1(\lambda)$ has only infinite poles and $G_2(\lambda)$ has only finite poles;  \\
                 &  \texttt{'stable'} &--& $G_1(\lambda)$ has only stable poles and
$G_2(\lambda)$ has only unstable and infinite poles; \\
                 &  \texttt{'unstable'}&--& $G_1(\lambda)$ has only unstable and infinite poles and $G_2(\lambda)$ has only stable poles.  \\
                                        \hline
\end{longtable}}
\end{description}

\subsubsection*{Output data}
\begin{description}
\item
\texttt{SYS1} contains  the descriptor system state-space realization of the transfer function matrix $G_1(\lambda)$ in the form
\be\label{gsdec:sysss1}
\begin{aligned}
 E_1\lambda x_1(t)  &=   A_1 x_1(t)+ B_1 u(t)  ,\\
y_1(t) &=  C_1 x_1(t)+ D u(t)  .
\end{aligned}
\ee
The pair $(A_1,E_1)$ is in a GRSF.
\item
\texttt{SYS2} contains  the descriptor system state-space realization of the transfer function matrix $G_2(\lambda)$ in the form
\be\label{gsdec:sysss2}
\begin{aligned}
 E_2\lambda x_2(t)  &=   A_2 x_2(t)+ B_2 u(t)  ,\\
y_2(t) &=  C_2 x_2(t)  .
\end{aligned}
\ee
The pair $(A_2,E_2)$ is in a GRSF.
\item
\texttt{Q} is the employed left transformation matrix used to reduce the pole pencil $A-\lambda E$ to a block-diagonal form (see \textbf{Method}).
\item
\texttt{Z} is the employed right transformation matrix used to reduce the pole pencil $A-\lambda E$ to a block-diagonal form (see \textbf{Method}).
\end{description}

\subsubsection*{Method}
The employed additive decomposition approach of the transfer function matrix $G(\lambda)$ in the form (\ref{gsdec:tfm_add}) is described in Section \ref{app:tfm-add}, and is based on the method proposed in \cite{Kags89}. For the  computation of spectral separations of the descriptor system \texttt{SYS}, the mex-function \url{sl_gsep} is employed. \index{MEX-functions!\url{sl_gsep}}%

For a certain domain of interest $\mathds{C}_g$, the basic computation consist in determining two invertible matrices $Q$ and $Z$, to obtain an equivalent descriptor system representation of
$G(\lambda)$ with a block-diagonal pole pencil, in the form
\be\label{gsdec:desc-specdec} G(\lambda) =  \ba{c|c} QAZ-\lambda QEZ & QB \\ \hline \\[-3mm] CZ & D \ea
= \ba{cc|c}  A_1 -\lambda E_1 & 0 &  B_{1} \\ 0 &  A_2 -\lambda E_2 &  B_{2} \\ \hline \\[-4mm]  C_1 &  C_2 & D \ea \, ,\ee
where $\Lambda(A_1 -\lambda E_1) \subset \mathds{C}_g$ and $\Lambda(A_2 -\lambda E_2) \subset \mathds{C}\setminus\mathds{C}_g$.
This leads to the additive decomposition of $G(\lambda)$ as
\be\label{gsdec:dss_add}  G(\lambda) = G_1(\lambda) + G_2(\lambda) ,\ee
where
\be\label{gsdec:dss_addterms} G_1(\lambda) = \ba{c|c}  A_1 -\lambda E_1 & B_{1} \\ \hline \\[-4mm]  C_1 &  D \ea, \quad G_2(\lambda) = \ba{c|c}  A_2 -\lambda E_2 & B_{2} \\ \hline \\[-4mm]  C_2 &  0 \ea \, .\ee

The computation of the descriptor realization (\ref{gsdec:desc-specdec}), with a block-diagonal pole pencil, is achieved in two steps. The first step involves the separation of the spectrum of $A-\lambda E$, using orthogonal transformations, in accordance with the selected option in \texttt{OPTIONS.job}, which defines the domain of interest $\mathds{C}_g$. In all cases, a preliminary finite-infinite or infinite-finite separation of the eigenvalues of the pole pencil $A-\lambda E$ is performed (only for descriptor systems) using the orthogonal staircase reduction algorithm proposed in \cite{Misr94} (for the computation of system zeros). The finite part of the reduced pole pencil is further reduced to a GRSF, and if necessary,  the finite eigenvalues are additionally separated into stable-unstable or unstable-stable blocks using eigenvalue reordering techniques (see \cite{Golu13}). In the second step, the block-diagonalization of the reduced pole pencil is performed using the approach of \cite{Kags89}.

Four spectral separations of the poles of $G(\lambda)$   can be selected via the option parameter \texttt{OPTIONS.job}.  The resulting pole  pencils of the corresponding descriptor system realizations (\ref{gsdec:sysss1}) of $G_1(\lambda)$ and (\ref{gsdec:sysss2}) of $G_2(\lambda)$,  exhibit several particular features, which are shortly addressed below:
\begin{description}
\item If \texttt{OPTIONS.job = 'finite'}, the resulting $A_2$ is nonsingular and upper triangular, while the resulting $E_2$ is nilpotent and upper triangular.
\item If \texttt{OPTIONS.job = 'infinite'}, the resulting $A_1$ is nonsingular and upper triangular, while the resulting $E_1$ is nilpotent and upper triangular. In this case, \texttt{SYS2} contains the strictly proper part of \texttt{SYS}.
\item If \texttt{OPTIONS.job = 'stable'}, the resulting pair $(A_2,E_2)$, in a GRSF, contains also all infinite eigenvalues of the pole pencil $A-\lambda E$ (the trailing part of $E_2$ contains a nilpotent matrix).
\item If \texttt{OPTIONS.job = 'unstable'}, the resulting pair $(A_1,E_1)$, in a GRSF, contains also all infinite eigenvalues of the pole pencil $A-\lambda E$ (the leading part of $E_1$ contains a nilpotent matrix).
\end{description}

\subsubsection*{Example}
\begin{example}\label{ex:gsdec}
Consider the $2\times 2$ improper transfer function matrix $G(s)$ of a continuous-time system
\[ G(s)  = \ba{cc} s^2 & \displaystyle\frac{s}{s+1} \\ \\[-3mm] 0 & \displaystyle\frac{1}{s} \ea . \]
To compute the proper-polynomial additive decomposition of $G(s)$,  we can use the following command sequence:
\begin{verbatim}
s = tf('s');              % define the complex variable s
Gc = [s^2 s/(s+1); 0 1/s] % define the 2-by-2 improper Gc(s)
sysc = ss(Gc);            % build continuous-time descriptor system realization

% compute the separation of proper and polynomial parts of Gc(s)
[sysf,sysi] = gsdec(sysc);

% for checking the results, convert the terms to zeros/poles/gain form
Gf = zpk(sysf)   % proper part
Gp = zpk(sysi)   % polynomial part
\end{verbatim}
The resulting transfer function matrices of the proper part $G_f(s)$ and polynomial part $G_p(s)$ are
\[ G_f(s)  = \ba{cc} 0 & \displaystyle\frac{s}{s+1} \\ \\[-3mm] 0 & \displaystyle\frac{1}{s} \ea , \qquad
G_p(s)  = \ba{cc} s^2 & 0 \\ \\[-3mm] 0 & 0 \ea . \]
\fine
\end{example}

\begin{example}\label{ex:gsdec2} For the transfer function matrix employed in Example \ref{ex:gsdec}, we can compute a polynomial-strictly proper additive decomposition of $G(s)$,  using the following command sequence:
\begin{verbatim}
s = tf('s');              % define the complex variable s
Gc = [s^2 s/(s+1); 0 1/s] % define the 2-by-2 improper Gc(s)
sysc = ss(Gc);            % build continuous-time descriptor system realization

% compute the separation of polynomial and strictly proper parts of Gc(s)
[syspol,syssp] = gsdec(sysc,struct('job','infinite'));

% for checking the results, convert the terms to zeros/poles/gain form
Gsp  = zpk(syssp)   % strictly proper part
Gpol = zpk(syspol)  % polynomial part
\end{verbatim}
The resulting strictly proper part $G_{sp}(s)$ and polynomial part $G_{pol}(s)$ are
\[ G_{sp}(s)  = \ba{cc} 0 & \displaystyle-\frac{1}{s+1} \\ \\[-3mm] 0 & \displaystyle\frac{1}{s} \ea , \qquad
G_{pol}(s)  = \ba{cc} s^2 & 1 \\ \\[-3mm] 0 & 0 \ea . \]
\fine
\end{example}

\subsubsection{\texttt{\bfseries grmcover1}}
\index{M-functions!\texttt{\bfseries grmcover1}}

\subsubsection*{Syntax}
\begin{verbatim}
[SYSX,INFO,SYSY] = grmcover1(SYS1,SYS2,TOL)
[SYSX,INFO,SYSY] = grmcover1(SYS,M1,TOL)
\end{verbatim}

\subsubsection*{Description}
\texttt{\bfseries grmcover1} computes, for given proper transfer function matrices $X_1(\lambda)$ and $X_2(\lambda)$, a proper $X(\lambda)$ and a strictly proper $Y(\lambda)$, both with minimal McMillan degrees, which satisfy
\be\label{grmcover1:syssscov} X(\lambda) = X_1(\lambda)+X_2(\lambda)Y(\lambda) \ee
and represent the solution of a right minimum cover problem. An approach based on the computation of a minimum dynamic cover of Type 1 is used to determine $X(\lambda)$ and $Y(\lambda)$, such that $\delta(X(\lambda)) \leq \delta(X_1(\lambda))$. If $X_1(\lambda)$ is improper, then a maximum reduction of $\delta(X(\lambda))$ is achieved, by maximally reducing the order of the proper part of $X_1(\lambda)$ and keeping unaltered its polynomial part.
\index{descriptor system!right minimal cover problem!Type 1}%
\index{transfer function matrix (TFM)!right minimal cover problem}%

\subsubsection*{Input data}
For the usage with
\begin{verbatim}
[SYSX,INFO,SYSY] = grmcover1(SYS1,SYS2,TOL)
\end{verbatim}
the input parameters are as follows:
\begin{description}
\item
\texttt{SYS1} is a LTI system, whose transfer function matrix is $X_1(\lambda)$, and is in a  descriptor system state-space form
\be\label{grmcover1:sysssG}
\begin{aligned}
E_1\lambda x_1(t)  &=   A_1x_1(t)+ B_1 u(t)  ,\\
y_1(t) &=  C_1 x_1(t)+ D_1 u(t)  ,
\end{aligned}
\ee
where $y_1(t) \in \mathds{R}^{p}$ and $u(t) \in \mathds{R}^{m_1}$.
\item
\texttt{SYS2} is a proper LTI system, whose transfer function matrix is $X_2(\lambda)$, and is in a  descriptor system state-space form
\be\label{grmcover1:sysssF}
\begin{aligned}
E_2\lambda x_2(t)  &=   A_2x_2(t)+ B_2 v(t)  ,\\
y_2(t) &=  C_2 x_2(t)+ D_2 v(t)  ,
\end{aligned}
\ee
where $y_2(t) \in \mathds{R}^{p}$.
\item
 \texttt{TOL} is a relative tolerance used for rank determinations. If \texttt{TOL} is not specified as input or if \texttt{TOL} = 0, an internally computed default value is used.
\end{description}
For the usage with
\begin{verbatim}
[SYSX,INFO,SYSY] = grmcover1(SYS,M1,TOL)
\end{verbatim}
the input parameters are as follows:
\begin{description}
\item
\texttt{SYS} is an input concatenated compound LTI system, \texttt{SYS = [ SYS1 SYS2 ]},  in a  descriptor system state-space form
\be\label{grmcover1:sysssGF}
\begin{aligned}
E\lambda x(t)  &=   Ax(t)+ B_1 u(t)+ B_2 v(t)  ,\\
y(t) &=  Cx(t)+ D_1 u(t)+ D_2 v(t)  ,
\end{aligned}
\ee
where \texttt{SYS1} has the transfer function matrix $X_1(\lambda)$, with the descriptor system realization $(A-\lambda E,B_1,C,D_1)$, and
\texttt{SYS2} has the proper transfer function matrix $X_2(\lambda)$, with the descriptor system realization $(A-\lambda E,B_2,C,D_2)$.
\item
\texttt{M1} is the dimension $m_1$ of the input vector $u(t)$ of the system \texttt{SYS1}.
\item
 \texttt{TOL} is a relative tolerance used for rank determinations. If \texttt{TOL} is not specified as input or if \texttt{TOL} = 0, an internally computed default value is used.
\end{description}

\subsubsection*{Output data}
\begin{description}
\item
\texttt{SYSX} contains, in the case when both $X_1(\lambda)$ and $X_2(\lambda)$ are proper,  the descriptor system state-space realization of the resulting reduced order $X(\lambda)$ in (\ref{grmcover1:syssscov}), in the form
\be\label{grmcover1:sysssX}
\begin{aligned}
E_r\lambda  x_{r,1}(t)  &=    A_r  x_{r,1}(t)+  B_r v(t)  ,\\
y_{r,1}(t) &= C_{r,1} x_{r,1}(t)+ D_1 v(t)  ,
\end{aligned}
\ee
with the pair $(A_r-\lambda E_r,B_r)$ in a controllability staircase form with $[\, B_r\; A_r\,]$ as in (\ref{cscf-defab}) and $E_r$ upper triangular and nonsingular, as in (\ref{cscf-defe}). If $X_1(\lambda)$ is improper,  but $X_2(\lambda)$ is proper, then \texttt{SYSX} contains the descriptor system realization corresponding to the sum of the reduced proper part of $X_1(\lambda)$ (in the form (\ref{grmcover1:sysssX})) and the polynomial part of $X_1(\lambda)$.
\item
 \texttt{INFO} is a MATLAB structure containing additional information, as follows:\\
{\begin{longtable}{|l|p{12cm}|} \hline
\textbf{\texttt{INFO} fields} & \textbf{Description} \\ \hline
 \texttt{stdim}   & vector which contains the dimensions of the diagonal
        blocks of $A_r-\lambda E_r$, which are the row dimensions of the full
        row rank diagonal blocks of the pencil $[\,B_r \; A_r-\lambda E_r\,]$ in
        controllability staircase form;\\ \hline
 \texttt{tcond}   & maximum of the Frobenius-norm condition numbers of
        the employed non-orthogonal transformation matrices (large values indicate possible  loss of numerical stability); \\ \hline
 \texttt{fnorm}   & the norm of the employed state-feedback $F$ used
       for minimum dynamic cover computation (see \textbf{Method}) (large values indicate possible  loss of numerical stability).   \\ \hline
\end{longtable}}
\item
\texttt{SYSY} contains the descriptor system state-space realization of the resulting (strictly proper) $Y(\lambda)$ in (\ref{grmcover1:syssscov}), in the form
\be\label{grmcover1:sysssY}
\begin{aligned}
E_r\lambda  x_{r,2}(t)  &=    A_r  x_{r,2}(t)+  B_r v(t)  ,\\
y_{r,2}(t) &= C_{r,2} x_{r,2}(t) ,
\end{aligned}
\ee
with the pair $(A_r-\lambda E_r,B_r)$ in a controllability staircase form with $[\, B_r\; A_r\,]$ as in (\ref{cscf-defab}) and $E_r$ upper triangular and nonsingular, as in (\ref{cscf-defe}).
\end{description}

\subsubsection*{Method}
The  approach to determine $X(\lambda)$ and $Y(\lambda)$, with least McMillan degrees, satisfying   (\ref{grmcover1:syssscov}) with both $X_1(\lambda)$ and $X_2(\lambda)$ proper, is based on computing a Type 1 minimum dynamic cover \cite{Kimu77} and is described in Section \ref{sec:dyncov}. This approach is based on a realization of the form (\ref{grmcover1:sysssGF}) of the input concatenated proper compound system \texttt{SYS = [ SYS1 SYS2 ]}, with invertible $E$. A detailed computational procedure to determine $X(\lambda)$ in   (\ref{grmcover1:syssscov})  is described in  \cite{Varg04a} (see also \textbf{Procedure GRMCOVER1} in \cite[Section 10.4.2]{Varg17} for more details). For the intervening reduction of the partitioned pair $(A-\lambda E,[\, B_1\;B_2\,]$ to a special controllability staircase form (see  \textbf{Procedure GSCSF} in \cite[Section 10.4.1]{Varg17}), the mex-function \url{sl_gstra} is employed. \index{MEX-functions!\url{sl_gstra}}%
If the descriptor realizations of \texttt{SYS1} and \texttt{SYS2} are separately provided, then an irreducible realization of the input concatenated compound system \texttt{[ SYS1 SYS2 ]} is internally computed. For orthogonal transformation based order reduction purposes, the mex-function \url{sl_gminr} is employed. \index{MEX-functions!\url{sl_gminr}}%

For the given descriptor system realization of the transfer function matrix $[\, X_1(\lambda) \; X_2(\lambda)\,]$ in the form (\ref{grmcover1:sysssGF}), a state-feedback matrix $F$ is determined to obtain the resulting reduced order $X(\lambda)$ and $Y(\lambda)$ in (\ref{grmcover1:syssscov}) with the descriptor realization
\be\label{grmcover1:Xtildegen2} \ba{c} X(\lambda) \\ Y(\lambda)\ea := \ba{c|c} A+B_2F-\lambda E & B_1 \\ \hline C+D_2F & D_1 \\
F & 0 \ea \, .\ee
The  matrix $F$ is determined such that the pair
$(A+B_2F-\lambda E,B_1)$ is {maximally uncontrollable}. Then, the resulting realizations of $X(\lambda)$ and $Y(\lambda)$ contain a maximum number of uncontrollable eigenvalues, which are eliminated by the Type 1 dynamic cover computation algorithm \cite{Varg04a}. This algorithm involves the use of non-orthogonal similarity transformations, whose maximum Frobenius-norm condition number is provided in \texttt{INFO.tcond}.  The resulting minimal realizations of $X(\lambda)$ and $Y(\lambda)$ have the forms (\ref{glmcover1:sysssX}) and (\ref{glmcover1:sysssY}), respectively.

As already mentioned, the underlying \textbf{Procedure GRMCOVER1} of
\cite[Section 10.4.2]{Varg17} is based on the algorithm proposed in \cite{Varg04a} for descriptor systems with invertible $E$. However, the function \texttt{\bfseries grmcover1} also works if the original descriptor system realization
       has $E$ singular, but \texttt{SYS2} is proper. In this case, the order
       reduction is performed working only with the proper part of \texttt{SYS1}.
       The polynomial part of \texttt{SYS1} is included, without modification, in the resulting realization of \texttt{SYSX}. To compute the intervening finite-infinite spectral separation, the mex-function \url{sl_gsep} is employed. \index{MEX-functions!\url{sl_gsep}}%

\subsubsection*{Examples}
\begin{example}\label{ex:simple-basis}
For a given transfer function matrix $G(\lambda)$, a simple minimal proper right nullspace basis can be computed in two steps: first, compute a minimal proper right nullspace basis $N_r(\lambda)$ of $G(\lambda)$, and then in a second step, determine successively the basis vectors of a minimal simple   proper basis $N_r^s(\lambda)$ by solving a sequence of right minimal cover problems. Specifically, if col$_i(N_r(\lambda))$ is the $i$-th basis vector (i.e., column) of $N_r(\lambda)$ and $V_i^c(\lambda)$ is the matrix formed from the rest of basis vectors, then the $i$-th basis vector col$_i(N_r^s(\lambda))$ of a minimal simple proper basis $N_r^s(\lambda)$ is the solution of the right cover problem
\[ \text{col}_i(N_r^s(\lambda)) = \text{col}_i(N_r(\lambda))+V_i^c(\lambda)Y_i(\lambda) ,\]
where $Y_i(\lambda)$ is a suitable strictly proper vector of least McMillan degree.

The following MATLAB commands illustrate this approach:
\begin{verbatim}
% computation of a minimal simple proper right nullspace basis
sys = rss(6,2,6);  % generate a random system with a 2x6 TFM

% compute a minimal proper right nullspace basis Nr with a 6x4 TFM
[Nr,info] = grnull(sys);
info.degs        % the expected orders of vectors of a minimal simple basis
nb = size(Nr,2); % number of basis vectors
% let Vi be the i-th basis vector and Vci the rest of basis vectors in Nr
% to compute the i-th basis vector Vsi of a simple basis Nrs,
% solve the right minimum cover problem:
% Vsi = Vi + Vci*Yi (for a suitable strictly proper Yi)
%
Nrs = ss(zeros(size(Nr)));   % initialize Nrs
for i = 1:nb;
    % apply the cover computation to Nr with permuted columns
    Vsi = grmcover1(Nr(:,[i,1:i-1,i+1:nb]),1);
    % check orders
    if order(Vsi) ~= info.degs(i)
       warning('Expected order not achieved')
    end
    Nrs(:,i) = Vsi;
end

% check nullspace condition sys*Nrs = 0
gminreal(sys*Nrs)
\end{verbatim}

The same approach can be used by calling the function \texttt{\bfseries grmcover2} instead \texttt{\bfseries grmcover1}.  \fine

\end{example}

\begin{example}\label{ex:polynomial-basis}
For a given transfer function matrix $G(\lambda)$, a minimal polynomial right nullspace basis can be computed in three steps: (1) compute a minimal proper right nullspace basis $N_r(\lambda)$ of $G(\lambda)$; (2) determine successively the basis vectors of a minimal simple proper basis $N_r^s(\lambda)$ by solving a sequence of right minimal cover problems; and (3) determine the polynomial numerator of each vector of the simple basis. The first two steps have been already described in Example \ref{ex:simple-basis}.

The following MATLAB commands illustrates the three step approach to compute polynomial bases:
\begin{verbatim}
% minimal polynomial nullspace computation
sys = rss(6,2,6);  % generate a random system with a 2x6 TFM

% compute a proper right nullspace basis Nr with a 6x4 TFM
[Nr,info] = grnull(sys);
info.degs        % the degrees of vectors of a minimal polynomial basis
nb = size(Nr,2); % number of basis vectors
% let Vi the i-th basis vector and Vci the rest of basis vectors in Nr
% to compute the i-th basis vector Vsi of a simple basis Nrs,
% solve the right minimum cover problem:
% Vsi = Vi + Vci*Yi (for a suitable strictly proper Yi)
% to obtain the corresponding polynomial basis vector, cancel all finite poles
% of Vsi
%
Nrp = tf(zeros(6,nb));   % initialize Nrs
for i = 1:nb;
    % apply the cover computation to Nr with permuted columns
    Vsi = grmcover1(Nr(:,[i,1:i-1,i+1:nb]),1);
    % check orders
    if order(Vsi) ~= info.degs(i)
       warning('Expected order not achieved')
    end
    Nrp(:,i) = minreal(tf(Vsi*tf(poly(eig(Vsi)),1)));
end

% check nullspace condition sys*Nrp = 0
gminreal(sys*Nrp,1.e-7)
\end{verbatim}

The same approach can be used by calling the function \texttt{\bfseries grmcover2} instead \texttt{\bfseries grmcover1}.  \fine

\end{example}

\subsubsection{\texttt{\bfseries glmcover1}}
\index{M-functions!\texttt{\bfseries glmcover1}}

\subsubsection*{Syntax}
\begin{verbatim}
[SYSX,INFO,SYSY] = glmcover1(SYS1,SYS2,TOL)
[SYSX,INFO,SYSY] = glmcover1(SYS,P1,TOL)
\end{verbatim}

\subsubsection*{Description}

\texttt{\bfseries glmcover1} computes, for given proper transfer function matrices $X_1(\lambda)$ and $X_2(\lambda)$, a proper $X(\lambda)$ and a strictly proper $Y(\lambda)$, both with minimal McMillan degrees, which satisfy
\be\label{glmcover1:syssscov} X(\lambda) = X_1(\lambda)+Y(\lambda)X_2(\lambda), \ee
and represent the solution of a left minimum cover problem. An approach based on the computation of a minimum dynamic cover of Type 1 is used to determine $X(\lambda)$ and $Y(\lambda)$, such that $\delta(X(\lambda)) \leq \delta(X_1(\lambda))$. If $X_1(\lambda)$ is improper, then a maximum reduction of $\delta(X(\lambda))$ is achieved, by maximally reducing the order of the proper part of $X_1(\lambda)$ and keeping unaltered its polynomial part.
\index{descriptor system!left minimal cover problem!Type 1}%
\index{transfer function matrix (TFM)!left minimal cover problem}%

\subsubsection*{Input data}
For the usage with
\begin{verbatim}
[SYSX,INFO,SYSY] = glmcover1(SYS1,SYS2,TOL)
\end{verbatim}
the input parameters are as follows:
\begin{description}
\item
\texttt{SYS1} is a LTI system, whose transfer function matrix $X_1(\lambda)$ has a  descriptor system state-space realization of the form
\be\label{glmcover1:sysssG}
\begin{aligned}
E_1\lambda x_1(t)  &=   A_1x_1(t)+ B_1 u(t)  ,\\
y_1(t) &=  C_1 x_1(t)+ D_1 u(t)  ,
\end{aligned}
\ee
where $y_1(t) \in \mathds{R}^{p_1}$ and $u(t) \in \mathds{R}^{m}$.
\item
\texttt{SYS2} is a proper LTI system, whose transfer function matrix $X_2(\lambda)$ has a  descriptor system state-space realization of the form
\be\label{glmcover1:sysssF}
\begin{aligned}
E_2\lambda x_2(t)  &=   A_2x_2(t)+ B_2 v(t)  ,\\
y_2(t) &=  C_2 x_2(t)+ D_2 v(t)  ,
\end{aligned}
\ee
where $v(t) \in \mathds{R}^{m}$.
\item
 \texttt{TOL} is a relative tolerance used for rank determinations. If \texttt{TOL} is not specified as input or if \texttt{TOL} = 0, an internally computed default value is used.
\end{description}
For the usage with
\begin{verbatim}
[SYSX,INFO,SYSY] = glmcover1(SYS,P1,TOL)
\end{verbatim}
the input parameters are as follows:
\begin{description}
\item
\texttt{SYS} is an output concatenated compound LTI system, \texttt{SYS = [ SYS1; SYS2 ]},  in a  descriptor system state-space form
\be\label{glmcover1:sysssGF}
\begin{aligned}
E\lambda x(t)  &=   Ax(t)+ B u(t)  ,\\
y_1(t) &=  C_1x(t)+ D_1 u(t)  , \\
y_2(t) &=  C_2x(t)+ D_2 u(t)  ,
\end{aligned}
\ee
where \texttt{SYS1} has the transfer function matrix $X_1(\lambda)$, with the descriptor system realization $(A-\lambda E,B,C_1,D_1)$, and
\texttt{SYS2} has the proper transfer function matrix $X_2(\lambda)$, with the descriptor system realization $(A-\lambda E,B,C_2,D_2)$.
\item
\texttt{P1} is the dimension $p_1$ of the output vector $y_1(t)$ of the system \texttt{SYS1}.
\item
 \texttt{TOL} is a relative tolerance used for rank determinations. If \texttt{TOL} is not specified as input or if \texttt{TOL} = 0, an internally computed default value is used.
\end{description}

\subsubsection*{Output data}
\begin{description}
\item
\texttt{SYSX} contains, in the case when both $X_1(\lambda)$ and $X_2(\lambda)$ are proper,  the descriptor system state-space realization of the resulting reduced order $X(\lambda)$  in (\ref{glmcover1:syssscov}), in the form
\be\label{glmcover1:sysssX}
\begin{aligned}
E_l\lambda  x_{l,1}(t)  &=    A_l  x_{l,1}(t)+  B_{l,1} v(t)  ,\\
y_{l,1}(t) &= C_{l} x_{l,1}(t)+ D_1 v(t)  ,
\end{aligned}
\ee
with the pair $(A_l-\lambda E_l,C_l)$ in an observability staircase form with $\left[\begin{smallmatrix} A_l\\ C_l \end{smallmatrix}\right]$ as in (\ref{oscf-defac}) and $E_l$ upper triangular and nonsingular, as in (\ref{oscf-defe}). If $X_1(\lambda)$ is improper,  but $X_2(\lambda)$ is proper, then \texttt{SYSX} contains the descriptor system realization corresponding to the sum of the reduced proper part of $X_1(\lambda)$ (in the form (\ref{glmcover1:sysssX})) and the polynomial part of $X_1(\lambda)$.
\item
 \texttt{INFO} is a MATLAB structure containing additional information, as follows:\\
{\begin{longtable}{|l|p{12cm}|} \hline
\textbf{\texttt{INFO} fields} & \textbf{Description} \\ \hline
 \texttt{stdim}   & vector containing the dimensions of the diagonal
        blocks of $A_l-\lambda E_l$, which are the column dimensions of the full
        column rank diagonal blocks of the pencil $\left[\begin{smallmatrix} A_l-\lambda E_l\\ C_l \end{smallmatrix}\right]$ in
        observability staircase form;\\ \hline
 \texttt{tcond}   & maximum of the Frobenius-norm condition numbers of
        the employed non-orthogonal transformation matrices (large values indicate possible  loss of numerical stability); \\ \hline
 \texttt{fnorm}   & the norm of the employed state-feedback $F$ used
       for minimum dynamic cover computation (see \textbf{Method}) (large values indicate possible  loss of numerical stability).   \\ \hline
\end{longtable}}
\item
\texttt{SYSY} contains the descriptor system state-space realization of the resulting (strictly proper) $Y(\lambda)$ in (\ref{glmcover1:syssscov}), in the form
\be\label{glmcover1:sysssY}
\begin{aligned}
E_l\lambda  x_{l,2}(t)  &=    A_l  x_{l,2}(t)+  B_{l,2} v(t)  ,\\
y_{l,2}(t) &= C_{l} x_{l,2}(t) ,
\end{aligned}
\ee
with the pair $(A_l-\lambda E_l,C_l)$ in an observability staircase form with $\left[\begin{smallmatrix} A_l\\ C_l \end{smallmatrix}\right]$ as in (\ref{oscf-defac}) and $E_l$ upper triangular and nonsingular, as in (\ref{oscf-defe}).
\end{description}

\subsubsection*{Method}

To determine $X(\lambda)$ and $Y(\lambda)$, with least McMillan degrees, satisfying  (\ref{glmcover1:syssscov}), the function \texttt{\bfseries glmcover1} calls \texttt{\bfseries grmcover1} to determine the dual
quantities $\widetilde X(\lambda)$ and $\widetilde Y(\lambda)$ satisfying
\[ \widetilde X(\lambda) = X_1^T(\lambda) + X_2^T(\lambda)\widetilde Y(\lambda) \]
and obtain $X(\lambda) = \widetilde X^T(\lambda)$ and $Y(\lambda) = \widetilde Y^T(\lambda)$.
\index{M-functions!\texttt{\bfseries grmcover1}}%
Thus, the employed solution method corresponds to the dual of the approach sketched in Section \ref{sec:dyncov} (see also \cite{Varg04a} and \textbf{Procedure GRMCOVER1} in \cite[Section 10.4.2]{Varg17} for more details). The function \texttt{\bfseries grmcover1} relies on the mex-functions \url{sl_gstra} to compute a special controllability staircase form (see  \textbf{Procedure GSCSF} in \cite[Section 10.4.1]{Varg17}) and \url{sl_gminr} to compute irreducible realizations.
\index{MEX-functions!\url{sl_gstra}}%
\index{MEX-functions!\url{sl_gminr}}%

For the given descriptor system realization of the transfer function matrix $\left[\begin{smallmatrix} X_1(\lambda) \\ X_2(\lambda)\end{smallmatrix}\right]$ in the form (\ref{glmcover1:sysssGF}), an output injection $F$ is determined to obtain the resulting reduced order $X(\lambda)$ and $Y(\lambda)$ in (\ref{glmcover1:syssscov}) with the descriptor realization
\be\label{glmcover1:Xtildegen2} [\, X(\lambda) \;\; Y(\lambda)\,] := \ba{c|cc} A+FC_2-\lambda E & B+FD_2 & F\\ \hline C_1 & D_1 & 0 \ea \, .\ee
The matrix $F$ is determined such that the pair
 $(A+FC_2-\lambda E,C_1)$ is {maximally unobservable}. Then, the resulting realizations of $X(\lambda)$ and $Y(\lambda)$ contain a maximum number of unobservable eigenvalues, which are eliminated by the Type 1 dynamic cover computation algorithm \cite{Varg04a}. This algorithm involves the use of non-orthogonal similarity transformations, whose maximum Frobenius-norm condition number is provided in \texttt{INFO.tcond}.  The resulting minimal realizations of $X(\lambda)$ and $Y(\lambda)$ have the forms (\ref{glmcover1:sysssX}) and (\ref{glmcover1:sysssY}), respectively.

The underlying \textbf{Procedure GRMCOVER1} of
\cite[Section 10.4.2]{Varg17}, used in a dual setting, is based on the algorithm proposed in \cite{Varg04a} for descriptor systems with invertible $E$. However, the function \texttt{\bfseries glmcover1} also works if the original descriptor system realization
       has $E$ singular, but \texttt{SYS2} is proper. In this case, the order
       reduction is performed working only with the proper part of \texttt{SYS1}.
       The polynomial part of \texttt{SYS1} is included, without modification, in the resulting realization of \texttt{SYSX}.

\subsubsection{\texttt{\bfseries grmcover2}}
\index{M-functions!\texttt{\bfseries grmcover2}}
\subsubsection*{Syntax}
\begin{verbatim}
[SYSX,INFO,SYSY] = grmcover2(SYS1,SYS2,TOL)
[SYSX,INFO,SYSY] = grmcover2(SYS,M1,TOL)
\end{verbatim}

\subsubsection*{Description}

\texttt{\bfseries grmcover2} computes, for given proper transfer function matrices $X_1(\lambda)$ and $X_2(\lambda)$, a proper $X(\lambda)$ and a proper $Y(\lambda)$, both with minimal McMillan degrees, which satisfy
\be\label{grmcover2:syssscov} X(\lambda) = X_1(\lambda)+X_2(\lambda)Y(\lambda) \ee
 and represent the solution of a right minimum cover problem. An approach based on the computation of a minimum dynamic cover of Type 2 is used to determine $X(\lambda)$ and $Y(\lambda)$, such that $\delta(X(\lambda)) \leq \delta(X_1(\lambda))$. If $X_1(\lambda)$ is improper, then a maximum reduction of $\delta(X(\lambda))$ is achieved, by maximally reducing the order of the proper part of $X_1(\lambda)$ and keeping unaltered its polynomial part.
\index{descriptor system!right minimal cover problem!Type 2}%
\index{transfer function matrix (TFM)!right minimal cover problem}%

\subsubsection*{Input data}
For the usage with
\begin{verbatim}
[SYSX,INFO,SYSY] = grmcover2(SYS1,SYS2,TOL)
\end{verbatim}
the input parameters are as follows:
\begin{description}
\item
\texttt{SYS1} is a LTI system, whose transfer function matrix $X_1(\lambda)$ has a  descriptor system state-space realization of the form
\be\label{grmcover2:sysssG}
\begin{aligned}
E_1\lambda x_1(t)  &=   A_1x_1(t)+ B_1 u(t)  ,\\
y_1(t) &=  C_1 x_1(t)+ D_1 u(t)  ,
\end{aligned}
\ee
where $y_1(t) \in \mathds{R}^{p}$ and $u(t) \in \mathds{R}^{m_1}$.
\item
\texttt{SYS2} is a proper LTI system, whose transfer function matrix $X_2(\lambda)$ has a  descriptor system state-space realization of the form
\be\label{grmcover2:sysssF}
\begin{aligned}
E_2\lambda x_2(t)  &=   A_2x_2(t)+ B_2 v(t)  ,\\
y_2(t) &=  C_2 x_2(t)+ D_2 v(t)  ,
\end{aligned}
\ee
where $y_2(t) \in \mathds{R}^{p}$.
\item
 \texttt{TOL} is a relative tolerance used for rank determinations. If \texttt{TOL} is not specified as input or if \texttt{TOL} = 0, an internally computed default value is used.
\end{description}
For the usage with
\begin{verbatim}
[SYSX,INFO,SYSY] = grmcover2(SYS,M1,TOL)
\end{verbatim}
the input parameters are as follows:
\begin{description}
\item
\texttt{SYS} is an input concatenated compound LTI system, \texttt{SYS = [ SYS1 SYS2 ]},  in a  descriptor system state-space form
\be\label{grmcover2:sysssGF}
\begin{aligned}
E\lambda x(t)  &=   Ax(t)+ B_1 u(t)+ B_2 v(t)  ,\\
y(t) &=  Cx(t)+ D_1 u(t)+ D_2 v(t)  ,
\end{aligned}
\ee
where \texttt{SYS1} has the transfer function matrix $X_1(\lambda)$ with the descriptor system realization $(A-\lambda E,B_1,C,D_1)$ and
\texttt{SYS2} has the proper transfer function matrix $X_2(\lambda)$ with the descriptor system realization $(A-\lambda E,B_2,C,D_2)$.
\item
\texttt{M1} is the dimension $m_1$ of the input vector $u(t)$ of the system \texttt{SYS1}.
\item
 \texttt{TOL} is a relative tolerance used for rank determinations. If \texttt{TOL} is not specified as input or if \texttt{TOL} = 0, an internally computed default value is used.
\end{description}

\subsubsection*{Output data}
\begin{description}
\item
\texttt{SYSX} contains, in the case when both $X_1(\lambda)$ and $X_2(\lambda)$ are proper,  the descriptor system state-space realization of the resulting reduced order $X(\lambda)$ in (\ref{grmcover2:syssscov}), in the form
\be\label{grmcover2:sysssX}
\begin{aligned}
E_r\lambda  x_{r,1}(t)  &=    A_r  x_{r,1}(t)+  B_r v(t)  ,\\
y_{r,1}(t) &= C_{r,1} x_{r,1}(t)+ D_1 v(t)  ,
\end{aligned}
\ee
with the pair $(A_r-\lambda E_r,B_r)$ in a controllability staircase form with $[\, B_r\; A_r\,]$ as in (\ref{cscf-defab}) and $E_r$ upper triangular and nonsingular, as in (\ref{cscf-defe}). If $X_1(\lambda)$ is improper,  but $X_2(\lambda)$ is proper, then \texttt{SYSX} contains the descriptor system realization corresponding to the sum of the reduced proper part of $X_1(\lambda)$ (in the form (\ref{grmcover2:sysssX})) and the polynomial part of $X_1(\lambda)$.
\item
 \texttt{INFO} is a MATLAB structure containing additional information, as follows:\\
{\begin{longtable}{|l|p{12cm}|} \hline
\textbf{\texttt{INFO} fields} & \textbf{Description} \\ \hline
 \texttt{stdim}   & vector which contains the dimensions of the diagonal
        blocks of $A_r-\lambda E_r$, which are the row dimensions of the full
        row rank diagonal blocks of the pencil $[\,B_r \; A_r-\lambda E_r\,]$ in
        controllability staircase form;\\ \hline
 \texttt{tcond}   & maximum of the Frobenius-norm condition numbers of
        the employed non-orthogonal transformation matrices (large values indicate possible  loss of numerical stability); \\ \hline
 \texttt{fnorm}   & the norm of the employed state-feedback $F$ used
       for minimum dynamic cover computation (see \textbf{Method}) (large values indicate possible  loss of numerical stability);   \\ \hline
 \texttt{gnorm}   & the norm of the employed feedforward matrix $G$ used
       for minimum dynamic cover computation (see \textbf{Method}) (large values indicate possible  loss of numerical stability).   \\ \hline
\end{longtable}}
\item
\texttt{SYSY} contains the descriptor system state-space realization of the resulting $Y(\lambda)$ in (\ref{grmcover2:syssscov}), in the form
\be\label{grmcover2:sysssY}
\begin{aligned}
E_r\lambda  x_r(t)  &=    A_r  x_{r,2}(t)+  B_r v(t)  ,\\
y_{r,2}(t) &= C_{r,2} x_{r,2}(t) + D_{r,2} v(t),
\end{aligned}
\ee
with the pair $(A_r-\lambda E_r,B_r)$ in a controllability staircase form with $[\, B_r\; A_r\,]$ as in (\ref{cscf-defab}) and $E_r$ upper triangular and nonsingular, as in (\ref{cscf-defe}).
\end{description}

\subsubsection*{Method}
The  approach to determine $X(\lambda)$ and $Y(\lambda)$, with least McMillan degrees, satisfying   (\ref{grmcover1:syssscov}),  with both $X_1(\lambda)$ and $X_2(\lambda)$ proper,  is based on computing a Type 2 minimum dynamic cover \cite{Kimu77} and is described in Section \ref{sec:dyncov}. This approach is based on a realization of the form (\ref{grmcover2:sysssGF}) of the input concatenated compound proper system \texttt{SYS = [ SYS1 SYS2 ]}, with invertible $E$. A detailed computational procedure to determine $X(\lambda)$ in   (\ref{grmcover2:syssscov})  is described in  \cite{Varg04a} (see also \textbf{Procedure GRMCOVER2} in \cite[Section 10.4.2]{Varg17} for more details). For the intervening reduction of the partitioned pair $(A-\lambda E,[\, B_2\;B_1\,]$ to a special controllability staircase form (see  \textbf{Procedure GSCSF} in \cite[Section 10.4.1]{Varg17}), the mex-function \url{sl_gstra} is employed. \index{MEX-functions!\url{sl_gstra}}%
If the descriptor realizations of \texttt{SYS1} and \texttt{SYS2} are separately provided, then an irreducible realization of the input concatenated compound system \texttt{[ SYS1 SYS2 ]} is internally computed. For orthogonal transformation based order reduction purposes, the mex-function \url{sl_gminr} is employed. \index{MEX-functions!\url{sl_gminr}}%

For the given descriptor system realization of the transfer function matrix $[\, X_1(\lambda) \; X_2(\lambda)\,]$ in the form (\ref{grmcover2:sysssGF}), a state-feedback matrix $F$ and a feedforward  gain $G$ are determined to obtain the resulting reduced order $X(\lambda)$ and $Y(\lambda)$ in (\ref{grmcover2:syssscov}) with the descriptor realization
\be\label{grmcover2:Xtildegen2} \ba{c} X(\lambda) \\ Y(\lambda)\ea := \ba{c|c} A+B_2F-\lambda E & B_1+B_2G \\ \hline C+D_2F & D_1+D_2G \\
F & G \ea .\ee
The feedback matrix $F$ and feedforward matrix $G$ are determined such that the descriptor pair
 $(A+B_2F-\lambda E,B_1+B_2G)$ is {maximally uncontrollable}. Then, the resulting realizations of $X(\lambda)$ and $Y(\lambda)$ contain a maximum number of uncontrollable eigenvalues, which are eliminated by the Type 2 dynamic cover computation algorithm of \cite{Varg04a} for descriptor systems and of \cite{Varg03e} for standard systems. This algorithm involves the use of non-orthogonal similarity transformations, whose maximum Frobenius-norm condition number is provided in \texttt{INFO.tcond}.  The resulting minimal realizations of $X(\lambda)$ and $Y(\lambda)$ have the forms (\ref{grmcover2:sysssX}) and (\ref{grmcover2:sysssY}), respectively.

As already mentioned, the underlying \textbf{Procedure GRMCOVER2} of
\cite[Section 10.4.2]{Varg17} is based on the algorithm proposed in \cite{Varg04a} for descriptor systems with invertible $E$. However, the function \texttt{\bfseries grmcover2} also works if the original descriptor system realization
       has $E$ singular, but \texttt{SYS2} is proper. In this case, the order
       reduction is performed working only with the proper part of \texttt{SYS1}.
       The polynomial part of \texttt{SYS1} is included, without modification, in the resulting realization of \texttt{SYSX}. To compute the intervening finite-infinite spectral separation, the mex-function \url{sl_gsep} is employed. \index{MEX-functions!\url{sl_gsep}}%

\subsubsection{\texttt{\bfseries glmcover2}}
\index{M-functions!\texttt{\bfseries glmcover2}}
\subsubsection*{Syntax}
\begin{verbatim}
[SYSX,INFO,SYSY] = glmcover2(SYS1,SYS2,TOL)
[SYSX,INFO,SYSY] = glmcover2(SYS,P1,TOL)
\end{verbatim}

\subsubsection*{Description}

\texttt{\bfseries glmcover2} computes, for given proper transfer function matrices $X_1(\lambda)$ and $X_2(\lambda)$, a proper $X(\lambda)$ and a proper $Y(\lambda)$, both with minimal McMillan degrees, which satisfy
\be\label{glmcover2:syssscov} X(\lambda) = X_1(\lambda)+Y(\lambda)X_2(\lambda), \ee
and represent the solution of a left minimum cover problem. An approach based on the computation of a minimum dynamic cover of Type 2 is used to determine $X(\lambda)$ and $Y(\lambda)$, such that $\delta(X(\lambda)) \leq \delta(X_1(\lambda))$. If $X_1(\lambda)$ is improper, then a maximum reduction of $\delta(X(\lambda))$ is achieved, by maximally reducing the order of the proper part of $X_1(\lambda)$ and keeping unaltered its polynomial part.
\index{descriptor system!left minimal cover problem!Type 2}%
\index{transfer function matrix (TFM)!left minimal cover problem}%

\subsubsection*{Input data}
For the usage with
\begin{verbatim}
[SYSX,INFO,SYSY] = glmcover2(SYS1,SYS2,TOL)
\end{verbatim}
the input parameters are as follows:
\begin{description}
\item
\texttt{SYS1} is a LTI system, whose transfer function matrix $X_1(\lambda)$ has a  descriptor system state-space realization of the form
\be\label{glmcover2:sysssG}
\begin{aligned}
E_1\lambda x_1(t)  &=   A_1x_1(t)+ B_1 u(t)  ,\\
y_1(t) &=  C_1 x_1(t)+ D_1 u(t)  ,
\end{aligned}
\ee
where $y_1(t) \in \mathds{R}^{p_1}$ and $u(t) \in \mathds{R}^{m}$.
\item
\texttt{SYS2} is a proper LTI system, whose transfer function matrix $X_2(\lambda)$ has a  descriptor system state-space realization of the form
\be\label{glmcover2:sysssF}
\begin{aligned}
E_2\lambda x_2(t)  &=   A_2x_2(t)+ B_2 v(t)  ,\\
y_2(t) &=  C_2 x_2(t)+ D_2 v(t)  ,
\end{aligned}
\ee
where $v(t) \in \mathds{R}^{m}$.
\item
 \texttt{TOL} is a relative tolerance used for rank determinations. If \texttt{TOL} is not specified as input or if \texttt{TOL} = 0, an internally computed default value is used.
\end{description}
For the usage with
\begin{verbatim}
[SYSX,INFO,SYSY] = glmcover2(SYS,P1,TOL)
\end{verbatim}
the input parameters are as follows:
\begin{description}
\item
\texttt{SYS} is an output concatenated compound LTI system, \texttt{SYS = [ SYS1; SYS2 ]},  in a  descriptor system state-space form
\be\label{glmcover2:sysssGF}
\begin{aligned}
E\lambda x(t)  &=   Ax(t)+ B u(t)  ,\\
y_1(t) &=  C_1x(t)+ D_1 u(t)  , \\
y_2(t) &=  C_2x(t)+ D_2 u(t)  ,
\end{aligned}
\ee
where \texttt{SYS1} has the transfer function matrix $X_1(\lambda)$ with the descriptor system realization $(A-\lambda E,B,C_1,D_1)$ and
\texttt{SYS2} has the proper transfer function matrix $X_2(\lambda)$ with the descriptor system realization $(A-\lambda E,B,C_2,D_2)$.
\item
\texttt{P1} is the dimension $p_1$ of the output vector $y_1(t)$ of the system \texttt{SYS1}.
\item
 \texttt{TOL} is a relative tolerance used for rank determinations. If \texttt{TOL} is not specified as input or if \texttt{TOL} = 0, an internally computed default value is used.
\end{description}

\subsubsection*{Output data}
\begin{description}
\item
\texttt{SYSX} contains, in the case when both $X_1(\lambda)$ and $X_2(\lambda)$ are proper,  the descriptor system state-space realization of the resulting reduced order $X(\lambda)$  in (\ref{glmcover2:syssscov}), in the form
\be\label{glmcover2:sysssX}
\begin{aligned}
E_l\lambda  x_{l,1}(t)  &=    A_l  x_{l,1}(t)+  B_{l,1} v(t)  ,\\
y_{l,1}(t) &= C_{l} x_{l,1}(t)+ D_1 v(t)  ,
\end{aligned}
\ee
with the pair $(A_l-\lambda E_l,C_l)$ in an observability  staircase form with $\left[\begin{smallmatrix} A_l\\ C_l \end{smallmatrix}\right]$ as in (\ref{oscf-defac}) and $E_l$ upper triangular and nonsingular, as in (\ref{oscf-defe}). If $X_1(\lambda)$ is improper,  but $X_2(\lambda)$ is proper, then \texttt{SYSX} contains the descriptor system realization corresponding to the sum of the reduced proper part of $X_1(\lambda)$ (in the form (\ref{glmcover2:sysssX})) and the polynomial part of $X_1(\lambda)$.
\item
 \texttt{INFO} is a MATLAB structure containing additional information, as follows:\\
{\begin{longtable}{|l|p{12cm}|} \hline
\textbf{\texttt{INFO} fields} & \textbf{Description} \\ \hline
 \texttt{stdim}   & vector containing the dimensions of the diagonal
        blocks of $A_l-\lambda E_l$, which are the column dimensions of the full
        column rank diagonal blocks of the pencil $\left[\begin{smallmatrix} A_l-\lambda E_l\\ C_l \end{smallmatrix}\right]$ in
        observability staircase form;\\ \hline
 \texttt{tcond}   & maximum of the Frobenius-norm condition numbers of
        the employed non-orthogonal transformation matrices (large values indicate possible  loss of numerical stability); \\ \hline
 \texttt{fnorm}   & the norm of the employed state-feedback $F$ used
       for minimum dynamic cover computation (see \textbf{Method}) (large values indicate possible  loss of numerical stability).   \\ \hline
\end{longtable}}
\item
\texttt{SYSY} contains the descriptor system state-space realization of the resulting $Y(\lambda)$ in (\ref{glmcover2:syssscov}), in the form
\be\label{glmcover2:sysssY}
\begin{aligned}
E_l\lambda  x_{l,2}(t)  &=    A_l  x_{l,2}(t)+  B_{l,2} v(t)  ,\\
y_{l,2}(t) &= C_{l} x_{l,2}(t) +  D_{l,2} v(t) ,
\end{aligned}
\ee
with the pair $(A_l-\lambda E_l,C_l)$ in an observability staircase form with $\left[\begin{smallmatrix} A_l\\ C_l \end{smallmatrix}\right]$ as in (\ref{oscf-defac}) and $E_l$ upper triangular and nonsingular, as in (\ref{oscf-defe}).
\end{description}

\subsubsection*{Method}
To determine $X(\lambda)$ and $Y(\lambda)$, with least McMillan degrees, satisfying  (\ref{glmcover1:syssscov}), the function \texttt{\bfseries glmcover2} calls \texttt{\bfseries grmcover2} to determine the dual
quantities $\widetilde X(\lambda)$ and $\widetilde Y(\lambda)$ satisfying
\[ \widetilde X(\lambda) = X_1^T(\lambda) + X_2^T(\lambda)\widetilde Y(\lambda) \]
and obtain $X(\lambda) = \widetilde X^T(\lambda)$ and $Y(\lambda) = \widetilde Y^T(\lambda)$.
\index{M-functions!\texttt{\bfseries grmcover1}}%
Thus, the employed solution method corresponds to the dual of the approach sketched in Section \ref{sec:dyncov} (see also \cite{Varg04a} and \textbf{Procedure GRMCOVER2} in \cite[Section 10.4.2]{Varg17} for more details). The function \texttt{\bfseries grmcover2} relies on the mex-functions \url{sl_gstra} to compute a special controllability staircase form (see  \textbf{Procedure GSCSF} in \cite[Section 10.4.1]{Varg17}) and \url{sl_gminr} to compute irreducible realizations.
\index{MEX-functions!\url{sl_gstra}}%
\index{MEX-functions!\url{sl_gminr}}%

For the given descriptor system realization of the transfer function matrix $\left[\begin{smallmatrix} X_1(\lambda) \\ X_2(\lambda)\end{smallmatrix}\right]$ in the form (\ref{glmcover2:sysssGF}), an output injection $F$ is determined to obtain the resulting reduced order $X(\lambda)$ and $Y(\lambda)$ in (\ref{glmcover2:syssscov}) with the descriptor realization
\be\label{glmcover2:Xtildegen2} [\, X(\lambda) \;\; Y(\lambda)\,] := \ba{c|cc} A+FC_2-\lambda E & B+FD_2 & F\\ \hline C_1+GC_2 & D_1+GD_2 & G\ea \, .\ee
The matrices $F$ and $G$ are determined such that the pair
 $(A+FC_2-\lambda E,C_1+GC_2)$ is {maximally unobservable}. Then, the resulting realizations of $X(\lambda)$ and $Y(\lambda)$ contain a maximum number of unobservable eigenvalues, which are eliminated by the Type 2 dynamic cover computation algorithm \cite{Varg04a}. This algorithm involves the use of non-orthogonal similarity transformations, whose maximum Frobenius-norm condition number is provided in \texttt{INFO.tcond}.  The resulting minimal realizations of $X(\lambda)$ and $Y(\lambda)$ have the forms (\ref{glmcover2:sysssX}) and (\ref{glmcover2:sysssY}), respectively.

The underlying \textbf{Procedure GRMCOVER2} of
\cite[Section 10.4.2]{Varg17}, used in a dual setting, is based on the algorithm proposed in \cite{Varg04a} for descriptor systems with invertible $E$. However, the function \texttt{\bfseries glmcover2} also works if the original descriptor system realization
       has $E$ singular, but \texttt{SYS2} is proper. In this case, the order
       reduction is performed working only with the proper part of \texttt{SYS1}.
       The polynomial part of \texttt{SYS1} is included, without modification, in the resulting realization of \texttt{SYSX}.

\subsubsection*{Example}
\begin{example}\label{ex:Wang-Davison2}
This example, taken from \cite{Wang73}, has been already considered in Example \ref{ex:Wang-Davison} to illustrate the computation of a left inverse of least McMillan degree of a full column rank transfer function matrix $G(s)$, using the function \texttt{\bfseries glsol}. In this example, we illustrates an alternative way to compute a left inverse of least McMillan degree, using the generators of all solution of the equation $X(\lambda)G(\lambda) = I$. A generator of all solutions is formed from a pair $(X_0(s),N_l(s))$, where $X_0(\lambda)$ is any particular solution (i.e., any particular left inverse) and $N_l(s)$ is a proper basis of the left nullspace of $G(s)$. Such a generator can be also computed using
\texttt{\bfseries glsol}.

The determination of a least order left inverse can be alternatively formulated in terms of the solution of the following left minimal cover problem: for a given generator $(X_0(s),N_l(s))$, compute least order $X(s)$ and $Y(s)$ such that
\[ X(s) = X_0(s)+N_l(s)Y(s). \]
Recall transfer function matrix used in Example \ref{ex:Wang-Davison}
\[ G(s) = \frac{1}{s^2+3s+2}\ba{cc} s+1 & s+2\\ s+3 & s^2+2s\\ s^2+3s & 0 \ea, \]
which has a third order minimal state-space realization, which can be computed using the function \texttt{\bfseries gir} (or \texttt{\bfseries minreal}). This realization is used to compute, using \texttt{\bfseries glsol}, a pair $(X_0(s),N_l(s))$ representing a generator of all solutions of $X(s)G(s) = I$. The function \texttt{\bfseries glmcover2} is then used to solve the above left minimal cover problem, to determine a least McMillan order proper left inverse $X(s)$. The following MATLAB commands illustrate this approach:
\begin{verbatim}
% Wang and Davison Example (1973)
s = tf('s');
g = [ s+1 s+2; s+3 s^2+2*s; s^2+3*s 0 ]/(s^2+3*s+2);
sysg = gir(ss(g));

% compute a generator (X0,Nl) of all solutions of X(s)G(s) = I
[~,~,sysgen] = glsol(sysg,ss(eye(2)));  % X0 = sysgen(:,1:2), Nl = sysgen(:,3)
gpole(sysgen)   % the generator is proper (no infinite poles)

% compute a least order inverse as X(s) = X0(s)+NL(s)*Y(s), by using
% order reduction based on a left minimal dynamic cover
[sysx,~,sysy] = glmcover2(sysgen,2);
gpole(sysx)     % resulting 2nd (least order) left inverse is unstable

% check solution applying gminreal to sysx*sysg-eye(2) or computing its norm
rez = gminreal(sysx*sysg-eye(2))
norm(sysx*sysg-eye(2),inf)
\end{verbatim}

The computed solution is unstable. Since the resulting $Y(s)$ is strictly proper, the same approach can be used by calling the function \texttt{\bfseries glmcover1} instead \texttt{\bfseries glmcover2}.  \fine

\end{example}

\subsubsection{\texttt{\bfseries gbilin}}
\index{M-functions!\texttt{\bfseries gbilin}}
\subsubsection*{Syntax}
\begin{verbatim}
[SYST,SYSI1] = gbilin(SYS,SYS1)
\end{verbatim}

\subsubsection*{Description}

\noindent \texttt{gbilin} computes for a LTI descriptor  system realization $(A-\lambda E,B,C,D)$ and a first order transfer function $g(\delta)$, a transformed descriptor system realization $(\widetilde A-\delta \widetilde E,\widetilde B,\widetilde C,\widetilde D)$ corresponding to the bilinear transformation $\lambda = g(\delta)$.
\index{descriptor system!bilinear transformation}

\subsubsection*{Input data}

\begin{description}
\item
\texttt{SYS} is a LTI system,  in a  descriptor system state-space form
\be\label{gbilin:sysss}
\begin{aligned}
E\lambda x(t)  &=   Ax(t)+ B u(t) ,\\
y(t) &=  C x(t)+ D u(t) .
\end{aligned}
\ee
\item
\texttt{SYS1} is a LTI system of McMillan degree one,  given as a real transfer function of the form  \be\label{gbilin:sys1io}
\begin{aligned}
g(\delta) = \frac{a\delta+b}{c\delta+d} ,
\end{aligned}
\ee
where, for a continuous-time system \texttt{SYS1}, $\delta = s$, the complex variable in
    the Laplace transform, while for a discrete-time system \texttt{SYS1},
   $\delta = z$, the complex variable in the Z-transform. The parameters $a$, $b$, $c$ and $d$ must be real and must satisfy $ad-bd \neq 0$ (to prevent cancellation). The transformation (\ref{gbilin:sys1io}) is a particular case of the \emph{M\"{o}bius transformation} (with complex parameters).
\item
 \texttt{OPTIONS} is a MATLAB structure to specify user options and has the following fields:\\
\pagebreak[4]
{\tabcolsep=1mm
\begin{longtable}{|l|lcp{11cm}|} \hline
\textbf{\texttt{OPTIONS} fields} & \multicolumn{3}{l|}{\textbf{Description}} \\ \hline
\texttt{tol}   & \multicolumn{3}{p{11cm}|}{tolerance for rank determinations (Default: internally computed)} \\ \hline
 \texttt{compact}   & \multicolumn{3}{p{12cm}|}{option to compute a compact descriptor system realization for the transformed system, without non-dynamic modes:}\\
                 &  \texttt{true} &--& determine a compact descriptor realization (default); \\
                 &  \texttt{false}&--& disable elimination of non-dynamic modes.   \\
                                        \hline
 \texttt{minimal}   & \multicolumn{3}{p{12cm}|}{option to compute a minimal descriptor system realization for the transformed system:}\\
                 &  \texttt{true} &--& determine a standard state-space realization; \\
                 &  \texttt{false}&--& no minimal realization computed (default)  \\
                                        \hline
 \texttt{ss}   & \multicolumn{3}{p{12cm}|}{option to compute a standard state-space
                       realization (if possible) for the transformed system:} \\
                 &  \texttt{true} &--& determine a standard state-space realization; \\
                 &  \texttt{false}&--& determine a descriptor system realization (default)  \\
                                        \hline
\end{longtable}}

\end{description}

\subsubsection*{Output data}

\begin{description}
\item
\texttt{SYST} contains the resulting transformed system in a descriptor system state-space form
\be\label{gbilin:systss}
\begin{aligned}
\widetilde E\delta \widetilde x(t)  &=   \widetilde A\widetilde x(t)+ \widetilde B u(t) ,\\
y(t) &=  \widetilde C x(t)+ \widetilde D u(t) .
\end{aligned}
\ee
corresponding to the bilinear transformation $\lambda = g(\delta)$. It follows, that if \texttt{SYS} has the TFM $G(\lambda)$, then \texttt{SYST} has the TFM $G(g(\delta))$. If the original system (\ref{gbilin:sysss}) has a standard state-space representation with $E = I$ and $g(\delta)$ is a first order polynomial, then the resulting system  (\ref{gbilin:systss}) is also determined in a standard state-space form with $\widetilde E = I$.

\item
\texttt{SYSI1} is a LTI system of McMillan degree one,  obtained as a transfer function of the form
\be\label{gbilin:sysi1io}
\begin{aligned}
g^{-1}(\lambda) = \frac{\phantom{-}d\lambda-b}{-c\lambda+a} \, ,
\end{aligned}
\ee
representing the inverse of the bilinear transformation $\lambda = g(\delta)$.

\end{description}

\subsubsection*{Method}
The following cases are explicitly considered:
\begin{enumerate}
\item
If $g(\delta)$ is a rational transfer function of the form (\ref{gbilin:sys1io}), then the resulting matrices of the descriptor realization $(\widetilde A-\delta \widetilde E,\widetilde B,\widetilde C,\widetilde D)$ are determined as follows:
\[ \widetilde A-\delta \widetilde E = \ba{cc} dA-bE -\delta (aE-cA) & dB+\delta cB\\
0 & -I \ea, \quad \widetilde B = \ba{c} 0\\ I \ea, \quad \widetilde C = [\, C \; D\,], \quad \widetilde D = 0 . \]
This realization is generally not minimal, even if the original realization (\ref{gbilin:sysss}) is minimal, and therefore, by default, a more compact equivalent descriptor realization, without non-dynamic modes, is determined, unless \texttt{OPTION.compact = false}. The elimination of non-dynamic modes is performed using the approach implemented in the function \texttt{gss2ss} (see equations (\ref{gss2ss:triu})) and the resulting $\widetilde E$ is upper triangular.  A minimal realization can be obtained by using \texttt{OPTIONS.minimal = true}.  A standard state-space realization of \texttt{SYST} (if exists) can be obtained using \texttt{OPTIONS.ss = true}.
\item
If $g(\delta)$ is a polynomial transfer function of the form $g(\delta) = a\delta+b$ and $E\neq I$, then the resulting matrices of the descriptor realization $(\widetilde A-\delta \widetilde E,\widetilde B,\widetilde C,\widetilde D)$ are determined as follows:
\[ \widetilde A-\delta \widetilde E = A-bE -\delta aE, \quad \widetilde B = B, \quad \widetilde C = C, \quad \widetilde D = D . \]
This realization is minimal, if the original realization (\ref{gbilin:sysss}) is minimal. A standard state-space realization of \texttt{SYST} (if exists) can be obtained using \texttt{OPTIONS.ss = true}.
\item
If $g(\delta)$ is a polynomial transfer function of the form $g(\delta) = a\delta+b$ and $E = I$, then the resulting matrices of the standard state-space realization $(\widetilde A-\delta I,\widetilde B,\widetilde C,\widetilde D)$ with $\widetilde E = I$, are determined as follows:
\[ \widetilde A = (A-bI)/a, \quad \widetilde B = B/a, \quad \widetilde C = C, \quad \widetilde D = D . \]
This realization is minimal, if the original realization (\ref{gbilin:sysss}) is minimal.
\end{enumerate}

\noindent \emph{Note:} The function \texttt{gbilin1} (not documented) can be employed to generate the transfer function $g(\delta)$ and its inverse $g^{-1}(\lambda)$ for several commonly used bilinear transformations.

\subsubsection*{Example}
\begin{example}\label{ex:gbilin}
Consider the $2\times 2$  transfer function matrix $G(s)$ of a continuous-time system
\[ G(s)  = \ba{cc} s^2 & \displaystyle\frac{s}{s+1} \\ \\[-3mm] 0 & \displaystyle\frac{1}{s} \ea . \]
It is straightforward to observe that $G(s)$ is improper and has poles and zeros in the origin and at infinity. We can try to perturb $G(s)$ to simultaneously move the poles and zeros in the origin into the stable domain and make the infinite pole and zero finite by using a bilinear transformation $s \leftarrow g(s) = \frac{s+0.01}{1+0.01s}$. Here, $g(s)$ is formed as a composition of a \emph{translation} to shift the poles and zeros (i.e., $s \leftarrow s+0.01$) and of a \emph{reflection} and \emph{inversion} with respect to the real axis to make infinite poles finite (i.e., $s \leftarrow \frac{1}{1+0.01s}$).
To compute the descriptor realization of the proper and stable $G(g(s))$,  we can use the following command sequence:
\begin{verbatim}
s = tf('s');              % define the complex variable s
G  = [s^2 s/(s+1); 0 1/s] % define the 2-by-2 improper G(s)
sys = ss(G);              % build continuous-time descriptor system realization
[p,m] = size(sys);        % get system dimensions

% pole-zero analysis
pol = gpole(sys), zer = gzero(sys)

% make all poles and zeros stable and finite using g(s) = (s+0.01)/(1+0.01*s)
g = (s+0.01)/(1+0.01*s);

% compute the transformed system
syst = gbilin(sys,g,struct('tol',1.e-7,'minimal',true));
minreal(zpk(syst),1.e-5)

% check resulting poles and zeros
pol_new = gpole(syst), zer_new = gzero(syst)

% compute the Vinnicombe's  nugap distance between models
nugap = gnugap(sys,syst)

% plot the Bode magnitude plot
bodemag(sys,syst,{0.01 100})

\end{verbatim}
The resulting proper transfer function matrix $G\big(g(s)\big)$, with stable and finite poles and zeros, is
\[ G\big(g(s)\big)  = \ba{cc} \displaystyle\frac{(s+0.01)^2}{(0.01s+1)^2} & 0.9901\displaystyle\frac{s+0.01}{s+1} \\ \\[-3mm] 0 & \displaystyle\frac{0.01s+1}{s+0.01} \ea  \]
and the resulting $\nu$-gap distance  \cite{Vinn93} between the transfer function matrices $G(s)$ (improper) and $G\big(g(s)\big)$ (proper),   is $\delta_\nu\big(G(s),G\big(g(s)\big)\big) = 0.0192$ (see Section \ref{sec:nugap}). The Bode-magnitude plots provide an alternative way to compare the frequency responses of $G(s)$ and $G(g(s))$ on a relevant frequency range.
\fine
\end{example}

\subsection{Functions for Factorizations}\label{dstools:fact}
These functions cover the computation of several factorizations of rational matrices, such as the coprime factorizations, inner-outer factorizations, and special spectral factorizations.

\subsubsection{\texttt{\bfseries grcf}}
\index{M-functions!\texttt{\bfseries grcf}}

\subsubsection*{Syntax}
\begin{verbatim}
[SYSN,SYSM] = grcf(SYS,OPTIONS)
\end{verbatim}

\subsubsection*{Description}
\texttt{\bfseries grcf} computes, for the transfer function matrix $G(\lambda)$ of a LTI descriptor state-space system, a right coprime factorization in the form
\be\label{grcf:def} G(\lambda) = N(\lambda)M^{-1}(\lambda) \, , \ee
such that $N(\lambda)$ and $M(\lambda)$ are proper transfer function matrices with poles in a specified stability  region $\mathds{C}_g \subset \mathds{C}$.
\index{factorization!right coprime (RCF)}%
\index{factorization!right coprime (RCF)!minimum-degree denominator}%

\subsubsection*{Input data}
\begin{description}
\item
\texttt{SYS} is a LTI system, whose transfer function matrix is $G(\lambda)$, and is in a  descriptor system state-space form
\be\label{grcf:sysss}
\begin{aligned}
E\lambda x(t)  &=   Ax(t)+ Bu(t) ,\\
y(t) &=  C x(t)+ D u(t) ,
\end{aligned}
\ee
with $x(t) \in \mathds{R}^n$ and $u(t) \in \mathds{R}^m$.
\item
 \texttt{OPTIONS} is a MATLAB structure to specify user options and has the following fields:\\
{\tabcolsep=1mm
\begin{longtable}{|l|lcp{11cm}|} \hline
\textbf{\texttt{OPTIONS} fields} & \multicolumn{3}{l|}{\textbf{Description}} \\ \hline
\texttt{tol}   & \multicolumn{3}{p{11cm}|}{tolerance for the singular values
                        based rank determination of $E$ (Default: $n^2\|E\|_1$\texttt{eps})} \\ \hline
\texttt{tolmin}   & \multicolumn{3}{p{10cm}|}{tolerance for the singular values
                        based controllability tests (Default: $nm\|B\|_1$\texttt{eps})} \\ \hline
 \texttt{smarg}   & \multicolumn{3}{p{12cm}|}{stability margin which specifies the stability region $\mathds{C}_g$ of the eigenvalues of the pole pencil as follows:
                  in the continuous-time case, the stable eigenvalues
                  have real parts less than or equal to \texttt{OPTIONS.smarg}, and
                  in the discrete-time case, the stable eigenvalues
                  have moduli less than or equal to \texttt{OPTIONS.smarg}. \newline
                  (Default: \texttt{-sqrt(eps)} for a continuous-time system \texttt{SYS}; \newline
                  \phantom{(Default:} \texttt{1-sqrt(eps}) for a discrete-time system \texttt{SYS}.)} \\ \hline
\texttt{sdeg}   & \multicolumn{3}{p{12.3cm}|}{prescribed stability degree for the poles of  the factors assigned within $\mathds{C}_g$
                    (Default: \texttt{[ ]}) } \\ \hline
 \texttt{poles}   & \multicolumn{3}{p{12cm}|}{complex conjugated set of desired poles  to be assigned for the factors
                     (Default: \texttt{[ ]})}\\
                                        \hline
 \texttt{mindeg}   & \multicolumn{3}{p{11cm}|}{option to compute a minimum degree denominator:}\\
                 &  \texttt{true} &--& determine a minimum degree denominator; \\
                 &  \texttt{false}&--& determine both factors with the same order (default)  \\
                                        \hline
 \texttt{mininf}   & \multicolumn{3}{p{12.3cm}|}{option for the removal of simple infinite eigenvalues (non-dynamic modes) of the factors:}\\
                 &  \texttt{true} &--& remove simple infinite eigenvalues; \\
                 &  \texttt{false}&--& keep simple infinite eigenvalues (default)  \\
                                        \hline
\end{longtable}}

\end{description}

\subsubsection*{Output data}
\begin{description}
\item
\texttt{SYSN} contains the descriptor system state-space realization of the numerator factor $N(\lambda)$ in the form
\be\label{grcf:sysN}
\begin{aligned}
E_N\lambda x_N(t)  &=   A_N x_N(t)+ B_N v(t)  ,\\
y_N(t) &=  C_N x_N(t)+ D_N v(t)  ,
\end{aligned}
\ee
where the pair $(A_N,E_N)$ is in a GRSF. The eigenvalues of $A_N-\lambda E_N$ include all eigenvalues of $A-\lambda E$ in $\mathds{C}_g$, and, additionally, all assigned eigenvalues in accordance with the specified stability degree in \texttt{OPTIONS.sdeg} and poles in \texttt{OPTIONS.poles}.  The resulting $E_N$ is invertible if \texttt{OPTIONS.mininf = true}. \index{condensed form!generalized real Schur (GRSF)}
\item
\texttt{SYSM}  contains the descriptor system state-space realization of the denominator factor $M(\lambda)$ in the form
\be\label{grcf:sysM}
\begin{aligned}
E_M\lambda x_M(t)  &=   A_M x_M(t)+ B_M w(t)  ,\\
y_M(t) &=  C_M x_M(t)+ D_M w(t)  ,
\end{aligned}
\ee
where the pair $(A_M,E_M)$ is in a GRSF. If \texttt{OPTIONS.mindeg = false}, then $N(\lambda)$ and $M(\lambda)$ have realizations of the same order  with $E_M = E_N$, $A_M = A_N$, and $B_M = B_N$ and the eigenvalues of $A_M-\lambda E_M$ include all eigenvalues of $A-\lambda E$ in $\mathds{C}_g$, which are however unobservable. Additionally, the eigenvalues of $A_M-\lambda E_M$ include the assigned eigenvalues in accordance with the specified stability degree in \texttt{OPTIONS.sdeg} and poles in \texttt{OPTIONS.poles}.  The resulting $E_M$ is invertible if \texttt{OPTIONS.mininf = true}. If \texttt{OPTIONS.mindeg = true} then the eigenvalues of $A_M-\lambda E_M$ include only the assigned eigenvalues in accordance with the specified stability degree in \texttt{OPTIONS.sdeg} and poles in \texttt{OPTIONS.poles}. In this case, the resulting $E_M$ is always invertible.
\end{description}

\subsubsection*{Method}
For the definitions related to coprime factorizations of transfer function matrices see Section \ref{appsec:fact}.
The implemented computational methods to compute the right coprime factorizations of general rational matrices  rely on a preliminary orthogonal reduction of the pole pencil $A-\lambda E$
to a special GRSF, which allows to preserve all eigenvalues of $A-\lambda E$ in $\mathds{C}_g$ in the resulting factors. The underlying reduction is described in \cite{Varg17d}. The function \texttt{grcf} implements the \textbf{Procedure GRCF} of \cite{Varg17d}, which represents
  an extension of the recursive factorization approach of \cite{Varg98a} to cope with
  infinite poles. In this procedure, all infinite poles are first assigned to finite real values.

If \texttt{OPTIONS.poles} is empty, then a stabilization oriented factorization is performed (see \cite{Varg17d}), where the infinite poles are assigned to real values specified by  \texttt{OPTIONS.sdeg} and all finite poles lying outside $\mathds{C}_g$ are assigned to the nearest values having a stability margin specified by \texttt{OPTIONS.sdeg}.
  If \texttt{OPTIONS.poles} is not empty, then a pole assignment oriented factorization is performed, by assigning first all infinite poles to real values specified in \texttt{OPTIONS.poles}. If \texttt{OPTIONS.poles} does not contain a sufficient
  number of real values, then a part or all of infinite poles are
  assigned to the value specified by \texttt{OPTIONS.sdeg}. Then, all finite poles lying outside $\mathds{C}_g$ are assigned to values specified in \texttt{OPTIONS.poles}  or assigned to the values having a stability margin specified by \texttt{OPTIONS.sdeg}.

\subsubsection*{Example}
\begin{example}\label{ex:grcf}
Consider the continuous-time improper TFM
\be\label{grcf:ex1}  G(s) = {\def\arraystretch{2}\left[\begin{array}{cc} s^2 & \displaystyle\frac{s}{s +1}\\[2mm]0 & \displaystyle\frac{1}{s} \end{array}\right]},
\ee
which has the following set of poles: $\{-1, 0, \infty, \infty\}$. To compute a stable and proper right coprime factorization of $G(s)$, we employed the pole assignment oriented factorization, with a stability degree of $-1$, a desired set of poles $\{ -1, -2, -3\}$, and with the option to eliminate the simple infinite eigenvalues. The resulting factors have the transfer function matrices
\[ N(s) = \ba{cc} -\displaystyle\frac{s^2}{(s+1)(s+2)} & \displaystyle\frac{s^2}{(s+1)(s+3)} \\[3mm] 0 & \displaystyle\frac{1}{s+3} \ea , \quad
M(s) = \ba{cc} -\displaystyle\frac{1}{(s+1)(s+2)} & 0 \\[2mm] 0 & \displaystyle\frac{s}{s+3} \ea  .
\]
The McMillan degree of $M(s)$ is three, thus the least possible one.
The above factors have been computed with the following sequence of commands:
\begin{verbatim}
% Varga (2017), Example 1
s = tf('s');  % define the complex variable s
% enter G(s) and determine a minimal state-space realization
G = [s^2 s/(s+1);
    0 1/s];
sys = ss(G);
gpole(sys)    % the system is unstable and improper

% compute the right coprime factorization G(s) = N(s)*inv(M(s))
[sysn,sysm] = grcf(sys,struct('poles',[-1,-2,-3,-4],'sdeg',-1,'mininf',true));

% check the factorization ||G(s)*M(s)-N(s)||_inf = 0
norm(gminreal(sys*sysm-sysn),inf)

% check the poles of the factors
gpole(sysm), gpole(sysn)

% check coprimeness of the factors
gzero(gir([sysn;sysm]))   % [N(s); M(s)] has no zeros
\end{verbatim}
\fine\end{example}

\subsubsection{\texttt{\bfseries glcf}}
\index{M-functions!\texttt{\bfseries glcf}}

\subsubsection*{Syntax}
\begin{verbatim}
[SYSN,SYSM] = glcf(SYS,OPTIONS)
\end{verbatim}

\subsubsection*{Description}
\texttt{\bfseries glcf} computes, for the transfer function matrix $G(\lambda)$ of a LTI descriptor state-space system, a left coprime factorization in the form
\be\label{glcf:def} G(\lambda) = M^{-1}(\lambda)N(\lambda) \, , \ee
such that $N(\lambda)$ and $M(\lambda)$ are proper transfer function matrices with poles in a specified stability  region $\mathds{C}_g \subset \mathds{C}$.
\index{factorization!left coprime (LCF)}%
\index{factorization!left coprime (LCF)!minimum-degree denominator}%

\subsubsection*{Input data}
\begin{description}
\item
\texttt{SYS} is a LTI system, whose transfer function matrix is $G(\lambda)$, and is in a  descriptor system state-space form
\be\label{glcf:sysss}
\begin{aligned}
E\lambda x(t)  &=   Ax(t)+ Bu(t) ,\\
y(t) &=  C x(t)+ D u(t) ,
\end{aligned}
\ee
with $x(t) \in \mathds{R}^n$ and $y(t) \in \mathds{R}^p$.
\item
 \texttt{OPTIONS} is a MATLAB structure to specify user options and has the following fields:\\
{\tabcolsep=1mm
\begin{longtable}{|l|lcp{11cm}|} \hline
\textbf{\texttt{OPTIONS} fields} & \multicolumn{3}{l|}{\textbf{Description}} \\ \hline
\texttt{tol}   & \multicolumn{3}{p{11cm}|}{tolerance for the singular values
                        based rank determination of $E$ (Default: $n^2\|E\|_1$\texttt{eps})} \\ \hline
\texttt{tolmin}   & \multicolumn{3}{p{10cm}|}{tolerance for the singular values
                        based observability tests (Default: $np\|C\|_\infty$\texttt{eps})} \\ \hline
 \texttt{smarg}   & \multicolumn{3}{p{12cm}|}{stability margin which specifies the stability region $\mathds{C}_g$ of the eigenvalues of the pole pencil as follows:
                  in the continuous-time case, the stable eigenvalues
                  have real parts less than or equal to \texttt{OPTIONS.smarg}, and
                  in the discrete-time case, the stable eigenvalues
                  have moduli less than or equal to \texttt{OPTIONS.smarg}. \newline
                  (Default: \texttt{-sqrt(eps)} for a continuous-time system \texttt{SYS}; \newline
                  \phantom{(Default:} \texttt{1-sqrt(eps}) for a discrete-time system \texttt{SYS}.)} \\ \hline
\texttt{sdeg}   & \multicolumn{3}{p{12.3cm}|}{prescribed stability degree for the poles of  the factors assigned within $\mathds{C}_g$
                    (Default: \texttt{[ ]}) } \\ \hline
 \texttt{poles}   & \multicolumn{3}{p{12cm}|}{complex conjugated set of desired poles  to be assigned for the factors
                     (Default: \texttt{[ ]})}\\
                                        \hline
 \texttt{mindeg}   & \multicolumn{3}{p{11cm}|}{option to compute a minimum degree denominator:}\\
                 &  \texttt{true} &--& determine a minimum degree denominator; \\
                 &  \texttt{false}&--& determine both factors with the same order (default)  \\
                                        \hline
 \texttt{mininf}   & \multicolumn{3}{p{12.3cm}|}{option for the removal of simple infinite eigenvalues (non-dynamic modes) of the factors:}\\
                 &  \texttt{true} &--& remove simple infinite eigenvalues; \\
                 &  \texttt{false}&--& keep simple infinite eigenvalues (default)  \\
                                        \hline
\end{longtable}}

\end{description}

\subsubsection*{Output data}
\begin{description}
\item
\texttt{SYSN} contains the descriptor system state-space realization of the numerator factor $N(\lambda)$ in the form
\be\label{glcf:sysN}
\begin{aligned}
E_N\lambda x_N(t)  &=   A_N x_N(t)+ B_N u(t)  ,\\
y_N(t) &=  C_N x_N(t)+ D_N u(t)  ,
\end{aligned}
\ee
where the pair $(A_N,E_N)$ is in a GRSF. The eigenvalues of $A_N-\lambda E_N$ include all eigenvalues of $A-\lambda E$ in $\mathds{C}_g$, and, additionally, all assigned eigenvalues in accordance with the specified stability degree in \texttt{OPTIONS.sdeg} and poles in \texttt{OPTIONS.poles}.  The resulting $E_N$ is invertible if \texttt{OPTIONS.mininf = true}. \index{condensed form!generalized real Schur (GRSF)}
\item
\texttt{SYSM}  contains the descriptor system state-space realization of the denominator factor $M(\lambda)$ in the form
\be\label{glcf:sysM}
\begin{aligned}
E_M\lambda x_M(t)  &=   A_M x_M(t)+ B_M w(t)  ,\\
y_M(t) &=  C_M x_M(t)+ D_M w(t)  ,
\end{aligned}
\ee
where the pair $(A_M,E_M)$ is in a GRSF. If \texttt{OPTIONS.mindeg = false}, then $N(\lambda)$ and $M(\lambda)$ have realizations of the same order with $E_M = E_N$, $A_M = A_N$, and $C_M = C_N$ and the eigenvalues of $A_M-\lambda E_M$ include all eigenvalues of $A-\lambda E$ in $\mathds{C}_g$, which are however uncontrollable. Additionally, the eigenvalues of $A_M-\lambda E_M$ include the assigned eigenvalues in accordance with the specified stability degree in \texttt{OPTIONS.sdeg} and poles in \texttt{OPTIONS.poles}.  The resulting $E_M$ is invertible if \texttt{OPTIONS.mininf = true}. If \texttt{OPTIONS.mindeg = true} then the eigenvalues of $A_M-\lambda E_M$ include only the assigned eigenvalues in accordance with the specified stability degree in \texttt{OPTIONS.sdeg} and poles in \texttt{OPTIONS.poles}. In this case, the resulting $E_M$ is always invertible.
\end{description}

\subsubsection*{Method}
For the definitions related to coprime factorizations of transfer function matrices see Section \ref{appsec:fact}.
To compute the left coprime factorization (\ref{glcf:def}), the function \texttt{glcf} calls \texttt{grcf} to compute the right coprime factorization of $G^T(\lambda)$ in the form
\[ G^T(\lambda) = \widetilde N(\lambda) \widetilde M^{-1}(\lambda) \]
and obtain the factors as $N(\lambda) = \widetilde N^T(\lambda)$ and $M(\lambda) = \widetilde M^T(\lambda)$. The function \texttt{grcf} implements the \textbf{Procedure GRCF} of \cite{Varg17d}, which represents
  an extension of the recursive factorization approach of \cite{Varg98a} to cope with
  infinite poles.

If \texttt{OPTIONS.poles} is empty, then a stabilization oriented factorization is performed (see \cite{Varg17d}), where the infinite poles are assigned to real values specified by  \texttt{OPTIONS.sdeg} and all finite poles lying outside $\mathds{C}_g$ are assigned to the nearest values having a stability margin specified by \texttt{OPTIONS.sdeg}.
  If \texttt{OPTIONS.poles} is not empty, then a pole assignment oriented factorization is performed, by assigning first all infinite poles to real values specified in \texttt{OPTIONS.poles}. If \texttt{OPTIONS.poles} does not contain a sufficient
  number of real values, then a part or all of infinite poles are
  assigned to the value specified by \texttt{OPTIONS.sdeg}. Then, all finite poles lying outside $\mathds{C}_g$ are assigned to values specified in \texttt{OPTIONS.poles}  or assigned to the values having a stability margin specified by \texttt{OPTIONS.sdeg}.

\subsubsection*{Example}
\begin{example}\label{ex:glcf}
Consider the continuous-time improper TFM
\be\label{glcf:ex1}  G(s) = {\def\arraystretch{2}\left[\begin{array}{cc} s^2 & \displaystyle\frac{s}{s +1}\\[2mm]0 & \displaystyle\frac{1}{s} \end{array}\right]},
\ee
which has the following set of poles: $\{-1, 0, \infty, \infty\}$. To compute a stable and proper left coprime factorization of $G(s)$, we employed the pole assignment oriented factorization, with a stability degree of $-1$, a desired set of poles $\{ -1, -2, -3\}$, and with the option to eliminate the simple infinite eigenvalues. The resulting factors have the transfer function matrices
\[ N(s) = \ba{cc} -\displaystyle\frac{s^2}{(s+1)(s+2)} & -\displaystyle\frac{s}{(s+1)^2(s+2)} \\[3mm] 0 & \displaystyle\frac{1}{s+3} \ea , \quad
M(s) = \ba{cc} -\displaystyle\frac{1}{(s+1)(s+2)} & 0 \\[2mm] 0 & \displaystyle\frac{s}{s+3} \ea  .
\]
The McMillan degree of $M(s)$ is three, thus the least possible one.
The above factors have been computed with the following sequence of commands:
\begin{verbatim}
% Varga (2017), Example 1
s = tf('s');  % define the complex variable s
% enter G(s) and determine a minimal state-space realization
G = [s^2 s/(s+1);
    0 1/s];
sys = ss(G);
gpole(sys)    % the system is unstable and improper

% compute the left coprime factorization G(s) = inv(M(s))*N(s)
[sysn,sysm] = glcf(sys,struct('poles',[-1,-2,-3,-4],'sdeg',-1,'mininf',true));

% check the factorization ||M(s)*G(s)-N(s)||_inf = 0
norm(gminreal(sysm*sys-sysn,1.e-7),inf)

% check the poles of the factors
gpole(sysm), gpole(sysn)

% check coprimeness of the factors
gzero(gir([sysn sysm]))  % [N(s) M(s)] has no zeros
\end{verbatim}

\fine\end{example}

\subsubsection{\texttt{\bfseries grcfid}}
\index{M-functions!\texttt{\bfseries grcfid}}

\subsubsection*{Syntax}
\begin{verbatim}
[SYSN,SYSM] = grcfid(SYS,OPTIONS)
\end{verbatim}

\subsubsection*{Description}
\texttt{\bfseries grcfid} computes, for the transfer function matrix $G(\lambda)$ of a LTI descriptor state-space system, a right coprime factorization with inner denominator in the form
\be\label{grcfid:def} G(\lambda) = N(\lambda)M^{-1}(\lambda) \, , \ee
such that $N(\lambda)$ and $M(\lambda)$ are proper and stable transfer function matrices, and $M(\lambda)$ is inner.
\index{factorization!right coprime (RCF)!with inner denominator}%

\subsubsection*{Input data}
\begin{description}
\item
\texttt{SYS} is a LTI system, whose transfer function matrix is $G(\lambda)$, and is in a  descriptor system state-space form
\be\label{grcfid:sysss}
\begin{aligned}
E\lambda x(t)  &=   Ax(t)+ Bu(t) ,\\
y(t) &=  C x(t)+ D u(t) ,
\end{aligned}
\ee
with $x(t) \in \mathds{R}^n$ and $u(t) \in \mathds{R}^m$. $G(\lambda)$ must not have poles in $\partial\mathds{C}_s$.
\item
 \texttt{OPTIONS} is a MATLAB structure to specify user options and has the following fields:\\
{\tabcolsep=1mm
\begin{longtable}{|l|lcp{11cm}|} \hline
\textbf{\texttt{OPTIONS} fields} & \multicolumn{3}{l|}{\textbf{Description}} \\ \hline
\texttt{tol}   & \multicolumn{3}{p{11cm}|}{tolerance for the singular values
                        based rank determination of $E$ (Default: $n^2\|E\|_1$\texttt{eps})} \\ \hline
\texttt{tolmin}   & \multicolumn{3}{p{10cm}|}{tolerance for the singular values
                        based controllability tests (Default: $nm\|B\|_1$\texttt{eps})} \\ \hline
 \texttt{mindeg}   & \multicolumn{3}{p{11cm}|}{option to compute a minimum degree denominator:}\\
                 &  \texttt{true} &--& determine a minimum degree denominator; \\
                 &  \texttt{false}&--& determine both factors with the same order (default)  \\
                                        \hline
 \texttt{mininf}   & \multicolumn{3}{p{12.3cm}|}{option for the removal of simple infinite eigenvalues (non-dynamic modes) of the factors:}\\
                 &  \texttt{true} &--& remove simple infinite eigenvalues; \\
                 &  \texttt{false}&--& keep simple infinite eigenvalues (default)  \\
                                        \hline
\end{longtable}}

\end{description}

\subsubsection*{Output data}
\begin{description}
\item
\texttt{SYSN} contains the descriptor system state-space realization of the numerator factor $N(\lambda)$ in the form
\be\label{grcfid:sysN}
\begin{aligned}
E_N\lambda x_N(t)  &=   A_N x_N(t)+ B_N v(t)  ,\\
y_N(t) &=  C_N x_N(t)+ D_N v(t)  ,
\end{aligned}
\ee
where the pair $(A_N,E_N)$ is in a GRSF. The eigenvalues of $A_N-\lambda E_N$ include all stable eigenvalues of $A-\lambda E$ (i.e., eigenvalues located in $\mathds{C}_s$). Additionally,  to each unstable eigenvalue of $A-\lambda E$  corresponds a stable eigenvalue of $A_N-\lambda E_N$ located in a symmetric location with respect to the imaginary axis, in the continuous-time case, or with respect to the unit circle centered in the origin, in the discrete-time case.  The resulting $E_N$ is invertible if \texttt{OPTIONS.mininf = true}. \index{condensed form!generalized real Schur (GRSF)}
\item
\texttt{SYSM}  contains the descriptor system state-space realization of the inner denominator factor $M(\lambda)$ in the form
\be\label{grcfid:sysM}
\begin{aligned}
E_M\lambda x_M(t)  &=   A_M x_M(t)+ B_M w(t)  ,\\
y_M(t) &=  C_M x_M(t)+ D_M w(t)  ,
\end{aligned}
\ee
where the pair $(A_M,E_M)$ is in a GRSF. The resulting $E_M$ is invertible if \texttt{OPTIONS.mininf = true}. If \texttt{OPTIONS.mindeg = false}, then $N(\lambda)$ and $M(\lambda)$ have realizations of the same order with $E_M = E_N$, $A_M = A_N$, and $B_M = B_N$ and the eigenvalues of $A_M-\lambda E_M$ include all stable eigenvalues of $A-\lambda E$, which are however unobservable. Additionally,  to each unstable eigenvalue of $\lambda_u \in \Lambda(A-\lambda E)$  corresponds a stable eigenvalue of $A_M-\lambda E_M$ located in a symmetric location with respect to the boundary of the appropriate stability domain (i.e.,  $-\bar\lambda_u \in \Lambda(A_M-\lambda E_M)$, in the continuous-time case, or $1/\lambda_u \in \Lambda(A_M-\lambda E_M)$ in the discrete-time case). If \texttt{OPTIONS.mindeg = true}, only the latter eigenvalues are present and the resulting $E_M$ is always invertible.
\end{description}

\subsubsection*{Method}
For the definitions related to coprime factorizations of transfer function matrices see Section \ref{appsec:fact}.
The implemented computational methods to compute the right coprime factorizations with inner denominators of transfer function matrices  rely on a preliminary orthogonal reduction of the pole pencil $A-\lambda E$
to a special GRSF, which allows to isolate all eigenvalues of $A-\lambda E$ lying outside of the stability domain $\mathds{C}_s$. The underlying reduction is described in \cite{Varg17d}. The function \texttt{grcfid} implements the \textbf{Procedure GRCFID} of \cite{Varg17d}, which represents
  an extension of the corresponding recursive factorization approach of \cite{Varg98a} to cope with
  infinite poles in the discrete-time case. In this procedure, all infinite poles of a discrete-time system are reflected into poles in the origin of the factors.

\subsubsection*{Example}
\begin{example}\label{ex:grcfid}
Consider the discrete-time improper TFM
\be\label{grcfid:ex1}  G(z) = {\def\arraystretch{2}\left[\begin{array}{cc} z^2 & \displaystyle\frac{z}{z-2}\\[2mm]0 & \displaystyle\frac{1}{z} \end{array}\right]},
\ee
which has the following set of poles: $\{2, 0, \infty, \infty\}$ and therefore, the right coprime factorization with inner denominator exists. With the option to eliminate the simple infinite eigenvalues, the function \texttt{grcfid}  computes the following factors having the transfer function matrices
\[ N(z) = \ba{cc} 1 & \displaystyle\frac{1}{2z-1} \\[3mm] 0 & \displaystyle\frac{z-2}{z(2z-1)} \ea , \quad
M(z) = \ba{cc} \displaystyle\frac{1}{z^2} & 0 \\[2mm] 0 & \displaystyle\frac{z-2}{2z-1} \ea  .
\]
The McMillan degree of $M(z)$ is three, thus the least possible one. Interestingly, the McMillan degree of $N(z)$ is only two, because two unobservable eigenvalues in 0 have been removed. These eigenvalues are the zeros of the (improper) all-pass factor $\diag(z^2,1)$ with two infinite poles, which is contained in $G(z)$.

The above factors have been computed with the following sequence of commands:
\begin{verbatim}
% Varga (2017), Example 2
z = tf('z');  % define the complex variable z
% enter G(z) and determine a minimal state-space realization
G = [z^2 z/(z-2);
    0 1/z];
sys = ss(G);
gpole(sys)    % the system is unstable and improper

% compute the right coprime factorization G(z) = N(z)*inv(M(z)),
% with inner denominator M(z)
[sysn,sysm] = grcfid(sys,struct('mininf',true));

% check the factorization ||G(z)*M(z)-N(z)||_inf = 0
norm(gminreal(sys*sysm-sysn),inf)

% check the innerness of M(z): ||conj(M(z))*M(z)-I||_inf = 0
norm(sysm'*sysm-eye(2),inf)

% check the poles of the factors
gpole(sysm), gpole(sysn)

% check coprimeness of the factors
gzero(gir([sysn;sysm]))   % [N(z); M(z)] has no zeros
\end{verbatim}

\fine\end{example}

\subsubsection{\texttt{\bfseries glcfid}}
\index{M-functions!\texttt{\bfseries glcfid}}

\subsubsection*{Syntax}
\begin{verbatim}
[SYSN,SYSM] = glcfid(SYS,OPTIONS)
\end{verbatim}

\subsubsection*{Description}
\texttt{\bfseries glcfid} computes, for the transfer function matrix $G(\lambda)$ of a LTI descriptor state-space system, a left coprime factorization with inner denominator in the form
\be\label{glcfid:def} G(\lambda) = M^{-1}(\lambda)N(\lambda) \, , \ee
such that $N(\lambda)$ and $M(\lambda)$ are proper and stable transfer function matrices, and $M(\lambda)$ is inner.
\index{factorization!left coprime (LCF)!with inner denominator}%

\subsubsection*{Input data}
\begin{description}
\item
\texttt{SYS} is a LTI system, whose transfer function matrix is $G(\lambda)$, and is in a  descriptor system state-space form
\be\label{glcfid:sysss}
\begin{aligned}
E\lambda x(t)  &=   Ax(t)+ Bu(t) ,\\
y(t) &=  C x(t)+ D u(t) ,
\end{aligned}
\ee
with $x(t) \in \mathds{R}^n$ and $y(t) \in \mathds{R}^p$. $G(\lambda)$ must not have poles in $\partial\mathds{C}_s$.
\item
 \texttt{OPTIONS} is a MATLAB structure to specify user options and has the following fields:\\
\pagebreak[4]
{\tabcolsep=1mm
\begin{longtable}{|l|lcp{11cm}|} \hline
\textbf{\texttt{OPTIONS} fields} & \multicolumn{3}{l|}{\textbf{Description}} \\ \hline
\texttt{tol}   & \multicolumn{3}{p{11cm}|}{tolerance for the singular values
                        based rank determination of $E$ (Default: $n^2\|E\|_1$\texttt{eps})} \\ \hline
\texttt{tolmin}   & \multicolumn{3}{p{10cm}|}{tolerance for the singular values
                        based observability tests (Default: $np\|C\|_\infty$\texttt{eps})} \\ \hline
 \texttt{mindeg}   & \multicolumn{3}{p{11cm}|}{option to compute a minimum degree denominator:}\\
                 &  \texttt{true} &--& determine a minimum degree denominator; \\
                 &  \texttt{false}&--& determine both factors with the same order (default)  \\
                                        \hline
 \texttt{mininf}   & \multicolumn{3}{p{12.3cm}|}{option for the removal of simple infinite eigenvalues (non-dynamic modes) of the factors:}\\
                 &  \texttt{true} &--& remove simple infinite eigenvalues; \\
                 &  \texttt{false}&--& keep simple infinite eigenvalues (default)  \\
                                        \hline
\end{longtable}}

\end{description}

\subsubsection*{Output data}
\begin{description}
\item
\texttt{SYSN} contains the descriptor system state-space realization of the numerator factor $N(\lambda)$ in the form
\be\label{glcfid:sysN}
\begin{aligned}
E_N\lambda x_N(t)  &=   A_N x_N(t)+ B_N u(t)  ,\\
y_N(t) &=  C_N x_N(t)+ D_N u(t)  ,
\end{aligned}
\ee
where the pair $(A_N,E_N)$ is in a GRSF. The eigenvalues of $A_N-\lambda E_N$ include all stable eigenvalues of $A-\lambda E$ (i.e., eigenvalues located in $\mathds{C}_s$). Additionally,  to each unstable eigenvalue of $A-\lambda E$  corresponds a stable eigenvalue of $A_N-\lambda E_N$ located in a symmetric location with respect to the imaginary axis, in the continuous-time case, or with respect to the unit circle centered in the origin, in the discrete-time case.  The resulting $E_N$ is invertible if \texttt{OPTIONS.mininf = true}. \index{condensed form!generalized real Schur (GRSF)}
\item
\texttt{SYSM}  contains the descriptor system state-space realization of the inner denominator factor $M(\lambda)$ in the form
\be\label{glcfid:sysM}
\begin{aligned}
E_M\lambda x_M(t)  &=   A_M x_M(t)+ B_M w(t)  ,\\
y_M(t) &=  C_M x_M(t)+ D_M w(t)  ,
\end{aligned}
\ee
where the pair $(A_M,E_M)$ is in a GRSF. The resulting $E_M$ is invertible if \texttt{OPTIONS.mininf = true}. If \texttt{OPTIONS.mindeg = false}, then $N(\lambda)$ and $M(\lambda)$ have realizations of the same order with $E_M = E_N$, $A_M = A_N$, and $B_M = B_N$ and the eigenvalues of $A_M-\lambda E_M$ include all stable eigenvalues of $A-\lambda E$, which are however unobservable. Additionally,  to each unstable eigenvalue of $\lambda_u \in \Lambda(A-\lambda E)$  corresponds a stable eigenvalue of $A_M-\lambda E_M$ located in a symmetric location with respect to the boundary of the appropriate stability domain (i.e.,  $-\bar\lambda_u \in \Lambda(A_M-\lambda E_M)$, in the continuous-time case, or $1/\lambda_u \in \Lambda(A_M-\lambda E_M)$ in the discrete-time case). If \texttt{OPTIONS.mindeg = true}, only the latter eigenvalues are present and the resulting $E_M$ is always invertible.
\end{description}

\subsubsection*{Method}
For the definitions related to coprime factorizations of transfer function matrices see Section \ref{appsec:fact}.
To compute the left coprime factorization (\ref{glcfid:def}), the function \texttt{glcfid} calls \texttt{grcfid} to compute the right coprime factorization with inner denominator of $G^T(\lambda)$ in the form
\[ G^T(\lambda) = \widetilde N(\lambda) \widetilde M^{-1}(\lambda) \]
and obtain the factors as $N(\lambda) = \widetilde N^T(\lambda)$ and $M(\lambda) = \widetilde M^T(\lambda)$.
The function \texttt{grcfid} implements the \textbf{Procedure GRCFID} of \cite{Varg17d}, which represents
  an extension of the corresponding recursive factorization approach of \cite{Varg98a} to cope with
  infinite poles in the discrete-time case.

\subsubsection*{Example}
\begin{example}\label{ex:glcfid}
Consider the discrete-time improper TFM
\be\label{glcfid:ex1}  G(z) = {\def\arraystretch{2}\left[\begin{array}{cc} z^2 & \displaystyle\frac{z}{z-2}\\[2mm]0 & \displaystyle\frac{1}{z} \end{array}\right]},
\ee
which has the following set of poles: $\{2, 0, \infty, \infty\}$ and therefore, the left coprime factorization with inner denominator exists. With the option to eliminate the simple infinite eigenvalues, the function \texttt{glcfid}  computes the following factors having the transfer function matrices
\[ N(z) = \ba{cc} \displaystyle\frac{z-2}{2z-1} & \displaystyle\frac{1}{z(2z-1)} \\[3mm] 0 & \displaystyle\frac{1}{z} \ea , \quad
M(z) = \ba{cc} \displaystyle\frac{z-2}{z^2(2z-1)}  & 0 \\[2mm] 0 & 1 \ea  .
\]
The McMillan degree of $M(z)$ is three, thus the least possible one. Interestingly, the McMillan degree of $N(z)$ is only two, because two unobservable eigenvalues in 0 have been removed. These eigenvalues are the zeros of the (improper) all-pass factor $\diag(z^2,1)$ with two infinite poles, which is contained in $G(z)$.

The above factors have been computed with the following sequence of commands:
\begin{verbatim}
% Varga (2017), Example 2
z = tf('z');  % define the complex variable z
% enter G(z) and determine a minimal state-space realization
G = [z^2 z/(z-2);
    0 1/z];
sys = ss(G);
gpole(sys)    % the system is unstable and improper

% compute the right coprime factorization G(z) = inv(M(z))*N(z),
% with inner denominator M(z)
[sysn,sysm] = glcfid(sys,struct('mininf',true));

% check the factorization ||M(z)*G(z)-N(z)||_inf = 0
norm(gminreal(sysm*sys-sysn),inf)

% check the innerness of M(z): ||conj(M(z))*M(z)-I||_inf = 0
norm(sysm'*sysm-eye(2),inf)

% check the poles of the factors
gpole(sysm), gpole(sysn)

% check coprimeness of the factors
gzero(gir([sysn sysm]))   % [N(z) M(z)] has no zeros
\end{verbatim}

\fine\end{example}

\subsubsection{\texttt{\bfseries gnrcf}}
\index{M-functions!\texttt{\bfseries gnrcf}}

\subsubsection*{Syntax}
\begin{verbatim}
[SYSN,SYSM] = gnrcf(SYS,OPTIONS)
\end{verbatim}

\subsubsection*{Description}
\texttt{\bfseries gnrcf} computes, for the transfer function matrix $G(\lambda)$ of a LTI descriptor state-space system, a normalized right coprime factorization in the form
\be\label{gnrcf:def} G(\lambda) = N(\lambda)M^{-1}(\lambda) \, , \ee
such that $N(\lambda)$ and $M(\lambda)$ are proper and stable transfer function matrices and $\left[\begin{smallmatrix} N(\lambda) \\ M(\lambda) \end{smallmatrix}\right]$ is inner.
\index{factorization!right coprime (RCF)!normalized}%

\subsubsection*{Input data}
\begin{description}
\item
\texttt{SYS} is a LTI system, whose transfer function matrix is $G(\lambda)$, and is in a  descriptor system state-space form
\be\label{gnrcf:sysss}
\begin{aligned}
E\lambda x(t)  &=   Ax(t)+ Bu(t) ,\\
y(t) &=  C x(t)+ D u(t) ,
\end{aligned}
\ee
with $x(t) \in \mathds{R}^n$ and $u(t) \in \mathds{R}^m$.
\item
 \texttt{OPTIONS} is a MATLAB structure to specify user options and has the following fields:\\
{\tabcolsep=1mm
\begin{longtable}{|l|lcp{11cm}|} \hline
\textbf{\texttt{OPTIONS} fields} & \multicolumn{3}{l|}{\textbf{Description}} \\ \hline
\texttt{tol}   & \multicolumn{3}{p{11cm}|}{tolerance for rank determinations (Default: internally computed)} \\ \hline
 \texttt{ss}   & \multicolumn{3}{p{11cm}|}{option to compute standard state-space
                       realizations of the factors:}\\
                 &  \texttt{true} &--& determine standard state-space realizations; \\
                 &  \texttt{false}&--& determine descriptor system realizations (default)  \\
                                        \hline
 \texttt{balance}   & \multicolumn{3}{p{11cm}|}{balancing option for the Riccati
                    equation solvers (see functions \texttt{care} and \texttt{dare} of the Control System Toolbox):}\\
                  &  \texttt{true} &--& apply balancing (default); \\
                  &  \texttt{false}&--& disable balancing.   \\
                                        \hline
\end{longtable}}

\end{description}

\subsubsection*{Output data}
\begin{description}
\item
\texttt{SYSN} contains the descriptor system state-space realization of the numerator factor $N(\lambda)$ in the form
\be\label{gnrcf:sysN}
\begin{aligned}
E_N\lambda x_N(t)  &=   A_N x_N(t)+ B_N v(t)  ,\\
y_N(t) &=  C_N x_N(t)+ D_N v(t)  .
\end{aligned}
\ee
The resulting $E_N = I_n$  if \texttt{OPTIONS.ss = true}.
\item
\texttt{SYSM}  contains the descriptor system state-space realization of the denominator factor $M(\lambda)$ in the form
\be\label{gnrcf:sysM}
\begin{aligned}
E_N\lambda x_M(t)  &=   A_N x_M(t)+ B_N w(t)  ,\\
y_M(t) &=  C_M x_M(t)+ D_M w(t)  .
\end{aligned}
\ee
The resulting $E_N = I_n$  if \texttt{OPTIONS.ss = true}.
\end{description}

\subsubsection*{Method}
The factors $N(\lambda)$ and $M(\lambda)$ are computed from a minimal inner basis $R(\lambda)$ of the range space of $\left[ \begin{smallmatrix} G(\lambda) \\ I_m \end{smallmatrix}\right]$ satisfying
\[ \ba{c} G(\lambda) \\ I_m \ea = R(\lambda)X(\lambda) , \]
with
\[ R(\lambda) = \ba{c} N(\lambda) \\ M(\lambda) \ea, \qquad X(\lambda) = M^{-1}(\lambda) . \]
For the computation of an inner range space of a rational matrix the method discussed in \cite{Varg17f} is employed.

\begin{example}\label{ex:Oara2001} %
For the polynomial transfer function matrix $G(s)$ considered in the \textbf{Example 3} of \cite{Oara00} with
\[ G(s) =
\left[\begin{array}{ccc} s^2 + s + 1 & 4\, s^2 + 3\, s + 2 & 2\, s^2 - 2\\ s & 4\, s - 1 & 2\, s - 2\\ s^2 & 4\, s^2 - s & 2\, s^2 - 2\, s \end{array}\right],
\]
the following MATLAB code can be used to compute a normalized right coprime factorization of $G(s)$:
\begin{verbatim}

% Oara and Varga (2000), Example 3
s = tf('s');  % define the complex variable s
% enter G(s) and determine a minimal state-space realization
G = [s^2+s+1 4*s^2+3*s+2 2*s^2-2;
    s 4*s-1 2*s-2;
    s^2 4*s^2-s 2*s^2-2*s];
sys = gir(ss(G));

% compute the normalized right coprime factorization
[N,M] = gnrcf(sys,1.e-7);

% check the coprime factorization
norm(gir(N*inv(M)-sys),inf)               % ||N*inv(M)-G||_inf = 0

% checking the innerness of [N;M]
norm([N;M]'*[N;M]-eye(size(sys,2)),inf)   % ||[N;M]'*[N;M]-I||_inf = 0
gpole(N), gpole(M)                        % N and M are stable
\end{verbatim}

\fine\end{example}

\subsubsection{\texttt{\bfseries gnlcf}}
\index{M-functions!\texttt{\bfseries gnlcf}}

\subsubsection*{Syntax}
\begin{verbatim}
[SYSN,SYSM] = gnlcf(SYS,OPTIONS)
\end{verbatim}

\subsubsection*{Description}
\texttt{\bfseries gnlcf} computes, for the transfer function matrix $G(\lambda)$ of a LTI descriptor state-space system, a normalized left coprime factorization in the form
\be\label{gnlcf:def} G(\lambda) = M^{-1}(\lambda)N(\lambda) \, , \ee
such that $N(\lambda)$ and $M(\lambda)$ are proper and stable transfer function matrices and $[\,N(\lambda) \; M(\lambda)\,]$ is coinner.
\index{factorization!left coprime (LCF)!normalized}%

\subsubsection*{Input data}
\begin{description}
\item
\texttt{SYS} is a LTI system, whose transfer function matrix is $G(\lambda)$, and is in a  descriptor system state-space form
\be\label{gnlcf:sysss}
\begin{aligned}
E\lambda x(t)  &=   Ax(t)+ Bu(t) ,\\
y(t) &=  C x(t)+ D u(t) ,
\end{aligned}
\ee
with $x(t) \in \mathds{R}^n$ and $y(t) \in \mathds{R}^p$.
\item
 \texttt{OPTIONS} is a MATLAB structure to specify user options and has the following fields:\pagebreak[3]
{\tabcolsep=1mm
\begin{longtable}{|l|lcp{11cm}|} \hline
\textbf{\texttt{OPTIONS} fields} & \multicolumn{3}{l|}{\textbf{Description}} \\ \hline
\texttt{tol}   & \multicolumn{3}{p{11cm}|}{tolerance for rank determinations (Default: internally computed)} \\ \hline
 \texttt{ss}   & \multicolumn{3}{p{11cm}|}{option to compute standard state-space
                       realizations of the factors:}\\
                 &  \texttt{true} &--& determine standard state-space realizations; \\
                 &  \texttt{false}&--& determine descriptor system realizations (default)  \\
                                        \hline
 \texttt{balance}   & \multicolumn{3}{p{11cm}|}{balancing option for the Riccati
                    equation solvers (see functions \texttt{care} and \texttt{dare} of the Control System Toolbox):}\\
                  &  \texttt{true} &--& apply balancing (default); \\
                  &  \texttt{false}&--& disable balancing.   \\
                                        \hline
\end{longtable}}

\end{description}

\subsubsection*{Output data}
\begin{description}
\item
\texttt{SYSN} contains the descriptor system state-space realization of the numerator factor $N(\lambda)$ in the form
\be\label{gnlcf:sysN}
\begin{aligned}
E_N\lambda x_N(t)  &=   A_N x_N(t)+ B_N u(t)  ,\\
y_N(t) &=  C_N x_N(t)+ D_N u(t)  .
\end{aligned}
\ee
The resulting $E_N = I_n$  if \texttt{OPTIONS.ss = true}.
\item
\texttt{SYSM}  contains the descriptor system state-space realization of the denominator factor $M(\lambda)$ in the form
\be\label{gnlcf:sysM}
\begin{aligned}
E_N\lambda x_M(t)  &=   A_N x_M(t)+ B_M w(t)  ,\\
y_M(t) &=  C_N x_M(t)+ D_M w(t)  ,
\end{aligned}
\ee
The resulting $E_N = I_n$  if \texttt{OPTIONS.ss = true}.

\end{description}

\subsubsection*{Method}

The factors $N(\lambda)$ and $M(\lambda)$ are computed from a minimal coinner basis $R(\lambda)$ of the co-image space of $[\, G(\lambda) \; I_p \,]$ satisfying
\[ [\, G(\lambda) \; I_p \,] = X(\lambda)R(\lambda) , \]
with
\[ R(\lambda) = [\, N(\lambda) \; M(\lambda) \,], \qquad X(\lambda) = M^{-1}(\lambda) . \]
For the computation of a coinner basis of the co-image space of a rational matrix, the inner range space basis of the transposed rational matrix is computed, for which the computational method discussed in \cite{Varg17f} is employed.

\begin{example}\label{ex:Oara2001disc} %
For the polynomial transfer function matrix $G(z)$ considered in the \textbf{Example 3} of \cite{Oara00}, obtained by replacing the complex Laplace transform variable $s$ with the complex $Z$-transform variable $z$,
\[ G(z) =
\left[\begin{array}{ccc} z^2 + z + 1 & 4\, z^2 + 3\, z + 2 & 2\, z^2 - 2\\ z & 4\, z - 1 & 2\, z - 2\\ z^2 & 4\, z^2 - z & 2\, z^2 - 2\, z \end{array}\right],
\]
the following MATLAB code can be used to compute a normalized right coprime factorization of $G(z)$:
\begin{verbatim}
% Oara and Varga (2000), Example 3
z = tf('z');  % define the complex variable z
% enter G(s) and determine a minimal state-space realization
G = [z^2+z+1 4*z^2+3*z+2 2*z^2-2;
    z 4*z-1 2*z-2;
    z^2 4*z^2-z 2*z^2-2*z];
sys = gir(ss(G));

% compute the normalized left coprime factorization
[N,M] = gnlcf(sys,struct('tol',1.e-7));

% check the coprime factorization
norm(gir(inv(M)*N-sys),inf)             % ||inv(M)*N-G||_inf = 0

% checking the coinnerness of [N M]
norm([N M]*[N M]'-eye(size(sys,1)),inf) % ||[N M]*[N M]'-I||_inf = 0
gpole(N), gpole(M)                      % N and M are stable
\end{verbatim}
\fine\end{example}

\subsubsection{\texttt{\bfseries giofac}}
\index{M-functions!\texttt{\bfseries giofac}}

\subsubsection*{Syntax}
\begin{verbatim}
[SYSI,SYSO,INFO] = giofac(SYS,OPTIONS)
\end{verbatim}

\subsubsection*{Description}
\texttt{\bfseries giofac} computes, for the transfer function matrix $G(\lambda)$ of a LTI descriptor state-space system, the extended inner--quasi-outer or the extended QR-like factorization in the form
\be\label{giofac:def} G(\lambda) = G_i(\lambda) \ba{c} G_o(\lambda) \\ 0 \ea \, , \ee
where $G_i(\lambda)$ is square and inner, and $G_o(\lambda)$ is quasi-outer or full row rank, respectively.
\index{factorization!inner--quasi-outer!extended}%
\index{factorization!QR-like!extended}%

\subsubsection*{Input data}
\begin{description}
\item
\texttt{SYS} is a LTI system, whose transfer function matrix is $G(\lambda)$, and is in a  descriptor system state-space form
\be\label{giofac:sysss}
\begin{aligned}
E\lambda x(t)  &=   Ax(t)+ Bu(t) ,\\
y(t) &=  C x(t)+ D u(t) .
\end{aligned}
\ee
\item
 \texttt{OPTIONS} is a MATLAB structure to specify user options and has the following fields:\\
{\tabcolsep=1mm
\begin{longtable}{|l|lcp{11cm}|} \hline
\textbf{\texttt{OPTIONS} fields} & \multicolumn{3}{l|}{\textbf{Description}} \\ \hline
 \texttt{tol}   & \multicolumn{3}{p{11cm}|}{relative tolerance for rank computations and observability tests (Default: internally computed)} \\ \hline
 \texttt{offset}   & \multicolumn{3}{p{12.5cm}|}{stability
 boundary offset $\beta$, to be used  to assess the finite zeros which belong to $\partial\mathds{C}_s$ (the boundary of the stability domain) as follows: in the
 continuous-time case these are the finite
  zeros having real parts in the interval $[-\beta, \beta]$, while in the
 discrete-time case these are the finite zeros having moduli in the
 interval $[1-\beta, 1+\beta]$ (Default: $\beta = 1.4901\cdot 10^{-08}$).} \\ \hline
 \texttt{minphase}   & \multicolumn{3}{p{12cm}|}{option to compute a minimum-phase quasi-outer factor:}\\
                  &  \texttt{true} &--& compute a minimum phase quasi-outer factor,
                            with all zeros stable, excepting possibly
                            zeros on the boundary of the appropriate
                            stability domain (default); \\
                  &  \texttt{false}&--& compute a full row rank factor, which
                            includes all zeros of $G(\lambda)$.   \\
                                        \hline
 \texttt{balance}   & \multicolumn{3}{p{12cm}|}{balancing option for the Riccati
                    equation solvers (see functions \texttt{care} and \texttt{dare} of the Control System Toolbox):}\\
                  &  \texttt{true} &--& apply balancing (default); \\
                  &  \texttt{false}&--& disable balancing.   \\
                                        \hline
\end{longtable}}
\end{description}

\subsubsection*{Output data}
\begin{description}
\item
\texttt{SYSI} contains  the descriptor system state-space realization of the square inner transfer function matrix $G_i(\lambda)$ in the form
\be\label{giofac:sysi}
\begin{aligned}
 E_i\lambda x_i(t)  &=   A_i x_i(t)+ B_i v(t)  ,\\
y_i(t) &=  C_i x_i(t)+ D_i v(t)  ,
\end{aligned}
\ee
where $E_i$ is invertible and $\Lambda(A_i-\lambda E_i) \subset \mathds{C}_s$. The realization of the inner factor is a standard system with $E_i =I$ if the original system (\ref{giofac:sysss}) is a standard system with $E = I$. If \texttt{OPTIONS.minphase = false}, then $G_i(\lambda)$ has the least possible McMillan degree.
\item
\texttt{SYSO} contains the descriptor system state-space realization of the transfer function matrix $G_o(\lambda)$ in the form
\be\label{giofac:syso}
\begin{aligned}
 E\lambda x_o(t)  &=   A x_o(t)+ B u(t)  ,\\
y_o(t) &=  C_o x_o(t)+ D_o u(t)  ,
\end{aligned}
\ee
where the dimension $r$ of $y_o(t)$ is the normal rank of $G(\lambda)$. If \texttt{OPTIONS.minphase = true}, then $G_o(\lambda)$ is quasi-outer and all zeros of $G_o(\lambda)$ lie in $\overline{\mathds{C}}_s$. If \texttt{OPTIONS.minphase = false}, then $G_o(\lambda)$ is full row rank and contains all zeros of $G(\lambda)$.
\item
 \texttt{INFO} is a MATLAB structure containing additional information, as follows:\\
{\begin{longtable}{|l|p{12cm}|} \hline
\textbf{\texttt{INFO} fields} & \textbf{Description} \\ \hline
 \texttt{nrank}   & normal rank of the transfer function matrix $G(\lambda)$; \\ \hline
 \texttt{nfuz}   & number of finite unstable zeros of \texttt{SYSO}; these are the finite zeros of
              \texttt{SYS} lying on the boundary of the stability region $\partial\mathds{C}_s$ within the offset specified
by \texttt{OPTION.offset};\\ \hline
 \texttt{niuz}   & number of infinite unstable zeros of \texttt{SYSO}; these are the infinite zeros of
              \texttt{SYS}, in the continuous-time case, and 0 in the discrete-time case; \\ \hline
 \texttt{ricrez}   & diagnosis flag, as provided provided by the generalized
               Riccati equation solvers \texttt{care} and \texttt{dare}; if non-negative,
               this value represents the Frobenius norm of relative
               residual of the Riccati equation, while a negative value
               indicates failure of solving the Riccati equation.  \\ \hline
\end{longtable}}
\end{description}

\subsubsection*{Method}
For the definitions related to inner-outer and QR-like factorizations of transfer function matrices see Section \ref{app_IOF}. Assume that the transfer function matrix $G(\lambda)$ has normal rank $r$ and the inner factor $G_i(\lambda)$ in (\ref{giofac:def}) is partitioned as $G_i(\lambda) = [\, G_{i,1}(\lambda) \; G_{i,2}(\lambda) \,]$, where $G_{i,1}(\lambda)$ has $r$ columns and is inner. Then $G_{i,2}(\lambda)$ represents the complementary inner factor $G_{i,1}^\bot(\lambda)$, which  is the inner orthogonal complement of $G_{i,1}(\lambda)$. Thus, the full rank inner--quasi-outer or QR-like factorization of $G(\lambda)$ has the form
\be\label{giofac:inoutMIMO} G(\lambda) = G_{i,1}(\lambda)G_o(\lambda) \, .\ee
If \texttt{OPTIONS.minphase = true}, the resulting factor $G_o(\lambda)$ has full row rank $r$ and  is minimum phase, excepting possibly zeros in $\partial \mathds{C}_s$, the boundary of the appropriate stability domain. If \texttt{OPTIONS.minphase = false}, $G_o(\lambda)$ has full row rank $r$ and contains all zeros of $G(\lambda)$. In this case, the resulting inner factor has the least possible McMillan degree. If $G(\lambda)$ contains a so-called free inner factor, then the resulting realization (\ref{giofac:syso}) of $G_o(\lambda)$  is unobservable. The unobservable eigenvalues of the pencil $\left[\begin{smallmatrix} A-\lambda E\\ C_o \end{smallmatrix}\right]$ are precisely the (stable) poles of the free inner factor and can be readily eliminated (e.g., by using the function \texttt{gir}).

The implemented computational methods to determine the inner--quasi-outer or QR-like factorizations of general rational matrices  rely on a preliminary orthogonal reduction of the system matrix pencil
\[ S(\lambda) = \ba{cc} A-\lambda E & B\\ C & D \ea \]
to a special Kronecker-like form (see (\ref{spec-klf})), which allows to reduce the original computational problem to a standard inner-outer factorization problem for a system with full column rank transfer function matrix and without zeros in $\partial \mathds{C}_s$. The underlying reduction is described in \cite{Oara00} and involves the use of the mex-function \url{sl_klf} to compute the appropriate Kronecker-like form.
\index{MEX-functions!\url{sl_klf}}%
For the computation of the inner and outer factors for the reduced problem, extensions of the standard factorization methods of \cite{Zhou96} are used. These methods involve the solution of appropriate (continuous- or discrete-time) generalized algebraic Riccati equations.
The overall factorization procedures are described in \cite{Oara00}
for continuous-time systems and in \cite{Oara05} for discrete-time systems.
The formulas for the determination of the complementary inner factors have been derived extending the results of \cite{Zhou96}. The function \texttt{\bfseries giofac} employs the mex-function
\url{sl_gminr} to obtain the inner factor with a standard state-space realization with $E_i = I$.
\index{MEX-functions!\url{sl_gminr}}%

\subsubsection*{Examples}
\begin{example}\label{ex:Oara-Varga-IOF}
This is Example 1 from \cite{Oara00} of the transfer function matrix of a continuous-time proper system:
\be\label{giofac:ex1} G(s) = {\def\arraystretch{2}
\left[\begin{array}{ccc} \displaystyle\frac{s - 1}{s + 2} & \displaystyle\frac{s}{s + 2} & \displaystyle\frac{1}{s + 2}\\ 0 & \displaystyle\frac{s - 2}{{\left(s + 1\right)}^2} & \displaystyle\frac{s - 2}{{\left(s + 1\right)}^2}\\ \displaystyle\frac{s - 1}{s + 2} & \displaystyle\frac{s^2 + 2\, s - 2}{\left(s + 1\right)\, \left(s + 2\right)} & \displaystyle\frac{2\, s - 1}{\left(s + 1\right)\, \left(s + 2\right)} \end{array}\right]} \, .
\ee
$G(s)$ has zeros at $\{1, 2, \infty\}$, poles at $\{-1, -1, -2, -2 \}$, and normal rank $r = 2$.
The extended inner--quasi-outer factorization of $G(s)$ can be computed with the following command sequence:
\begin{verbatim}
% Oara and Varga (2000), Example 1
s = tf('s');  % define the complex variable s
% enter G(s) and determine a minimal state-space realization
G = [(s-1)/(s+2) s/(s+2) 1/(s+2);
    0 (s-2)/(s+1)^2 (s-2)/(s+1)^2;
    (s-1)/(s+2) (s^2+2*s-2)/(s+1)/(s+2) (2*s-1)/(s+1)/(s+2)];
sys = minreal(ss(G));
gpole(sys)    % the system is stable
gzero(sys)    % the system has 2 unstable zeros and an infinite zero
nrank(sys)    % the normal rank of G(s) is 2

% compute the extended inner-quasi-outer factorization G(s) = Gi(s)*[Go(s);0]
[sysi,syso,info] = giofac(sys,struct('tol',1.e-7));  % use tolerance 1.e-7

% check the factorization
nr = size(syso,1);                %  nr = 2 is also the normal rank of G(s)
norm(sysi(:,1:nr)*syso-sys,inf)   %  ||Gi(:,1:nr)(s)*Go(s)-G(s)||_inf = 0

% checking the innerness of Gi(s)
norm(sysi'*sysi-eye(3),inf)       %  ||conj(Gi(s))*Gi(s)-I||_inf = 0

syso = gir(syso);                 % a free inner factor is present in G(s)
gzero(syso)                       % Go(s) has no zeros in open right-half plane
info.niuz                         % Go(s) has an infinite zero
info.nfuz                         % Go(s) has no zeros on the imaginary axis
\end{verbatim}

\fine\end{example}

\subsubsection{\texttt{\bfseries goifac}}
\index{M-functions!\texttt{\bfseries goifac}}

\subsubsection*{Syntax}
\begin{verbatim}
[SYSI,SYSO,INFO] = goifac(SYS,OPTIONS)
\end{verbatim}

\subsubsection*{Description}
\texttt{\bfseries goifac} computes, for the transfer function matrix $G(\lambda)$ of a LTI descriptor state-space system,  the extended quasi-co-outer--inner or the extended RQ-like factorization in the form
\be\label{goifac:def} G(\lambda) = [\, G_o(\lambda) \; 0 \,] G_i(\lambda) \, , \ee
where $G_i(\lambda)$ is square and inner, and $G_o(\lambda)$ is quasi-co-outer or full column rank, respectively.
\index{factorization!quasi-co-outer--coinner!extended}%
\index{factorization!RQ-like!extended}%

\subsubsection*{Input data}
\begin{description}
\item
\texttt{SYS} is a LTI system, whose transfer function matrix is $G(\lambda)$, and is in a  descriptor system state-space form
\be\label{goifac:sysss}
\begin{aligned}
E\lambda x(t)  &=   Ax(t)+ Bu(t) ,\\
y(t) &=  C x(t)+ D u(t) .
\end{aligned}
\ee
\item
 \texttt{OPTIONS} is a MATLAB structure to specify user options and has the following fields:\\
\pagebreak[3]
{\tabcolsep=1mm
\begin{longtable}{|l|lcp{11cm}|} \hline
\textbf{\texttt{OPTIONS} fields} & \multicolumn{3}{l|}{\textbf{Description}} \\ \hline
 \texttt{tol}   & \multicolumn{3}{p{11cm}|}{relative tolerance for rank computations and observability tests (Default: internally computed)} \\ \hline
 \texttt{offset}   & \multicolumn{3}{p{12.5cm}|}{stability
 boundary offset $\beta$, to be used  to assess the finite zeros which belong to $\partial\mathds{C}_s$ (the boundary of the stability domain) as follows: in the
 continuous-time case these are the finite
  zeros having real parts in the interval $[-\beta, \beta]$, while in the
 discrete-time case these are the finite zeros having moduli in the
 interval $[1-\beta, 1+\beta]$ (Default: $\beta = 1.4901\cdot 10^{-08}$).}  \\ \hline
 \texttt{minphase}   & \multicolumn{3}{p{12cm}|}{option to compute a minimum-phase quasi-co-outer factor:}\\
                  &  \texttt{true} &--& compute a minimum phase quasi-co-outer factor,
                            with all zeros stable, excepting possibly
                            zeros on the boundary of the appropriate
                            stability domain (default); \\
                  &  \texttt{false}&--& compute a full row rank factor, which
                            includes all zeros of $G(\lambda)$.   \\
                                        \hline
 \texttt{balance}   & \multicolumn{3}{p{12cm}|}{balancing option for the Riccati
                    equation solvers (see functions \texttt{care} and \texttt{dare} of the Control System Toolbox):}\\
                  &  \texttt{true} &--& apply balancing (default); \\
                  &  \texttt{false}&--& disable balancing.   \\
                                        \hline
\end{longtable}}
\end{description}

\subsubsection*{Output data}
\begin{description}
\item
\texttt{SYSI} contains  the descriptor system state-space realization of the square inner transfer function matrix $G_i(\lambda)$ in the form
\be\label{goifac:sysi}
\begin{aligned}
 E_i\lambda x_i(t)  &=   A_i x_i(t)+ B_i u(t)  ,\\
y_i(t) &=  C_i x_i(t)+ D_i u(t)  ,
\end{aligned}
\ee
where $E_i$ is invertible and $\Lambda(A_i-\lambda E_i) \subset \mathds{C}_s$. The realization of the inner factor is a standard system with $E_i =I$ if the original system (\ref{goifac:sysss}) is a standard system with $E = I$. If \texttt{OPTIONS.minphase = false}, then $G_i(\lambda)$ has the least possible McMillan degree.
\item
\texttt{SYSO} contains  the descriptor system state-space realization of the transfer function matrix $G_o(\lambda)$ in the form
\be\label{goifac:syso}
\begin{aligned}
 E\lambda x_o(t)  &=   A x_o(t)+ B_o v(t)  ,\\
y_o(t) &=  C x_o(t)+ D_o v(t)  ,
\end{aligned}
\ee
where the dimension $r$ of $v(t)$ is the normal rank of $G(\lambda)$.
If \texttt{OPTIONS.minphase = true}, then $G_o(\lambda)$ is quasi-co-outer and all zeros of $G_o(\lambda)$ lie in $\overline{\mathds{C}}_s$. If \texttt{OPTIONS.minphase = false}, then $G_o(\lambda)$ is full column rank and contains all zeros of $G(\lambda)$.
\item
 \texttt{INFO} is a MATLAB structure containing additional information, as follows:\\
{\begin{longtable}{|l|p{12cm}|} \hline
\textbf{\texttt{INFO} fields} & \textbf{Description} \\ \hline
 \texttt{nrank}   & normal rank of the transfer function matrix $G(\lambda)$; \\ \hline
 \texttt{nfuz}   & number of finite unstable zeros of \texttt{SYSO}; these are the finite zeros of
              \texttt{SYS} lying on the boundary of the stability region $\partial\mathds{C}_s$ within the offset specified
by \texttt{OPTION.offset};\\ \hline
 \texttt{niuz}   & number of infinite unstable zeros of \texttt{SYSO}; these are the infinite zeros of
              \texttt{SYS}, in the continuous-time case, and 0 in the discrete-time case; \\ \hline
 \texttt{ricrez}   & diagnosis flag, as provided provided by the generalized
               Riccati equation solvers \texttt{care} and \texttt{dare}; if non-negative,
               this value represents the Frobenius norm of relative
               residual of the Riccati equation, while a negative value
               indicates failure of solving the Riccati equation.  \\ \hline
\end{longtable}}
\end{description}

\subsubsection*{Method}
For the definitions related to co-outer--coinner or RQ-like  factorizations of transfer function matrices see Section \ref{app_IOF}. Assume that the transfer function matrix $G(\lambda)$ has normal rank $r$ and the inner factor $G_i(\lambda)$ in (\ref{goifac:def}) is partitioned as $G_i(\lambda) = \left[\begin{smallmatrix} G_{i,1}(\lambda) \\ G_{i,2}(\lambda) \end{smallmatrix}\right]$, where $G_{i,1}(\lambda)$ has $r$ rows and is coinner. Then $G_{i,2}(\lambda)$ represents the complementary coinner factor $G_{i,1}^\bot(\lambda)$, which  is the coinner orthogonal complement of $G_{i,1}(\lambda)$. Thus, the full rank quasi-co-outer--coinner or full rank RQ-like factorization of $G(\lambda)$ has the form
\be\label{goifac:inoutMIMO} G(\lambda) = G_o(\lambda)G_{i,1}(\lambda) \, .\ee
If \texttt{OPTIONS.minphase = true}, the resulting factor $G_o(\lambda)$ has full row rank $r$ and  is minimum phase, excepting possibly zeros in $\partial \mathds{C}_s$, the boundary of the appropriate stability domain. If \texttt{OPTIONS.minphase = false}, $G_o(\lambda)$ has full row rank $r$ and contains all zeros of $G(\lambda)$. In this case, the resulting inner factor has the least possible McMillan degree. If $G(\lambda)$ contains a so-called free inner factor, then the resulting realization (\ref{goifac:syso}) of $G_o(\lambda)$  is uncontrollable. The uncontrollable eigenvalues of the pencil $[\, A-\lambda E\; B_o \,]$ are precisely the (stable) poles of the free inner factor and can be readily eliminated (e.g., by using the function \texttt{gir}).

To compute the extended quasi-co-outer--inner or the extended RQ-like factorization (\ref{goifac:def}), the function \texttt{goifac} calls \texttt{giofac} to compute the extended inner--quasi-outer or extended QR-like factorization of $G^T(\lambda)$ in the form
\[ G^T(\lambda) = \widetilde G_i(\lambda) \ba{c} \widetilde G_o(\lambda) \\ 0 \ea \]
and obtain the inner and quasi-co-outer/full column rank factors as $G_i(\lambda) = \widetilde G_i^T(\lambda)$ and $G_o(\lambda) = \widetilde G_o^T(\lambda)$, respectively.
The factorization procedures underlying the function \texttt{giofac} are described in \cite{Oara00}
for continuous-time systems and in \cite{Oara05} for discrete-time systems.
The function \texttt{\bfseries giofac} relies on the mex-functions
\url{sl_klf} and \url{sl_gminr}.
 \index{MEX-functions!\url{sl_klf}}%
\index{MEX-functions!\url{sl_gminr}}%

\subsubsection*{Examples}
\begin{example}\label{ex:Oara05}
This is Example 1 from \cite{Oara05} of the transfer function matrix of a discrete-time proper system:
\[ G(z) = {\def\arraystretch{2} \arraycolsep=1mm
\left[\begin{array}{ccc} \displaystyle\frac{ z^4 - \displaystyle\frac{z^3}{2} - 16\, z^2 - \displaystyle\frac{29\, z}{2} + 18}{z^4 + \displaystyle\frac{5\, z^3}{2} + 2\, z^2 + \displaystyle\frac{z}{2}} & \displaystyle\frac{z^4 + 5\, z^3 - z^2 - 11\, z + 6}{z^4 + \displaystyle\frac{5\, z^3}{2} + 2\, z^2 + \displaystyle\frac{z}{2}} & \displaystyle\frac{\displaystyle\frac{11\, z^3}{2} + 15\, z^2 + \displaystyle\frac{7\, z}{2} - 12}{z^4 + \displaystyle\frac{5\, z^3}{2} + 2\, z^2 + \displaystyle\frac{z}{2}}\\ \displaystyle\frac{- 3\, z^2 + 12}{z^4 + \displaystyle\frac{5\, z^3}{2} + 2\, z^2 + \displaystyle\frac{z}{2}} & \displaystyle\frac{  z^3 - z^2 - 4\, z + 4}{z^4 + \displaystyle\frac{5\, z^3}{2} + 2\, z^2 + \displaystyle\frac{z}{2}} & \displaystyle\frac{  z^3 + 2\, z^2 - 4\, z - 8}{z^4 + \displaystyle\frac{5\, z^3}{2} + 2\, z^2 + \displaystyle\frac{z}{2}}\\ \displaystyle\frac{  z^4 - \displaystyle\frac{z^3}{2} - 19\, z^2 -\displaystyle\frac{23\, z}{2} + 24}{z^4 + \displaystyle\frac{5\, z^3}{2} + 2\, z^2 + \displaystyle\frac{z}{2}} & \displaystyle\frac{z^4 + 6\, z^3 - 3\, z^2 - 12\, z + 8}{z^4 + \displaystyle\frac{5\, z^3}{2} + 2\, z^2 + \displaystyle\frac{z}{2}} & \displaystyle\frac{  \displaystyle\frac{13\, z^3}{2} + 16\, z^2 - \displaystyle\frac{z}{2} - 16}{z^4 + \displaystyle\frac{5\, z^3}{2} + 2\, z^2 + \displaystyle\frac{z}{2}} \end{array}\right]} .
\]
$G(z)$ has the zeros $\{1, 2,\infty\}$, the poles $\{0, -0.5,-1,-1\}$, and normal rank $r = 2$.
The extended quasi-co-outer--inner factorization of $G(z)$ can be computed with the following sequence of commands:
\begin{verbatim}
% Oara (2005), Example 1
z = tf('z');  % define the complex variable z
% enter G(z) and determine a minimal state-space realization
G = 1/(z^4+5/2*z^3+2*z^2+z/2)*...
    [z^4-z^3/2-16*z^2-29/2*z+18 z^4+5*z^3-z^2-11*z+6 11/2*z^3+15*z^2+7/2*z-12
    -3*z^2+12 z^3-z^2-4*z+4 z^3+2*z^2-4*z-8;
    z^4-z^3/2-19*z^2-23/2*z+24 z^4+6*z^3-3*z^2-12*z+8 13/2*z^3+16*z^2-z/2-16];
sys = gir(ss(G),1.e-7);
gpole(sys)    % the system is marginally stable
gzero(sys)    % the system has an unstable zero and an infinite zero
nrank(sys)    % the normal rank of G(z) is 2

% compute the extended quasi-co-outer-inner factorization
% G(z) = [Go(z) 0]*Gi(z)
[sysi,syso,info] = goifac(sys,struct('tol',1.e-7));  % use tolerance 1.e-7

% check the factorization
nr = size(syso,2);                %  nr = 2 is also the normal rank of G(z)
%  check that ||Go(z)*Gi(1:nr,:)(z)-G(z)||_inf = 0
norm(gir(syso*sysi(1:nr,:)-sys,1.e-7),inf)

% check the innerness of Gi(z): ||conj(Gi(z))*Gi(z)-I||_inf = 0
norm(sysi'*sysi-eye(3),inf)

gzero(syso)                       % Go(z) has no zeros outside the unit circle
info.nfuz                         % Go(z) has one zero on the unit circle
\end{verbatim}

\fine\end{example}

\subsubsection{\texttt{\bfseries grsfg}}
\index{M-functions!\texttt{\bfseries grsfg}}

\subsubsection*{Syntax}
\begin{verbatim}
SYSF = grsfg(SYS,GAMMA,OPTIONS)
\end{verbatim}

\subsubsection*{Description}
\texttt{\bfseries grsfg} solves, for the transfer function matrix $G(\lambda)$ of a LTI descriptor state-space system, and a given $\gamma$ satisfying $\gamma > \|G(\lambda)\|_\infty$,  the right stable and minimum-phase spectral factorization problem
\be\label{grsfg:def} \gamma^2 I - G^\sim(\lambda)G(\lambda) = F^\sim(\lambda)F(\lambda) \, , \ee
 such that the resulting spectral factor $F(\lambda)$ is proper, stable and minimum-phase. \index{factorization!spectral!special, stable minimum-phase right}

\subsubsection*{Input data}
\begin{description}
\item
\texttt{SYS} is a LTI system, whose transfer function matrix is $G(\lambda)$, and is in a  descriptor system state-space form
\be\label{grsfg:sysss}
\begin{aligned}
E\lambda x(t)  &=   Ax(t)+ Bu(t) ,\\
y(t) &=  C x(t)+ D u(t) ,
\end{aligned}
\ee
with $x(t) \in \mathds{R}^n$ and $y(t) \in \mathds{R}^p$. $G(\lambda)$ must not have poles in $\partial\mathds{C}_s$.
\item
\texttt{GAMMA} is a given scalar $\gamma$, which must satisfy $\gamma > \|G(\lambda)\|_\infty$.
\item
 \texttt{OPTIONS} is a MATLAB structure to specify user options and has the following fields:\\
{\tabcolsep=1mm
\begin{longtable}{|l|lcp{11cm}|} \hline
\textbf{\texttt{OPTIONS} fields} & \multicolumn{3}{l|}{\textbf{Description}} \\ \hline
\texttt{tol}   & \multicolumn{3}{p{11cm}|}{tolerance for the singular values
                        based rank determination of $E$ (Default: $n^2\|E\|_1$\texttt{eps})} \\ \hline
\texttt{tolmin}   & \multicolumn{3}{p{10cm}|}{tolerance for the singular values
                        based observability tests (Default: $np\|C\|_\infty$\texttt{eps})} \\ \hline
 \texttt{stabilize}   & \multicolumn{3}{p{11cm}|}{stabilization option:}\\
                 &  \texttt{true} &--& perform a preliminary stabilization using a left coprime factorization with inner denominator of $G(\lambda)$ (see \textbf{Method}) (default); \\
                 &  \texttt{false}&--& no preliminary stabilization is performed. \\
                                        \hline
\end{longtable}}

\end{description}

\subsubsection*{Output data}
\begin{description}
\item
\texttt{SYSF} contains the descriptor system state-space realization of the spectral factor $F(\lambda)$ in the form
\be\label{grsfg:sysF}
\begin{aligned}
E_F\lambda x_F(t)  &=   A_F x_F(t)+ B_F v(t)  ,\\
y_F(t) &=  C_F x_F(t)+ D_F v(t)  .
\end{aligned}
\ee
\end{description}

\subsubsection*{Method}
For the computation of the right spectral factorization (\ref{grsfg:def}) the dual of the two-step approach sketched in Section \ref{app_IOF} is employed. In the first step,
a preliminary left coprime factorization with inner denominator of $G(\lambda)$ is computed such that $G(\lambda) = M^{-1}(\lambda)N(\lambda)$, with both $N(\lambda)$ and $M(\lambda)$ stable, and $M(\lambda)$ inner. For this purpose, the dual algorithm to compute right coprime factorizations with inner denominators, given in \textbf{Procedure GRCFID} of \cite{Varg17d}, is employed. This step is not performed if \texttt{OPTIONS.stabilize = false}, in which case $N(\lambda) := G(\lambda)$ and $M(\lambda) = I_p$. In the second step, the spectral factorization problem is solved
\[ \gamma^2 I - G^\sim(\lambda)G(\lambda) = \gamma^2 I - N^\sim(\lambda)N(\lambda) = F^\sim(\lambda)F(\lambda)  \]
for the minimum-phase phase factor $F(\lambda)$. For this computation, the formulas provided by the dual versions of Lemma~\ref{L:special-specfacc} and Lemma~\ref{L:special-specfacd} are employed. These lemmas extend to proper descriptor system the formulas developed in  \cite{Zhou96}.

\subsubsection*{Example}
\begin{example}\label{ex:grsfg}
Consider the discrete-time improper TFM
\be\label{grsfg:ex1}  G(z) = {\left[\begin{array}{ccc} z^2 + z + 1 & 4\, z^2 + 3\, z + 2 & 2\, z^2 - 2\\ z & 4\, z - 1 & 2\, z - 2\\ z^2 & 4\, z^2 - z & 2\, z^2 - 2\, z \end{array}\right]
},
\ee
which has two infinite poles (i.e., McMillan-degree of $G(z)$ is equal to 2) and has a minimal descriptor state-space realization of order 4. Therefore, the spectral factorization problem (\ref{grsfg:def}) has a solution for all $\gamma > \|G(z)\|_\infty = 10.4881$. With $\gamma = 1.1\|G(z)\|_\infty$, the function \texttt{grsfg}  computes the proper minimal-phase spectral factor $F(z)$, having two poles in 0 and two stable zeros in $\{-0.2908,0.4188\}$.

The spectral factor $F(z)$ can be computed with the following sequence of commands:
\begin{verbatim}
z = tf('z');  % define the complex variable z
% enter G(z) and determine a minimal state-space realization
G = [z^2+z+1 4*z^2+3*z+2 2*z^2-2;
    z 4*z-1 2*z-2;
    z^2 4*z^2-z 2*z^2-2*z];
sys = gir(ss(G));
gpole(sys)                % the system is unstable and improper
gamma = 1.1*norm(sys,inf) % set gamma = 1.1*||G||_inf

% compute the minimum-phase stable spectral factor F(z) satisfying
% gamma^2*I-conj(G(z))*G(z) = conj(F(z))*F(z)
sysf = grsfg(sys,gamma);

% check the factorization ||F'(z)*F(z)+G'(z)*G(z)-gamma^2*I||_inf = 0
norm(gir(sysf'*sysf+sys'*sys,1.e-7)-gamma^2*eye(size(sys,2)),inf)

% check the stability of poles and zeros of F(z)
gpole(sysf), gzero(sysf)  % F(z) is stable and minimum-phase
\end{verbatim}

\fine\end{example}

\subsubsection{\texttt{\bfseries glsfg}}
\index{M-functions!\texttt{\bfseries glsfg}}

\subsubsection*{Syntax}
\begin{verbatim}
SYSF = glsfg(SYS,GAMMA,OPTIONS)
\end{verbatim}

\subsubsection*{Description}
\texttt{\bfseries glsfg} solves, for the transfer function matrix $G(\lambda)$ of a LTI descriptor state-space system, and a given $\gamma$ satisfying $\gamma > \|G(\lambda)\|_\infty$,  the left stable and minimum-phase spectral factorization problem
\be\label{glsfg:def} \gamma^2 I - G(\lambda)G^\sim(\lambda) = F(\lambda)F^\sim(\lambda) \, , \ee
such that the resulting spectral factor $F(\lambda)$ is proper, stable and minimum-phase. \index{factorization!spectral!special, stable minimum-phase left}

\subsubsection*{Input data}
\begin{description}
\item
\texttt{SYS} is a LTI system, whose transfer function matrix is $G(\lambda)$, and is in a  descriptor system state-space form
\be\label{glsfg:sysss}
\begin{aligned}
E\lambda x(t)  &=   Ax(t)+ Bu(t) ,\\
y(t) &=  C x(t)+ D u(t) ,
\end{aligned}
\ee
with $x(t) \in \mathds{R}^n$ and $u(t) \in \mathds{R}^m$. $G(\lambda)$ must not have poles in $\partial\mathds{C}_s$.
\item
\texttt{GAMMA} is a given scalar $\gamma$, which must satisfy $\gamma > \|G(\lambda)\|_\infty$.
\item
 \texttt{OPTIONS} is a MATLAB structure to specify user options and has the following fields:
{\tabcolsep=1mm
\begin{longtable}{|l|lcp{11cm}|} \hline
\textbf{\texttt{OPTIONS} fields} & \multicolumn{3}{l|}{\textbf{Description}} \\ \hline
\texttt{tol}   & \multicolumn{3}{p{11cm}|}{tolerance for the singular values
                        based rank determination of $E$ (Default: $n^2\|E\|_1$\texttt{eps})} \\ \hline
\texttt{tolmin}   & \multicolumn{3}{p{10cm}|}{tolerance for the singular values
                        based controllability tests (Default: $nm\|B\|_1$\texttt{eps})} \\ \hline
 \texttt{stabilize}   & \multicolumn{3}{p{11cm}|}{stabilization option:}\\
                 &  \texttt{true} &--& perform a preliminary stabilization using a right coprime factorization with inner denominator of $G(\lambda)$ (see \textbf{Method}) (default); \\
                 &  \texttt{false}&--& no preliminary stabilization is performed. \\
                                        \hline
\end{longtable}}

\end{description}

\subsubsection*{Output data}
\begin{description}
\item
\texttt{SYSF} contains the descriptor system state-space realization of the spectral factor $F(\lambda)$ in the form
\be\label{glsfg:sysF}
\begin{aligned}
E_F\lambda x_F(t)  &=   A_F x_F(t)+ B_F v(t)  ,\\
y_F(t) &=  C_F x_F(t)+ D_F v(t)  .
\end{aligned}
\ee
\end{description}

\subsubsection*{Method}
For the computation of the left spectral factorization (\ref{glsfg:def}) the two-step approach sketched in Section \ref{app_IOF} is employed. In the first step,
a preliminary right coprime factorization with inner denominator of $G(\lambda)$ is computed such that $G(\lambda) = N(\lambda)M^{-1}(\lambda)$, with both $N(\lambda)$ and $M(\lambda)$ stable, and $M(\lambda)$ inner.  For this purpose, the algorithm to compute right coprime factorizations with inner denominators, given in \textbf{Procedure GRCFID} of \cite{Varg17d}, is employed. This step is not performed if \texttt{OPTIONS.stabilize = false}, in which case $N(\lambda) := G(\lambda)$ and $M(\lambda) = I_m$. In the second step, the left spectral factorization problem is solved
\[ \gamma^2 I - G(\lambda)G^\sim(\lambda) = \gamma^2 I - N(\lambda)N^\sim(\lambda) = F(\lambda)F^\sim(\lambda)  \]
for the minimum-phase phase factor $F(\lambda)$. For this computation, the formulas provided by  Lemma~\ref{L:special-specfacc} and Lemma~\ref{L:special-specfacd} are employed. These lemmas extend to proper descriptor system the formulas developed in  \cite{Zhou96}.

\subsubsection*{Example}
\begin{example}\label{ex:glsfg}
Consider the discrete-time improper TFM
\be\label{glsfg:ex1}  G(z) = {\left[\begin{array}{ccc} z^2 + z + 1 & 4\, z^2 + 3\, z + 2 & 2\, z^2 - 2\\ z & 4\, z - 1 & 2\, z - 2\\ z^2 & 4\, z^2 - z & 2\, z^2 - 2\, z \end{array}\right]
},
\ee
which has two infinite poles (i.e., McMillan-degree of $G(z)$ is equal to 2) and has a minimal descriptor state-space realization of order 4. Therefore, the spectral factorization problem (\ref{glsfg:def}) has a solution for all $\gamma > \|G(z)\|_\infty = 10.4881$. With $\gamma = 1.1\|G(z)\|_\infty$, the function \texttt{glsfg}  computes the proper minimal-phase spectral factor $F(z)$, having two poles in 0 and two stable zeros in $\{-0.2908,0.4188\}$.

The spectral factor $F(z)$ can be computed with the following sequence of commands:
\begin{verbatim}

z = tf('z');  % define the complex variable z
% enter G(z) and determine a minimal state-space realization
G = [z^2+z+1 4*z^2+3*z+2 2*z^2-2;
    z 4*z-1 2*z-2;
    z^2 4*z^2-z 2*z^2-2*z];
sys = gir(ss(G));
gpole(sys)                % the system is unstable and improper
gamma = 1.1*norm(sys,inf) % set gamma = 1.1*||G||_inf

% compute the minimum-phase stable spectral factor F(z) satisfying
% gamma^2*I-G(z)*conj(G(z)) = F(z)*conj(F(z))
sysf = glsfg(sys,gamma);

% check the factorization ||F(z)*F'(z)+G(z)*G'(z)-gamma^2*I||_inf = 0
norm(gir(sysf*sysf'+sys*sys',1.e-7)-gamma^2*eye(size(sys,1)),inf)

% check the stability of poles and zeros of F(z)
gpole(sysf), gzero(sysf)  % F(z) is stable and minimum-phase
\end{verbatim}

\fine\end{example}

\subsection{Functions for Approximations}\label{dstools:approx}
These functions cover the computation of Nehari approximations and the solution of the 1-block and 2-block least distance problems.

\subsubsection{\texttt{\bfseries gnehari}}
\index{M-functions!\texttt{\bfseries gnehari}}

\subsubsection*{Syntax}
\begin{verbatim}
[SYSX,S1] = gnehari(SYS)
[SYSX,S1] = gnehari(SYS,GAMMA)
[SYSX,S1] = gnehari(SYS,GAMMA,TOL)
\end{verbatim}

\subsubsection*{Description}
\texttt{\bfseries gnehari} computes an optimal or suboptimal stable Nehari approximation of the transfer function matrix $G(\lambda)$ of a LTI descriptor state-space system. The optimal Nehari approximation $X(\lambda)$ satisfies
\be\label{gnehari:opt} \| G(\lambda)-X(\lambda)\|_\infty = \|G_u^\sim(\lambda)\|_H \, ,\ee
\index{transfer function matrix (TFM)!Nehari approximation!optimal}%
where $G_u(\lambda)$ is the anti-stable part of $G(\lambda)$. For a given $\gamma > \|G_u^\sim(\lambda)\|_H$, the suboptimal approximation satisfies
\be\label{gnehari:subopt} \| G(\lambda)-X(\lambda)\|_\infty <\gamma \,.\ee
\index{transfer function matrix (TFM)!Nehari approximation!suboptimal}%

\subsubsection*{Input data}
\begin{description}
\item
\texttt{SYS} is a LTI system, whose transfer function matrix is $G(\lambda)$, and is in a  descriptor system state-space form
\be\label{gnehari:sysss}
\begin{aligned}
E\lambda x(t)  &=   Ax(t)+ B u(t) ,\\
y(t) &=  C x(t)+ D u(t) .
\end{aligned}
\ee
$G(\lambda)$ must not have poles in $\partial\mathds{C}_s$.
\item
\texttt{GAMMA}, if specified,  is the desired suboptimality level $\gamma$ for the suboptimal Nehari approximation problem (\ref{gnehari:subopt}) and must satisfy $\gamma > \|G_u^\sim(\lambda)\|_H$, where $G_u(\lambda)$ is the anti-stable part of $G(\lambda)$. If \texttt{GAMMA = [ ]}, then the optimal Nehari approximation problem (\ref{gnehari:opt}) is solved.
\item
 \texttt{TOL}, if specified, is a relative tolerance used for rank computations \newline (Default: internally computed).
\end{description}

\subsubsection*{Output data}
\begin{description}
\item
\texttt{SYSX} contains the descriptor system state-space realization of the optimal or suboptimal stable Nehari approximation $X(\lambda)$ in the form
\be\label{gnehari:sysssred}
\begin{aligned}
E_X\lambda  x_X(t)  &=    A_X x_X(t)+ B_X u(t) ,\\
y_X(t) &=  C_X x_X(t)+ D_X u(t) .
\end{aligned}
\ee
\item
\texttt{S1} is the Hankel-norm of the antistable part of $G(\lambda)$ (also the $\mathcal{L}_\infty$-norm of the optimal approximation error).
\end{description}

\subsubsection*{Method}
The case when $G(\lambda)$ is anstistable, is discussed Section \ref{app:nehari}. For a general $G(\lambda)$ without poles in $\partial\mathds{C}_s$, a preliminary spectral separation is performed as
\be\label{gnehari:sdec} G(\lambda) = G_s(\lambda) + G_u(\lambda) , \ee
where $G_s(\lambda)$ is the stable part (i.e., all poles of $G_s(\lambda)$ are in $\mathds{C}_s$) and $G_u(\lambda)$ is the antistable part (i.e., all poles of $G_u(\lambda)$ are in $\mathds{C}_u$).
The optimal or suboptimal Nehari approximation of $G_u(\lambda)$ is then computed, by determining $Y^\sim(\lambda)$, the optimal zeroth-order Hankel-norm approximation or the suboptimal Hankel-norm approximation of $G_u^\sim(\lambda)$, respectively, using the methods proposed in \cite{Glov84} (see also \cite{Safo90}) with straightforward
extensions for proper descriptor systems. For the computation of the optimal Nehari approximation, the system balancing-based \textbf{Procedure GNEHARI} in \cite{Varg17} (extended to non-square systems) is used.
The solution of the Nehari approximation problem for the original problem is obtained as
\[ X(\lambda) = G_s(\lambda)+Y(\lambda) .\]
Explicit approximation formulas developed in \cite{Glov84} are employed for continuous-time systems, while for discrete-time systems, the bilinear transformation based approach suggested in  \cite{Glov84} is used. For the  computation of additive spectral separation (\ref{gnehari:sdec}) of the descriptor system \texttt{SYS}, the mex-function \url{sl_gsep} is employed. \index{MEX-functions!\url{sl_gsep}}%
In the computation of the balancing transformations,  the intervening Lyapunov and Stein equations, satisfied by the gramians, have been solved directly for the Cholesky factors of the gramians using the mex-function \url{sl_glme}.
\index{MEX-functions!\url{sl_glme}}%

\begin{example}\label{ex:gnehari}
Consider the discrete-time improper TFM
\be\label{gnehari:ex1}  G(z) = {\left[\begin{array}{ccc} z^2 + z + 1 & 4\, z^2 + 3\, z + 2 & 2\, z^2 - 2\\ z & 4\, z - 1 & 2\, z - 2\\ z^2 & 4\, z^2 - z & 2\, z^2 - 2\, z \end{array}\right]
},
\ee
which has two infinite poles (i.e., $G(z)$ is antistable and its McMillan-degree is equal to 2) and has a minimal descriptor state-space realization of order 4. The Hankel-norm of  $G^\sim(z)$ is $\|G^\sim(z)\|_H = 8.6622$, and therefore the optimal stable Nehari approximation $X(z)$ must satisfy $\|G(z)-X(z)\|_\infty = 8.6622$.
Indeed, the optimal stable Nehari approximation, computed with the function \texttt{gnehari}, achieves exactly this approximation error, with $X(z)$ having the McMillan degree equal to 1.
In contrast, the suboptimal stable Nehari approximation computed for $\gamma = 10$, has McMillan degree 2 and the achieved approximation error is 9.8207.

The optimal and suboptimal Nehari approximations can be computed with the following sequence of commands:
\begin{verbatim}
z = tf('z');  % define the complex variable z
% enter G(z) and determine a minimal state-space realization
G = [z^2+z+1 4*z^2+3*z+2 2*z^2-2;
    z 4*z-1 2*z-2;
    z^2 4*z^2-z 2*z^2-2*z];
sys = gir(ss(G));
gpole(sys)                % the system is unstable and improper
ghanorm(sys')             % the achievable optimal approximation error

% compute the optimal Nehari approximation
[sysx,s1] = gnehari(sys);

% check the approximation error ||G(z)-X(z)||_inf = s1
norm(gminreal(sys-sysx,1.e-7),inf)-s1

% check the stability of poles of X(z)
gpole(sysx)  % X(z) is stable and has order 1

% compute the suboptimal Nehari approximation for gamma = 10
[sysxsub,s1] = gnehari(sys,10);

% compute the approximation error ||G(z)-Xsub(z)||_inf
norm(gminreal(sys-sysxsub,1.e-7),inf)

% check the stability of poles of Xsub(z)
gpole(sysxsub)  % Xsub(z) is stable and has order 2

\end{verbatim}
\fine\end{example}

\subsubsection{\texttt{\bfseries glinfldp}}
\index{M-functions!\texttt{\bfseries glinfldp}}

\subsubsection*{Syntax}
\begin{verbatim}
[SYSX,MINDIST] = glinfldp(SYS1,SYS2,OPTIONS)
[SYSX,MINDIST] = glinfldp(SYS,M2,OPTIONS)
\end{verbatim}

\subsubsection*{Description}
\texttt{\bfseries glinfldp} solves the 2-block optimal least distance problem to find a stable $X(\lambda)$ such that
\be\label{glinfldp:opt}  \|[\, G_1(\lambda)-X(\lambda) \;\; G_2(\lambda)\,] \|_\infty = \min \, , \ee
or the 2-block suboptimal least distance problem to find a stable $X(\lambda)$ such that
\be\label{glinfldp:subopt}  \|[\, G_1(\lambda)-X(\lambda) \;\; G_2(\lambda)\,] \|_\infty < \gamma \, , \ee
where $G_1(\lambda)$ and $G_2(\lambda)$ are the transfer function matrices of LTI descriptor state-space systems and $\gamma > \|G_2(\lambda)\|_\infty$.

\subsubsection*{Input data}
For the usage with
\begin{verbatim}
[SYSX,MINDIST] = glinfldp(SYS1,SYS2,OPTIONS)
\end{verbatim}
the input parameters \texttt{SYS1} and \texttt{SYS2} are as follows:
\begin{description}
\item
\texttt{SYS1} is a LTI system, whose transfer function matrix is $G_1(\lambda)$, and is in a descriptor system state-space form
\be\label{glinfld:sysssG1}
\begin{aligned}
E_1\lambda x_1(t)  &=   A_1x_1(t)+ B_1 u(t)  ,\\
y_1(t) &=  C_1 x_1(t)+ D_1 u(t)  ,
\end{aligned}
\ee
where $y_1(t) \in \mathds{R}^{p}$ and $u(t) \in \mathds{R}^{m_1}$. $G_1(\lambda)$ must not have poles in $\partial\mathds{C}_s$.
\item
\texttt{SYS2} is a LTI system, whose transfer function matrix is $G_2(\lambda)$, and is in a  descriptor system state-space  form
\be\label{glinfld:sysssG2}
\begin{aligned}
E_2\lambda x_2(t)  &=   A_2x_2(t)+ B_2 v(t)  ,\\
y_2(t) &=  C_2 x_2(t)+ D_2 v(t)  ,
\end{aligned}
\ee
where $y_2(t) \in \mathds{R}^{p}$ and $v(t) \in \mathds{R}^{m_2}$. If \texttt{SYS2} is empty, a 1-block least distance problem is solved. $G_2(\lambda)$ must not have poles in $\partial\mathds{C}_s$.
\item
\end{description}
\noindent For the usage with
\begin{verbatim}
[SYSX,MINDIST] = glinfldp(SYS,M2,OPTIONS)
\end{verbatim}
the input parameters \texttt{SYS} and \texttt{M2} are as follows:
\begin{description}
\item
\texttt{SYS} is an input concatenated compound LTI system, \texttt{SYS = [ SYS1 SYS2 ]},  in a  descriptor system state-space form
\be\label{glinfldp:sysssG12}
\begin{aligned}
E\lambda x(t)  &=   Ax(t)+ B_1 u(t)+ B_2 v(t)  ,\\
y(t) &=  Cx(t)+ D_1 u(t)+ D_2 v(t)  ,
\end{aligned}
\ee
where \texttt{SYS1} has the transfer function matrix $G_1(\lambda)$ with the descriptor system realization $(A-\lambda E,B_1,C,D_1)$ and
\texttt{SYS2} has the transfer function matrix $G_2(\lambda)$ with the descriptor system realization $(A-\lambda E,B_2,C,D_2)$. $[\,G_1(\lambda)\;G_2(\lambda)\,]$ must not have poles in $\partial\mathds{C}_s$.
\item
\texttt{M2} is the dimension $m_2 \geq 0$ of the input vector $v(t)$ of the system \texttt{SYS2}. If $m_2 = 0$, a 1-block least distance problem is solved.
\end{description}
\noindent For both usages:
\begin{description}
\item
 \texttt{OPTIONS} is a MATLAB structure to specify user options and has the following fields:\\
{\tabcolsep=1mm
\begin{longtable}{|l|lcp{11cm}|} \hline
\textbf{\texttt{OPTIONS} fields} & \multicolumn{3}{l|}{\textbf{Description}} \\ \hline
\texttt{tol}   & \multicolumn{3}{p{11cm}|}{tolerance for rank determinations (Default: internally computed)} \\ \hline
\texttt{reltol}   & \multicolumn{3}{p{11cm}|}{specifies the relative tolerance $rtol$
                    for the desired accuracy of the gamma-iteration.
                    The iterations are performed until the current
                    estimates of the maximum distance $\gamma_u$ and minimum distance $\gamma_l$, which bound the optimal distance $\gamma_{opt}$ (i.e.,  $\gamma_l \leq \gamma_{opt} \leq \gamma_u$), satisfy
                    $\gamma_u-\gamma_l < rtol (\|[\,G_1(\lambda)\; G_2(\lambda)\,]\|_\infty-\|G_2(\lambda)\|_\infty)$.
                    (Default: $rtol = 10^{-4}$)    } \\ \hline
 \texttt{gamma}   & \multicolumn{3}{p{11cm}|}{desired suboptimality level $\gamma$ for the suboptimal least distance problem (\ref{glinfldp:subopt})
 (Default: \texttt{[ ]}, i.e., the optimal least distance problem (\ref{glinfldp:opt}) is solved)}\\
                                        \hline
\end{longtable}}

\end{description}

\subsubsection*{Output data}
\begin{description}
\item
\texttt{SYSX} contains the descriptor system state-space realization of the optimal or suboptimal solution $X(\lambda)$ in the form
\be\label{glinfldp:sysssred}
\begin{aligned}
E_X\lambda  x_X(t)  &=    A_X x_X(t)+ B_X u(t) ,\\
y_X(t) &=  C_X x_X(t)+ D_X u(t) .
\end{aligned}
\ee
\item
\texttt{MINDIST} is the achieved distance
by the computed (optimal or suboptimal) solution $X(\lambda)$.
\end{description}

\subsubsection*{Method}
The solution approach is sketched in Section \ref{appsec:LDP} and corresponds to extensions of the method proposed in \cite{Chu86} to descriptor system state-space representations.

\subsubsection*{Example}
\begin{example}\label{ex:glinfldp}
\index{model-matching problem}
The formulation of many so-called  {approximate model-matching problems}  can be done as  error minimization problems, where {approximate} solutions of  rational equations of the form $X(\lambda)G(\lambda)=F(\lambda)$ are determined by minimizing the $\mathcal{L}_\infty$-norm of the error
$\mathcal{E}(\lambda) := F(\lambda)-X(\lambda)G(\lambda)$.  For example, the standard formulation of the optimal $\mathcal{H}_\infty$ \emph{model-matching problem} ($\mathcal{H}_\infty$-MMP) is: given $G(\lambda), F(\lambda) \in \mathcal{H}_\infty$,  find $X(\lambda) \in \mathcal{H}_\infty$ which minimizes $\|\mathcal{E}(\lambda)\|_\infty$. The optimal solution is typically computed by solving a sequence of suboptimal $\mathcal{H}_\infty$ {model-matching problems}, such that $\|\mathcal{E}(\lambda)\|_\infty < \gamma$, for suitably chosen $\gamma$ values.

For the solution of the $\mathcal{H}_\infty$-MMPs we can often assume that $G(\lambda)$ has full row rank for all $\lambda \in \partial\mathds{C}_s$ and employ the (extended) co-outer--coinner factorization of $G(\lambda)$ to reduce this problem to a $\mathcal{H}_\infty$ least distance problem (LDP). Consider the extended factorization
\[ G(\lambda) = \ba{cc} G_o(\lambda) & 0 \ea G_i(\lambda) = \ba{cc} G_o(\lambda) & 0 \ea \ba{c} G_{i,1}(\lambda)\\ G_{i,2}(\lambda) \ea = G_o(\lambda)G_{i,1}(\lambda), \] where $G_i(\lambda) := \ba{c} G_{i,1}(\lambda)\\ G_{i,2}(\lambda) \ea$ is square and inner and $G_o(\lambda)$ is square and outer (therefore invertible in $\mathcal{H}_\infty$).
This allows to write successively
\[
\begin{array}{ll}
\|\mathcal{E}(\lambda)\|_{\infty} &= \|F(\lambda) -X(\lambda)G(\lambda)\|_{\infty} \\ & = \left\|\left(F(\lambda) G_i^{\sim}(\lambda)-X(\lambda)\ba{cc} G_o(\lambda) & 0 \ea\right)  G_i(\lambda)\right\|_{\infty} \\ & = \left\|\ba{cc} \widetilde F_1(\lambda)-Y(\lambda) & ~~\widetilde F_2(\lambda) \ea\right\|_{\infty} \, ,\end{array}\]
where $Y(\lambda) := X(\lambda)G_o(\lambda) \in \mathcal{H}_\infty$ and
\[ F(\lambda) G_i^{\sim}(\lambda) = {\arraycolsep=1mm\ba{c|c} F(\lambda) G_{i,1}^{\sim}(\lambda) & F(\lambda) G_{i,2}^{\sim}(\lambda) \ea := \ba{c|c} \widetilde F_1(\lambda) & \widetilde F_2(\lambda) \ea \, .} \]
Thus, the problem of computing a stable $X(\lambda)$ which minimizes the error norm $\|\mathcal{E}(\lambda)\|_{\infty}$ has been reduced to a LDP to compute the stable solution $Y(\lambda)$ which minimizes the distance $\big\|[\, \widetilde F_1(\lambda)-Y(\lambda) \;\; \widetilde F_2(\lambda)\,]\big\|_{\infty}$. A $\gamma$-iteration based approach is used for this purpose, as described in \cite{Fran87}. The solution of the original MMP is given by
\[ X(\lambda) = Y(\lambda)G_o^{-1}(\lambda)\, . \]

We apply this approach to solve a MMP discussed in the book of Francis \cite[Example 1, p.~112]{Fran87} with
\[ G(s) = \ba{cc} \displaystyle -\frac{s-1}{s^2+s+1} & \displaystyle\frac{s(s-1)}{s^2+s+1} \ea W(s) , \quad F(s) = [\, W(s) \; \; 0 \,], \]
where $W(s) = (s+1)/(10s+1)$ is a suitable weighting factor.
The optimal solution
\[ X(s) =   \frac{2.3144 (s+0.4569) (s+2.189) (s^2 + s + 1)}{(s+3.095) (s+2.189) (s+1) (s+0.4569)} \]
leads to the optimal error norm $\|F(\lambda)-X(\lambda)G(\lambda)\|_\infty = 0.2521$, which fully agrees with the computed minimum distance \texttt{mindist} (see bellow).
The solution in Francis' book \cite[p.~114]{Fran87} corresponds to a suboptimal solution for $\gamma = 0.2729$ and is
\[ X_{sub}(s) =   \frac{ 2.2731 (s+0.4732) (s+2.183) (s^2 + s + 1)}{(s+3.108) (s+2.189) (s+1) (s+0.4569)} .\]
The corresponding suboptimal error norm $\|F(\lambda)-X_{sub}(\lambda)G(\lambda)\|_\infty = 0.2536$. The solutions of the optimal and suboptimal $\mathcal{H}_\infty$-MMPs
can be computed using the following MATLAB code:

\begin{verbatim}

s = tf('s');  % define the complex variable s
% enter W(s), G(s) and F(s)
W = (s+1)/(10*s+1);    % weighting function
G = [ -(s-1)/(s^2+s+1) (s^2-2*s)/(s^2+s+1)]*W;
F = [ W 0 ];

sys = ss(G);

% compute the extended outer-coinner factorization
[Gi,Go] = goifac(sys,struct('tol',1.e-7));

% define the LDP
Fbar = F*Gi'; m2 = 1;

% compute the optimal solution of the LDP
[Y,mindist] = glinfldp(Fbar,m2); mindist

% compute the optimal solution of the MMP
X = minreal(zpk(Y/Go))

% compute the error norm of the optimal solution
norm(X*G-F,inf)

% compute the suboptimal solution of the LDP
[Ysub,mindistsub] = glinfldp(Fbar,m2,struct('gamma',0.2729)); mindistsub

% compute the suboptimal solution of the MMP
Xsub = minreal(zpk(Ysub/Go))

% compute the error norm of the suboptimal solution
norm(Xsub*G-F,inf)
\end{verbatim}

\fine
\end{example}

\subsubsection{\texttt{\bfseries grasol}}
\index{M-functions!\texttt{\bfseries grasol}}

\subsubsection*{Syntax}
\begin{verbatim}
[SYSX,INFO] = grasol(SYSG,SYSF,OPTIONS)
[SYSX,INFO] = grasol(SYSGF,MF,OPTIONS)
\end{verbatim}

\subsubsection*{Description}
\texttt{\bfseries grasol} computes for the (rational) transfer function matrices $G(\lambda)$ and $F(\lambda)$ of  LTI descriptor systems, an approximate solution $X(\lambda)$ of the linear rational matrix equation $G(\lambda)X(\lambda) = F(\lambda)$.
The optimal solution $X(\lambda)$ minimizes the error norm such that
\be\label{grasol:opt} \|G(\lambda)X(\lambda) - F(\lambda)\|_{\infty/2} = \min .\ee
Optionally, for a given suboptimality level $\gamma$, a suboptimal solution $X(\lambda)$ is determined such that
\be\label{grasol:subopt} \|G(\lambda)X(\lambda) - F(\lambda)\|_{\infty} \leq \gamma .\ee
The resulting $X(\lambda)$ has all poles stable or  lying on the boundary of the stability domain.
\index{transfer function matrix (TFM)!model-matching problem!approximate}

\subsubsection*{Input data}
For the usage with
\begin{verbatim}
[SYSX,INFO] = grasol(SYSG,SYSF,OPTIONS)
\end{verbatim}
the input parameters are as follows:
\begin{description}
\item
\texttt{SYSG} is a LTI system, whose transfer function matrix is $G(\lambda)$, and is in   a descriptor system state-space form
\be\label{grasol:sysssG}
\begin{aligned}
E_G\lambda x_G(t)  &=   A_Gx_G(t)+ B_G u(t)  ,\\
y_G(t) &=  C_G x_G(t)+ D_G u(t)  ,
\end{aligned}
\ee
where $y_G(t) \in \mathds{R}^{p}$.
\item
\texttt{SYSF} is a stable LTI system, whose transfer function matrix is $F(\lambda)$, and is in a  descriptor system state-space form
\be\label{grasol:sysssF}
\begin{aligned}
E_F\lambda x_F(t)  &=   A_Fx_F(t)+ B_F v(t)  ,\\
y_F(t) &=  C_F x_F(t)+ D_F v(t)  ,
\end{aligned}
\ee
where $y_F(t) \in \mathds{R}^{p}$.
\item
 \texttt{OPTIONS} is a MATLAB structure to specify user options and has the following fields:\\
{\tabcolsep=1mm
\begin{longtable}{|l|lcp{11cm}|} \hline
\textbf{\texttt{OPTIONS} fields} & \multicolumn{3}{l|}{\textbf{Description}} \\ \hline
 \texttt{tol}   & \multicolumn{3}{l|}{relative tolerance for rank computations (Default: internally computed);} \\ \hline
\texttt{offset}   & \multicolumn{3}{p{12.5cm}|}{stability
 boundary offset $\beta$, to be used  to assess the finite zeros which belong to $\partial\mathds{C}_s$ (the boundary of the stability domain) as follows: in the
 continuous-time case these are the finite
  zeros having real parts in the interval $[-\beta, \beta]$, while in the
 discrete-time case these are the finite zeros having moduli in the
 interval $[1-\beta, 1+\beta]$ (Default: $\beta = 1.4901\cdot 10^{-08}$). } \\ \hline
 \texttt{sdeg}   & \multicolumn{3}{p{12cm}|}{prescribed stability degree for the free
                    poles of the solution $X(\lambda)$ \newline
                    (Default: \texttt{[ ]}, in which case:
                    \begin{tabular}[t]{l} $-0.05$, in the continuous-time case;\\
                     \hspace*{0.75em}0.95, in the discrete-time case )\end{tabular} } \\ \hline
 \texttt{poles}   & \multicolumn{3}{p{11cm}|}{a complex conjugated set of desired poles  to be assigned for the free poles of the solution $X(\lambda)$
                     (Default: \texttt{[ ]}).}\\
                                        \hline
 \texttt{mindeg}   & \multicolumn{3}{p{12cm}|}{option to compute a minimum degree solution:}\\
                 &  \texttt{true} &--& determine, if possible, a  minimum order
                             solution with all poles in $\mathds{C}_s \cup \partial\mathds{C}_s$; \\
                 &  \texttt{false}&--& determine a particular solution which has
                             possibly non-minimal order and all its poles in $\mathds{C}_s \cup \partial\mathds{C}_s$ (default).  \\
                                        \hline
 \texttt{H2sol}   & \multicolumn{3}{p{12.5cm}|}{option to compute an $\mathcal{H}_2$-norm optimal solution :}\\
                 &  \texttt{true} &--& compute a $\mathcal{H}_2$-norm optimal solution; \\
                 &  \texttt{false}&--& compute a $\mathcal{H}_\infty$-norm optimal or suboptimal solution (default).  \\
                                        \hline
\texttt{reltol}   & \multicolumn{3}{p{12.3cm}|}{specifies the relative tolerance $rtol$
                    for the desired accuracy of the $\gamma$-iteration to solve the $\mathcal{H}_\infty$ least distance problem (see \textbf{Method}).
                    The iterations are performed until the current
                    estimates of the maximum distance $\gamma_u$ and minimum distance $\gamma_l$, which bound the optimal distance $\gamma_{opt}$ (i.e.,  $\gamma_l \leq \gamma_{opt} \leq \gamma_u$), satisfy
                    $\gamma_u-\gamma_l < rtol \cdot (\gamma_u^0-\gamma_l^0)$, where $\gamma_u^0$ and $\gamma_l^0$ are the initial estimates of $\gamma_u$ and $\gamma_l$, respectively. 
                    (Default: $rtol = 10^{-4}$)    } \\ \hline
 \texttt{gamma}   & \multicolumn{3}{p{12.3cm}|}{desired suboptimality level $\gamma$ for solving the suboptimal model-matching problem (\ref{grasol:subopt})\newline
 (Default: \texttt{[ ]}, i.e., the optimal model-matching problem (\ref{grasol:opt}) is solved)}\\
                                        \hline
\end{longtable}}
\end{description}
\noindent For the usage with
\begin{verbatim}
[SYSX,INFO] = grasol(SYSGF,MF,OPTIONS)
\end{verbatim}
the input parameters are as follows:
\begin{description}
\item
\texttt{SYSGF} is an input concatenated compound LTI system, \texttt{SYSGF = [ SYSG SYSF ]},  in a  descriptor system state-space form
\be\label{grasol:sysssGF}
\begin{aligned}
E\lambda x(t)  &=   Ax(t)+ B_G u(t)+ B_F v(t)  ,\\
y(t) &=  Cx(t)+ D_G u(t)+ D_F v(t)  ,
\end{aligned}
\ee
where \texttt{SYSG} has the transfer function matrix $G(\lambda)$, with the descriptor system realization $(A-\lambda E,B_G,C,D_G)$, and
\texttt{SYSF} has the transfer function matrix $F(\lambda)$, with the descriptor system realization $(A-\lambda E,B_F,C,D_F)$.
\item
\texttt{MF} is the dimension of the input vector $v(t)$ of the system \texttt{SYSF}. \item
\texttt{OPTIONS} is a MATLAB structure to specify user options and has the same fields as described previously.
\end{description}

\subsubsection*{Output data}
\begin{description}
\item
\texttt{SYSX} contains  the descriptor system state-space realization of the solution $X(\lambda)$ in the form
\be\label{grasol:sysssX}
\begin{aligned}
\widetilde E\lambda \widetilde x(t)  &=   \widetilde A \widetilde x(t)+ \widetilde B v(t)  ,\\
u(t) &=  \widetilde C \widetilde x(t)+ \widetilde D v(t)  .
\end{aligned}
\ee
\item
 \texttt{INFO} is a MATLAB structure containing additional information, as follows:\\
{\begin{longtable}{|l|p{12cm}|} \hline
\textbf{\texttt{INFO} fields} & \textbf{Description} \\ \hline
 \texttt{nrank}   & normal rank of the transfer function matrix $G(\lambda)$; \\ \hline
 \texttt{mindist}   & achieved approximation error norm $\|G(\lambda)X(\lambda)-F(\lambda)\|_{\infty/2}$ ;\\ \hline
 \texttt{nonstandard}   & logical value, which is set to \texttt{true} for a \emph{non-standard problem} (when $G(\lambda)$ has zeros in $\partial\mathds{C}_s$ within the stability offset \texttt{OPTIONS.offset} for finite zeros), and to \texttt{false} for a \emph{standard problem} (when $G(\lambda)$ has no zeros in $\partial\mathds{C}_s$); \\ \hline
\end{longtable}}
\end{description}

\subsubsection*{Method}
The employed solution approach is sketched in Section \ref{appsec:AMMP} for the \emph{left} linear rational matrix equation $X(\lambda)G(\lambda) = F(\lambda)$ and extends the approach proposed in \cite{Fran87} to arbitrary $G(\lambda)$. In what follows, we give the three main steps of a similar approach to solve the \emph{right} linear rational matrix equation $G(\lambda)X(\lambda) = F(\lambda)$:
\begin{enumerate}
\item Compute the (extended) inner--quasi-outer factorization of $G(\lambda)$ as
\be\label{grasol:iofac}
 G(\lambda) = G_i(\lambda)\ba{c} G_o(\lambda) \\ 0 \ea  = [\, G_{i,1}(\lambda)\; G_{i,2}(\lambda)\,] \ba{cc} G_o(\lambda)\\ 0 \ea = G_{i,1}(\lambda)G_o(\lambda), \ee
 where $G_i(\lambda) := [\, G_{i,1}(\lambda) \; G_{i,2}(\lambda) \,]$ is square and inner, and $G_o(\lambda)$ is full row rank (i.e., is right invertible), has the same poles as $G(\lambda)$ and has only zeros  in $\mathds{C}_s \cup \partial\mathds{C}_s$.
\item Compute
\[ G_i^{\sim}(\lambda)F(\lambda)  = {\arraycolsep=1mm\ba{cc} G_{i,1}^{\sim}(\lambda)F(\lambda) \\ \hline  G_{i,2}^{\sim}(\lambda)F(\lambda)  \ea := \ba{cc} \widetilde F_1(\lambda) \\ \hline \\[-4mm] \widetilde F_2(\lambda) \ea } \]
and compute the stable solution $Y(\lambda)$ of the least distance problem
\[  \gamma_{opt}:= \left\|\ba{c}\widetilde F_1(\lambda)-Y(\lambda) \\ \widetilde F_2(\lambda)\ea\right\|_{\infty/2} = \min  \]
or of the $\gamma$-suboptimal distance problem
\[  \left\|\ba{c}\widetilde F_1(\lambda)-Y(\lambda) \\ \widetilde F_2(\lambda)\ea\right\|_{\infty/2} \leq \gamma .  \]
\item Compute an exact solution $X(\lambda)$ of the linear rational equation $G_o(\lambda)X(\lambda) = Y(\lambda)$, where $G_o(\lambda)$ is left invertible.
    \end{enumerate}

If \texttt{OPTIONS.mindeg = false}, the computed solution $X(\lambda)$ at Step 3 may not have the least achievable McMillan degree (excepting the case when $G_o(\lambda)$ is invertible), but its only unstable poles are (part of) the unstable zeros of $G_o(\lambda)$. These poles are also (part of) the zeros of $G(\lambda)$ lying in $\partial\mathds{C}_s$. If $G_o(\lambda)$ is non-square, $X(\lambda)$ has a number of free poles, which  are assigned to values specified via the option parameters \texttt{OPTIONS.sdeg} and \texttt{OPTIONS.poles}.

If \texttt{OPTIONS.mindeg = true}, the least order solution $X(\lambda)$ of the linear rational equation $G_o(\lambda)X(\lambda) = Y(\lambda)$ has the least achievable McMillan degree, but besides possible unstable poles which are the unstable zeros of $G_o(\lambda)$ lying in $\partial\mathds{C}_s$, $X(\lambda)$ may also have additional unstable poles (e.g., in the case when $G_o(\lambda)$ has no least order stable right inverse). In this case, the solution corresponding to \texttt{OPTIONS.mindeg = false} is computed.

The resulting value of the achieved distance $\left\|\left[\begin{smallmatrix}\widetilde F_1(\lambda)-Y(\lambda) \\ \widetilde F_2(\lambda)\end{smallmatrix}\right]\right\|_{\infty/2}$ at Step 2, is returned in \texttt{INFO.mindist}.
If \texttt{OPTIONS.H2sol = true}, an $\mathcal{H}_2$-LDP is solved (see Section \ref{appsec:AMMP}), in which case the distance may be infinite if $\widetilde F_2(\lambda)$ is not strictly proper.

\subsubsection*{Example}

\begin{example}\label{ex:grasol1}
This example is taken from \cite{Gao89} and has been also considered in  \emph{Example} \ref{ex:Gao-Antsaklis1989}, where an exact model matching problem is solved for a stable solution  of  $G(s)X(s) = F(s)$, with
\[ G(s) =
\left[\begin{array}{cc} \displaystyle\frac{s - 1}{s\,  \left(s + 1\right)} &  \displaystyle\frac{s - 1}{s\, \left(s + 2\right)} \end{array}\right], \quad
F(s) =
\left[\begin{array}{cc}  \displaystyle\frac{s - 1}{\left(s + 1\right)\, \left(s + 3\right)} & \displaystyle\frac{s - 1}{\left(s + 1\right)\, \left(s + 4\right)} \end{array}\right] .
\]
Both $G(s)$ and $[\, G(s)\; F(s)\,]$ have rank equal to 1 and zeros $\{1,\infty\}$. It follows, according to Lemma \ref{L-SEMMP}, that the linear rational matrix equation $G(s)X(s) = F(s)$ has a stable and proper solution. A fourth order stable and proper solution has been computed in \cite{Gao89}. A least order solution, with McMillan degree equal to 2, has been computed with \texttt{grasol} as
\[ X(s) = {\def\arraystretch{2}\ba{cc} \displaystyle\frac{0.75 (s-0.3333)}{s+3} & \displaystyle\frac{0.75 (s-0.3333)}{s+4} \\
\displaystyle\frac{0.25  (s+2)}{s+3} & \displaystyle\frac{0.25  (s+2)}{s+4} \ea } .\]
Note that this solution is different (but of same order) from the exact solution computed with \texttt{grsol} in \emph{Example} \ref{ex:Gao-Antsaklis1989}.

To compute a least order solution $X(s)$, the following sequence of commands can be used:
\begin{verbatim}
% Gao & Antsaklis (1989)
s = tf('s');
G = [(s-1)/(s*(s+1))  (s-1)/(s*(s+2))];
F = [(s-1)/((s+1)*(s+3))  (s-1)/((s+1)*(s+4))];

% solve G(s)*X(s) = F(s) for the least order solution
[X,info] = grasol(ss(G),ss(F),struct('mindeg',true,'tol',1.e-7)); info
minreal(zpk(X))

% check  solution
minreal(G*X-F)
\end{verbatim}
\fine\end{example}

\begin{example}\label{ex:grasol2}
\index{model-matching problem}
We solve the  $\mathcal{H}_\infty$-MMP discussed in the book of Francis \cite[Example 1, p.~112]{Fran87} with
\[ G(s) = \ba{c} \displaystyle -\frac{s-1}{s^2+s+1} \\ \\[-4mm]\displaystyle\frac{s(s-1)}{s^2+s+1} \ea W(s) , \quad F(s) = \ba{c}  W(s) \\ 0  \ea, \]
where $W(s) = (s+1)/(10s+1)$ is a suitable weighting factor.
The optimal solution
\[ X(s) =   \frac{2.3144 (s+0.4569) (s+2.189) (s^2 + s + 1)}{(s+3.095) (s+2.189) (s+1) (s+0.4569)} \]
leads to the optimal error norm $\|F(\lambda)-G(\lambda)X(\lambda)\|_\infty = 0.2521$.
The solution in Francis' book \cite[p.~114]{Fran87} corresponds to a suboptimal solution for $\gamma = 0.2729$ and is
\[ X_{sub}(s) =   \frac{ 2.2731 (s+0.4732) (s+2.183) (s^2 + s + 1)}{(s+3.108) (s+2.189) (s+1) (s+0.4569)} .\]
The corresponding suboptimal error norm $\|F(\lambda)-G(\lambda)X_{sub}(\lambda)\|_\infty = 0.2536$.

The solutions of the optimal and suboptimal $\mathcal{H}_\infty$-MMPs
can be computed using the following MATLAB code:

\begin{verbatim}
s = tf('s');  % define the complex variable s
% enter W(s), G(s) and F(s)
W = (s+1)/(10*s+1);    % weighting function
G = [ -(s-1)/(s^2+s+1); (s^2-2*s)/(s^2+s+1)]*W;
F = [ W; 0 ];

% solve G(s)*X(s) = F(s) for the least order solution
[X,info] = grasol(ss(G),ss(F),struct('mindeg',true,'tol',1.e-7)); info
minreal(zpk(X))

% compute the error norm of the optimal solution
norm(G*X-F,inf)


% compute the suboptimal solution for gamma = 0.2729
opts = struct('mindeg',true,'tol',1.e-7,'gamma',0.2729);
[Xsub,info] = grasol(ss(G),ss(F),opts); info
minreal(zpk(Xsub))

% compute the error norm of the suboptimal solution
norm(G*Xsub-F,inf)
\end{verbatim}

\fine
\end{example}

\subsubsection{\texttt{\bfseries glasol}}
\index{M-functions!\texttt{\bfseries glasol}}

\subsubsection*{Syntax}
\begin{verbatim}
[SYSX,INFO] = glasol(SYSG,SYSF,OPTIONS)
[SYSX,INFO] = glasol(SYSGF,MF,OPTIONS)
\end{verbatim}

\subsubsection*{Description}
\texttt{\bfseries glasol} computes for the (rational) transfer function matrices $G(\lambda)$ and $F(\lambda)$ of  LTI descriptor systems, an approximate solution $X(\lambda)$ of the linear rational matrix equation $X(\lambda)G(\lambda) = F(\lambda)$.
The optimal solution $X(\lambda)$ minimizes the error norm such that
\be\label{glasol:opt} \|X(\lambda)G(\lambda) - F(\lambda)\|_{\infty/2} = \min .\ee
Optionally, for a given suboptimality level $\gamma$, a suboptimal solution $X(\lambda)$ is determined such that
\be\label{glasol:subopt} \|X(\lambda)G(\lambda) - F(\lambda)\|_{\infty} \leq \gamma .\ee
The resulting $X(\lambda)$ has all poles stable or  lying on the boundary of the stability domain.
\index{transfer function matrix (TFM)!model-matching problem!approximate}

\subsubsection*{Input data}
For the usage with
\begin{verbatim}
[SYSX,INFO] = glasol(SYSG,SYSF,OPTIONS)
\end{verbatim}
the input parameters are as follows:
\begin{description}
\item
\texttt{SYSG} is a LTI system, whose transfer function matrix is $G(\lambda)$, and is in   a descriptor system state-space form
\be\label{glasol:sysssG}
\begin{aligned}
E_G\lambda x_G(t)  &=   A_Gx_G(t)+ B_G u(t)  ,\\
y_G(t) &=  C_G x_G(t)+ D_G u(t)  ,
\end{aligned}
\ee
where $u(t) \in \mathds{R}^{m}$.
\item
\texttt{SYSF} is a stable LTI system, whose transfer function matrix is $F(\lambda)$, and is in a  descriptor system state-space form
\be\label{glasol:sysssF}
\begin{aligned}
E_F\lambda x_F(t)  &=   A_Fx_F(t)+ B_F v(t)  ,\\
y_F(t) &=  C_F x_F(t)+ D_F v(t)  ,
\end{aligned}
\ee
where $v(t) \in \mathds{R}^{m}$.
\item
 \texttt{OPTIONS} is a MATLAB structure to specify user options and has the following fields:\\
\pagebreak[4]
{\tabcolsep=1mm
\begin{longtable}{|l|lcp{11cm}|} \hline
\textbf{\texttt{OPTIONS} fields} & \multicolumn{3}{l|}{\textbf{Description}} \\ \hline
 \texttt{tol}   & \multicolumn{3}{l|}{relative tolerance for rank computations (Default: internally computed);} \\ \hline
\texttt{offset}   & \multicolumn{3}{p{12.5cm}|}{stability
 boundary offset $\beta$, to be used  to assess the finite zeros which belong to $\partial\mathds{C}_s$ (the boundary of the stability domain) as follows: in the
 continuous-time case these are the finite
  zeros having real parts in the interval $[-\beta, \beta]$, while in the
 discrete-time case these are the finite zeros having moduli in the
 interval $[1-\beta, 1+\beta]$ (Default: $\beta = 1.4901\cdot 10^{-08}$). }  \\ \hline
 \texttt{sdeg}   & \multicolumn{3}{p{12cm}|}{prescribed stability degree for the free
                    poles of the solution $X(\lambda)$ \newline
                    (Default: \texttt{[ ]}, in which case:
                    \begin{tabular}[t]{l} $-0.05$, in the continuous-time case;\\
                     \hspace*{0.75em}0.95, in the discrete-time case )\end{tabular} } \\ \hline
 \texttt{poles}   & \multicolumn{3}{p{11cm}|}{a complex conjugated set of desired poles  to be assigned for the free poles of the solution $X(\lambda)$
                     (Default: \texttt{[ ]}).}\\
                                        \hline
 \texttt{mindeg}   & \multicolumn{3}{p{12cm}|}{option to compute a minimum degree solution:}\\
                 &  \texttt{true} &--& determine, if possible, a  minimum order
                             solution with all poles in $\mathds{C}_s \cup \partial\mathds{C}_s$; \\
                 &  \texttt{false}&--& determine a particular solution which has
                             possibly non-minimal order and all its poles in $\mathds{C}_s \cup \partial\mathds{C}_s$ (default).  \\
                                        \hline
 \texttt{H2sol}   & \multicolumn{3}{p{12.5cm}|}{option to compute an $\mathcal{H}_2$-norm optimal solution :}\\
                 &  \texttt{true} &--& compute a $\mathcal{H}_2$-norm optimal solution; \\
                 &  \texttt{false}&--& compute a $\mathcal{H}_\infty$-norm optimal or suboptimal solution (default).  \\
                                        \hline
\texttt{reltol}   & \multicolumn{3}{p{12.3cm}|}{specifies the relative tolerance $rtol$
                    for the desired accuracy of the $\gamma$-iteration to solve the $\mathcal{H}_\infty$ least distance problem (see \textbf{Method}).
                    The iterations are performed until the current
                    estimates of the maximum distance $\gamma_u$ and minimum distance $\gamma_l$, which bound the optimal distance $\gamma_{opt}$ (i.e.,  $\gamma_l \leq \gamma_{opt} \leq \gamma_u$), satisfy
                    $\gamma_u-\gamma_l < rtol \cdot (\gamma_u^0-\gamma_l^0)$, where $\gamma_u^0$ and $\gamma_l^0$ are the initial estimates of $\gamma_u$ and $\gamma_l$, respectively. 
                    (Default: $rtol = 10^{-4}$)    } \\ \hline
 \texttt{gamma}   & \multicolumn{3}{p{12.3cm}|}{desired suboptimality level $\gamma$ for solving the suboptimal model-matching problem (\ref{glasol:subopt})\newline
 (Default: \texttt{[ ]}, i.e., the optimal model-matching problem (\ref{glasol:opt}) is solved)}\\
                                        \hline
\end{longtable}}
\end{description}
\noindent For the usage with
\begin{verbatim}
[SYSX,INFO] = glasol(SYSGF,MF,OPTIONS)
\end{verbatim}
the input parameters are as follows:
\begin{description}
\item
\texttt{SYSGF} is an output concatenated compound LTI system, \texttt{SYSGF = [ SYSG; SYSF ]},  in a  descriptor system state-space form
\be\label{glasol:sysssGF}
\begin{aligned}
E\lambda x(t)  &=   Ax(t)+ B u(t)  ,\\
y_G(t) &=  C_Gx(t)+ D_G u(t)  , \\
y_F(t) &=  C_Fx(t)+ D_F u(t)  ,
\end{aligned}
\ee
where \texttt{SYSG} has the transfer function matrix $G(\lambda)$, with the descriptor system realization $(A-\lambda E,B,C_G,D_G)$, and
\texttt{SYSF} has the transfer function matrix $F(\lambda)$, with the descriptor system realization $(A-\lambda E,B,CF,D_F)$.
\item
\texttt{MF} is the dimension of the output vector $y_F(t)$ of the system \texttt{SYSF}.
\item
\texttt{OPTIONS} is a MATLAB structure to specify user options and has the same fields as described previously.
\end{description}

\subsubsection*{Output data}
\begin{description}
\item
\texttt{SYSX} contains  the descriptor system state-space realization of the solution $X(\lambda)$ in the form
\be\label{glasol:sysssX}
\begin{aligned}
\widetilde E\lambda \widetilde x(t)  &=   \widetilde A \widetilde x(t)+ \widetilde B y_G(t)  ,\\
y(t) &=  \widetilde C \widetilde x(t)+ \widetilde D y_G(t)  .
\end{aligned}
\ee
\item
 \texttt{INFO} is a MATLAB structure containing additional information, as follows:\\
{\begin{longtable}{|l|p{12cm}|} \hline
\textbf{\texttt{INFO} fields} & \textbf{Description} \\ \hline
 \texttt{nrank}   & normal rank of the transfer function matrix $G(\lambda)$; \\ \hline
 \texttt{mindist}   & achieved approximation error norm $\|X(\lambda)G(\lambda)-F(\lambda)\|_{\infty/2}$ ;\\ \hline
 \texttt{nonstandard}   & logical value, which is set to \texttt{true} for a \emph{non-standard problem} (when $G(\lambda)$ has zeros in $\partial\mathds{C}_s$ within the stability offset \texttt{OPTIONS.offset} for finite zeros), and to \texttt{false} for a \emph{standard problem} (when $G(\lambda)$ has no zeros in $\partial\mathds{C}_s$); \\ \hline
\end{longtable}}
\end{description}

\subsubsection*{Method}
The employed solution approach to determine the approximate solution of the \emph{left} linear rational matrix equation $X(\lambda)G(\lambda) = F(\lambda)$ is sketched in Section \ref{appsec:AMMP} and extends the approach proposed in \cite{Fran87}  to arbitrary $G(\lambda)$. The three main steps of this approach are:
\begin{enumerate}
\item Compute the (extended) quasi-co-outer--coinner  factorization of $G(\lambda)$ as
\be\label{glasol:iofac}
G(\lambda) = \ba{cc} G_o(\lambda) & 0 \ea G_i(\lambda) = \ba{cc} G_o(\lambda) & 0 \ea \ba{c} G_{i,1}(\lambda)\\ G_{i,2}(\lambda) \ea = G_o(\lambda)G_{i,1}(\lambda), \ee
 where $G_i(\lambda) := \left[\begin{smallmatrix} G_{i,1}(\lambda) \\ G_{i,2}(\lambda) \end{smallmatrix}\right]$ is square and inner, and $G_o(\lambda)$ is full column rank (i.e., is left invertible), has the same poles as $G(\lambda)$ and has only zeros  in $\mathds{C}_s \cup \partial\mathds{C}_s$.
\item Compute
\[ F(\lambda) G_i^{\sim}(\lambda) = {\arraycolsep=1mm\ba{c|c} F(\lambda) G_{i,1}^{\sim}(\lambda) & F(\lambda) G_{i,2}^{\sim}(\lambda) \ea := \ba{c|c} \widetilde F_1(\lambda) & \widetilde F_2(\lambda) \ea \, .} \]
and compute the stable solution $Y(\lambda)$ of the least distance problem
\[  \gamma_{opt}:= \big\|\big[\, \widetilde F_1(\lambda)-Y(\lambda) \; \widetilde F_2(\lambda)\, \big]\big\|_{\infty/2} = \min  \]
or of the $\gamma$-suboptimal distance problem
\[  \big\|\big[\, \widetilde F_1(\lambda)-Y(\lambda) \; \widetilde F_2(\lambda)\, \big]\big\|_{\infty/2} \leq \gamma .  \]
\item Compute an exact solution $X(\lambda)$ of the linear rational equation $X(\lambda)G_o(\lambda) = Y(\lambda)$, where $G_o(\lambda)$ is left invertible.
    \end{enumerate}

If \texttt{OPTIONS.mindeg = false}, the computed solution $X(\lambda)$ at Step 3 may not have the least achievable McMillan degree (excepting the case when $G_o(\lambda)$ is invertible), but its only unstable poles are (part of) the unstable zeros of $G_o(\lambda)$. These poles are also (part of) the zeros of $G(\lambda)$ lying in $\partial\mathds{C}_s$. If $G_o(\lambda)$ is non-square, $X(\lambda)$ has a number of free poles, which  are assigned to values specified via the option parameters \texttt{OPTIONS.sdeg} and \texttt{OPTIONS.poles}.

If \texttt{OPTIONS.mindeg = true}, the least order solution $X(\lambda)$ of the linear rational equation $X(\lambda)G_o(\lambda) = Y(\lambda)$ has the least achievable McMillan degree, but besides possible unstable poles which are the unstable zeros of $G_o(\lambda)$ lying in $\partial\mathds{C}_s$, $X(\lambda)$ may also have additional unstable poles (e.g., in the case when $G_o(\lambda)$ has no least order stable left inverse). In this case, the solution corresponding to \texttt{OPTIONS.mindeg = false} is computed.

The resulting value of the achieved distance $\big\|\big[\, \widetilde F_1(\lambda)-Y(\lambda) \; \widetilde F_2(\lambda)\, \big]\big\|_{\infty/2}$ at Step 2, is returned in \texttt{INFO.mindist}.
If \texttt{OPTIONS.H2sol = true}, an $\mathcal{H}_2$-LDP is solved (see Section \ref{appsec:AMMP}), in which case the distance may be infinite if $\widetilde F_2(\lambda)$ is not strictly proper.

\subsubsection*{Example}

\begin{example}\label{ex:glasol1}
This example illustrates the computation of a stable left inverse for the example  used in \cite{Wang73} (see also \emph{Example} \ref{ex:Wang-Davison}).
The transfer function matrices $G(s)$ and $F(s)$ are
\[ G(s) = \frac{1}{s^2+3s+2}\ba{cc} s+1 & s+2\\ s+3 & s^2+2s\\ s^2+3s & 0 \ea, \quad F(s) = \ba{cc} 1 & 0 \\ 0 & 1 \ea. \]
The solution $X(s)$ of the rational equation $X(s)G(s) = I$ is a left inverse of $G(s)$.
Since $G(s)$ has no zeros, a stable left inverse of $G(s)$ of McMillan degree equal to three exists. However, as illustrated in \emph{Example} \ref{ex:Wang-Davison}, any left inverse of least McMillan order equal to two is always unstable. The solution computed with \texttt{glasol} results always stable, even if the option for least order is selected. For the three free poles of the left inverse we can assign the poles to $\{-1,-2,-3\}$. For the computation of a stable left inverse with poles assigned to $\{-1,-2,-3\}$, the following MATLAB commands can be used:
\begin{verbatim}
% Wang and Davison Example (1973)
s = tf('s');
g = [ s+1 s+2; s+3 s^2+2*s; s^2+3*s 0 ]/(s^2+3*s+2); f = eye(2);
sysg = gir(ss(g)); sysf = ss(f);

% compute a stable solution of X(s)G(s) = I
[sysx,info] = glasol(sysg,sysf,struct('mindeg',true,'poles',[-1 -2 -3])); info
gpole(sysx)   % the left inverse is stable

% check solution
gir(sysx*sysg-sysf,1.e-7)
\end{verbatim}
The computed solution is stable and the resulting minimal realization of the left inverse \texttt{sysx} has the poles equal to $\{-1,-2,-3\}$.  \fine

\end{example}

\begin{example}\label{ex:glasol2}
\index{model-matching problem}
We solve the  $\mathcal{H}_\infty$-MMP discussed in the book of Francis \cite[Example 1, p.~112]{Fran87} with
\[ G(s) = \ba{cc} \displaystyle -\frac{s-1}{s^2+s+1} & \displaystyle\frac{s(s-1)}{s^2+s+1} \ea W(s) , \quad F(s) = [\,  W(s) \;\; 0  \,], \]
where $W(s) = (s+1)/(10s+1)$ is a suitable weighting factor.
The optimal solution
\[ X(s) =   \frac{2.3144 (s+0.4569) (s+2.189) (s^2 + s + 1)}{(s+3.095) (s+2.189) (s+1) (s+0.4569)} \]
leads to the optimal error norm $\|F(\lambda)-X(\lambda)G(\lambda)\|_\infty = 0.2521$.
The solution in Francis' book \cite[p.~114]{Fran87} corresponds to a suboptimal solution for $\gamma = 0.2729$ and is
\[ X_{sub}(s) =   \frac{ 2.2731 (s+0.4732) (s+2.183) (s^2 + s + 1)}{(s+3.108) (s+2.189) (s+1) (s+0.4569)} .\]
The corresponding suboptimal error norm $\|F(\lambda)-X_{sub}(\lambda)G(\lambda)\|_\infty = 0.2536$.

The solutions of the optimal and suboptimal $\mathcal{H}_\infty$-MMPs
can be computed using the following MATLAB code:

\begin{verbatim}
s = tf('s');  % define the complex variable s
% enter W(s), G(s) and F(s)
W = (s+1)/(10*s+1);    % weighting function
G = [ -(s-1)/(s^2+s+1) (s^2-2*s)/(s^2+s+1)]*W;
F = [ W 0 ];

% solve G(s)*X(s) = F(s) for the least order solution
[X,info] = glasol(ss(G),ss(F),struct('mindeg',true,'tol',1.e-7)); info
minreal(zpk(X))

% compute the error norm of the optimal solution
norm(X*G-F,inf)


% compute the suboptimal solution for gamma = 0.2729
opts = struct('mindeg',true,'tol',1.e-7,'gamma',0.2729);
[Xsub,info] = glasol(ss(G),ss(F),opts); info
minreal(zpk(Xsub))

% compute the error norm of the suboptimal solution
norm(Xsub*G-F,inf)
\end{verbatim}

\fine
\end{example}

\subsection{Functions for Matrix Pencils and Stabilization}\label{dstools:pencil}
These functions cover the reduction of linear matrix pencils to several Kronecker-like forms, the computation of specially ordered generalized real Schur forms, and the stabilization using state feedback.

\subsubsection{\texttt{\bfseries gklf}}
\index{M-functions!\texttt{\bfseries gklf}}

\subsubsection*{Syntax}
\begin{verbatim}
[AT,ET,INFO,Q,Z] = gklf(A,E,TOL,JOBOPT)
[AT,ET,INFO,Q,Z] = gklf(A,E,TOL,...,QOPT)
\end{verbatim}

\subsubsection*{Description}

\noindent \texttt{gklf} computes several Kronecker-like forms $\widetilde A-\lambda \widetilde E$ of a linear matrix pencil $A-\lambda E$.
\index{condensed form!Kronecker-like}

\subsubsection*{Input data}

\begin{description}
\item
\texttt{A,E} are the $m\times n$ real matrices $A$ and $E$, which define the linear matrix pencil $A-\lambda E$.
\item
 \texttt{TOL} is a relative tolerance used for rank determinations. If \texttt{TOL} is not specified as input or if \texttt{TOL} = 0, an internally computed default value is used.
\item
 \texttt{JOBOPT}  is a character option variable to specify various options to compute Kronecker-like forms. The valid options are: \\[-7mm]
 {\tabcolsep=1mm
\begin{longtable}{lcp{12cm}}
   ~~~~~~~~\texttt{'standard'} &--& compute the standard Kronecker-like form  (\ref{gklf:standard}), with infinite-finite ordering of the blocks of the regular part (see \textbf{Method}) (default); \\
   ~~~~~~~~\texttt{'reverse'}&--& compute the standard Kronecker-like form  (\ref{gklf:reverse}), with reverse ordering of the blocks of the regular part (see \textbf{Method}); \\
   ~~~~~~~~\texttt{'right'}&--& compute the Kronecker-like form    (\ref{gklf:right}), which exhibits the right and infinite Kronecker structures (see \textbf{Method}); \\
   ~~~~~~~~\texttt{'left'}&--& compute the Kronecker-like form (\ref{gklf:left}), which exhibits the left and infinite Kronecker structures (see \textbf{Method}).\\
   \end{longtable}
}
\item
 \texttt{QOPT}  is a character option variable to specify the options to accumulate or not the left orthogonal transformations in $Q$: \\[-7mm]
 {\tabcolsep=2mm
\begin{longtable}{lcp{12cm}}
   ~~\texttt{'Q'} &--& accumulate $Q$ (default); \\
   ~~\texttt{'noQ'}&--& do not accumulate $Q$. \\
   \end{longtable}
}

\end{description}

\subsubsection*{Output data}

\begin{description}
\item
\texttt{AT,ET} contain the matrices $\widetilde A$ and $\widetilde E$ which define the resulting pencil $\widetilde A -\lambda \widetilde E$ in a Kronecker-like form, satisfying
\be\label{gklf:AE}
 \widetilde A -\lambda \widetilde E = Q^T(A-\lambda E)Z , \ee
where $Q$ and $Z$ are orthogonal transformation matrices.
\item
\texttt{INFO} is a MATLAB structure, which provides
information on the  structure of the pencil $\widetilde A -\lambda \widetilde E$ (see \textbf{Method}) as follows: \\
\hspace*{-3mm}{\tabcolsep=1mm\begin{tabular}{llp{11.3cm}}
    ~~\texttt{INFO.mr,INFO.nr} & -- & the dimensions of the full row rank diagonal blocks of $[\, B_r \; A_r-\lambda E_r\,]$, which characterize the right Kronecker structure  (see \textbf{Method}); \texttt{INFO.mr} and \texttt{INFO.nr} are empty if no right structure exists; if \texttt{JOBOPT = 'left'}, \texttt{INFO.mr}
and \texttt{INFO.nr} are set to the negative values of the row and column dimensions of the pencil $A_{r,f}-\lambda E_{r,f}$ in (\ref{gklf:left}); \\
    ~~\texttt{INFO.minf}& -- &the dimensions of the square diagonal blocks of $A_\infty-\lambda E_\infty$, which characterize the infinite Kronecker structure; \texttt{INFO.minf} is empty if no infinite structure exists; \\
    ~~\texttt{INFO.mf}& -- &the dimension of the square regular pencil $A_f-\lambda E_f$, which contains the finite eigenvalues; \\
    ~~\texttt{INFO.ml,INFO.nl} & -- & the dimensions of the full column rank diagonal blocks of $\left[\begin{smallmatrix} A_l-\lambda E_l\\C_l\end{smallmatrix}\right]$, which characterize the left Kronecker structure  (see \textbf{Method}); \texttt{INFO.ml} and \texttt{INFO.nl} are empty if no left structure exists; if \texttt{JOBOPT = 'right'}, \texttt{INFO.ml}
and \texttt{INFO.nl} are set to the negative values of the row and column dimensions of the pencil $A_{f,l}-\lambda E_{f,l}$ in (\ref{gklf:right}). \\
    \end{tabular}}
\item
\texttt{Q,Z} contain the orthogonal matrices $Q$ and $Z$ used to compute the Kronecker-like form $\widetilde A -\lambda \widetilde E$ in (\ref{gklf:AE}). $Q$ and $Z$ are not accumulated if both output variables \texttt{Q} and \texttt{Z} are not specified. If \texttt{QOPT = 'noQ'}, $Q$ is not accumulated and, if specified, set to \texttt{Q = [ ]}.
\end{description}

\subsubsection*{Method}
The Kronecker-like forms, obtainable by using orthogonal transformations,  contains most of the relevant structural information provided by the potentially highly sensitive Kronecker canonical form (see Section \ref{sec:matrix-pencils}).
The computation of  Kronecker-like forms is based on the method proposed in \cite{Beel88}, which underlies the implementation of the mex-function \url{sl_klf}, called by \texttt{gklf}.
\index{MEX-functions!\url{sl_klf}}%

The (standard) Kronecker-like form has the following structure
\be\label{gklf:standard}
\widetilde A -\lambda \widetilde E =
\ba{ccccc} B_r & A_r-\lambda E_r & \ast & \ast & \ast \\
0 & 0 & A_\infty-\lambda E_\infty & \ast & \ast \\
0 & 0 & 0 & A_f -\lambda E_f & \ast \\
0 & 0 & 0 & 0 & A_l-\lambda E_l\\
0 & 0 & 0 & 0 & C_l \ea ,\ee
where
\begin{itemize}
 \item[(1)] $[\, B_r \; A_r-\lambda E_r\,]$, with $E_r$ invertible and upper triangular, contains the right Kronecker structure and is in a controllability staircase form with $[\, B_r \; A_r\,]$ as in (\ref{cscf-defab}) and $E_r$ as in (\ref{cscf-defe}); the dimensions of the full row rank diagonal blocks of $[\, B_r \; A_r-\lambda E_r\,]$ are provided in \texttt{INFO.mr} and \texttt{INFO.nr}; \index{condensed form!controllability staircase form}%
 \item[(2)] $A_\infty-\lambda E_\infty$, with $A_\infty$ invertible and upper triangular, and $E_\infty$ nilpotent and upper triangular, contains the
     infinite structure and is in a block upper triangular form; the dimensions of the square diagonal blocks of $A_\infty-\lambda E_\infty$ are provided in \texttt{INFO.minf};
 \item[(3)] $A_f-\lambda E_f$, with $E_f$ invertible,  contains the finite structure; the dimension of the square regular pencil $A_f-\lambda E_f$ is provided in \texttt{INFO.mf}.
 \item[(4)] $\left[\begin{smallmatrix} A_l-\lambda E_l\\C_l\end{smallmatrix}\right]$, with $E_l$ invertible and upper triangular, contains the left Kronecker structure and is in an observability staircase form with $\left[\begin{smallmatrix} A_l \\C_l\end{smallmatrix}\right]$ as in (\ref{oscf-defac}) and $E_l$ as in (\ref{oscf-defe}); the dimensions of the full column rank subdiagonal blocks of $\left[\begin{smallmatrix} A_l-\lambda E_l\\C_l\end{smallmatrix}\right]$ are provided in \texttt{INFO.ml} and \texttt{INFO.nl}.
     \index{condensed form!observability staircase form}%
\end{itemize}
Depending on the option selected via the option parameter \texttt{JOBOPT}, several Kronecker-like forms can be determined, which contains basically the same blocks, however ordered differently, or contains only a part of the main structural blocks, which are relevant for particular applications (e.g., nullspace computation).

The following Kronecker-like forms can be computed:
\begin{enumerate}
\item
\texttt{JOBOPT = 'standard'}: to determine the (standard) Kronecker-like form (\ref{gklf:standard}).
\item
\texttt{JOBOPT = 'reverse'}: to determine a Kronecker-like form with a reversed order of the regular blocks
\be\label{gklf:reverse}
\widetilde A -\lambda \widetilde E =
\ba{ccccc} B_r & A_r-\lambda E_r & \ast & \ast & \ast \\
0 & 0 & A_f -\lambda E_f & \ast & \ast \\
0 & 0 & 0 & A_\infty-\lambda E_\infty & \ast \\
0 & 0 & 0 & 0 & A_l-\lambda E_l\\
0 & 0 & 0 & 0 & C_l \ea .\ee
\item
\texttt{JOBOPT = 'right'}: to determine a Kronecker-like form emphasizing the right and infinite structures
\be\label{gklf:right}
\widetilde A -\lambda \widetilde E =
\ba{cccc} B_r & A_r-\lambda E_r & \ast & \ast  \\
0 & 0 & A_\infty-\lambda E_\infty & \ast  \\
0 & 0 & 0 & A_{f,l} -\lambda E_{f,l}
\ea, \ee
where $A_{f,l} -\lambda E_{f,l}$ contains the finite and left Kronecker structures.
 In this case, \texttt{INFO.ml} and \texttt{INFO.nl} are set to the negative values of the
 row and column dimensions of the pencil $A_{f,l} -\lambda E_{f,l}$, respectively, and
 \texttt{INFO.mf} = 0.
\item
\texttt{JOBOPT = 'left'}: to determine a Kronecker-like form emphasizing the left and infinite structures
\be\label{gklf:left}
\widetilde A -\lambda \widetilde E =
\ba{ccc} A_{r,f}-\lambda E_{r,f} & \ast & \ast  \\
0 & A_\infty-\lambda E_\infty & \ast \\
0 & 0 & A_l-\lambda E_l\\
0 & 0 & C_l \ea ,\ee
 where $A_{r,f}-\lambda E_{r,f}$ contains the right and finite Kronecker structures.
 In this case, \texttt{INFO.mr} and \texttt{INFO.nr} are set to the negative values of the
 row and column dimensions of the pencil $A_{r,f}-\lambda E_{r,f}$, respectively, and
 \texttt{INFO.mf} = 0.

\end{enumerate}

\subsubsection{\texttt{\bfseries gsklf}}
\index{M-functions!\texttt{\bfseries gsklf}}

\subsubsection*{Syntax}
\begin{verbatim}
[AT,ET,DIMSC,Q,Z] = gsklf(SYS,TOL,ZEROSEL)
[AT,ET,DIMSC,Q,Z] = gsklf(SYS,TOL,ZEROSEL,OFFSET)
[AT,ET,DIMSC,Q,Z] = gsklf(SYS,TOL,...,QOPT)
\end{verbatim}

\subsubsection*{Description}

\noindent \texttt{gsklf} computes several special Kronecker-like forms of the system matrix pencil of a LTI descriptor system.
\index{condensed form!Kronecker-like!special}

\subsubsection*{Input data}

\begin{description}
\item
\texttt{SYS} is a LTI system,  in a  descriptor system state-space form
\be\label{gsklf:sysss}
\begin{aligned}
E\lambda x(t)  &=   Ax(t)+ B u(t) ,\\
y(t) &=  C x(t)+ D u(t) .
\end{aligned}
\ee
\item
 \texttt{TOL} is a relative tolerance used for rank determinations. If \texttt{TOL} is not specified as input or if \texttt{TOL} = 0, an internally computed default value is used.\\[-3mm]
\item
 \texttt{ZEROSEL}  is a character option variable to specify various zero selection options for the computation of the special Kronecker-like form (\ref{gsklf:spec-klf}) using various partitions of the form  (\ref{gsklf:Cgoodbad}) of the extended complex plane (see \textbf{Method}). The valid options are:  \\[-7mm]
 {\tabcolsep=1mm
\begin{longtable}{lcp{12cm}}
   ~~~~~~~~\texttt{'none'} &--& use $\mathds{C}_g = \mathds{C} \cup \{\infty\}$ and $\mathds{C}_b = \emptyset$  (default); \\
   ~~~~~~~~\texttt{'unstable'}&--& use $\mathds{C}_g = \overline{\mathds{C}}_s$ and $\mathds{C}_b = \mathds{C}_u$; \\
   ~~~~~~~~\texttt{'s-unstable'}&--& use $\mathds{C}_g = {\mathds{C}}_s$ and $\mathds{C}_b = \overline{\mathds{C}}_u$; \\
   ~~~~~~~~\texttt{'stable'}&--& use $\mathds{C}_g = \overline{\mathds{C}}_u$ and $\mathds{C}_b = \mathds{C}_s$; \\
   ~~~~~~~~\texttt{'all'}&--& use $\mathds{C}_g = \emptyset$ and $\mathds{C}_b = \mathds{C} \cup \{\infty\}$;\\
   ~~~~~~~~\texttt{'finite'}&--& use $\mathds{C}_g = \{\infty\}$ and $\mathds{C}_b = \mathds{C}$;\\
   ~~~~~~~~\texttt{'infinite'}&--& use $\mathds{C}_g = \mathds{C}$ and $\mathds{C}_b =  \{\infty\}$.\\
   \end{longtable}
}
\item
 \texttt{OFFSET}  is the stability
 boundary offset $\beta$, to be used  to assess the finite zeros which belong to $\partial\mathds{C}_s$ (the boundary of the stability domain) as follows: in the
 continuous-time case these are the finite
  zeros having real parts in the interval $[-\beta, \beta]$, while in the
 discrete-time case these are the finite zeros having moduli in the
 interval $[1-\beta, 1+\beta]$. \newline (Default: $\beta = 1.4901\cdot 10^{-08}$). \item
 \texttt{QOPT}  is a character option variable to specify the options to accumulate or not the left orthogonal transformations in $Q$: \\[-7mm]
 {\tabcolsep=2mm
\begin{longtable}{lcp{12cm}}
   ~~\texttt{'Q'} &--& accumulate $Q$ (default); \\
   ~~\texttt{'noQ'}&--& do not accumulate $Q$. \\
   \end{longtable}
}

\end{description}

\subsubsection*{Output data}

\begin{description}
\item
\texttt{AT,ET} contain the matrices $\widetilde A$ and $\widetilde E$ which define the resulting system matrix pencil $\widetilde A -\lambda \widetilde E$ in a special Kronecker-like form (\ref{gsklf:spec-klf}), satisfying
\be\label{gsklf:AE}
 \widetilde A -\lambda \widetilde E = \ba{cc} Q^T & 0 \\ 0 & I_p \ea \ba{cc} A - \lambda E & B \\ \hline C & D \ea Z , \ee
where $Q$ and $Z$ are orthogonal transformation matrices.
\item
\texttt{DIMSC} is a five-dimensional (integer) vector. \texttt{DIMSC(1:4)} contain the column dimensions of the matrices $A_{rg}$, $A_{b\ell}$, $B_{b\ell}$ and $B_n$, respectively, in the special Kronecker-like form (\ref{gsklf:spec-klf}). The column dimension \texttt{DIMSC(3)} of $B_{b\ell}$ represents the normal rank of the transfer function matrix $G(\lambda)$ of the system (\ref{gsklf:sysss}). \texttt{DIMSC(5)}, if non-negative, contains the number of finite zeros of \texttt{SYS} on the boundary of the
  stability region $\partial\mathds{C}_s$, within the offset $\beta$ specified by \texttt{OFFSET}. \texttt{DIMSC(6)}, if nonnegative, contains the number of infinite zeros of \texttt{SYS} in the continuous-time case and is set to 0 in the discrete-time case.  Nonnegative values of \texttt{DIMSC(5)} and \texttt{DIMSC(6)} may only result for \texttt{OPTIONS.ZEROSEL} set to \texttt{'unstable'}, \texttt{'stable'} or \texttt{'s-unstable'}.
\item
\texttt{Q,Z} contain the orthogonal matrices $Q$ and $Z$ used to compute the special Kronecker-like form $\widetilde A -\lambda \widetilde E$ in (\ref{gsklf:spec-klf}). $Q$ and $Z$ are not accumulated if both outputs \texttt{Q} and \texttt{Z} are not specified. If \texttt{QOPT = 'noQ'}, $Q$ is not accumulated and, if specified, set to \texttt{Q = [ ]}.
\end{description}

\subsubsection*{Method}
Consider a disjunct partition of the extended complex plane $\mathds{C}_e = \mathds{C} \cup \{\infty\}$ as
\be\label{gsklf:Cgoodbad}  \mathds{C}_e = \mathds{C}_g \cup \mathds{C}_b, \quad \mathds{C}_g \cap \mathds{C}_b = \emptyset \, ,\ee
where $\mathds{C}_g$ and $\mathds{C}_b$ are symmetric with respect to the real axis. Assume that the descriptor system realization (\ref{gsklf:sysss}) is infinite controllable (but $\mathds{C}_b$-{stabilizability is not assumed)} and let $G(\lambda)$ be the $p\times m$  transfer function matrix of the system (\ref{gsklf:sysss}) having normal rank $r$.   Then, there exist two orthogonal matrices $Q$ and $Z$ such that
\be\label{gsklf:spec-klf}
\ba{cc} Q^T & 0 \\ 0 & I \ea \ba{cc} A - \lambda E & B \\ \hline C & D \ea Z =
\ba{cccc} A_{rg}-\lambda E_{rg} & \ast & \ast & \ast \\
0 & A_{b\ell}-\lambda E_{b\ell} & B_{b\ell} & \ast \\
0 & 0 & 0 & B_n \\ \hline
0 & C_{b\ell} & D_{b\ell} & \ast \ea , \ee
where
\begin{itemize}
\item[(a)] The pencil $A_{rg}-\lambda E_{rg}$ has full row rank for $\lambda \in \mathds{C}_g$ and $E_{rg}$ has full row rank.
\item[(b)]  $E_{b\ell}$ and $B_n$ are invertible, the pencil
\be\label{bl_syspencil} S_{bl}(\lambda) := \ba{cc}   A_{b\ell}-\lambda E_{b\ell} & B_{b\ell} \\ C_{b\ell} & D_{b\ell} \ea \ee
has full column rank $n_{b\ell}+r$ in $\mathds{C}_g$ (the pair $(A_{b\ell}-\lambda E_{b\ell}, B_{b\ell})$ is $\mathds{C}_b$-stabilizable if the descriptor system realization (\ref{gsklf:sysss}) is $\mathds{C}_b$-stabilizable).
\end{itemize}
The pencil $S_{bl}(\lambda)$ is the system matrix pencil of a descriptor system $(A_{b\ell}-\lambda E_{b\ell},B_{b\ell},C_{b\ell},D_{b\ell})$, whose  proper transfer function matrix $R(\lambda)$ is a range space basis matrix of $G(\lambda)$ \cite{Varg17f}.
By construction, the zeros of $R(\lambda)$ are those zeros of $G(\lambda)$ which are contained in $\mathds{C}_b$, and can be selected via the option parameter \texttt{ZEROSEL}.

The computation of  the special Kronecker-like form (\ref{gsklf:spec-klf}) is based on the method proposed in \cite{Oara05}. The function \texttt{gsklf} calls the more general function \texttt{gklf} to compute standard Kronecker-like forms, which is based on the mex-function \url{sl_klf}.
\index{MEX-functions!\url{sl_klf}}%

\subsubsection{\texttt{\bfseries gsorsf}}
\index{M-functions!\texttt{\bfseries gsorsf}}

\subsubsection*{Syntax}
\begin{verbatim}
[AT,ET,Q,Z,DIMS,NI] = gsorsf(A,E,OPTIONS)
\end{verbatim}

\subsubsection*{Description}

\noindent \texttt{gsorsf} computes, for a pair of square matrices $(A,E)$, an orthogonally similar pair $(\widetilde A,\widetilde E)$ in a specially ordered GRSF.
\index{condensed form!generalized real Schur (GRSF)!specially ordered}%

\subsubsection*{Input data}

\begin{description}
\item
\texttt{A,E} are $n\times n$ real matrices $A$ and $E$, which define a regular linear matrix pencil $A-\lambda E$.
\item
 \texttt{OPTIONS} is a MATLAB structure to specify user options and has the following fields:\\[5mm]
{\tabcolsep=1mm
\begin{longtable}{|l|lcp{11cm}|} \hline
\textbf{\texttt{OPTIONS} fields} & \multicolumn{3}{l|}{\textbf{Description}} \\ \hline
\texttt{tol}   & \multicolumn{3}{p{11cm}|}{tolerance for the singular values
                        based rank determination of $E$ (Default: $n^2\|E\|_1$\texttt{eps})} \\ \hline
 \texttt{disc}   & \multicolumn{3}{p{11cm}|}{disc option for the ``good'' region $\mathds{C}_g$:}\\
                 &  \texttt{true}&--& $\mathds{C}_g$ is a disc centered in the origin; \\                  &  \texttt{false} &--& $\mathds{C}_g$ is a left half complex plane (default). \\
                                        \hline
 \texttt{smarg}   & \multicolumn{3}{p{12.5cm}|}{stability margin $\beta$, which defines the stability region $\mathds{C}_g$ of the eigenvalues of $A-\lambda E$, as follows:
                  if \texttt{OPTIONS.disc = false}, the stable eigenvalues
                  have real parts less than or equal to $\beta$, and
                  if \texttt{OPTIONS.disc = true}, the stable eigenvalues
                  have moduli less than or equal to $\beta$. \newline
                  (Default: \texttt{-sqrt(eps)} if \texttt{OPTIONS.disc = false}; \newline
                  \phantom{(Default:} \texttt{1-sqrt(eps}) if \texttt{OPTIONS.disc = true}.)} \\ \hline
 \texttt{reverse}   & \multicolumn{3}{p{11cm}|}{option for reverse ordering of diagonal blocks:}\\
                 &  \texttt{true} &--& the diagonal blocks are as in (\ref{bad-good-inf-spec}) (see \textbf{Method}); \\
                 &  \texttt{false}&--& the diagonal blocks are as in (\ref{inf-good-bad-spec}) (see \textbf{Method})  (default).  \\
                                        \hline
  \texttt{sepinf}   & \multicolumn{3}{p{12.3cm}|}{option for the separation of higher order infinite eigenvalues:}\\
                 &  \texttt{true} &--& separate higher order generalized infinite
                            eigenvalues in the trailing positions if
                            \texttt{OPTIONS.reverse = false}, or in the leading
                            positions if \texttt{OPTIONS.reverse = true} (default); \\
                 &  \texttt{false}&--& no separation of higher order infinite
 eigenvalues. \\
                                        \hline
\texttt{fast}   & \multicolumn{3}{p{12.3cm}|}{option for fast separation of higher
                    order infinite eigenvalues (to be used in conjuntion
                    with the \texttt{OPTIONS.sepinf = true}):}\\
                 &  \texttt{true} &--& fast separation of higher order infinite
                            eigenvalues using orthogonal pencil
                            manipulation techniques with QR decompositon
                            based rank determinations (default); \\
                 &  \texttt{false}&--& separation of higher order infinite
                            eigenvalues, by using SVD-based rank determinations
                            (potentially more reliable, but slower).  \\
                                        \hline
\end{longtable}}

\end{description}

\subsubsection*{Output data}

\begin{description}
\item
\texttt{AT,ET} contain the transformed matrices $\widetilde A$ and $\widetilde E$, obtained as
\be\label{gsorsf:AE}
 \widetilde A  = Q^TAZ, \qquad \widetilde E = Q^TEZ , \ee
where $Q$ and $Z$ are orthogonal transformation matrices. The pair $(\widetilde A,\widetilde E)$ is in a specially ordered GRSF, with the regular pencil $\widetilde A -\lambda \widetilde E$ in the form (\ref{inf-good-bad-spec}), if \texttt{OPTIONS.reverse = false}, or in the form (\ref{bad-good-inf-spec}), if \texttt{OPTIONS.reverse = true}.
\item
\texttt{Q,Z} contain the orthogonal matrices $Q$ and $Z$ used to compute the pair $(\widetilde A, \widetilde E)$ in (\ref{gsorsf:AE}), in a specially ordered GRSF.
\item
\texttt{DIMS(1:4)} contain the orders of the diagonal blocks $(A_\infty,A_g,A_{b,f},A_{b,\infty})$,
    if \texttt{OPTIONS.reverse = false}, or of the diagonal blocks $(A_{b,\infty},A_{b,f},A_g,A_\infty)$,
    if \texttt{OPTIONS.reverse = true} (see \textbf{Method}).
\item
\texttt{NI} contains the dimensions of the square diagonal blocks of $A_{b,\infty}-\lambda E_{b,\infty}$, which characterize the higher order infinite eigenvalues of the pair $(A,E)$. \texttt{NI} is empty if no higher order infinite eigenvalues exist.
\end{description}

\subsubsection*{Method}
Consider a disjunct partition of the complex plane $\mathds{C}$ as
\be\label{gsorsf:Cgoodbad}  \mathds{C} = \mathds{C}_g \cup \mathds{C}_b, \quad \mathds{C}_g \cap \mathds{C}_b = \emptyset \, ,\ee
where both $\mathds{C}_g$ and $\mathds{C}_b$ are symmetrically located with respect to the real axis, and $\infty \in \mathds{C}_b$. The complex domains $\mathds{C}_g$ and $\mathds{C}_b$  are typically associated with the ``good'' and ``bad'' generalized eigenvalues of the matrix pair $(A,E)$, respectively. Using orthogonal similarity transformations as in (\ref{gsorsf:AE}), the pair $(A,E)$ can be reduced to the specially ordered GRSF
\be\label{inf-good-bad-spec}\hspace*{-1.5mm} \widetilde A -\lambda \widetilde E = {\arraycolsep=0.2mm\ba{cccc} A_\infty & \ast &\ast&\ast \\ 0 & A_g\!-\!\lambda E_g & \ast &\ast\\ 0 & 0 & A_{b,f}\! -\! \lambda E_{b,f} &\ast \\ 0 & 0 & 0 & A_{b,\infty}\! -\! \lambda E_{b,\infty}
\ea} , \ee
where:  $(i)$ $A_{\infty}$ is an $(n-r)\times (n-r)$ invertible (upper triangular) matrix, with $r = \rank  E$; the leading pair $(A_\infty,0)$ contains all infinite eigenvalues of $A-\lambda E$ corresponding to first-order eigenvectors; $(ii)$ $A_g$ and $E_g$ are $n_g\times n_g$ matrices, such that the pair $(A_g,E_g)$,  with $E_g$ invertible, is in a GRSF (i.e., $A_g$ upper quasi-triangular and $E_g$ upper triangular)  and $\Lambda(A_g-\lambda E_g)$ are the finite eigenvalues lying in $\mathds{C}_g$; $(iii)$ $A_{b,f}$ and $E_{b,f}$ are $n_b^f\times n_b^f$ matrices, such that the pair $(A_{b,f},E_{b,f})$,  with $E_{b,f}$ invertible,  is in a GRSF and  $\Lambda(A_{b,f}-\lambda E_{b,f})$ are the finite eigenvalues lying in $\mathds{C}_b$; and $(iv)$ $A_{b,\infty}$ and $E_{b,\infty}$ are $n_b^\infty\times n_b^\infty$ upper triangular matrices, with $A_{b,\infty}$ invertible and $E_{b,\infty}$ nilpotent,  and $\Lambda(A_{b,\infty}-\lambda E_{b,\infty})$ are the higher order infinite eigenvalues.  The orders $[n-r, n_g, n_b^f, n_b^\infty]$ of the diagonal blocks $(A_\infty,A_g,A_{b,f},A_{b,\infty})$, respectively, are provided in \texttt{DIMS(1:4)}. $A_{b,\infty}-\lambda E_{b,\infty}$ has a block upper triangular form with $k$ diagonal blocks, whose increasingly ordered dimensions are provided in $\texttt{NI}(1\!:\!k)$.
The dimensions $\texttt{NI}(1\!:\!k)$ define the multiplicities of the higher order infinite eigenvalues as follows: for $i = 1, \ldots, k$, there are $\texttt{NI}(k-i+1)-\texttt{NI}(k-i)$ infinite eigenvalues of multiplicity $i$, where \texttt{NI}$(0) := 0$. The multiplicity of simple infinite eigenvalues is $\texttt{DIMS}(1)-\texttt{NI}(k)$.

To obtain the specially ordered GRSF (\ref{inf-good-bad-spec}), the
\textbf{Procedure GSORSF}, described in \cite{Varg17d}, can be used. For the separation of the higher order infinite generalized eigenvalues, the  mex-function \url{sl_klf} is called by \texttt{gsorsf}, if \texttt{OPTIONS.fast = true}. This function employs rank determinations based on QR-decompositions with column pivoting and, therefore, is more efficient than an alternative, potentially more reliable approach, based on SVD-based rank determination. This latter approach is performed if \texttt{OPTIONS.fast = false}.
\index{MEX-functions!\url{sl_klf}}%

The same procedure, applied to the transposed pair $(A^T,E^T)$, is used to obtain  the specially ordered GRSF with a reverse ordering of the blocks
\be\label{bad-good-inf-spec}\hspace*{-1.5mm} \widetilde A -\lambda \widetilde E = {\arraycolsep=0.2mm\ba{cccc} A_{b,\infty}\! -\! \lambda E_{b,\infty} & \ast &\ast&\ast \\ 0 & A_{b,f}\! -\! \lambda E_{b,f}& \ast &\ast\\ 0 & 0 & A_g\!-\!\lambda E_g  &\ast \\ 0 & 0 & 0 & A_\infty
\ea} . \ee
The orders $[n_b^\infty,n_b^f, n_g, n-r]$ of the diagonal blocks $(A_{b,\infty},A_{b,f},A_g,A_\infty)$, respectively, are provided in \texttt{DIMS(1:4)}. $A_{b,\infty}-\lambda E_{b,\infty}$ has a block upper triangular form with $k$ diagonal blocks, whose decreasingly ordered dimensions are provided in $\texttt{NI}(1\!:\!k)$.
The dimensions $\texttt{NI}(1\!:\!k)$ define the multiplicities of the higher order infinite eigenvalues as follows: for $i = 1, \ldots, k$, there are $\texttt{NI}(i)-\texttt{NI}(i+1)$ infinite eigenvalues of multiplicity $i$, where $\texttt{NI}(k+1) := 0$. The multiplicity of simple infinite eigenvalues is $\texttt{DIMS}(4)-\texttt{NI}(1)$.

\subsubsection{\texttt{\bfseries gsfstab}}
\index{M-functions!\texttt{\bfseries gsfstab}}

\subsubsection*{Syntax}
\begin{verbatim}
[F,INFO] = gsfstab(A,E,B,POLES,SDEG,OPTIONS)
\end{verbatim}

\subsubsection*{Description}

\noindent \texttt{gsfstab} computes, for a descriptor system pair $(A-\lambda E,B)$, a state-feedback matrix $F$ such that the controllable finite generalized eigenvalues of the closed-loop pair $(A+BF,E)$ lie in the stability domain $\mathds{C}_s$.

\subsubsection*{Input data}

\begin{description}
\item
\texttt{A,E} are $n\times n$ real matrices $A$ and $E$, which define a regular linear pencil $A-\lambda E$. $E = I$ is assumed, for \texttt{E = [ ]}.
\item
\texttt{B} is a $n\times m$ real matrix $B$.
\item
\texttt{POLES} specifies a complex conjugated set of desired
                       eigenvalues to be assigned for the pair $(A+BF,E)$.
  The specified values are intended to replace the finite generalized eigenvalues of $(A,E)$ lying outside
   of the specified stability domain $\mathds{C}_s$. If the number of specified
   eigenvalues in \texttt{POLES} is less than the number of controllable
   generalized eigenvalues of $(A,E)$ outside of $\mathds{C}_s$, then the rest of
   generalized eigenvalues of $(A+BF,E)$ are assigned to the nearest values
   on the boundary of $\mathds{C}_s$ (defined by the parameter \texttt{SDEG}, see below) or are kept unmodified if
   \texttt{SDEG = [ ]}. (Default: \texttt{POLES = [ ]})
\item
\texttt{SDEG}  specifies a prescribed stability degree for the
                       eigenvalues of the pair $(A+BF,E)$. For a continuous-time setting, with $\texttt{SDEG} < 0$, the stability domain
   $\mathds{C}_s$ is the set of complex numbers with real parts at most \texttt{SDEG},
   while for a discrete-time setting, with $0 \leq \texttt{SDEG} < 1$, $\mathds{C}_s$ is the set
   of complex numbers with moduli at most \texttt{SDEG} \newline
     (Default: $-0.2$, used if \texttt{SDEG = [ ]} and \texttt{POLES = [ ]}).
\item
 \texttt{OPTIONS} is a MATLAB structure to specify user options and has the following fields:\\
{\tabcolsep=1mm
\begin{longtable}{|l|lcp{11cm}|} \hline
\textbf{\texttt{OPTIONS} fields} & \multicolumn{3}{l|}{\textbf{Description}} \\ \hline
\texttt{tol}   & \multicolumn{3}{p{11cm}|}{tolerance for rank determinations (Default: internally computed)}\\ \hline
  \texttt{sepinf}   & \multicolumn{3}{p{12.3cm}|}{option for a preliminary separation
                    of the infinite eigenvalues:}\\
                 &  \texttt{true} &--& perform preliminary separation of the infinite
                            generalized eigenvalues from the finite ones ; \\
                 &  \texttt{false}&--& no separation of infinite generalized
 eigenvalues (default) \\
                                        \hline
\end{longtable}}

\end{description}

\subsubsection*{Output data}

\begin{description}
\item
\texttt{F} contains the resulting $m\times n$ state-feedback gain $F$.
\item
\texttt{INFO} is a MATLAB structure, which provides additional information on the closed-loop matrices
   $A_{cl} := Q(A+BF)Z$, $E_{cl} := QEZ$, and $B_{cl} := QB$, where $Q$ and $Z$ are the
   orthogonal matrices used to obtain the pair $(A_{cl},E_{cl})$ in a GRSF. The fields of the INFO structure are: \\
\hspace*{3mm}{\begin{longtable}{|l|p{12cm}|} \hline
\textbf{\texttt{INFO} fields} & \textbf{Description} \\ \hline
    \texttt{Acl} & contains $A_{cl}$ in a quasi-upper triangular form; \\  \hline
    \texttt{Ecl}& contains $E_{cl}$ in an upper triangular form; \\ \hline
    \texttt{Bcl}&  contains $B_{cl}$ ; \\ \hline
    \texttt{Q,Z} &  contains the orthogonal transformation matrices $Q$ and $Z$. \\ \hline
    \texttt{ninf}&  number of infinite generalized eigenvalues of $(A,E)$; \\ \hline
    \texttt{nfg}&  number of finite generalized eigenvalues of $(A,E)$
                  lying in the stability domain $\mathds{C}_s$; \\ \hline
    \texttt{naf}&  number of assigned finite generalized eigenvalues in $\mathds{C}_s$; \\ \hline
    \texttt{nuf}&  number of uncontrollable finite generalized eigenvalues
                   lying outside $\mathds{C}_s$. \\ \hline
    \end{longtable}}
\end{description}

\subsubsection*{Method}
For a standard system pair $(A,I)$ (for \texttt{E = [ ]}), the Schur method
  of \cite{Varg81b} is used, while for a generalized system pair $(A,E)$ the
  generalized Schur method of \cite{Varg95} is used. The resulting
  closed-loop matrices
   $A_{cl} := Q(A+BF)Z$, $E_{cl} := QEZ$, and $B_{cl} := QB$, where $Q$ and $Z$ are the
   orthogonal matrices used to obtain the pair $(A_{cl},E_{cl})$ in a GRSF, have the forms
\be\label{gsorsf:cl} {\arraycolsep=1mm A_{cl} = \ba{cccc}A_\infty & \ast & \ast & \ast \\
0 & A_{f,g} & \ast & \ast\\ 0 & 0 & A_{f,a} & \ast\\ 0 & 0 & 0 & A_{f,u} \ea, \quad
E_{cl} = \ba{cccc}E_\infty & \ast & \ast & \ast \\
0 & E_{f,g} & \ast & \ast\\ 0 & 0 & E_{f,a} & \ast\\ 0 & 0 & 0 & E_{f,u} \ea, \quad
B_{cl} = \ba{c} \ast\\ \ast \\ \ast \\ 0 \ea }, \ee
   where: $(i)$ the pair $(A_\infty,E_\infty)$, with $A_\infty$ upper triangular and invertible and $E_\infty$ upper
   triangular and nilpotent, contains the \texttt{INFO.ninf} infinite generalized
   eigenvalues of $(A,E)$; $(ii)$ the pair $(A_{f,g},E_{f,g})$, in GRSF, contains the \texttt{INFO.nfg} finite
   generalized eigenvalues of $(A,E)$ in $\mathds{C}_s$; $(iii)$ the pair $(A_{f,a},E_{f,a})$, in GRSF, contains
   the \texttt{INFO.naf} assigned finite generalized eigenvalues in $\mathds{C}_s$; and, $(iv)$ the pair $(A_{f,u},E_{f,u})$, in GRSF, contains the uncontrollable finite generalized
   eigenvalues of $(A,E)$ lying outside $\mathds{C}_s$. For the separation of the infinite generalized eigenvalues, the  mex-function \url{sl_klf} is called by \texttt{gsfstab}. \index{MEX-functions!\url{sl_klf}}%

\bookmarksetup{startatroot}
\cleardoublepage
\phantomsection

\appendix
\newpage
\section{Installing DSTOOLS} \label{appA}

\textbf{DSTOOLS} runs  with MATLAB R2015b (or later versions) under 64-bit Windows 7 (or later). Additionally, the \emph{Control System Toolbox} (Version 9.10 or later) is necessary to be installed.
To install \textbf{DSTOOLS},  perform the following steps:
\begin{itemize}
\item  download \textbf{DSTOOLS} as a zip file from Bitbucket\footnote{\url{https://bitbucket.org/DSVarga/dstools}}
\item create on your computer the directory \texttt{dstools}
\item  extract, using any unzip utility, the functions of the \textbf{DSTOOLS} collection in the corresponding directory  \texttt{dstools}
\item  start MATLAB and put the directory \texttt{dstools} on the MATLAB path, by using the \texttt{pathtool} command; for repeated use, save the new MATLAB search path, or alternatively, use the \texttt{addpath} command to set new path entries in \texttt{startup.m}
\item try out the installation by running the demonstration script \texttt{DSToolsdemo.m}
\end{itemize}

\newpage
\section{Current \texttt{Contents.m} File }\label{app:contents}

The M-functions available in the current version of \textbf{FDITOOLS}  are listed in the current version of the \texttt{Contents.m} file, given below:
{\small
\begin{verbatim}

% DSTOOLS - Descriptor System Tools.
% Version 0.71            30-Sep-2018
% Copyright (c) 2016-2018 by A. Varga
%
% Demonstration.
%   DSToolsdemo - Demonstration of DSTOOLS.
%
% System analysis.
%   gpole     - Poles of a LTI descriptor system.
%   gzero     - Zeros of a LTI descriptor system.
%   gnrank    - Normal rank of the transfer function matrix of a LTI system.
%   ghanorm   - Hankel norm of a proper and stable LTI descriptor system.
%   gnugap    - Nu-gap distance between two LTI systems.
%
% System order reduction.
%   gir       - Reduced order realizations of LTI descriptor systems.
%   gminreal  - Minimal realization of a LTI descriptor system.
%   gbalmr    - Balancing-based model reduction of a LTI descriptor system.
%   gss2ss    - Conversions to SVD-like forms without non-dynamic modes.
%
% Operations on transfer function matrices.
%   grnull    - Right nullspace basis of a transfer function matrix.
%   glnull    - Left nullspace basis of a transfer function matrix.
%   grange    - Range space basis of a transfer function matrix.
%   gcrange   - Coimage space basis of a transfer function matrix.
%   grsol     - Solution of the linear rational matrix equation G*X = F.
%   glsol     - Solution of the linear rational matrix equation X*G = F.
%   gsdec     - Generalized additive spectral decompositions.
%   grmcover1 - Right minimum dynamic cover of Type 1 based order reduction.
%   glmcover1 - Left minimum dynamic cover of Type 1 based order reduction.
%   grmcover2 - Right minimum dynamic cover of Type 2 based order reduction.
%   glmcover2 - Left minimum dynamic cover of Type 2 based order reduction.
%   gbilin    - Generalized bilinear transformation.
%   gbilin1   - Transfer functions of commonly used bilinear transformations.
%
% Factorizations of transfer function matrices.
%   grcf      - Right coprime factorization with proper and stable factors.
%   glcf      - Left coprime factorization with proper and stable factors.
%   grcfid    - Right coprime factorization with inner denominator.
%   glcfid    - Left coprime factorization with inner denominator.
%   gnrcf     - Normalized right coprime factorization.
%   gnlcf     - Normalized left coprime factorization.
%   giofac    - Inner-outer/QR-like factorization.
%   goifac    - Co-outer-coinner/RQ-like factorization.
%   grsfg     - Right spectral factorization of gamma^2*I-G'*G.
%   glsfg     - Left spectral factorization of gamma^2*I-G*G'.
%
% Model-matching problem.
%   grasol    - Approximate solution of the linear rational matrix equation G*X = F.
%   glasol    - Approximate solution of the linear rational matrix equation X*G = F.
%   glinfldp  - Solution of the least distance problem min||G1-X G2||_inf.
%   gnehari   - Generalized Nehari approximation.
%
% Feedback stabilization.
%   gsfstab    - Generalized state-feedback stabilization.
%   eigselect1 - Selection of a real eigenvalue to be assigned.
%   eigselect2 - Selection of a pair of eigenvalues to be assigned.
%   galoc2     - Generalized pole allocation for second order systems.
%
% Pencil similarity transformations
%   gklf      - Kronecker-like staircase forms of a linear matrix pencil.
%   gsklf     - Special Kronecker-like form of a system matrix pencil.
%   gsorsf    - Specially ordered generalized real Schur form.
%
% SLICOT-based mex-functions.
%   sl_gstra  - Descriptor system coordinate transformations.
%   sl_gminr  - Minimal realization of descriptor systems.
%   sl_gsep   - Descriptor system additive spectral decompositions.
%   sl_gzero  - Computation of system zeros and Kronecker structure.
%   sl_klf    - Pencil reduction to Kronecker-like forms.
%   sl_glme   - Solution of generalized linear matrix equations.
\end{verbatim}
}

\newpage
\section{DSTOOLS Release Notes} \label{appB}

The \textbf{DSTOOLS} Release Notes describe the changes introduced in the successive versions of the \textbf{DSTOOLS} collection, as new features, enhancements to functions, or major bug fixes. The following versions of \textbf{DSTOOLS} have been released:\\

\begin{tabular}{llp{9cm}}
Version & Release date & Comments\\ \hline
V0.5 &  December 31, 2016 & Initial version accompanying the book \cite{Varg17}.\\
V0.6 &  July 31, 2017 & New function for range computation, many enhancements of existing functions. \\
V0.61 &  January 29, 2018 & New function for coimage computation\\
V0.64 &  July 31, 2018 & New functions for approximate solutions of linear rational equations.\\
V0.7 &  August 23, 2018 & New functions for normalized coprime factorizations and bilinear transformation, and several enhancements of existing functions.\\
V0.71 &  September 30, 2018 & New function for the computation of $\nu$-gap metric, and several enhancements of existing functions related to reliably detecting poles and zeros on the stability domain boundary. \\
\hline
\end{tabular}

\subsection{Release Notes V0.5 }
This is the initial version of the \textbf{DSTOOLS} collection of M- and MEX-functions, which accompanies the book \cite{Varg17}. All numerical results presented in this book have been obtained using this version of \textbf{DSTOOLS}.

\subsubsection{New Features}
The M-functions and MEX-functions available in the Version 0.5 of \textbf{DSTOOLS} are listed below, where we kept the  originally employed descriptions of the functions:

\begin{verbatim}
% DSTOOLS - Descriptor System Tools.
% Version 0.5             31-Dec-2016
% Copyright (c) 2016 by A. Varga
%
% Demonstration.
%   DSToolsdemo - Demonstration of DSTOOLS.
%
% System analysis.
%   gpole     - Poles of a LTI descriptor system.
%   gzero     - Invariant zeros and Kronecker structure of a system pencil.
%   nrank     - Normal rank of a transfer function matrix of a LTI system.
%   ghanorm   - Hankel norm of a proper and stable LTI descriptor system.
%
% Order reduction.
%   gir       - Irreducible realizations of LTI descriptor systems.
%   gminreal  - Minimal realization of a LTI descriptor system.
%   gbalmr    - Balancing-based model reduction of a stable descriptor system.
%   gss2ss    - Conversions to SVD-like forms without non-dynamic modes.
%
% Operations on generalized LTI systems.
%   glnull    - Left nullspace basis of a rational matrix.
%   grnull    - Right nullspace basis of a rational matrix.
%   glsol     - Solution of linear rational equation X(s)*G(s)=F(s).
%   grsol     - Solution of linear rational equation G(s)*X(s)=F(s).
%   gsdec     - Generalized additive spectral decompositions.
%   glmcover1 - Left minimum dynamic cover of Type 1 based order reduction.
%   grmcover1 - Right minimum dynamic cover of Type 1 based order reduction.
%   glmcover2 - Left minimum dynamic cover of Type 2 based order reduction.
%   grmcover2 - Right minimum dynamic cover of Type 2 based order reduction.
%

% Factorizations.
%   giofac    - Generalized inner-outer factorization of descriptor systems.
%   goifac    - Generalized co-outer-inner factorization of descriptor systems.
%   glcf      - Generalized left coprime factorization.
%   grcf      - Generalized right coprime factorization.
%   glcfid    - Generalized left coprime factorization with inner denominator.
%   grcfid    - Generalized right coprime factorization with inner denominator.
%   glsfg     - Generalized left spectral factorization of g^2-G*G'.
%   grsfg     - Generalized right spectral factorization of g^2-G'*G.
%
% Approximations.
%   gnehari   - Generalized Nehari approximation.
%   glinfldp  - Solution of the least distance problem min||F1-X F2||_inf.
%
% Feedback stabilization.
%   gsfstab    - Generalized state-feedback stabilization.
%   eigselect1 - Selection of a real eigenvalue to be assigned.
%   eigselect2 - Selection of a pair of eigenvalues to be assigned.
%   galoc2     - Generalized pole allocation for second order systems.
%
% Pencil similarity transformations
%   gklf      - Generalized Kronecker-like staircase form of a linear pencil.
%   gsorsf    - Specially ordered generalized real Schur form.
%
% SLICOT-based mex-functions.
%   sl_gstra  - Descriptor system coordinate transformations.
%   sl_gminr  - Minimal realization of descriptor systems.
%   sl_gsep   - Descriptor system spectral separations.
%   sl_gzero  - Computation of system zeros and Kronecker structure.
%   sl_klf    - Pencil reduction to Kronecker-like forms.
%   sl_glme   - Solution of generalized linear matrix equations.
\end{verbatim}

\subsection{Release Notes V0.6 }
This version of the \textbf{DSTOOLS} includes minor revisions of most functions, by adding exhaustive input parameter checks and performing several simplifications in the codes. Besides this, two new functions have been implemented and the underlying mex-functions have been replaced with new ones, which use an enhanced rank determination strategy.

\subsubsection{New Features}
Two new functions have been added to \textbf{DSTOOLS}:
\begin{verbatim}
% Operations on generalized LTI systems.
%   grange    - Range space basis of a transfer function matrix.
% Pencil similarity transformations
%   gsklf     - Special Kronecker-like form of a system matrix pencil.
\end{verbatim}
The function \texttt{\bfseries grange} to compute a range space basis of a transfer function matrix is based on a special Kronecker-like form of the system matrix pencil, which is computed by  the function \texttt{\bfseries gsklf}.

Several additional changes have been performed in the following functions:
\begin{description}
\item \texttt{\bfseries gpole}: the computations for standard state-space realizations have been separated from those for descriptor system realizations;
\item \texttt{\bfseries nrank}: functionality extended to handle all LTI system objects (i.e., \texttt{ss}, \texttt{tf}, \texttt{zpk});
\item \texttt{\bfseries gir}: \\
-- explicit handling of standard state-space representations added;\\
-- option structure removed and replaced with a character string option parameter;
\item \texttt{\bfseries gminreal}: \\
-- explicit handling of standard state-space representations added;\\
-- logical option parameter replaced with a character string option parameter;
\item \texttt{\bfseries gbalmr}: \\ -- original realization preserved if no order reduction and no balancing take place;\\ -- checking the invertibility of $E$ added;
\item \texttt{\bfseries gss2ss}: \\ -- relies entirely on \url{sl_gstra} to compute the SVD-like coordinate forms;\\ -- exploits the generalized Hessenberg form of the pair $(A,E)$ if $E$ is invertible;
\item \texttt{\bfseries grnull}: stabilization and pole assignment options added;
\item \texttt{\bfseries glnull}: stabilization and pole assignment options added;
\item \texttt{\bfseries grsol}: \texttt{INFO} provides always full structural information in \texttt{INFO.nr}, \texttt{INFO.nf} and \texttt{INFO.ninf};
\item \texttt{\bfseries glsol}: \texttt{INFO} provides always full structural information in \texttt{INFO.ninf}, \texttt{INFO.nf} and \texttt{INFO.nl};
\item \texttt{\bfseries gsdec}: explicit handling of standard state-space representations added;
\item \texttt{\bfseries giofac}: \\
-- \texttt{OPTIONS} structure input parameter introduced to specify several user options;\\
-- new functionality added to compute the QR-like factorization of rational matrices; \\
-- the code sequence to compute a special Kronecker-like form replaced by a call of \texttt{\bfseries gsklf};
\item \texttt{\bfseries goifac}: \\
-- \texttt{OPTIONS} structure input parameter introduced to specify several user options;\\
-- new functionality added to compute the RQ-like factorization of rational matrices; \item \texttt{\bfseries gnehari}: functionality restricted to systems without poles on the stability domain boundary;
\item \texttt{\bfseries gsorsf}: \\ -- the output parameter \texttt{DIMS} contains now the orders of all four diagonal blocks;\\ -- the default option for \texttt{OPTIONS.sepinf} has been redefined;
\end{description}

\subsubsection{Bug Fixes}
Several minor bug fixes have been performed in the following functions:
\begin{description}
\item \texttt{\bfseries grnull}: \\ -- nullspace computation fixed for zero input dimensions;
\item \texttt{\bfseries glnull}: \\ -- nullspace computation fixed for zero output dimensions;
\end{description}

\subsection{Release Notes V0.61 }
This version of the \textbf{DSTOOLS} includes a new implemented function (\textbf{\texttt{gcrange}}), minor enhancements of two functions (\textbf{\texttt{grange}} and \textbf{\texttt{gsklf}}), and a few bug fixes.

\subsubsection{New Features}
A new function \textbf{\texttt{gcrange}} has been implemented for the computation of coimage space bases of  transfer function matrices. This function relies on the new, enhanced versions of the functions \textbf{\texttt{grange}} and \textbf{\texttt{gsklf}}, with new options to handle ``strictly unstable'' system zeros.
\subsubsection{Bug Fixes}
Several minor bug fixes have been performed in the following functions:
\begin{description}
\item \texttt{\bfseries glinfldp}: \\ -- bug fixes in initializing the $\gamma$-iteration;
\item \texttt{\bfseries gbalmr}: \\ -- bug fix to handle non-dynamic reduced systems;
\item \texttt{\bfseries gnehari}: \\ -- bug fix to handle non-dynamic systems.
\end{description}

\subsection{Release Notes V0.64 }
This version of the \textbf{DSTOOLS} includes
two new  functions \textbf{\texttt{grasol}} and \textbf{\texttt{glasol}}, to compute approximate solutions of linear rational equations, minor enhancements of the function \textbf{\texttt{gir}}
and a few bug fixes.

\subsubsection{New Features}
Two new functions have been added to \textbf{DSTOOLS}:
\begin{verbatim}
% Model-matching problem.
%  grasol  - Approximate solution of the linear rational matrix equation G*X = F.
%  glasol  - Approximate solution of the linear rational matrix equation X*G = F.
\end{verbatim}

The function \textbf{\texttt{gir}} has been extended to handle arbitrary two-dimensional arrays of state-space systems.

\subsubsection{Bug Fixes}
Several minor bug fixes have been performed in the following functions:
\begin{description}
\item \texttt{\bfseries gnehari}: \\ -- extra argument added to specify a tolerance for rank computations.
\item \texttt{\bfseries glinfldp}: \\ -- bug fix when early terminating the $\gamma$-iteration without performing any iteration;
\\ -- an explicit minimum value for a suboptimal $\gamma$ value is provided in the error message;
\\ -- tolerance added in several calls to the function \texttt{gnehari}
\end{description}

\subsection{Release Notes V0.7 }
This version of the \textbf{DSTOOLS} includes
four new  functions, several enhancements of existing functions
and a few bug fixes.

\subsubsection{New Features}
The following new functions have been added to \textbf{DSTOOLS}: \\

\noindent\begin{tabular}{lcp{13cm}}
\texttt{gnrcf} &--& generalized normalized right coprime factorization;\\
\texttt{gnlcf} &--& generalized normalized left coprime factorization; \\
\texttt{gbilin} &--& generalized bilinear transformations; \\
\texttt{gbilin1} &--& generation of first order transfer functions which describe several commonly used bilinear transformations.
\end{tabular} \\

\noindent The functions \textbf{\texttt{grnull}} and \textbf{\texttt{glnull}} have been enhanced to optionally compute inner, respectively, coinner nullspace bases.  \\

\noindent The functions \textbf{\texttt{grange}} and \textbf{\texttt{gcrange}} provide additional information in an INFO structure, as the normal rank of the transfer function matrix, the number of its zeros lying on the boundary of stability region, and the accuracy of the solution of the involved Riccati equation. \texttt{INFO.nrank} is now provided in both functions instead \texttt{INFO.rankG}.   \\

\noindent In the functions \textbf{\texttt{grsol}} and \textbf{\texttt{glsol}}, \texttt{INFO.nrank} is now provided instead \texttt{INFO.rankG}.   \\

\noindent The functions \textbf{\texttt{giofac}} and \textbf{\texttt{goifac}} provide additional information in an INFO structure, as the normal rank of the transfer function matrix, the number of its zeros lying on the boundary of stability region, and the accuracy of the solution of the involved Riccati equation. \\

\noindent The functions \textbf{\texttt{grasol}} and \textbf{\texttt{glasol}} use the new information provided by \textbf{\texttt{giofac}} and \textbf{\texttt{goifac}}, to detect nonstandard problems without performing additional computations. \texttt{INFO.nrank} is now provided in both functions instead \texttt{INFO.rankG}.   \\

\noindent The function \textbf{\texttt{gsklf}} provides additionally the number of finite zeros on the boundary of the stability region and the number of infinite zeros in the continuous-time case.

\subsubsection{Bug Fixes}
Several minor bug fixes have been performed in the following functions:
\begin{description}
\item \texttt{\bfseries gcrange}: \\ -- option \texttt{inner} changed in \texttt{coinner};
\item \texttt{\bfseries glasol}: \\ -- minor enhancements to properly work with empty systems;
\item \texttt{\bfseries grasol}: \\ -- minor enhancements to properly work with empty systems;
\item \texttt{\bfseries glinfldp}: \\ -- minor bug fixes to enhance termination issues of the $\gamma$-iteration;
\end{description}

\subsection{Release Notes V0.71 }
This version of the \textbf{DSTOOLS} includes a new function to evaluate the $\nu$-gap metric,
several enhancements of existing functions related to reliably detecting poles and zeros on the stability domain boundary and
a few bug fixes.

\subsubsection{New Features}
A new function \texttt{gnugap} has been implemented to compute the $\nu$-gap metric for LTI systems. This function is applicable to arbitrary (even improper) systems and relies on the new enhancements of the functions \textbf{\texttt{gpole}} and \textbf{\texttt{gzero}}. \\

\noindent The functions \textbf{\texttt{gpole}} and \textbf{\texttt{gzero}} have been substantially enhanced, by providing extensive additional information on the structure of the underlying pencils and properties of computed poles and zeros, respectively. An additional optional entry can be used  to specify an offset for the stability domain boundary. \\

\noindent The function \textbf{\texttt{gnrank}} replaces \textbf{\texttt{nrank}}. For compatibility purposes, the obsolete function \textbf{\texttt{nrank}} can be still used, but it will be removed in a future version of \texttt{DSTOOLS}. \\

\noindent The functions \textbf{\texttt{grasol}} and \textbf{\texttt{glasol}} have an additional option entry, to specify an offset for the stability domain boundary. \\

\noindent The functions \textbf{\texttt{grange}} and \textbf{\texttt{gcrange}} have an additional option entry, to specify an offset for the stability domain boundary. \\

\noindent The functions \textbf{\texttt{giofac}} and \textbf{\texttt{goifac}} have an additional option entry, to specify an offset for the stability domain boundary. \\

\noindent The function \textbf{\texttt{gsklf}} has an additional input to specify an offset for  the stability domain boundary. For compatibility purposes, the old calling sequence used up to version V0.7 can be still used.

\subsubsection{Bug Fixes}
Several bug fixes and minor improvements have been performed in the following functions:
\begin{description}
\item \texttt{\bfseries gpole}: \\ -- bug fixed in determining the number of eigenvalues to be set to \texttt{NaN} values;
\item \texttt{\bfseries gir}: \\ -- improved performance in handling standard systems;
\item \texttt{\bfseries grnull}: \\ -- stabilizability test added when computing inner basis;
\item \texttt{\bfseries glnull}: \\ -- detectability test added when computing coinner basis;
\item \texttt{\bfseries grange}: \\ -- controllability is enforced by eliminating all uncontrollable eigenvalues;
\item \texttt{\bfseries giofac}: \\ -- controllability is enforced by eliminating all uncontrollable eigenvalues;
\item \texttt{\bfseries gsklf}: \\ -- bug fixed in evaluating the number of unstable finite zeros; \\
    -- bug fixed in detecting the lack of impulse controllability;
\item \texttt{\bfseries grasol}: \\ -- minor bug fixed in handling sampling-time compatibility related issues;
\item \texttt{\bfseries glasol}: \\ -- minor bug fixed in handling sampling-time compatibility related issues;
\end{description}

\newpage
\pdfbookmark[1]{References}{Refs}

\newpage

\pdfbookmark[1]{Index}{Ind}
\begin{theindex}

  \item canonical form
    \subitem Jordan, \hyperpage{12}
    \subitem Kronecker, \hyperpage{14}
    \subitem Weierstrass, \hyperpage{12}
  \item coimage space
    \subitem basis
      \subsubitem coinner, \hyperpage{26}
      \subsubitem minimal proper, \hyperpage{25}
  \item condensed form
    \subitem controllability staircase form, \hyperpage{16},
		\hyperpage{161}
    \subitem generalized real Schur (GRSF), \hyperpage{13},
		\hyperpage{118}, \hyperpage{121}, \hyperpage{124},
		\hyperpage{127}
      \subsubitem specially ordered, \hyperpage{165}
    \subitem Kronecker-like, \hyperpage{16}, \hyperpage{160}
      \subsubitem special, \hyperpage{163}
    \subitem observability staircase form, \hyperpage{16},
		\hyperpage{162}
    \subitem real Schur (RSF), \hyperpage{14}

  \indexspace

  \item descriptor system
    \subitem $\nu$-gap metric, \hyperpage{63}
    \subitem additive decomposition, \hyperpage{26}, \hyperpage{97}
    \subitem bilinear transformation, \hyperpage{114}
    \subitem conjugate, \hyperpage{53}
    \subitem controllability, \hyperpage{12}
    \subitem controllable eigenvalue, \hyperpage{24}
    \subitem coprime factorization, \hyperpage{28}
    \subitem exponential stability, \hyperpage{24}
    \subitem finite controllability, \hyperpage{12}
    \subitem finite detectable, \hyperpage{24}
    \subitem finite observability, \hyperpage{12}
    \subitem finite stabilizable, \hyperpage{24}
    \subitem Hankel norm, \hyperpage{36}, \hyperpage{62}
    \subitem improper, \hyperpage{24}
    \subitem infinite controllability, \hyperpage{12}
    \subitem infinite observability, \hyperpage{12}
    \subitem inverse, \hyperpage{52}
    \subitem irreducible realization, \hyperpage{12}, \hyperpage{64},
		\hyperpage{66}
    \subitem left minimal cover problem, \hyperpage{35}
      \subsubitem Type 1, \hyperpage{105}
      \subsubitem Type 2, \hyperpage{110}
    \subitem linear rational matrix equation, \hyperpage{33},
		\hyperpage{88}, \hyperpage{93}
    \subitem minimal nullspace basis, \hyperpage{20}, \hyperpage{71},
		\hyperpage{76}
    \subitem minimal realization, \hyperpage{12}, \hyperpage{66}
    \subitem normal rank, \hyperpage{24}
    \subitem observability, \hyperpage{12}
    \subitem observable eigenvalue, \hyperpage{24}
    \subitem poles, \hyperpage{24}
    \subitem polynomial, \hyperpage{24}
    \subitem proper, \hyperpage{24}
    \subitem right minimal cover problem, \hyperpage{35}
      \subsubitem Type 1, \hyperpage{35}, \hyperpage{100}
      \subsubitem Type 2, \hyperpage{35}, \hyperpage{107}
    \subitem similarity transformation, \hyperpage{11}
    \subitem strongly detectable, \hyperpage{25}
    \subitem strongly stabilizable, \hyperpage{25}
    \subitem uncontrollable eigenvalue, \hyperpage{24}
    \subitem unobservable eigenvalue, \hyperpage{24}
    \subitem zeros, \hyperpage{24}

  \indexspace

  \item factorization
    \subitem co-outer--coinner, \hyperpage{28}, \hyperpage{30}
      \subsubitem extended, \hyperpage{29}
    \subitem fractional, \hyperpage{27}
    \subitem full rank, \hyperpage{25, 26}, \hyperpage{81},
		\hyperpage{85}
    \subitem inner--outer, \hyperpage{28}, \hyperpage{30}
      \subsubitem extended, \hyperpage{28}
    \subitem inner--quasi-outer, \hyperpage{28}
      \subsubitem extended, \hyperpage{28}, \hyperpage{133}
    \subitem left coprime (LCF), \hyperpage{27}, \hyperpage{120}
      \subsubitem minimum-degree denominator, \hyperpage{27},
		\hyperpage{120}
      \subsubitem normalized, \hyperpage{28}, \hyperpage{131}
      \subsubitem with inner denominator, \hyperpage{27},
		\hyperpage{126}
    \subitem QR-like
      \subsubitem extended, \hyperpage{29}, \hyperpage{133}
    \subitem quasi-co-outer--coinner, \hyperpage{28}
      \subsubitem extended, \hyperpage{29}, \hyperpage{136}
    \subitem right coprime (RCF), \hyperpage{27}, \hyperpage{117}
      \subsubitem minimum-degree denominator, \hyperpage{27},
		\hyperpage{117}
      \subsubitem normalized, \hyperpage{28}, \hyperpage{129}
      \subsubitem with inner denominator, \hyperpage{27},
		\hyperpage{123}
    \subitem RQ-like
      \subsubitem extended, \hyperpage{29}, \hyperpage{136}
    \subitem spectral, \hyperpage{29}
      \subsubitem minimum-phase left, \hyperpage{29}
      \subsubitem minimum-phase right, \hyperpage{29}
      \subsubitem special, stable minimum-phase left, \hyperpage{31},
		\hyperpage{39}, \hyperpage{142}
      \subsubitem special, stable minimum-phase right, \hyperpage{32},
		\hyperpage{140}
      \subsubitem stable left, \hyperpage{29}
      \subsubitem stable minimum-phase left, \hyperpage{30}
      \subsubitem stable minimum-phase right, \hyperpage{29}
      \subsubitem stable right, \hyperpage{29}

  \indexspace

  \item linear matrix pencil
    \subitem eigenvalues, \hyperpage{12}
    \subitem finite eigenvalues, \hyperpage{12}
    \subitem infinite eigenvalues, \hyperpage{12}
    \subitem Kronecker canonical form, \hyperpage{14}
    \subitem Kronecker indices, \hyperpage{15}
    \subitem strict equivalence, \hyperpage{12}, \hyperpage{23}
    \subitem Weierstrass canonical form, \hyperpage{12}
    \subitem zeros, \hyperpage{54}
      \subsubitem finite, \hyperpage{54}
      \subsubitem infinite, \hyperpage{54}

  \indexspace

  \item M-functions
    \subitem \texttt{\bfseries gbalmr}, \hyperpage{68}
    \subitem \texttt{\bfseries gbilin}, \hyperpage{114}
    \subitem \texttt{\bfseries gcrange}, \hyperpage{85}
    \subitem \texttt{\bfseries ghanorm}, \hyperpage{62}
    \subitem \texttt{\bfseries giofac}, \hyperpage{133}
    \subitem \texttt{\bfseries gir}, \hyperpage{64}
    \subitem \texttt{\bfseries gklf}, \hyperpage{160}
    \subitem \texttt{\bfseries glasol}, \hyperpage{155}
    \subitem \texttt{\bfseries glcfid}, \hyperpage{126}
    \subitem \texttt{\bfseries glcf}, \hyperpage{120}
    \subitem \texttt{\bfseries glinfldp}, \hyperpage{146}
    \subitem \texttt{\bfseries glmcover1}, \hyperpage{104}
    \subitem \texttt{\bfseries glmcover2}, \hyperpage{110}
    \subitem \texttt{\bfseries glnull}, \hyperpage{76}
    \subitem \texttt{\bfseries glsfg}, \hyperpage{142}
    \subitem \texttt{\bfseries glsol}, \hyperpage{92}
    \subitem \texttt{\bfseries gminreal}, \hyperpage{66}
    \subitem \texttt{\bfseries gnehari}, \hyperpage{144}
    \subitem \texttt{\bfseries gnlcf}, \hyperpage{131}
    \subitem \texttt{\bfseries gnrank}, \hyperpage{60}
    \subitem \texttt{\bfseries gnrcf}, \hyperpage{129}
    \subitem \texttt{\bfseries gnugap}, \hyperpage{63}
    \subitem \texttt{\bfseries goifac}, \hyperpage{136}
    \subitem \texttt{\bfseries gpole}, \hyperpage{54}
    \subitem \texttt{\bfseries grange}, \hyperpage{81}
    \subitem \texttt{\bfseries grasol}, \hyperpage{150}
    \subitem \texttt{\bfseries grcfid}, \hyperpage{123}
    \subitem \texttt{\bfseries grcf}, \hyperpage{117}
    \subitem \texttt{\bfseries grmcover1}, \hyperpage{100},
		\hyperpage{107}, \hyperpage{112}
    \subitem \texttt{\bfseries grmcover2}, \hyperpage{107}
    \subitem \texttt{\bfseries grnull}, \hyperpage{71}
    \subitem \texttt{\bfseries grsfg}, \hyperpage{140}
    \subitem \texttt{\bfseries grsol}, \hyperpage{88}, \hyperpage{95}
    \subitem \texttt{\bfseries gsdec}, \hyperpage{97}
    \subitem \texttt{\bfseries gsfstab}, \hyperpage{167}
    \subitem \texttt{\bfseries gsklf}, \hyperpage{162}
    \subitem \texttt{\bfseries gsorsf}, \hyperpage{165}
    \subitem \texttt{\bfseries gss2ss}, \hyperpage{69}
    \subitem \texttt{\bfseries gzero}, \hyperpage{57}
  \item matrix equation
    \subitem generalized algebraic Riccati
      \subsubitem continuous-time (GCARE), \hyperpage{30},
		\hyperpage{32}
      \subsubitem discrete-time (GDARE), \hyperpage{30}, \hyperpage{32}
    \subitem generalized Lyapunov, \hyperpage{36}
    \subitem generalized Stein, \hyperpage{36}
  \item MEX-functions
    \subitem \url{sl_glme}, \hyperpage{62}, \hyperpage{69},
		\hyperpage{145}
    \subitem \url{sl_gminr}, \hyperpage{66}, \hyperpage{68},
		\hyperpage{91}, \hyperpage{95}, \hyperpage{102},
		\hyperpage{107}, \hyperpage{110}, \hyperpage{112},
		\hyperpage{135}, \hyperpage{138}
    \subitem \url{sl_gsep}, \hyperpage{98}, \hyperpage{103},
		\hyperpage{110}, \hyperpage{145}
    \subitem \url{sl_gstra}, \hyperpage{70}, \hyperpage{91},
		\hyperpage{95}, \hyperpage{102}, \hyperpage{107},
		\hyperpage{109}, \hyperpage{112}
    \subitem \url{sl_gzero}, \hyperpage{57}, \hyperpage{59},
		\hyperpage{61}
    \subitem \url{sl_klf}, \hyperpage{74}, \hyperpage{80},
		\hyperpage{84}, \hyperpage{87}, \hyperpage{91},
		\hyperpage{95}, \hyperpage{135}, \hyperpage{138},
		\hyperpage{161}, \hyperpage{164}, \hyperpage{167},
		\hyperpage{169}
  \item minimal basis
    \subitem polynomial, \hyperpage{19}
    \subitem proper rational, \hyperpage{19}
    \subitem simple, proper rational, \hyperpage{19}
  \item model-matching problem, \hyperpage{148}, \hyperpage{154},
		\hyperpage{159}
    \subitem approximate (AMMP), $\mathcal{H}_2$-norm
      \subsubitem solvability, \hyperindexformat{\ii}{40}
    \subitem approximate (AMMP), $\mathcal{H}_\infty$-norm
      \subsubitem solvability, \hyperindexformat{\ii}{40}

  \indexspace

  \item nullspace
    \subitem basis, \hyperpage{18}
      \subsubitem minimal polynomial, left, \hyperpage{19}
      \subsubitem minimal proper, left, \hyperpage{20}, \hyperpage{76}
      \subsubitem minimal proper, right, \hyperpage{20}, \hyperpage{71}
      \subsubitem minimal rational, left, \hyperpage{20}
      \subsubitem minimal rational, right, \hyperpage{20}
      \subsubitem simple minimal proper, left, \hyperpage{20},
		\hyperpage{76}
      \subsubitem simple minimal proper, right, \hyperpage{20},
		\hyperpage{71}
    \subitem left, \hyperpage{19}
    \subitem right, \hyperpage{19}

  \indexspace

  \item polynomial basis
    \subitem irreducible, \hyperpage{19}
    \subitem minimal, \hyperpage{18}
    \subitem row reduced, \hyperpage{19}
  \item polynomial matrix
    \subitem invariant polynomials, \hyperpage{21}
    \subitem normal rank, \hyperpage{21}
    \subitem Smith form, \hyperpage{21}
    \subitem unimodular, \hyperpage{10}
    \subitem zeros, \hyperpage{22}
      \subsubitem finite, \hyperpage{22}
      \subsubitem infinite, \hyperpage{23}

  \indexspace

  \item range space
    \subitem basis
      \subsubitem inner, \hyperpage{25}
      \subsubitem minimal proper, \hyperpage{25}
  \item rational basis
    \subitem minimal proper, \hyperpage{19}
    \subitem simple minimal proper, \hyperpage{19}

  \indexspace

  \item transfer function
    \subitem anti-stable, \hyperpage{21}
    \subitem exponential stability, \hyperpage{21}
    \subitem minimum-phase, \hyperpage{21}
    \subitem poles, \hyperpage{20}
      \subsubitem finite, \hyperpage{20}
      \subsubitem infinite, \hyperpage{20}
      \subsubitem stability degree, \hyperpage{21}
      \subsubitem stable, \hyperpage{21}
      \subsubitem unstable, \hyperpage{21}
    \subitem stable, \hyperpage{21}
    \subitem winding number, \hyperpage{37}
    \subitem zeros, \hyperpage{20}
      \subsubitem finite, \hyperpage{20}
      \subsubitem infinite, \hyperpage{20}
      \subsubitem minimum-phase, \hyperpage{21}
      \subsubitem non-minimum-phase, \hyperpage{21}
  \item transfer function matrix (TFM)
    \subitem $\nu$-gap metric, \hyperpage{36}, \hyperpage{63}
    \subitem additive decomposition, \hyperpage{26}, \hyperpage{39},
		\hyperpage{97}
    \subitem biproper, \hyperpage{10}
    \subitem co-outer, \hyperpage{28}
    \subitem coinner, \hyperpage{27}
    \subitem conjugate, \hyperpage{27}
    \subitem Hankel norm, \hyperpage{36, 37}, \hyperpage{62}
    \subitem improper, \hyperpage{10}
    \subitem inner, \hyperpage{27}
    \subitem left minimal cover problem, \hyperpage{34},
		\hyperpage{105}, \hyperpage{110}
    \subitem linear rational matrix equation, \hyperpage{32},
		\hyperpage{88}, \hyperpage{93}
    \subitem McMillan degree, \hyperpage{23}
    \subitem minimum-phase, \hyperpage{23}
    \subitem model-matching problem
      \subsubitem approximate, \hyperpage{31}, \hyperpage{38},
		\hyperpage{40}, \hyperpage{150}, \hyperpage{155}
      \subsubitem exact, \hyperpage{33}
    \subitem Nehari approximation
      \subsubitem optimal, \hyperpage{38, 39}, \hyperpage{144}
      \subsubitem suboptimal, \hyperpage{38}, \hyperpage{144}
    \subitem non-minimum-phase, \hyperpage{23}
    \subitem normal rank, \hyperpage{18}, \hyperpage{57},
		\hyperpage{60}, \hyperpage{62}
    \subitem outer, \hyperpage{28}
    \subitem poles, \hyperpage{23}, \hyperpage{54}
      \subsubitem finite, \hyperpage{54}
      \subsubitem infinite, \hyperpage{23}, \hyperpage{54}
    \subitem proper, \hyperpage{10}
    \subitem quasi-co-outer, \hyperpage{28}
    \subitem quasi-outer, \hyperpage{28}
    \subitem right minimal cover problem, \hyperpage{34},
		\hyperpage{100}, \hyperpage{107}
    \subitem Smith-McMillan form, \hyperpage{22}
    \subitem stable, \hyperpage{23}
    \subitem strictly proper, \hyperpage{10}
    \subitem unstable, \hyperpage{23}
    \subitem zeros, \hyperpage{23}, \hyperpage{54}, \hyperpage{57}
      \subsubitem finite, \hyperpage{23}, \hyperpage{54},
		\hyperpage{57}
      \subsubitem infinite, \hyperpage{23}, \hyperpage{54},
		\hyperpage{57}

\end{theindex}


\end{document}